\documentclass[english,a4paper,12pt]{article}
\usepackage[utf8]{inputenc}
\usepackage[T1]{fontenc}
\usepackage{titling}
\usepackage{amsmath, amssymb}
\usepackage{pgf}
\usepackage{amsfonts}
\usepackage{dsfont}
\usepackage{mathrsfs}
\usepackage{mathabx}
\usepackage{stmaryrd}
\usepackage{makeidx}
\usepackage{amsbsy}
\usepackage{amsthm}
\usepackage{color}
\usepackage{fullpage}
\usepackage{enumitem}
\usepackage[vcentermath]{youngtab}
\usepackage{marginnote} %%% margin note
\usepackage{mdframed}	%%% implement corrections
\newmdenv[topline=false, bottomline=false, skipabove=\topsep, skipbelow=\topsep]{siderules}
\usepackage{tikz}
\usetikzlibrary{decorations.text,calc,arrows.meta}
\usepackage[vcentermath]{youngtab}
\usepackage{tkz-euclide}
\usetkzobj{all}
\usepackage{pgfplots}
\usetikzlibrary{shapes}
\usetikzlibrary{decorations.shapes}
\usetikzlibrary{positioning}

\newtheorem{theorem}{Theorem}
\newtheorem{corollary}{Corollary}

\newtheorem{proposition}{Proposition}
\newtheorem{lemma}{Lemma}
\newtheorem{remark}{Remark}
\newtheorem{definition}{Definition}
\newtheorem{example}{Example}

\def\sH{\mathscr{H}}
\def\sI{\mathscr{I}}

\def\sM{\mathscr{M}}

\def\bC{{\mathbb C}}

\def\gC{{\mathfrak C}}

\def\gM{{\mathfrak M}}

\newcommand{\W}{\mathfrak{W}}
\newcommand{\A}{\mathfrak{A}}
\newcommand{\B}{\mathfrak{B}}

\newcommand{\R}{\mathfrak{R}}
\newcommand{\D}{\mathfrak{D}}

\newcommand{\ca}[1]{{\cal #1}}
\newcommand{\ben}{\begin{equation}}
\newcommand{\een}{\end{equation}}
\def\bena{\begin{eqnarray}}
\def\eena{\end{eqnarray}}
\def\non{\nonumber}

\def\cC{{\ca C}}

\def\cE{{\ca E}}
\def\cF{{\ca F}}

\def\cP{{\ca P}}

\def\cS{{\ca S}}

\def\cX{{\ca X}}
\def\cY{{\ca Y}}

\def\C{{\cal C}}
\renewcommand{\H}{\mathcal{H}}
\newcommand{\F}{\mathcal{F}}
\renewcommand{\O}{{\mathcal O}}
\newcommand{\K}{\ensuremath{\mathcal{K}}}

\def\id{{\mathrm{id}}}
\def\1{{\mathds{1}}}

\def\supp{{\mathrm{supp} \,}}

\newcommand{\dd}{{\rm d}}
\newcommand{\tr}{\operatorname{Tr}}
\newcommand{\ctg}{\operatorname{ctg}}
\renewcommand{\log}{\operatorname{ln}}
\newcommand{\dist}{{\rm dist}}
\renewcommand{\epsilon}{\varepsilon}
\renewcommand{\slash}{ \, / \! \! \!}
\def\dom{{\mathrm{dom}}}

\newcommand{\RR}{\mathbb{R}}
\newcommand{\CC}{\mathbb{C}}

\newcommand{\ZZ}{\mathbb{Z}}

\newcommand{\half}{\tfrac{1}{2}}
\newcommand{\quarter}{\frac{1}{4}}

\begin{document}
\title{Entanglement measures and their properties in quantum field theory}

	\author{
	Stefan Hollands$^{1}$\thanks{\tt stefan.hollands@uni-leipzig.de} \and
	Ko Sanders$^{2}$\thanks{\tt jacobus.sanders@dcu.ie}
	\\ \\
%%%
{\it ${}^{1}$Institut f\" ur Theoretische Physik,% }\\
%{\it
Universit\" at Leipzig, }\\
{\it Br\" uderstrasse 16, D-04103 Leipzig, Germany} \\
{\it ${}^{2}$Dublin City University, School of Mathematical Sciences,} \\
{\it Glasnevin, Dublin 9, Ireland} \\
	}
\date{\today}
	
\maketitle

\newpage
\tableofcontents
\newpage
	
\begin{abstract}
An entanglement measure for a bipartite quantum system is a state functional that vanishes on separable states and that does not increase under separable (local) operations.
For pure states, essentially all entanglement measures are equal to the v. Neumann entropy of the reduced state, but for mixed states, this uniqueness is lost.
In quantum field theory, bipartite systems are associated with causally disjoint regions. But if these regions touch each other, there are no separable normal states to begin with, and one must hence leave a finite ``safety-corridor'' between the regions. Due to this corridor, the normal states of bipartite systems are necessarily mixed, so the v. Neumann entropy is not a good entanglement measure any more in this sense. In this volume, we study various good entanglement measures. In particular, we study the relative entanglement entropy, $E_R$, defined as the minimum relative entropy between the given state and an
arbitrary separable state. We establish upper and lower bounds on this quantity in several situations: 1) In arbitrary CFTs in $d+1$ dimensions, we
provide an upper bound for the entanglement measure of the vacuum state if the two regions of the bipartite system are a diamond and the complement of another diamond. The bound is given in terms of the
spins, dimensions of the CFT and the geometric invariants associated with the regions. 2) In integrable models in $1+1$ dimensions defined by a general analytic,
crossing symmetric 2-body scattering matrix, we give an upper bound for the entanglement measure of the vacuum state
for a pair of diamonds that are far apart, showing exponential decay
with the distance between the diamonds. The class of models includes e.g. the Sinh-Gordon field theory.
3) We give upper bounds for our entanglement measure
for a free Klein-Gordon/Dirac field in the ground state on an arbitrary static spacetime. Our upper bounds show exponential decay of the entanglement measure for large geodesic distance and an ``area law'' for small distances (modified by a logarithm).
4) We show that if we add charged particles to an arbitrary state, then $E_R$  decreases by a positive amount which is no more than the logarithm of the
quantum dimension of the charges (this dimension need not be integer). 5) We establish a lower bound on our entanglement measure for arbitrary regions that
get close to each other. This lower bound is of the type of an ``area law'' with the proportionality constant given by the number $N$ of free fields in the UV-fixed point times
a quantity $D_2$ that can be interpreted as the distillable entanglement of one ``Cbit-pair'' in the state.
\end{abstract}
	
\section{Introduction}

While correlations between different parts of a system can exist both in classical and quantum physics, there can exist in quantum systems
certain more subtle correlations that are absent in classical ones. Such correlations are nowadays referred to as the ``entanglement'' between the subsystems. Historically,
the first quantitative measure of entanglement were the Bell-inequalities~\cite{bell_1,bell_2} -- or rather, their violation.

Motivated not least by technological advances in controlling and manipulating quantum systems, there has by now emerged an
understanding of  certain types of operations that one can think of, in a definite way, as not increasing the entanglement originally
present in a bi-partite quantum system (see e.g.~\cite{plenio} for a review). The set of these operations, often called ``LOCC-operations''~\footnote{
This stands for ``local operations and classical communications''. In this volume, we will actually use an even broader class.} -- as well as various ``asymptotic'' generalizations thereof,
where one is allowed to access and manipulate arbitrarily many copies of the given bipartite system -- give the set of states on a bipartite quantum system
an ordering:
A state $\sigma_1$ is not more entangled than a state $\sigma_2$, if $\sigma_1$ can be obtained from $\sigma_2$ by LOCCs.

On the one extreme, one has states that are not entangled at all. These are called ``separable'' and are described by statistical operators $\sigma$ of the form
\ben\label{separable}
\sigma = p_1 \rho_{A 1} \otimes \rho_{B 1} + p_2 \rho_{A 2} \otimes \rho_{B 2} + \dots \ ,
\een
where, according to the usual principles of quantum theory, the total Hilbert space is the tensor product $\H = \H_A \otimes \H_B$, where
$\rho_{A 1}, \rho_{A 2}$ etc. respectively $\rho_{B 1}, \rho_{B 2}$ etc. are statistical operators for subsystem $A$ respectively $B$, and where $p_i \ge 0, \sum_i p_i=1$. In classical physics, all states are separable\footnote{In classical physics, if $\mu$ is a measure on a product phase space $X=X_A \times X_B$ which is, say, absolutely continuous relative to the Lebesgue measure, then
we can approximate it with arbitrary precision by sums of product measures $\sum_i \mu_{Ai} \times \mu_{Bi}$ (e.g. on the dense subspace of smooth observables).}.
On the other hand, in quantum physics, one has non-separable, i.e. entangled, states.
In particular, one has maximally entangled states.
In between these extremes, one has states that can neither be manipulated using LOCCs into a maximally entangled state, nor be obtained from separable states by such operations.
In general, the ordering is only partial: we cannot say for each and every pair of states whether one or the other is more entangled.

To understand better the structure of entangled states, it is useful to introduce entanglement measures. At a bare minimum\footnote{It may or may not be
possible/desirable to also have other
properties such as convexity under convex linear combinations.}, an entanglement measure $E(\rho)$ should
clearly satisfy the following properties:
\begin{enumerate}
\item[(I)] $E(\rho)$ should give
a number in $[0,\infty]$, returning 0 for separable states,
\end{enumerate}
and
\begin{enumerate}
\item[(II)] $E(\rho)$ should be monotonically decreasing under LOCCs.
\end{enumerate}
A great variety of entanglement measures has been introduced in the literature. While a classification seems presently out of reach, a particularly simple and satisfactory story emerges for the set of pure states $\rho = |\Psi \rangle \langle \Psi|$.
Here, one can show~\cite{donald_1} under moderate and reasonable technical
 assumptions, that every entanglement measure is, up to a change of normalization, equal to the v. Neumann entropy of
the reduced density matrix for subsystem $A$ (or equivalently $B$), i.e. $E(\rho) = H_{\rm vN} (\rho_B) = H_{\rm vN} (\rho_A)$.
As usual, the reduced density matrix $\rho_A=\tr_B \rho$ corresponds to the restriction to subsystem $A$, and similarly for $B$.
Furthermore, it can be shown~\cite{vedral} that if two pure states have the same $E(\rho)$, then they can be converted into each other ``asymptotically''
by LOCC operations.

Unfortunately, for general mixed states $\rho$,  uniqueness is lost and one can say under the same types of technical assumptions only that
a general entanglement measure $E$ must always yield values between two extreme, conceptually distinguished entanglement measures called ``entanglement
cost'' and ``distillable entanglement''. Furthermore, for mixed states, the v. Neumann entropy of the reduced density matrix does not provide a reasonable measure, as it
can for instance return the same value for separable and maximally entangled states.

\begin{figure}[h!]
\begin{center}
\begin{tikzpicture}
	% space
	\draw[color=black, very thick] (-4, -1) -- (1, -1) -- (2, 1) -- (-3, 1) -- cycle;
	% base
	\fill[
	      top color=red!60,
	      bottom color=red!10,
	      shading=axis,
	      opacity=0.25
	      ]
	    (-1,0) circle (2cm and 0.5cm);
	    % boundary
	    \draw[red, thick] (-3, 0) arc (180: 360: 2cm and 0.5cm) -- (1, 0);
	    \draw[red, thick] (-3, 0) arc (180: 0: 2cm and 0.5cm);
	
	% big thick arrow
	\draw[line width=2mm,>={Triangle[length=3mm,width=5mm]},->] (2, 0) -- (4, 0);
	
	% causal diamond
	% space
	%\draw[color=black, very thick] (4, -1) -- (9, -1) -- (10, 1) -- (5, 1) -- cycle;
	\draw[color=black, very thick] (4, -1) -- (9, -1) -- (10, 1) -- (8, 1);
	\draw[color=black, very thick] (4, -1) -- (5, 1) -- (6, 1);
	% upward cone
	\fill[
	      left color=red!0,
	      right color=red!60,
	      middle color=red!30,
	      shading=axis,
	      opacity=0.25
	    ]
	    (5, 0) -- (7, 2) -- (9, 0) arc (0: 0: 2cm and 0.5cm);
	% downward cone
	\fill[
	      left color=red!10,
	      right color=red!30,
	      middle color=red!60,
	      shading=axis,
	      opacity=0.25
	    ]
	    (5, 0) -- (7, -2) -- (9, 0) arc (0: 0: 2cm and 0.5cm);
	    % base
	\fill[
	      top color=red!60,
	      bottom color=red!10,
	      shading=axis,
	      opacity=0.5
	      ]
	    (7,0) circle (2cm and 0.5cm);
	% boarder
	\draw[red, thick] (5, 0) arc (180: 360: 2cm and 0.5cm) -- (7, 2) -- cycle;
	\draw[color=red, dashed, thick] (5, 0) arc (180: 0: 2cm and 0.5cm);
	\draw[color=red, dashed, thick] (5, 0) -- (6, -1);
	\draw[color=red, dashed, thick] (9, 0) -- (8, -1);
	\draw[color=red, thick] (6, -1) -- (7, -2);
	\draw[color=red, thick] (8, -1) -- (7, -2);
	
	% labelling
	\node at (-1, 0) {A};
	\node[right] at (-4, -1.5) {time slice $=$ Cauchy surface $\mathcal C$};
	\node[right] at (6, 2) {$O_{A}$};
	\node[right] at (9, -1.5) {$ \mathcal C$};
    \end{tikzpicture}

    \end{center}
    \caption{Causal diamond associated with $A$.}
\label{fig:diamond}
    \end{figure}
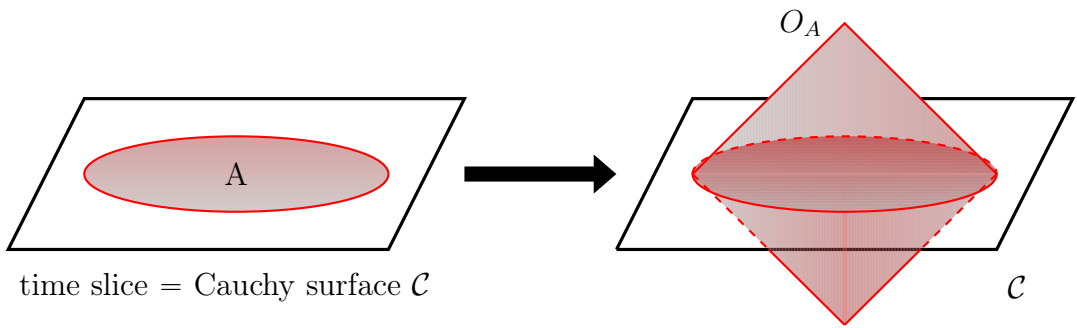

A considerable part of the literature in Quantum Information Theory has focussed on quantum systems with finitely many degrees of freedom -- leading at the technical
level mostly to (complicated) questions about algebras of matrices -- and, furthermore, to some extent, on an underlying assumption that the kinematics is
non-relativistic. It is therefore interesting to extend the analysis to relativistic quantum field theory (QFT) where both assumptions no longer hold. One may ask:
\begin{enumerate}
\item[(QI)] Whether a relativistic setup will lead to modified concepts of classical communication, distillation  protocols, etc. at a kinematical level.

\item[(QII)] Whether qualitatively new features can arise due to the presence of infinitely many degrees of freedom.
\end{enumerate}
In this volume, we will basically ignore (QI) and work with essentially the same concepts of LOCCs as in
the standard theory; see e.g.~\cite{chau, kaniewski,kent} and references therein for further developments in direction (QI).
Instead, we will focus on (QII). First of all, we note that, in QFT, the notion of subsystem is always tied to the
localization in spacetime. Thus, if $A$ is some subset of a Cauchy surface (i.e., a ``time slice'') in Minkowski spacetime (or, more generally, a
globally hyperbolic curved Lorentzian spacetime), then one ascribes~\cite{haag_1, haag_2} to $A$ a set of observables $\A_A$ localized in the
``causal diamond'' $O_A$ with base $A$, see fig.~\ref{fig:diamond}.

Informally speaking, $\A_A$ is the algebra generated by the quantum fields
localized at points in $O_A$. More precisely, $\A_A$ is the v. Neumann algebra generated by the spectral projections of the quantum fields
that are ``smeared'' against a test function supported in $O_A$.
It is a standard property of relativistic quantum field theories that if a region $B$ on the same time-slice as
$A$ is disjoint from $A$ -- so that there is no causal curve connecting $O_A$ with $O_B$ -- then
\ben
[\A_A , \A_B] = \{0\} \ .
\een
This relation of course also holds for non-relativistic quantum mechanical systems, since the algebra of
observables $\A = \A_A \vee \A_B$ generated by $\A_A$ and $\A_B$ is by definition set up in the form of a tensor product
$\A_A \otimes \A_B$ where the factors act on $\H_A \otimes \H_B$.

At this point, however, an -- at first sight seemingly academic -- difference arises in  QFT. This difference has its origins in
a mathematical fact about v. Neumann algebras. As is well-known, there are different ``types'' of v. Neumann algebras~\cite{murray_1, connes}.
The algebras appearing in non-relativistic systems, for instance matrix algebras, are typically of so-called type I, whereas the
algebras appearing in QFT are typically of type III~\cite{buchholz_2,fredenhagen_5}.
It is not so important for us what precisely these types mean (see e.g.~\cite{KadisonRingrose,BratteliRobinson} for general references). The key point for us
is rather that for type III, unlike type I,
\ben
[\A_A , \A_B] = \{0\} \quad \text{does {\em not} always imply} \quad \A \cong \A_A \otimes \A_B \ ,
\een
where ``$\cong$'' means ``up to unitary equivalence''. In fact, the conclusion -- which is closely related to the
``split property'' (see ~\cite{doplicher_4,buchholz_2,buchholz_4} and below) only holds if $A$ and $B$ are separated by a {\em finite} corridor, but
it does not hold for instance when $B$ is the complement of $A$. Thus, it fails in the usual and most natural
situation wherein we partition the total system into the union of two disjoint subsystems. For further discussion on this and related issues see~\cite{yngvason} and the nice review of~\cite{Witten}, which is directed at a wider theoretical physics audience.

This seemingly academic difference between QFT and  quantum systems with finitely many degrees of freedom
has the following important consequence. If $ \A \cong \A_A \otimes \A_B$ holds -- in which case we say that
$\A_A$ and $\A_B$ are ``statistically independent''~\cite{florig} -- then we can define separable states as in~\eqref{separable} for the total system. On the other hand,
if $\A_A$ and $\A_B$ only commute but are not statistically independent, then we cannot have (normal) separable states. Thus, we are entirely outside the
usual setup for discussing entanglement in Quantum Information Theory. In particular, we are outside this framework if $B$ is the complement of $A$
on a time slice. On the other hand, if we leave a finite safety corridor between $A$ and $B$,
then $\A_A$ and $\A_B$ are typically statistically independent, and the usual concepts from Quantum Information Theory such as separable states, LOCCs, etc.
carry over.

Thus, in QFT, we {\em should leave a finite safety corridor between $A$ and $B$}. But then it is clear that if we start with a state of the full quantum field theory, $\rho$
(for instance the vacuum state $\rho_0 = |0 \rangle \langle 0|$), then, since $A \cup B$ has an open complement $C$ (the corridor), the restriction of $\rho$ to $\A_A \otimes \A_B$ is never a pure state, as we shall prove rigorously below in sec.~\ref{sec:AQFT}. Therefore, following our general discussion about entanglement measures, we no longer have a unique
measure with which to quantify the entanglement of a state $\rho$ across $A$ and $B$. In particular, the v. Neumann entropy does {\em not} yield a satisfactory
entanglement measure.

We are thus forced in relativistic QFT to consider alternative entanglement measures with good properties [at least (I) and (II)] for mixed states. In this volume, we shall study
several such measures in the framework of algebraic QFT~\cite{haag_1,haag_2}. The measure which we shall focus on mostly is
the so-called ``relative entropy of entanglement'' proposed in~\cite{vedral}. This measure is based on Umegaki's relative entropy
functional~\cite{umegaki} $H(\rho,\sigma) = \tr(\rho \log \rho - \rho \log \sigma)$, or rather its generalization to v. Neumann algebras of general type due to Araki~\cite{araki_3a,araki_3b}. The relative entanglement entropy is given
by
\ben\label{eq:ER}
E_R(\rho) = \inf_{\sigma \ {\rm separable}} H(\rho,\sigma) \ .
\een
$E_R(\rho)$ may be interpreted as the information that we can expect to gain if we update our belief about the state of the system from being separable to
$\rho$ \cite{baez}.
Due to the variational definition [infimum over all separable states, cf.~\eqref{separable}], it is not even clear, a priori, whether $E_R(\rho) > 0$, nor is it clear that
$E_R(\rho)<\infty$ for {\rm any} state in the QFT setting.~\footnote{In fact, as shown in~\cite{narnhofer_1}, entanglement measures that are well-behaved in the type I-setting can
become ill-defined for type III, as is the case e.g. for the ``entanglement of formation''. \cite{narnhofer_1} has also shown that the entanglement entropy $E_R(\rho_0)$ behaves well under a ``nuclearity condition'', a technique to which we will come back in the body of the volume.} Our aim
is thus to investigate $E_R$ and provide upper and lower bounds in several situations.

\subsection{Summary of main results}
Our main results can be summarized as follows:
\begin{enumerate}
\item
Let $A$ be a ball of radius $r$ in a $t=0$ time slice of Minkowski space and $B$ the complement of a concentric ball with radius $R>r$ (see fig.~\ref{fig:concentric}).
\begin{figure}[h!]
\begin{center}

 \begin{tikzpicture}[scale=.7,transform shape]
        % drawing the boundaries
        \filldraw[color=green!60, fill=green!10, very thick] (7, 4) -- (7, -4) -- (-1, -4) -- (-1, 4)-- cycle;
	\filldraw[color=green, fill=white, very thick](3,0) circle (3);
	\filldraw[color=red!60, fill=red!10, very thick](3,0) circle (1);
	
	% radius of the circles
	\draw[line width=1pt, >={Triangle[length=1mm,width=2mm]}, ->] (3,0) -- (3, 1);
	\draw[line width=1pt, >={Triangle[length=1mm,width=2mm]}, ->] (3,0) -- (1.5, 2.65);
	
	% labelling
	\node[below] at (3, 0) {\textcolor{red}{$A$}};
	\node[above] at (-0.5, 1) {\textcolor{green}{$B$}};
	\node[right] at (3, 0.5) {$ r $};
	\node[right] at (1, 2) {$R$};
    \end{tikzpicture}

   \end{center}
    \caption{The regions $A$ and $B$.}
\label{fig:concentric}
    \end{figure}
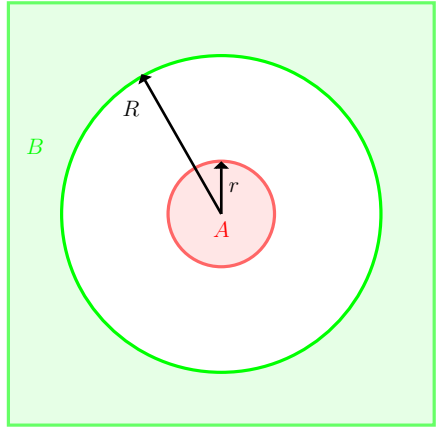

In any conformal field
theory in $d+1$ dimensions, we have
\ben
E_R(\rho_0) \le \log \ \sum_{\O} \left( \frac{r}{R} \right)^{d_\O} \ ,
\een
where $\rho_0 = |0 \rangle \langle 0|$ is the vacuum state\footnote{In the body of this volume we will distinguish, for technical reasons,
the expectation value function of a statistical operator $\omega( \ . \ ) = \tr (\ . \ \rho)$ and
the statistical operator $\rho$ itself.}, and where $d_\O$ are the dimensions of the local operators $\O$ in the theory.

If, more generally, the diamonds are not
necessarily concentric, we can introduce the conformally invariant cross ratio $u =\frac{
(x_{B+}-x_{B-})^2(x_{A+} - x_{A-})^2
}{
(x_{A-}-x_{B-})^2(x_{A+} - x_{B+})^2
}$ and similarly $v$, see~\eqref{xr},
associated with upper upper/lower tip of diamond $O_A$ and the upper lower tip of diamond $O_B'$, see fig.~\ref{fig:diamond1}.

\begin{figure}
\begin{center}
    \begin{tikzpicture}
        \filldraw[color=green, fill=green!20!white, very thick] (2, 0) arc (360: 180: 2cm and 0.4cm) -- (-1, 1) -- (-3, 1) -- (-5, -1) -- (3, -1) -- (5, 1) -- (1, 1);
        \filldraw[color=gray, fill=gray!20!white, thick] (-2, 0) -- (0, 2) -- (2, 0);
        \filldraw[color=gray, fill=gray!40!white, dashed, thick] (-2, 0) arc (180: 0: 2cm and 0.4cm);
        \filldraw[color=gray, fill=gray!40!white, thick] (2, 0) arc (360: 180: 2cm and 0.4cm);
        \filldraw[color=gray, fill=gray!20!white, thick] (-1, -1) -- (0, -2) -- (1, -1);
        \draw[gray, dashed, thick] (-1, -1) -- (-2, 0);
        \draw[gray, dashed, thick] (1, -1) -- (2, 0);
        % small causal diamond
        % base
        \fill[
        top color=red!70,
        bottom color=red!70,
        shading=axis,
        opacity=0.50, rotate around={ 20: (0, 1) }
        ]
         (0, 1) circle (0.5cm and 0.10cm) ;
        % upward cone
        \fill[
        left color=red!100,
        right color=red!100,
        middle color=red!100,
        shading=axis,
        opacity=0.25
        ]
        (-0.50, 0.83) -- (0.18, 1.50) -- (0.49, 1.18) arc (0: 0: 0.5cm and 0.10cm);
        % downward cone
        \fill[
        left color=red!100,
        right color=red!100,
        middle color=red!100,
        shading=axis,
        opacity=0.25
        ]
        (-0.50, 0.83) -- (-0.18, 0.55) -- (0.50, 1.18) arc (0: 0: 0.5cm and 0.10cm);
        % boarder
        \draw[red, thick, rotate around={ 20 : (-0.50, 0.83) }] (-0.50, 0.83) arc (180: 360: 0.52cm and 0.10cm);
        \draw[color=red, dashed, thick, rotate around={ 20 : (-0.50, 0.83) }] (-0.50, 0.83) arc (180: 0: 0.5cm and 0.10cm);
        \draw[color=red, thick] (-0.50, 0.83) -- (0.18, 1.50) -- (0.50, 1.18) -- (-0.18, 0.55) -- cycle;
        % labelling
	\node[right] at (2.2, 0) {\textcolor{green}{$B$}};
	\node[right] at (.4, 1.1) {\textcolor{red}{$A$}};
	\node[below] at (0, -1.9) {$x_{B-}$};
	\node[above] at (0, 1.9) {$x_{B+}$};
	\node[above] at (.4, 1.41) {$x_{A+}$};
	\node[below] at (0, .53) {$x_{A-}$};
  \end{tikzpicture}
      \caption{Nested causal diamonds.}
      \label{fig:diamond1}
\end{center}
\end{figure}
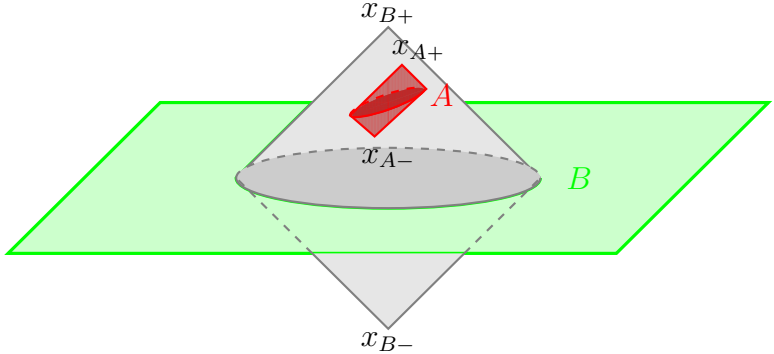

If $\tau, \theta \in \RR$ are the functions of these cross ratios $u,v$ defined below in eq.~\eqref{thtau}, then we get in $3+1$ dimensions
\ben\label{nubound2}
E_R(\rho_0) \le \log \ \sum_{\O} e^{-\tau d_\O} \left[ 2S_\O^R+1 \right]_\theta \left[ 2S_\O^L+1 \right]_\theta
%\frac{\sinh\frac12(s_\O^{}+1)\theta \  \sinh \frac12 (s'_\O + 1)\theta}{\sinh^2(\frac12 \theta)}
\ .
\een
Here, $S_\O^L, S_\O^R$ is the number of primed/unprimed spinor indices of the operator $\O$ and $[n]_\theta$ is
a suitably defined ``quantum deformed'' natural number $n$.
A similar bound is obtained also for chiral conformal field theories. When the outer diamond is much larger, $r/R \ll 1$, our bound gives for instance
\ben
E_R(\rho_0) \lesssim N_\O \, \left( \frac{r}{R} \right)^{d_\O}
\een
for concentric diamonds, where $\O$ is the operator with the smallest non-trivial dimension $d_\O$, and $N_\O$ the multiplicity of such operators.
This result is consistent  with the ``small $x$ expansion'' obtained by~\cite{Calabrese1,Calabrese2} in $1+1$ dimensions.
\item Let $\omega$ be any state of finite energy, and let $\chi^*\omega$ be a state obtained by adding ``charges'' $\chi=\prod \chi_i^{n_i}$
(in a generalized sense which may include charged
pseudo-particles with braid-group statistics~\cite{fredenhagen_2,longo_1,longo_2}) in region $A$ or $B$ as indicated in fig.~\ref{fig:charges}.

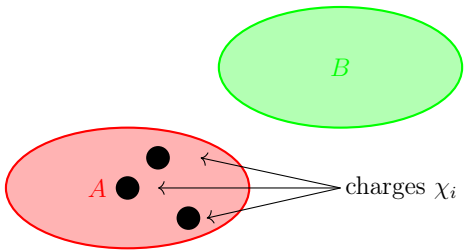
\begin{figure}[h!]
\begin{center}
\begin{tikzpicture}[scale=.8,transform shape]
            \draw[red, thick] (2.5, 0) arc (0: 360: 2cm and 1cm);
            \draw[green, thick] (6, 2) arc (0: 360: 2cm and 1cm);
            \filldraw[red=10!white, opacity=0.30] (2.5, 0) arc (0: 360: 2cm and 1cm);
            \filldraw[green=10!white, opacity=0.30] (6, 2) arc (0: 360: 2cm and 1cm);

            \node at (0, 0) {\textcolor{red}{$A$}};
        \node at (4, 2) {\textcolor{green}{$B$}};
        \node[circle,fill=black] at (0.5, 0) {};
        \node[circle,fill=black] at (1, 0.5) {};
        \node[circle,fill=black] at (1.5, -0.5) {};
        \node at (5,0) {charges $\chi_i$};

        \draw[->] (4, 0) -- (1, 0);
        \draw[->] (4, 0) -- (1.8, -0.5);
        \draw[->] (4, 0) -- (1.7, 0.5);
        \end{tikzpicture}
       \end{center}
    \caption{Charges $\chi_i$ added to $A$.}
\label{fig:charges}
    \end{figure}
We have
\ben
0\le  E_R(\omega) - E_R(\chi^*\omega)  \le \log \prod_i \dim (\chi_i)^{2n_i} \ ,
\een
irrespective of the nature of $\omega$, or the relative position of $A$ and $B$, or the dimension $d+1$ of spacetime.
Here, $n_i$ is the number of irreducible charges $\chi_i$ of type $i$, and $\dim(\chi_i)$ the ``quantum dimension'' of the charge. For instance, this dimension is $N$ if
the charge is created by a local operator transforming in the fundamental representation of $O(N)$, but can also be non-integer e.g. for
anyonic pseudo-particles in $1+1$ dimensions.
\item For the real Klein-Gordon scalar QFT with field equation
\ben
\Box \phi - m^2 \phi = 0 \ ,
\een
and positive mass $m$
on an arbitrary ultra-static, globally hyperbolic spacetime $\sM$ of dimension $d+1$ with metric $-\dd t^2 + h_{ij}(x) \dd x^i \dd x^j$, we show that the entanglement
of the ground state $\rho_0$ of the
theory decays for $mr \gg 1$ as\footnote{Formula~\eqref{ERintegr} below suggests that the upper bound can be
improved to $C_\infty(\delta) e^{-mr(1-\delta)}$ for each $\delta>0$.}
\ben
E_R(\rho_0) \lesssim C_\infty \, e^{-mr/2} \ ,
\een
where $r$ is the geodesic distance (with respect to $h_{ij}$) between $A$ and $B$ in a static slice, see fig.~\ref{fig:regionsAB} and where $C_\infty$
is some constant.

 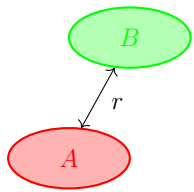
\begin{figure}[h!]
\begin{center}
\begin{tikzpicture}[scale=.8,transform shape]
            \draw[red, thick] (1, 0) arc (0: 360: 1cm and 0.50cm);
            \draw[green, thick] (2, 2) arc (0: 360: 1cm and 0.50cm);
            \filldraw[red=10!white, opacity=0.30] (1, 0) arc (0: 360: 1cm and 0.50cm);
            \filldraw[green=10!white, opacity=0.30] (2, 2) arc (0: 360: 1cm and 0.50cm);

            \draw[<->] (0.2, 0.5) -- (0.75, 1.5);

            \node at (0, 0) {\textcolor{red}{$A$}};
        \node at (1, 2) {\textcolor{green}{$B$}};
        \node at (0.8, 0.9) {$r$};
    \end{tikzpicture}
  \end{center}
    \caption{The regions $A$ and $B$.}
\label{fig:regionsAB}
\end{figure}
For the Majorana Dirac QFT with field equation
\ben
(\slash \nabla + m) \psi = 0 \ ,
\een
and non-vanishing mass, we also have an upper bound when the geodesic distance $r$ between $A$ and $B$ goes to zero in a static slice, in
a spacetime of the product form $\sM = \RR^{1,1} \times \Sigma$. More precisely,
let $A=\{(t,x,y) \mid x<0, t= 0\}$ and $B =\{ (t,x,y) \mid x>r, t=0\}$ (where $(t,x)$ are standard coordinates on $\RR^{1,1}$ and $y \in \Sigma$), and let $\rho_0$ be the
ground state. Then as $mr \to 0$, we have the upper bound
\ben
E_R(\rho_0) \le C_0 |\log(mr)| \sum_{j\le d-1} r^{-j} \int_{\partial A} a_j \lesssim c_0  |\log(mr)| \frac{|\partial A|}{r^{d-1}} \ ,
\een
where we assume $d\ge 2$ and the $a_j$ are geometric invariants associated with a heat kernel on
$\partial A \cong \Sigma$ and $C_0,c_0$ are constants.
We expect our methods to yield
similar results to hold generally on spacetimes with bifurcate Killing horizon, fig.~\ref{fig:KW}, see~\cite{wald_2} for this notion.

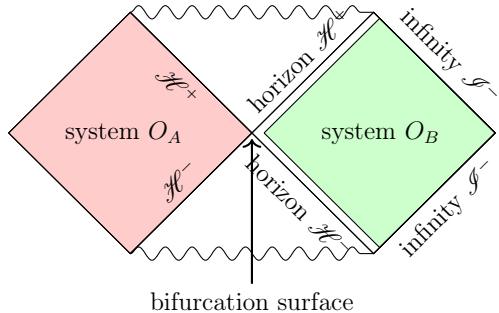
\begin{figure}[h!]
\begin{center}
\begin{tikzpicture}[scale=.80, transform shape]
\filldraw[fill=red!20] (0,0) -- (-2,2) -- (-4,0) -- (-2,-2) -- (0,0);
\filldraw[fill=green!20] (0.2,0) -- (2.1,1.9) -- (4,0) -- (2.1,-1.9) -- (0.2,0);
\draw (0,0) -- (2,2) -- (4,0) -- (2,-2) -- (0,0);
%\draw (-6,2) -- (-4,0) -- (-6,-2);
\draw (0,0) -- (-2,2) -- (-4,0) -- (-2,-2) --   (0,0);
%\draw[snake] (-2,-2) -- (-6,-2);
%\draw[dashed] (-6,-2)  --  (-6,2);
%\draw[snake] (-6,2) -- (-2,2);
\draw (0,0) -- node[black,above, sloped]{horizon $\sH^+$}(2,2) -- (4,0) -- (2,-2) -- node[black,below, sloped]{horizon $\sH^-$}(0,0);
\draw (-2,2) --node[black,below, sloped]{$\sH^+$}(0,0);
\draw (-2,-2) --node[black,above, sloped]{$\sH^-$}(0,0);
\draw (-2,2) decorate[decoration=snake] {--(2,2)};
\draw (1,0) node[black,right]{system $O_B$};
\draw (-1,0) node[black,left]{system $O_A$};
\draw (-2,-2)  decorate[decoration=snake] {-- (2,-2)};
\draw (2,-2) --node[black,below, sloped]{infinity $\sI^-$}(4,0);
%\draw (4,0) --node[black,below, sloped]{cosmic ${\mathscr H}_A$}(6,2);
%\draw (4,0) --node[black,above, sloped]{cosmic ${\mathscr H}_B$}(6,-2);
\draw (2,2) --node[black,above, sloped]{infinity $\sI^-$}(4,0);
%\draw[->, thick] (-6.1,1.5) node[black,left] {static chart} -- (2,0);
%\draw[->, thick] (-3.7,-3.5)  -- (-4,0);
\draw[->, thick] (0,-2.5)  node[black,below] {bifurcation surface}-- (0,-.1);
%\draw[->, thick] (-2.5,-3.5) node[black,below] {bifurcation surfaces} -- (4,0);
%\draw[blue] (2,2) .. controls (4,2) and (4,2) .. (6,2);
%\draw[blue] (2,2) .. controls (4,1.5) and (4,1.5) .. (6,2);
%\draw[blue] (2,2) .. controls (4,1) and (4,1) .. (6,2);
%\draw[blue] (2,2) .. controls (4,0.5) and (4,0.5) .. (6,2);
%\draw[blue] (2,2) .. controls (4,0) and (4,0) .. (6,2);
\end{tikzpicture}
\end{center}
    \caption{Spacetime with bifurcate Killing horizon.}
\label{fig:KW}
\end{figure}
\item
For the class of massive integrable models on $1+1$ dimensional Minkowski space with factorizable  two-body scattering
matrix $S_2$ of the general form \eqref{S2def}, and $A$ and $B$ given by two half-lines of the time
slice $\RR$ separated by $r>0$, the vacuum state $\rho_0 = | 0 \rangle \langle 0|$ satisfies the following bound.
For $mr \gg 1$ (here $m$ is the mass) we have for any $\kappa >0$ such that $S_2(\zeta)$ has no poles in the
strip $\{ \zeta \in \CC : -\kappa < \Im \zeta < \pi + \kappa\}$, and any small\footnote{We cannot put $\kappa$ or $\delta$ to zero, since
the asymptotic bound holds, roughly speaking, when $1/(\delta \kappa) \lesssim mr$.} $\delta>0$,
\ben\label{ERintegr}
E_R(\rho_0)  \lesssim \frac{4e}{\kappa} \sqrt{\frac{\| S_2 \|_\kappa}{\pi m r}} \, e^{-mr(1-\delta)} \ .
\een
where $\| S_2 \|_\kappa$ is the supremum of $|S_2(\zeta)|$ in the strip.
The constant in the bound diverges when the poles of $S_2$ approach the ``physical strip''. Since poles inside
the physical strip are characteristic for models with bound states, we can say that our upper bound on the entanglement
entropy deteriorates as we approach this situation.

Our bound applies in
particular to the {\em Sinh-Gordon model} with equation of motion
\ben\label{shG}
\Box \phi - m^2 \phi- g^2 \sinh \phi = 0,
\een
where
$S_2(\theta)=\frac{\sinh \theta - i\sin b}{\sinh \theta + i\sin b} \ , \ \ b = \frac{\pi g^2}{1 + g^2}$,
and the constants $C$ are given in terms of $b$ or $g$. In this case, the constants deteriorate for $g \to \infty$.
\item
Consider a massive QFT on $d+1$ dimensional Minkowski space satisfying a ``nuclearity condition'' in the sense of~\cite{buchholz_3}, and let $A$ and $B$
 regions separated by $r \gg$ than the size of $A$ and $B$, see fig.~\ref{fig:regionsAB}.

 We show that the entanglement in the vacuum state $\rho_0$ has sub-exponential decay, i.e.
 \ben
 E_R(\rho_0) \lesssim C \, e^{-(mr)^k} \ ,
 \een
 for any given $k<1$ (our $C$ diverges when $k \to 1$). Since such nuclearity conditions have been shown for massive free fields of spin $0,1/2$~\cite{buchholz_4,dantoni},
 the bound holds in particular for such theories.
\item
Consider a QFT in $d+1$ dimensions satisfying a suitable ``nuclearity condition'' for a thermal state $\rho_\beta$,
cf.~\eqref{Eqn_Theta_rz}, and
let $A$ and $B$ again be regions separated by $r \gg$ than the size of $A$ and $B$, see fig.~\ref{fig:regionsAB}. We show:
 \ben
 E_R(\rho_\beta) \lesssim C \, r^{-\alpha+1} \ ,
 \een
 where $\alpha$ is a parameter entering the nuclearity condition.
 \item
 For any conformal QFT in $d+1$ dimensions with vacuum state $\rho_0=|0 \rangle \langle 0|$, and
 for $A$ and $B$ regions separated by a thin corridor of diameter $\epsilon>0$, we show that asymptotically, as
 $\epsilon \to 0$
 \ben
 E_R(\rho_0) \gtrsim
 \begin{cases}
 D_2 \cdot |\partial A|/\epsilon^{d-1} & \text{$d>1$,}\\
 D_2 \cdot \log \frac{\min(|A|, |B|)}{\epsilon} & \text{$d=1$,}
 \end{cases}
 \een
 (``area law''), where $D_2$ is the distillable entropy of an elementary ``Cbit'' pair (defined in sec.~\ref{ssec:arealaw}).
 As we point out in the text, the same argument shows that
 for a non-conformal, asymptotically free theory and for any state $\omega$ with finite energy in $d>1$ spatial dimensions, one would get the bound
 \ben
  E_R(\omega) \gtrsim N \cdot D_2 \cdot |\partial A|/\epsilon^{d-1}
 \een
 where $N$ is the number of independent free fields in the short-distance scaling limit, for instance
 $N=n^2-1$ in ${\rm SU}(n)$ pure Yang-Mills theory.

 For a massive Dirac field, we have found a qualitatively similar upper bound in 3), so these lower bounds should be expected to be qualitatively sharp.
\end{enumerate}

In order to obtain these results, we use several other entanglement measures which give upper and lower bounds on
$E_R$ and which are often easier to estimate, such as $E_D$ (distillable entropy), $E_N$ (logarithmic dominance), $E_M$
(modular entanglement), and others. Some of these are of independent interest, and to our knowledge, new. A table comparing
these entanglement measures is presented in section~\ref{SSec_summarymeasures}. A key role is also played in our proofs by techniques from  Tomita-Takesaki modular theory for v. Neumann
algebras and their---to a large extent well-known---relation to quantum field theory. These come in on the one hand via
their relation with spacetime symmetries (Bisognano-Wichmann~\cite{bisognano} and Hislop-Longo~\cite{hislop} theorems)
and on the other hand via their relationship with ``nuclearity bounds''~\cite{buchholz_1,buchholz_3,buchholz_4,lechner_2}, both of
which are combined with methods from complex analysis. We note that the usefulness of nuclearity bounds  in the context of entanglement has been appreciated already
by~\cite{narnhofer_1}, and they have been used also more recently by~\cite{otani}, which appeared after our work was completed.
Among the other tools that we use is the theory of superselection sectors~\cite{doplicher_1,doplicher_2,doplicher_3,fredenhagen_2,longo_1,longo_2}.

\subsection{Comparison with other approaches to entanglement in QFT}
It is worth commenting on the conceptual and practical differences between our approach and the substantial body of literature on entanglement in QFT
based on the ``replica trick'', or ``holographic'' methods, see e.g.~\cite{solodukhin,nishioka} for reviews with many references.

In the ``replica trick''~\cite{calabrese1},
which applies most straightforwardly to ground states on static spacetimes, one ignores the problems discussed above with the v. Neumann entropy, and formally represents
$H_n(\omega_A) = \frac{1}{1-n} (\log Z(\sM_n) - n \log Z(\sM_1))$. Here, $\omega_A=\tr_A |0 \rangle \langle 0|$ is the reduced state,
$H_n$ is a regulated version of the v. Neumann entropy\footnote{It is defined as
$H_n(\rho) = \frac{1}{1-n} \log \tr \rho^n \ .$}, and
$\sM_n$ is an $n$-sheeted cover of $\sM$ obtained by gluing $n$ copies of $\sM$ across the boundary of $A$. $Z$ is the the partition function of the corresponding Euclidean QFT,
often represented in terms of a functional integral, or ``defect operators'' as in the pioneering paper~\cite{calabrese} on 1+1 dimensional conformal field theories.
In either case, the result is divergent due to the conical singularity
on $\sM_n$ along $\partial A$, but one can at this stage introduce a short distance (UV) cutoff $\epsilon$ of some sort, and get a finite answer,
$H_{\rm vN}^{\epsilon}(\omega_A)$.
The divergent terms are often found to be organized in a series in inverse powers of $\epsilon$. The most divergent term is usually $\propto |\partial A|/\epsilon^{d-1}, d>1$
(``area law''~\cite{bombelli,srednicki,susskind}), and the sub-leading ones are often -- though not always~\cite{marolf} -- given in terms of curvature invariants associated with $\partial A$.

Compared with our entanglement measure~\eqref{eq:ER}, one is tempted to perhaps expect a relationship of the form
\ben
H_{\rm vN}^{\epsilon}(\omega_A) \sim E_R(\omega)
\een
when $\epsilon \sim \dist(A,B)$ becomes small compared to the volume of $A$ and all other length scales in the QFT (including any scales introduced by the state $\omega = |\Psi \rangle \langle \Psi|$ if it is not the vacuum), and when $B$ approaches the complement of $A$.
Some of our results indicate that the above relation may indeed be roughly correct in many cases, but 3) and 4) indicate that this is perhaps so
only in massive theories, and perhaps only up to powers in $\ln(m\epsilon)$.

Compared to our approach, the replica trick has, at any rate, a rather different distribution of strengths and weaknesses. The strengths are that the basic formulas are, although formal, strikingly elegant, and in principle concrete, making rather non-trivial computations possible in many interesting examples, and establishing also an interesting link to other ideas in quantum field theory, such as e.g. the $c$-theorem~\cite{casini_1,casini_2}.

The weakness is that, in order to obtain a finite answer in the limit $\epsilon \to 0$, one must
either subtract by hand the divergent terms, or consider differences of quantities like for instance $H_n(\tr_A |0 \rangle \langle 0|) - H_n(\tr_A |\Psi \rangle \langle \Psi|)$, where
$|\Psi \rangle$ is some reference state, or like the ``mutual information'' for disjoint regions  $A,B$ -- hoping that the divergences are state-independent and cancel (this is not always the case~\cite{marolf}). At any rate, one does {\em not} obtain a quantity that satisfies the basic postulates (I) and (II), and furthermore, the assumption of a short distance cutoff at intermediate stages invalidates the basic assumptions of locality (if the cutoff is imposed in momentum space), or relativistic covariance (if the cutoff is implemented on a lattice), or it introduces an unwanted dependence on ``boundary conditions'' (if the cutoff is imposed by such conditions as e.g. in \cite{Cardy2}).

On the other hand, a strength of our approach -- apart from mathematical exactness -- is that postulates (I) and (II) are demonstrably satisfied for our entanglement measure $E_R$, but a weakness  is that the basic definitions are rather indirect, and do not lead to very explicit formulas that are amenable to straightforward computations or at least approximations. In fact, as is clear from
1)--7), we have only been able to compute upper and lower bounds. In the future, it would be interesting to establish a relation between the approaches. We conjecture that
the ``Buchholz free energy'' $\log Z_B(| 0 \rangle \langle 0|)$ introduced in the text may be seen as a regulated version of the formal quantity $\frac{1}{1-n} (\log Z(\sM_n) - n \log Z(\sM_1))$, but this remains to be investigated further. The relation between our approach and ``holographic methods'' based on the Ryu-Takayanagi
proposal~(\cite{hubeny,ryu}, and \cite{rang} for a recent textbook) is on the other hand less clear to us. Perhaps such a relation can be established via the intriguing relationship between entanglement and the geometry of the space of causal diamonds recently found by~\cite{deBoer_1,deBoer_2}.

\medskip

The organization of this volume is as follows. For the benefit of readers with a background in Quantum Information Theory, we first review
in sec. 2 mathematical definitions and results from operator algebra theory, and present important examples of algebraic QFTs. For the benefit of readers with a background in QFT, we then review in sec. 3 basic notions about LOCCs, entanglement measures etc., showing e.g. that (I) and (II) hold for $E_R$ in the QFT setting. In fact, besides $E_R$ we introduce several other such measures that serve as tools in deriving our results, and which may be of some independent interest. Our results 1)-7) are presented in detail and proven in sec. 4 and sec. 5.

\medskip
{\bf Notations and conventions:} Upper case Gothic letters $\A, \B, \dots$ denote v. Neumann or $C^*$-algebras. Lower case Greek letters such as $\omega, \sigma$
denote linear functionals or states on a v. Neumann or $C^*$-algebras. Hilbert spaces are always assumed (or manifestly) separable.
The dimension of spacetime is $d+1$, and our convention for the signature of the spacetime metric is $(-++...+)$.
Scalar products on Hilbert space are anti-linear in the first entry.
If $f(t), g(t)$ are non-negative functions of a real variable $t$, we write $f(t) \lesssim g(t)$ as $t\to\infty$ if for any $\delta>0$ there is a $t_0$ such that
$f(t) \le (1+\delta) g(t)$ for all $t \ge t_0$. We write $f(t) \sim g(t)$ when $f(t) \lesssim g(t)$ and $g(t) \lesssim f(t)$.

\section{Formalism for QFT}

\subsection{$C^*$-algebras and v. Neumann algebras}\label{SSec_algebras}

This section is intended to present the most important notions from the theory of operator algebras used in later sections. (See e.g. \cite{KadisonRingrose,BratteliRobinson}
for general references.) We begin with $C^*$-algebras:

\begin{definition}
A $C^*$-algebra is a complex, associative algebra $\A$ with a unit $1$, an involution $a \mapsto a^*$ and a norm $\| a \|$, such that for all $a,b \in \A$, one has
\ben
\|ab\| \le \|a\| \ \|b\|, \quad \|a^* \| = \| a\|, \quad \| a^* a \| = \| a\|^2 \ ,
\een
and such that $\A$ is complete with respect to this norm.
\end{definition}

The norm of a $C^*$-algebra is an intrinsic property of the algebra in the sense that there cannot be two different $C^*$-norms. This is a consequence of the fact that homomorphisms between $C^*$-algebras, i.e. linear maps $\phi$ satisfying $\phi(ab) = \phi(a)\phi(b), \phi(a^*) = \phi(a)^*$, are automatically continuous with $\|\phi(a)\|\le\|a\|$. The norm of a linear functional $\varphi: \A \to \CC$  is defined by
\ben
\| \varphi \| = \sup_{a \in \A, \|a\| \le 1} |\varphi(a)| \ .
\een
A linear functional on a $C^*$-algebra is called hermitian if $\varphi(a^*) = \overline{\varphi(a)}$ and positive if $\varphi(a^*a) \ge 0$ for all $a \in \A$. A positive functional is automatically hermitian and bounded, i.e. $\|\varphi\|<\infty$. In fact, it has $\|\varphi\|=\varphi(1)$.

\begin{definition}
A positive functional $\omega$ such that $\omega(1) = 1$ is called a ``state''.
\end{definition}

A state automatically has $\| \omega \| =1$, and vice-versa, any linear functional such that $\|\omega\| = 1, \omega(1) = 1$ is a state. A state is called ``pure'' if it cannot be written as a non-trivial  combination $\omega = \sum_i p_i \omega_i$ of other states, where $p_i > 0$. Otherwise it is called ``mixed''.

A standard example of a $C^*$-algebra is the set $\B(\H)$ of all bounded operators on a Hilbert space $\H$. The norm is defined concretely in this case by the usual operator norm,
\ben
\|a\| = \sup_{0 \neq |\Psi \rangle \in \H} \| \, a \Psi  \|/\|  \Psi  \| \ ,
\een
where the norm of a vector in Hilbert space is denoted by $\|  \Psi \|^2 = \langle \Psi | \Psi \rangle$.
The involution $*$ is concretely given by the hermitian adjoint, $\langle \Psi|a^* \Phi \rangle = \langle a \Psi | \Phi \rangle$ with respect to the inner product in $\H$. More generally, any linear subspace of $\B(\H)$ which is closed under products, hermitian adjoints, and limits, is a $C^*$-algebra. A *-homomorphism $\pi: \A \to \B(\H)$ is called a ``representation'' of $\A$ on $\H$ . The statistical operators $\rho$ on $\H$ (i.e. hermitian, positive semi-definite operators $\rho$ on $\H$ with $\tr_\H \rho =1$) automatically give rise to states $\omega_\rho$, in the algebraic sense of linear functionals on $\A$ described above, by the formula
\ben\label{omrho}
\omega_\rho(a) = \tr_\H (\rho \pi(a)) \ .
\een
The set of all such states is called the ``folium of $\pi$'', denoted $S_\pi(\A)$. One should be aware though that:
\begin{enumerate}
\item The set of states encompassed in this way by a given representation $\pi$ is in general very far from containing all states $\omega$. There can, and in general will be, disjoint folia.
\item It is in general not true that $\rho = |\Psi \rangle \langle \Psi|$ is equivalent to $\omega_\rho$ being pure!
\item It is in general not true that $\omega_\rho=\omega_{\rho'}$ implies $\rho=\rho'$.
\end{enumerate}
These issues are closely related to the existence of many, often inequivalent, representations. For this, it is useful to define the notion of intertwiner. An intertwiner between two representations $(\pi_i, \H_i), i=1,2$ is a bounded linear operator $T:\H_1 \to \H_2$ such that $T\pi_1(a) = \pi_2(a)T$ for all $a \in \A$. One says that
\begin{enumerate}
\item a representation $\pi$ is irreducible if there are no non-trivial self-intertwiners (other than multiples of the identity), or equivalently,  no invariant subspaces for $\pi(\A)$ other than $\{0\}$ and $\H$ itself.
\item two representations $\pi_i$ are unitarily equivalent if there is a unitary intertwiner.
\item two representations $\pi_i$ are quasi-equivalent if their folia coincide.
\item two representations $\pi_i$ are disjoint if there is no intertwiner $T\not=0$, i.e. if the folia $S_{\pi_i}(\A)$ have an empty intersection.
\end{enumerate}

If the representation $\pi$ is irreducible, then $\omega_\rho$ is pure if and only if $\rho = |\Psi \rangle \langle \Psi|$, and vice versa, but this is no longer true if $\pi$ is not irreducible. If there are several quasi-equivalence classes, then there exist representations and states which are not represented by density matrices in this representation. Nevertheless, it can be shown that given an algebraic state $\omega: \A \to \CC$, there is always {\em some} representation containing a vector $|\Omega \rangle$ such that $\omega$ is represented by this vector i.e. by the density matrix $\rho=|\Omega \rangle \langle \Omega|$. This is demonstrated by a simple, canonical, but conceptually very important construction called the ``GNS-construction''.

The starting point of this construction is the simple observation that the algebra $\A$ itself, as a linear space, always forms a representation $\pi$ by left multiplication, i.e. $\pi(a) b \equiv ab$. One would like to equip this representation with a Hilbert space structure, i.e. a positive definite inner product. It seems natural to define $\langle a|b\rangle = \omega(a^* b)$, but this will in general lead to non-zero vectors with vanishing norm. Introduce ${\mathfrak J}_\omega = \{ a \in \A \mid \omega(a^*a) = 0\}$. By the Cauchy-Schwarz inequality, $|\omega(a^*b)| \le \omega(a^*a)^{1/2} \omega(b^*b)^{1/2}$, we have ${\mathfrak J}_\omega = \{ a \in \A \mid \forall b\in \A, \omega(b^*a) = 0\}$, so it is a closed linear subspace and a left ideal of $\A$ containing precisely the null vectors. We can then define $\H_\omega = \A/{\mathfrak J}_\omega$ and complete it in the induced inner product. The left representation induces a representation on $\H_\omega$ which is called $\pi_\omega$. It is the desired GNS-representation. The vector $|\Omega_\omega \rangle \in \H_\omega$ representing $\omega$ is simply the equivalence class of the unit operator, $1$. It is by construction ``cyclic'' in the sense that the set $\pi_\omega(\A)|\Omega_\omega \rangle$ is dense in $\H_\omega$. We say that two states are quasi-equivalent if their GNS-representations are. Note that even mixed states are always represented by a vector in their GNS representation. Thus, in this case, the GNS-representation cannot be irreducible.

A mathematical concept related to that of a $C^*$-algebra is a v. Neumann algebra. Such algebras can be characterized in different ways. One way to characterize a v. Neumann algebra is as follows:

\begin{definition}
A v. Neumann algebra is a $C^*$-algebra $\A$ with a distinguished folium, the folium of ``normal states'', which spans a linear space $S_n(\A)$ of linear functionals on $\A$. This folium should satisfy the properties:
\begin{enumerate}
\item If  $a,b \in \A$ are such that $\omega(a)=\omega(b)$ for all $\omega \in S_n(\A)$, then $a=b$.
\item If $f: S_n(\A) \to \CC$ is a bounded, linear functional, then there exists an $a \in \A$ such that $f(\omega) = \omega(a)$ for all $\omega$ in the distinguished folium.
\end{enumerate}
\end{definition}

One sometimes also writes $\A_*$ for $S_n(\A)$ and calls it the ``predual''. States on a v. Neumann algebra which are not normal are called ``singular''. Given any normal state $\omega$, we can represent $\A$ on the Hilbert space $\H_\omega$ by the GNS-representation. The set of operators $\{\pi_\omega(a)\mid a\in\A, \|a\|\le1\}$ on $\H_\omega$ obtained in this way is always weakly closed, i.e. closed in the topology generated by the seminorms $N_\rho(a) = |\tr(\rho \pi_\omega(a))|$.

Furthermore, the v. Neumann bi-commutant theorem holds. This theorem says the following. Let $\A$ be a v. Neumann algebra represented by operators on a Hilbert space, $\H$, i.e. by Definition 3, $\A$ can be seen as a weakly closed *-subalgebra of $\B(\H)$. The commutant is the subalgebra $\A'$ of $\B(\H)$ given by all bounded operators commuting with all elements of $\A$,
$$
\A' = \{ a' \in \B(\H) \mid [a, a']=0 \ \ \text{for all $a \in \A$} \} \ .
$$
$\A'$ is again a v. Neumann algebra. The v. Neumann bi-commutant theorem states that $(\A')' = \A'' = \A$. In fact, one can show that v. Neumann algebras can actually be characterized in this way. A v. Neumann algebra represented by operators on a Hilbert space $\H$ is said to be in ``standard form'' if there exists a vector $|\Omega \rangle$ which is cyclic (see above) and separating, where ``separating'' means that $a|\Omega \rangle = 0$ implies $a=0$. In our applications to quantum field theory below, the v. Neumann algebras are almost always naturally presented in such a standard form\footnote{In general, a v. Neumann algebra is isomorphic to a v. Neumann algebra in standard form if it has a faithful representation which in turn is the case if it has a faithful normal state, i.e. a normal state
such that $\omega(a^*a) = 0$ implies $a=0$. In the following, we will always assume that this is the case.}.

We now come to an important construction for v. Neumann algebras in standard form due to Tomita and Takesaki. Let $S: \H \to \H$ be the anti-linear operator defined by $Sa|\Omega \rangle = a^*|\Omega \rangle$. It can be shown that the closure of $S$ has a polar decomposition $S = J \Delta^{1/2}$, where $J$ is anti-linear, (anti-)unitary and $\Delta^{1/2}$ is a self-adjoint, positive operator on $\H$. Furthermore, it can be shown that:
\begin{proposition}\label{Prop1}
For any v. Neumann algebra in standard form, the operators $J,\Delta^{1/2}$ satisfy:
\begin{enumerate}
\item $J \Delta^{\frac12} J = \Delta^{-\frac12}$, $J^* = J = J^{-1}$.
\item $\Delta |\Omega \rangle = |\Omega \rangle, J|\Omega \rangle = |\Omega \rangle$.
\item $a \mapsto \sigma_{t}(a) = \Delta^{it} a \Delta^{-it}$ is a 1-parameter group of automorphisms of $\A$.
\item $a\mapsto JaJ$ maps $\A$ onto $\A'$.
\item Let $\omega(a) = \langle \Omega | a \Omega \rangle$. Then $\omega$ is a {\bf KMS-state} with respect to $\sigma_t$ meaning the following. For each $a,b \in \A$, there exists a complex valued function $f_{a,b}(z)$ on the strip $\{ z \in \CC \mid 0<\Im z<1 \}$ which is bounded on the closure of the strip, analytic in the interior, and has continuous boundary values
\ben
\lim_{s \to 1^-} f_{a,b}(t+is) = \omega(\sigma_{-t}(b)a), \quad \lim_{s \to 0^+} f_{a,b}(t+is) = \omega(a \sigma_{-t}(b)) \ .
\een
\end{enumerate}
\end{proposition}
The operators $J, \Delta$ depend on $|\Omega \rangle$ and $\A$, which in principle should be included in the notation.

\medskip

{\bf Key example:}
The essence of the theorem is maybe easiest to understand in the case when $\A = M_N$ is the v. Neumann algebra of complex $N$ by $N$ matrices acting on $\CC^N$. Let $\omega$ be a state on $M_N$. Any such state can be represented by a unique density matrix, $\rho_\omega$, i.e. $\omega(a) = \tr_{\CC^N}(a\rho_\omega)$. Suppose that $\rho_{\omega}$ has no zero eigenvalues. The GNS-representation can then be described as follows: $\H_\omega$ is identified with the algebra $M_N$ itself. The GNS-vector is identified with $|\Omega_\omega\rangle = \sqrt{\rho_\omega} \in M_N$. The GNS-inner product is identified with $\langle\Psi|\Phi\rangle = \tr_{\CC^N}(\Psi^* \Phi)$. The representation acts by left-multiplication, i.e. $\pi_\omega(a)|\Psi \rangle= |a\Psi \rangle$, $\Psi \in M_N = \H_\omega$. Because $\rho_{\omega}$ has only positive eigenvalues, it immediately follows that $|\Omega_\omega\rangle$ is separating (and cyclic), hence standard. In order to describe the operators $J, \Delta$, we identify the Hilbert space $\H_\omega$ with $\CC^N \otimes \bar \CC^N$, where $\bar \CC^N$ is the dual space of $\CC^N$. Under this identification, we may also write $\pi_\omega(a) |\Psi \rangle = (a \otimes 1) |\Psi \rangle$. The commutant, $M_N'$ of $M_N$ in the representation $\pi_\omega$ is isomorphic to the opposite algebra of $M_N$ itself (with the products in reversed order). An element $b$ in the commutant $M_N'$ acts by\footnote{It is understood here that $b$ acts on $ \langle \psi |$ in $\bar \CC^N$ by $b \langle \psi| \equiv \langle b^* \psi|$.} $(1 \otimes b)|\Psi \rangle$ on $\H_\omega$. The operators $J, \Delta$ are given by
\ben\label{Eqn_DeltaExample}
\Delta^{\frac12} |\Psi \rangle = (\rho_{\omega}^{1/2} \otimes \rho_{\omega}^{-1/2}) |\Psi\rangle \ , \quad J|\Psi \rangle = |\Psi^*\rangle \ .
\een
The properties 1)-4) of the theorem are rather obvious in this example. To check 5), it is instructive to define $K = -\log \rho_{\omega}$. The state may then be written as $\omega(a) = \tr (a e^{-K})$, and the automorphism $\sigma_t$ as $\sigma_t(a) = e^{-itK} a e^{itK}$, i.e. it corresponds to the ``Heisenberg time evolution'', generated by the ``modular hamiltonian'' $K$. The state $\omega$ is then obviously a Gibbs state with respect to the modular hamiltonian. The notion of KMS-state in item 5) of the previous theorem encodes precisely this. Indeed,
\ben
f_{a,b}(t+is)=\tr (e^{-K} a e^{i(t+is)K} b e^{-i(t+is)K})
\een
has all the required properties.

\medskip

With any v. Neumann algebra $\A$ with standard vector $|\Omega \rangle$ there is associated a natural cone $\mathcal{P}^\sharp \subset \H$. It is defined by
\ben
\mathcal{P}^\sharp = \overline{\Delta^{\quarter} \A^+ |\Omega \rangle} \ ,
\een
where $\A^+$ denotes the set of non-negative elements of the v. Neumann algebra $\A$, and the overbar symbol means the closure. This cone has many beautiful properties. We will need (cf.~\cite{BratteliRobinson}):

\begin{proposition}\label{prop_cone}
\begin{enumerate}
\item Any normal state $\omega'$ has precisely one vector representative in the natural cone, i.e. $\omega'(a) =  \langle \Omega'| a \Omega' \rangle$ for a unique vector $|\Omega' \rangle \in \mathcal{P}^\sharp$.
\item $\mathcal{P}^\sharp$ is invariant under the modular group $\Delta^{it}$.
\item $\mathcal{P}^\sharp$ is the closed cone in $\H$ generated by vectors of the form $a (JaJ) |\Omega \rangle, a \in \A$.
\item Let $|\Phi \rangle, |\Phi' \rangle$ be the unique vector representatives in $\mathcal{P}^\sharp$ of normal states $\varphi, \varphi'$ on $\A$. Then
$\| \varphi - \varphi'\| \ge \|\Phi-\Phi'\|^2$.
\end{enumerate}
\end{proposition}

{\bf Key example continued:} The meaning of this proposition is maybe best understood in the case of the previous example. According to 3), the natural cone $\mathcal{P}^\sharp$ can be seen in this example to be the set of `vectors' in $\H=\CC^N \otimes \bar \CC^N$ of the form $\sum_j |\psi_j \rangle \langle \psi_j|$. A state on $M_N$ of the form $\varphi(a) = \tr_{\CC^N} (\rho_\varphi a)$ has the vector representative $|\Phi \rangle = \sqrt{\rho_\varphi}$, which is in the cone due to the Schmidt decomposition theorem. (Alternatively, it follows from (\ref{Eqn_DeltaExample}) that $\Delta^{\quarter}=\rho_{\omega}^{\quarter}\otimes\rho_{\omega}^{-\quarter}$ and hence $\mathcal{P}^\sharp=\overline{\rho_{\omega}^{\quarter}\A^+\rho_{\omega}^{\quarter}}$. Since $\rho_{\omega}$ is invertible, $\mathcal{P}^\sharp=\A^+$, and a state $\varphi$ is represented by $\sqrt{\rho_\varphi}$ in $\mathcal{P}^\sharp$.) Furthermore, in this example, the norm between two such states $\varphi, \varphi'$ may also be written as $\| \varphi - \varphi' \| = \| \rho_{\varphi'} - \rho_\varphi\|_1$ where the norm is defined by $\|a\|_1:=\tr\sqrt{a^*a}$.  By the Powers-St{\o}rmer-inequality \cite{powers}, we get
\ben
\| \varphi - \varphi'\| = \|\rho_{\varphi} - \rho_{\varphi'}\|_1 \ge \| \sqrt{\rho_{\varphi}} - \sqrt{\rho_{\varphi'}} \|_2^2 \ ,
\een
where the norm $\| a \ \|_2=\sqrt{\tr a^* a}$ is the Hilbert-Schmidt norm. Under our identification $|\Phi \rangle = \sqrt{\rho_\varphi}, |\Phi' \rangle = \sqrt{\rho_{\varphi'}} \in \mathcal{P}^{\sharp}$, the Hilbert-Schmidt-norm is nothing but the Hilbert space norm in $\H_{\omega}$, so we get 4), in the case of $\A = M_N$.

\medskip

It is sometimes too restrictive to demand that $|\Omega\rangle$ is separating for $\A$. To treat this more general situation, the above construction can be adapted as follows, see ~\cite{araki_3a,araki_3b} for details. First, one defines the subspace $\overline{\A'  |\Omega\rangle} = \H'$ with associated orthogonal projection $Q$ onto $\H'$. On $Q \A |\Omega\rangle \subset \H'$ the operator $S$ is now defined as $S Qa|\Omega\rangle = Qa^* |\Omega\rangle$ and extended by $0$ on $\H^{\prime \perp}$, so $Sa|\Omega\rangle = Qa^*|\Omega\rangle$. The closure of $S$ then has the decomposition $S=J\Delta^{1/2}$, and it follows that
${\rm ker} \Delta = \H^{\prime \perp}$ and $J^2 = Q$. As an example of this construction consider $\A=M_N$, $\H = \CC^N, |\Omega\rangle  \in \H$. This representation is obviously irreducible, $\A'= {\mathbb C}1$ and $|\Omega\rangle$ is obviously not separating. The Hilbert space $\H' = {\mathbb C} |\Omega\rangle$ and $\Delta  = |\Omega\rangle \langle \Omega |$.

\medskip

We finish this brief introduction with a subtle, but important point related to the ``statistical independence'' of two commuting v. Neumann algebras $\A_A$ and $\A_B$, represented on some common Hilbert space $\H$. (More precisely, we use $W^*$-independence in the product sense \cite{florig}.) Let $\A_A\vee\A_B=(\A_A'\cap\A_B')'$ be the v. Neumann algebra generated by $\A_A$ and $\A_B$ together, and let $\A_A\otimes\A_B$ be their v. Neumann algebraic tensor product, which we may identify with $(\A_A\otimes 1)\vee (1\otimes\A_B)$ on $\H\otimes\H$.

\begin{definition}
The algebras $\A_A$ and $\A_B$ are said to be statistically independent iff there is an isomorphism of the v. Neumann algebras $\A_A\vee\A_B \simeq \A_A\otimes\A_B$.
\end{definition}

When $\A_A$ and $\A_B$ are finite dimensional and $\A_A\cap\A_B=\bC 1$, then the algebras are always statistically independent. In the infinite dimensional case, however, statistical independence does not automatically follow. In particular, it does not follow in quantum field theory if the algebras correspond to two space-like regions which ``touch each other'' (see below).

\medskip
\noindent
{\bf Split property:} The notion of statistical independence is closely related to the ``split property''~\cite{doplicher_4, buchholz_4}: When local algebras in quantum field theory are statistically independent, there is typically a vector $|\Psi \rangle \in\H$ which is cyclic for $\A_A$ and $\A_B$ and separating for $\A_A\vee\A_B$. In this case, $|\Psi \rangle\otimes|\Psi \rangle$ is cyclic and separating for $\A_A\otimes\A_B$ and statistical independence then entails that there is a unitary map $W: \H \to \H \otimes \H$ such that
$(a \in \A_A, b \in \A_B)$
\ben
W a W^* = \pi_A(a) \otimes 1 \ , \quad W b W^* = 1 \otimes \pi_B(b) \ .
\een
We may identify the v. Neumann algebras $\A_A\simeq W\A_A W^*\subset \B(\H_A)\otimes 1$ and $\A_B\simeq W\A_BW^*\subset 1\otimes \B(\H_B)$. Furthermore, setting ${\mathfrak N} = W^*(\B(\H_A) \otimes 1)W$, one has the inclusion
\ben
\A_A \subset {\mathfrak N} \subset \A_B' \ ,
\een
which is also called the ``split''. The split and the unitary $W$ are unique (for given $|\Psi \rangle \in\H$) if we require that
$W^*(|\Psi \rangle\otimes|\Psi \rangle)$ is in the natural cone of $|\Psi\rangle$ for $\A_A\otimes\A_B$.

\subsection{Examples of $C^*$ and v. Neumann algebras}

We will now discuss some examples of $C^*$ and v. Neumann algebras which are relevant in quantum physics.

\subsubsection{The Weyl algebra}\label{ssec_Weyl}

The Weyl algebra encodes the canonical commutation relations. To define it we need a real vector space $K_{\RR}$ and a symplectic form $\sigma: K_{\RR} \times K_{\RR} \to \RR$. The Weyl algebra $\W(K_{\RR},\sigma)$ is generated by elements $W(F)$, $F \in K_{\RR}$, subject to the relations
\ben
W(F) W(F') = e^{\frac{-i}{2}\sigma(F,F')} W(F+F') \ , \quad W(F)^* = W(-F) \ .
\een
$\W(K_{\RR},\sigma)$ is turned into a $C^*$-algebra by introducing a (unique) norm and forming the completion in the norm topology, see e.g.~\cite{BratteliRobinson,binz}. We will continue to denote this completion by $\W(K_{\RR},\sigma)$. Due to the exponential nature of the Weyl-operators, the Weyl algebra behaves naturally under taking direct sums of symplectic vector spaces, in the sense that
\ben\label{tiso}
\W(K_{\RR,1} \oplus K_{\RR,2}, \sigma_1 \oplus \sigma_2) \cong \W(K_{\RR,1}, \sigma_1)
\otimes \W(K_{\RR,2}, \sigma_2) \ ,
\een
where the precise notion of the spatial tensor product between $C^*$-algebras is explained in Ch.~11 of \cite{KadisonRingrose}. In the finite dimensional case, we may take $K_{\RR} = \RR^{2n}$ and $\sigma$ to be the standard skew-symmetric form on $K_{\RR}$. This gives a $C^*$-version of the canonical commutation relations of $n$ positition variables and $n$ conjugate momenta. Informally, letting $F=(p,q)$ and $(P,Q)$ the corresponding operators with $[Q^j, P_k]=i\delta_k^j 1$, then ``$W(F)=\exp[ip \cdot Q- iq \cdot P]$'', and the Weyl relations formally follow by the Baker-Campbell-Hausdorff formula.

In order to obtain a v. Neumann algebra one takes a state $\omega$ and takes the weak closure in its GNS-representation, $\pi_{\omega}(\W(K_{\RR},\sigma))''$. The resulting v. Neumann algebra (and even its type) will in general depend on the choice of $\omega$. Quasifree (sometimes called ``Gaussian'') states of the Weyl algebra are in one-to-one correspondence with symmetric bilinear forms $\mu: K_{\RR} \times K_{\RR} \to \RR$ such that
\ben\label{sigboun}
\mu(F, F) \ge 0 \ , \qquad
\half|\sigma(F,F')| \le [\mu(F,F) \mu(F',F')]^{\half}
\een
for all $F,F' \in K_{\RR}$. The state corresponding to $\mu$ is defined by
\ben
\omega_{\mu}(W(F)) = e^{-\half \mu(F,F)} \ ,
\een
and one checks that this is indeed positive~\cite{kaersten}. We denote the GNS-representation by $(\pi_{\mu},\H_{\mu},\Omega)$.

The GNS-representations of quasi-free states can be described in terms of a Fock-space structure. In general, if $\H_1$ is a Hilbert space with inner product $( \ , \ )$, the bosonic Fock space over $\H_1$ is defined as the Hilbert space
\ben\label{FockW}
\F(\H_1) = \CC \oplus \bigoplus_{n>0} E_n \H^{\otimes n}_1 \ ,
\een
where $E_n$ is the projector onto the subspace of totally symmetric elements. The ''vacuum'' vector $|\Omega \rangle
= (1, 0 , 0 , 0 , \dots)$ corresponds to the first summand ``$\CC\ $''. The inner product $(  \ , \ )$ on this Fock space is the natural one inherited from $\H_1$. The summand $\H_1$ is called the ``1-particle subspace''. Creation and annihilation operators on $\F(\H_1)$ are denoted $a^*(\chi)$, $a(\chi)$, $|\chi)\in\H_1$, respectively, where, for any $|\Psi_n \rangle
=E_n|\psi_1)\otimes\ldots\otimes|\psi_n) \in E_n \H^{\otimes n}_1$
\bena\label{adef}
a^*(\chi) |\Psi_n \rangle &=& (n+1)^{\half} E_{n+1}|\chi \otimes \Psi_n \rangle\\
a(\chi) |\Psi_n \rangle &=& n^{-\half} \sum_{j=1}^n(\chi,\psi_j)
E_{n-1}|\psi_1)\otimes\ldots|\widehat{\psi_{j}} )\ldots\otimes |\psi_n).\non
\eena
These operators are closed and satisfy $a(\chi)=[a^*(\chi)]^*$ and $[a(\chi),a^*(\chi')]=(\chi,\chi')1$.

In order to describe the GNS-representations of quasi-free states as a Fock space, it is convenient to introduce the complexification $K$ of $K_{\RR}$ and to extend $\mu, \sigma$ to sesquilinear forms on $K$. Since $\sigma$ is non-degenerate, the inequality~\eqref{sigboun} implies that $\mu$ defines an inner product on $K$, which we write as $\langle \ | \ \rangle$. We denote the Hilbert space completion by $\text{clo}_\mu K$. The bound~\eqref{sigboun} in combination with Riesz' theorem shows that there exists a unique, bounded, self-adjoint operator $\Sigma$ on $\text{clo}_\mu K$ such that
\ben\label{Sdef}
\frac{i}{2}\sigma(F,F') = \langle F,\Sigma F'\rangle \quad \text{for all $F,F' \in K$.}
\een
We have $\|\Sigma\| \le 1$, $\Sigma^*=\Sigma$ and $\Gamma\Sigma \Gamma = -\Sigma$, where $\Gamma$ is the complex conjugation on $K$, which can be extended to an anti-unitary operator on $\text{clo}_\mu K$. We will write the polar decomposition of $\Sigma$ as $\Sigma=V|\Sigma|$, where we note that $V|\Sigma|=|\Sigma|V$ and the partial isometry $V$ satisfies $V^*=V$.

After these preliminaries, we can now describe the GNS-representations of quasi-free states. We set $\H_1:=\mathrm{ker}(1+\Sigma)^{\perp}\subset\text{clo}_\mu K$. The GNS-Hilbert space of $\omega_{\mu}$ is $\H_{\mu} = \F(\H_1)$ and the corresponding representation of the Weyl-operators $(F \in K_{\RR})$ is given by
\ben
\pi_{\mu}(W(F)) = \exp[i\{ a^*(\sqrt{1+\Sigma}F) + a(\sqrt{1+\Sigma}F) \}] \ .
\een
(We refer to~\cite{BratteliRobinson} for functional analytic details on how the exponential makes sense.) The GNS-vacuum vector is $|\Omega_\omega \rangle= (1, 0 , 0 , \dots)$, i.e. the Fock-space vacuum. One often informally defines the ``field operator'' $\phi(F) = -i\partial_t \pi_\mu(W(tF)) |_{t=0}$ for $F\in K_{\RR}$. Informally, the representation of the field operator is then
\ben
\phi(F) = a^*(\sqrt{1+\Sigma}F) + a(\sqrt{1+\Sigma}(\Gamma F)) \ .
\een
Here we inserted the operator $\Gamma$ to ensure that this field has a natural complex linear extension to $F\in K$.

Quasifree states can enjoy the following additional properties:
\begin{enumerate}
\item $\omega_{\mu}$ is pure if and only if $|\Sigma|=1$, see e.g.~\cite{wald_2, manuceau}, i.e.~ its GNS-representation is irreducible. When $\omega_{\mu}$ is pure, then $P_\pm = \half(1 \pm \Sigma)$ define projections in $\text{clo}_\mu K$ onto so-called positive and negative frequency subspaces. We then have $\H_1=P_+\text{clo}_\mu K$. Because $\Gamma P_{\pm}=P_{\mp}\Gamma$ we find $\phi(F) = \sqrt{2}\{ a^*(P_+F) + a(\Gamma P_-F) \}$.
\item $\omega_{\mu}$ is called primary iff $\mathrm{ker}(\Sigma)=\{0\}$. Note that this may fail in general, because we take a completion of $K$. Pure states are necessarily primary. More generally, $\omega_{\mu}$ is primary if and only if the v. Neumann algebra $\pi_{\mu}[\W(K_{\RR},\sigma)]''$ of the GNS-representation has a trivial centre (i.e.~it is a v. Neumann factor). In general, $V^2$ is the orthogonal projection onto the orthogonal complement $\mathrm{ker}(\Sigma)^{\perp}$, so $V=V^*=V^{-1}$ for primary states.
\item The GNS-vector representative $|\Omega \rangle$ is separating for $\pi_{\mu}[\W(K_{\RR},\sigma)]''$ if and only if $|\Sigma|<1$, i.e. $\mathrm{ker}(1-|\Sigma|)=\{0\}$ (\cite{araki_4}, thm.3.12, \cite{leyland} thm. I.3.2). In this case we have $\H_1=\text{clo}_\mu K$. Note that the 1-particle space $\H_1$ is ``twice as large'' by comparison with the case of a pure state. This corresponds to the fact that the representation of $\W(K_{\RR},\sigma)$ is now reducible.
\end{enumerate}

\subsubsection{The CAR algebra} \label{ssec_CAR}
The CAR algebra encodes the canonical anti-commutation relations. To define it we need a Hilbert space $K$ with inner product denoted by $( \ . \ , \ . \ )_K$, and an anti-linear involution
$\Gamma$ on $K$ satisfying $(\Gamma k_1, \Gamma k_2)_K = (k_2, k_1)_K$. The CAR algebra $\gC(K,\Gamma)$ is generated by the
elements $\psi(k)$, $k \in K$, subject to the relations
\ben
\psi(k_1) \psi(k_2) + \psi(k_2) \psi(k_1) = (\Gamma k_1, k_2)_K 1 \ , \quad \psi(k)^* = \psi(\Gamma k) \ .
\een
$\gC(K,\Gamma)$ is turned into a $C^*$-algebra by introducing the (unique) norm $\| \psi(k) \| = \| k \|_K/\sqrt{2}$ and forming the completion in the norm topology, see e.g.~\cite{araki_5,BratteliRobinson}. We will continue to denote this completion by $\gC(K,\Gamma)$. There is a *-automorphism $\alpha$ on $\gC(K,\Gamma)$ characterized uniquely by  $\alpha(\psi(k))=-\psi(k)$ which gives the CAR algebra a ${\mathbb Z}_2$-grading.

One has the functorial property
\ben\label{eq:functor}
\gC(K_1 \oplus K_2, \Gamma_1 \oplus \Gamma_2) \cong \gC(K_1, \Gamma_1) \hat \otimes \gC(K_2, \Gamma_2) \ ,
\een
where $\hat \otimes$ is the graded tensor product\footnote{
As a vector space, the graded tensor product is first defined to be the usual (algebraic) tensor product. The product is
defined as $(a_1 \hat \otimes b_1)(a_2 \hat \otimes b_2)= (-1)^{deg(a_2)deg(b_1)} \, a_1 a_2 \hat \otimes b_1 b_2$ and the *-operation
is $(a \hat \otimes b)^* = (-1)^{deg(a)deg(b)} a^* \hat \otimes \ b^*$, where the degree is defined to be $0$ resp. $1$ for even resp. odd elements under
$\alpha$. It is then shown that a natural $C^*$-norm compatible with these relations and the above isomorphism
can be defined which extends the $C^*$-norm of $\gC(K_i, \Gamma_i)$. The graded tensor product is the $C^*$-closure under this norm.
}.
As for the case of the Weyl algebra, one can develop a complete theory of
quasi-free states and describe their representations, see~\cite{araki_5} for details.

Here, we only describe pure, quasi-free states, and we do so by directly describing the associated GNS-representation. The input is an orthogonal projector, $P$, on $K$, obeying the relation $\Gamma P \Gamma = 1-P$. Then one sets $\H_1 = PK$ equipped with the restriction $( \ . \ , \ . \ )$ of the inner product on $K$. Next, the fermionic Fock space over $\H_1$ is defined exactly as in~\eqref{FockW}, with the only difference that $E_n$ now projects onto the subspace of totally anti-symmetric elements. Fermionic creation and annihilation operators are defined again as in~\eqref{adef}. We then define the desired representation associated with $P$ as
\ben\label{eq:fieldCAR}
\pi_P(\psi(k)) = a^*(Pk) + a(P\Gamma k) \ .
\een
The vector $|0\rangle$ in Fock space defines a state  on the CAR algebra called $\omega_P$.
The ``2-point'' function of the state is $\omega_P(\psi(k_1) \psi(k_2)) \equiv
\langle 0 | \pi_P(\psi(k_1)) \pi_P(\psi(k_2)) | 0 \rangle = (\Gamma k_1, Pk_2)$, and similar formulas can be derived for the higher ``$n$-point'' functions, see~\cite{araki_5} for details.

\subsubsection{The Cuntz algebra $\mathcal{O}_n$}\label{ssec_Cuntz}

A $C^*$-algebra arising naturally in the theory of superselection sectors (see below section~\ref{sec:charged}) is the Cuntz-algebra. Let $\H$ be a separable, infinite dimensional Hilbert space.
A partial isometry, $V$, on $\H$ is a linear operator such that $V^*V = 1$ and such that $VV^*$ is a projection. Now let $n>1$ a natural number. Then it is not hard to see that one can construct partial isometries $\psi_i, i=1, \dots, n$ on $\H$ satisfying the relations
\ben\label{cntz}
\sum_{i=1}^n\psi_i \psi_i^* = 1 \ ,\quad
\psi^*_i \psi_j = \delta_{ij}1 \quad \text{for all $i,j$}.
\een
Cuntz~\cite{cuntz} has shown that there exists a unique (up to $C^*$-isomorphism)
simple $C^*$-algebra generated by these elements and relations. It is denoted $\mathcal{O}_n$.

\subsection{The basic principles of quantum field theory} \label{sec:AQFT}

In algebraic quantum field theory, the algebraic relations between the quantum fields are encoded in a collection of $C^*$ or v. Neumann algebras associated with spacetime regions. The precise framework depends somewhat on the type of theory, spacetime background etc. one would like to consider. In the case of Minkowski space $\RR^{d,1}$, a standard set of assumptions, manifestly satisfied by many examples, and believed to be satisfied by all reasonable QFTs, is as follows. Call a ``causal diamond'' $O \subset \RR^{d,1}$ any set of the form $O = D(A)$, where $A$ is any relatively compact open set contained in a Cauchy surface $\cong \RR^d$ of Minkowski space, and $D(A)$ its domain of dependence, i.e. the set of points $x \in \RR^{d,1}$ such that any inextendible causal curve through $x$ must hit $A$ at least once, see fig.~\ref{fig:diamond}. Poincar\' e transformations $g=(\Lambda, a) \in
{\rm P} = {\rm SO}_+(d,1) \ltimes \RR^{d+1}$ act on points by $g\cdot x = \Lambda x + a$. Since Poincar\' e transformations are isometries of Minkowski spacetime, they map causal diamonds to causal diamonds, so we get  an action $O \mapsto g \cdot O$ on the set of causal diamonds.

In the algebraic approach, a quantum field theory is a collection (``net'') of $C^*$-algebras $\A(O)$ subject to the following conditions:

\begin{enumerate}
\item[a1)] (Isotony) $\A(O_1) \subset \A(O_2)$ if $O_1 \subset O_2$. We write $\A = \overline{\bigcup_O \A(O)}$ with completion in the $C^*$-norm.
\item[a2)] (Causality) $[\A(O_1),\A(O_2)]=\{0\}$ if $O_1$ is space-like related to $O_2$. In other words, algebras for space-like related causal diamonds commute. Denoting the causal complement of a set $O$ by $O'$, we may also write this more suggestively as
$$
\A(O') \subset \A(O)'
$$
where the prime on the right side is the commutant.
\item[a3)] (Relativistic covariance) For each transformation $g \in \widetilde{{\rm P}}$ covering\footnote{The covering group is needed to describe non-integer spin.} a Poincar\'e transformation $(\Lambda,a) \in {\rm P} = {\rm SO}_+(d,1) \ltimes \RR^{d+1}$, there is an automorphism $\alpha_g$ on $\A$ such that $\alpha_g \A(O) = \A(\Lambda O+a)$ for all causal diamonds $O$ and such that $\alpha_g \alpha_{g'} = \alpha_{gg'}$ and $\alpha_{(1,0)}=\id$ is the identity.
\item[a4)] (Vacuum) There is a unique state $\omega_0$ on $\A$ invariant under $\alpha_g$. On its GNS-representation $(\pi_0, \H_0, |0\rangle)$, $\alpha_g$ is implemented by a projective positive energy representation $U$ of the Poincar\' e group $\widetilde{{\rm P}}$ in the sense that $\pi_0(\alpha_g(a)) = U(g) \pi_0(a)U(g)^*$ for all $a \in \A, g \in \widetilde{{\rm P}}$. Positive energy means that the representation is strongly continuous and that, if $x \in \RR^{d,1} \subset {\rm P}$ is a translation by $x$, we can write
\ben
U(x) = \exp(-i P^\mu x_\mu),
\een
and the vector generator $P=(P^\mu)$ has spectral values $p=(p^\mu)$ in the forward lightcone
 $p \in \bar{V}^+ = \{ p \mid p^2 \ge 0, p^0>0\}$.
\end{enumerate}

For technical reasons, one often forms the weak closures $[\pi_0(\A(O))]''$, which gives a corresponding net of v. Neumann algebras (on $\H_0$). We will often work with this net.
Axioms a1) and a2) can be generalized straightforwardly to any general globally hyperbolic spacetime $(\sM,g)$. If such a spacetime has any symmetries, then a3) can be generalized too, with the Poincar\'e group replaced by the isometry group $\rm G$ of the spacetime. If this group $\rm G$ has a 1-parameter family of isometries with everywhere complete time-like orbits (as is the case e.g. in a static spacetime), then a version of a4) can be imposed, too. For more details, we refer the reader to~\cite{hollands_1}, and references therein.

A straightforward, but important, consequence of axioms a1)-a4) (and Araki's ``weak additivity axiom'' \cite{araki_6}, see below) is the Reeh-Schlieder theorem~\cite{reeh}, which is the following. We know by construction that $\pi_0(\A) |0 \rangle$ is dense in the entire Hilbert space, $\H_0$. One might guess at first that the subspace of states $\pi_0(\A(O)) |0 \rangle$, describing excitations relative to the vacuum localized in a causal diamond $O$, would depend on $O$. This expectation is usually incorrect, however, and instead the Reeh-Schlieder theorem holds:

\begin{theorem} (Reeh-Schlieder theorem)
Assume that any element of $\A$ can be approximated arbitrarily well
in the sense of matrix elements in the vacuum representation $\pi_0$ by finite sums of elements of the form $\alpha_{x_i}(a_i)$, where $a_i$
are in some arbitrarily small causal diamond, and where $\alpha_{x_i}$ denotes a translation (``weak addititivity\footnote{In terms of algebras:
$$
\pi_0(\A) = \left(
\bigcup_{x \in \RR^{d,1}}
\pi_0(\A(O+x))
\right)''
$$
for any causal diamond $O$.
}'').
For any causal diamond $O$, the set of vectors $\pi_0(\A(O)) |0\rangle$ is dense in the entire Hilbert space. The same statement remains true if $|0\rangle$ is
replaced with a vector with finite energy.
\end{theorem}

{\em Proof:} The proof of this theorem is standard. We drop the reference to $\pi_0$, so $a$ means $\pi_0(a)$, etc. Suppose that the closure of $\{ a | 0 \rangle \mid
a \in \pi_0(\A(\tilde O)) \}$ has a non-trivial orthogonal complement in $\H_0$ where $\tilde O$ is a causal diamond whose closure is contained in $O$ (i.e. ``inside'' $O$).
Let $|\Phi \rangle$ be a vector in the orthogonal complement, $a \in \A(\tilde O)$, and let $\RR^{d,1} \owns z \mapsto f(z) = \langle \Phi | U(z) a | 0 \rangle$. Due to the spectral properties of
$P$ postulated in a3), $f(z)$ has a holomorphic extension to the set $\{ z \in \CC^{d+1} \mid \Im (z) \in V^+\}$. By assumption $f(z) = 0$ if $z$ is in some sufficiently
small real neighborhood of $z=0$, since the translated element $U(z) a U(z)^*$ remains in $\A(O)$ for such $z$, by a3). By the edge-of-the-wedge theorem, see appendix~\ref{eow}, $f(z) = \langle \Phi| U(z) a U(z)^*| 0 \rangle$ therefore vanishes for all $z \in \RR^{d,1}$. Since an arbitrary $b \in \A$ can be approximated in the sense of matrix elements by a sum of translated elements from $\A(\tilde O)$ (by assumption), we have $\langle \Phi |  b | 0 \rangle=0$ for all $b \in \A$, proving
that also $|\Phi \rangle$ vanishes identically.

If $|\Psi \rangle$ is a vector with finite energy, then $U(z)^* |\Psi \rangle$ has an analytic continuation to $\{ z \in \CC^{d+1} \mid \Im (z) \in V^+\}$ and a similar
argument applies. \qed
\\

The Reeh-Schlieder theorem shows in particular that it is not straightforward to define ``localized states''.  To define these, and to have a mathematical counterpart of the intuitive idea that the set of states localized in $O$ with energy below $E$ should be approximately finite-dimensional, one has to go beyond the above axioms~\cite{haag_3}. Nowadays, this idea is phrased mathematically in terms of ``nuclearity conditions'', which will also play a role in this work. One such condition, due to Buchholz and Wichmann~\cite{buchholz_4}, is described now.
We formulate it as an additional axiom a5), although the precise form of this axiom is perhaps not so natural and irrefutable as the previous ones, a1)-a4), and in fact, we will
throughout this volume also consider variants of a5) [see a5') in sec.~\ref{sec:upper_bounds}].

\begin{enumerate}
\item[a5)] (BW-nuclearity) Let $A$ be a ball of radius $r$ in a time-slice, and let $O_r=D(A)$ be the corresponding double cone. Consider the map
\ben
\Theta_{\beta,r}: \A(O_r) \to \H_0 \ , \quad a \mapsto e^{-\beta H} \pi_0(a) |0 \rangle \ ,
\een
where $\beta>0$ and where $H=P^0$ is the Hamiltonian, i.e. the time-component of $P^\mu$ in item a4). It is required that there exist positive constants $n>0$ and $c = c(r)>0$ such that for $r>0, \beta>0$
\ben\label{BW}
\| \Theta_{\beta, r} \|_1 \le e^{(c/\beta)^n} \ .
\een
\end{enumerate}

Here we use the following~\cite{pietsch,fewster_1} definition/lemma:
\begin{definition}\label{def_1norm}
The 1-nuclear norm $\| \  . \ \|_1$ of a linear operator $T$ between two Banach spaces $\mathcal{X}, \mathcal{Y}$ is defined as $\|T\|_1 = \inf \sum_j \| y_j \| \ \|\psi_j\|$, where the infimum is taken over all possible ways (if any\footnote{If there are none, then the norm is set to infinity.}) of writing $T( \ . \ ) = \sum_j y_j \psi_j( \ . \ )$ as a norm convergent sum in terms of linear functionals $\psi_j \in \mathcal{X}^*$ and vectors $y_j \in \mathcal{Y}$.
\end{definition}
If both $\mathcal{X}, \mathcal{Y}$ are Hilbert spaces, then $\|T\|_1 = \tr |T|$. The $1$-nuclear norm satisfies the following properties:
\ben\label{pnorm}
\| ST \|_1 \le \| S \| \| T \|_1  \ , \quad
\| ST \|_1 \le \| S \|_1 \| T \| \ ,
\een
where $S$ is another linear operator between Banach spaces $\mathcal{Y}, \mathcal{Z}$.

The idea is that this 1-norm measures the ``size'' of the ellipsoid in Hilbert space given by the collection of vectors of the form $e^{-\beta H} \pi_0(a) |0\rangle$ with $\|a\| \le 1$. Such vectors represent states which are localized in space (by $r$) and have an exponentially suppressed energy (essentially by $1/\beta$). One typically expects $c \propto r$ and $n = d$ in $d$ spatial dimensions. (Note that in quantum mechanics, using $\mathcal{B}(\H)$ instead of $\A(O_r)$ and assuming that $H$ has an orthonormal eigenbasis, $\|\Theta_{\beta,r}\|_1$ reduces to the partition function.)

The BW-nuclearity condition a5) and its variants -- in particular the modular nuclearity condition a5') given below  --
can be shown to have a consequence which is of essential importance in this volume. If $A$ and $B$ are disjoint regions in a spatial slice with distance ${\rm dist}(A,B) > 0$ and corresponding causal diamonds $O_A$ and $O_B$, then a5) implies that the split property holds for $\A_A = \pi_0(\A(O_A))''$ and $\A_B = \pi_0(\A(O_B))''$ \cite{buchholz_2}. In other words, the v. Neumann algebras not only commute (by item a2)), but they are statistically independent. This fact will be crucial when defining relative entanglement entropies between such algebras.

A closely related consequence of a5) describes what happens when the regions $O_A$ and $O_B$ ``touch'' each other. It deals with a net of v. Neumann algebras $\A(O)$, and it assumes an additional asymptotic scale invariance at small length scales. Under these circumstances, the local v. Neumann algebras are direct sums of type III$_1$ factors~\cite{buchholz_2}. Furthermore, when we call a state $\omega$ on $\A$ locally normal when its restriction to each $\A(O)$ is normal, we have the following~\cite{fewster_2}:
\begin{theorem}\label{thm_split}
\begin{enumerate}
\item[a)] The restriction of a locally normal state $\omega$ on $\A$ to a local algebra $\A(O)$ is {\rm never pure}.
\item[b)] If $\omega_O$ is a pure normal state on some local algebra $\A(O)$ then it cannot be extended to a normal state on any larger local algebra.
\item[c)] Let $A$ and $B$ be disjoint subsets of a Cauchy surface touching each other in the sense that $\bar A$ intersects $\bar B$. Then there cannot exist a normal, separable state on $\A(O_A) \vee \A(O_B)$, i.e. one such that $\omega(ab) = \sum_j \varphi_j(a) \psi_j(b)$ for all $a \in \A(O_A), b \in \A(O_B)$, where $\varphi_j, \psi_j$ are normal positive functionals on $\A(O_A), \A(O_B)$, respectively. In particular, $\A(O_A), \A(O_B)$ cannot be statistically independent.
\end{enumerate}
\end{theorem}

{\em Proof:} a) follows immediately from b), which is proved in corollary 3.3 of~\cite{fewster_2}. To see that c) follows from a) we assume that a normal separable state $\omega$ exists, and we consider its separable decomposition. We can then find pure normal states $\omega_A$ and $\omega_B$ on $\A(O_A)$ and $\A(O_B)$, respectively, and a constant $c>0$ such that $\omega_A\le\sqrt{c}\varphi_1$ and $\omega_B\le\sqrt{c}\psi_1$. For all $a\in\A(O_A)$ and $b\in\A(O_B)$ we then have $0\le\omega_A(a^*a)\omega_B(b^*b)\le c\omega(a^*ab^*b)$, an estimate which can be extended to the algebraic tensor product $\A(O_A)\odot\A(O_B)$ and then to the entire algebra $\A(O_A)\vee\A(O_B)$. We denote the resulting state constructed out of $\omega_A$ and $\omega_B$ with a tensor product, so that $\omega_A\otimes\omega_B\le c\omega$. This estimate shows that $\omega_A\otimes\omega_B$ must be normal on $\A(O_A)\vee\A(O_B)$, which means in turn that it can be described by a density matrix $\rho_{AB}$ on the vacuum Hilbert space. Finally, $\rho_{AB}$ defines a locally normal state on all of $\A$, which restricts to the pure normal state $\omega_A$ on $\A(O_A)$, contradicting statement a). Hence, a normal separable state on $\A(O_A)\vee\A(O_B)$ cannot exist.\qed

The BW-nuclearity condition is designed specifically for Minkowski space, or more generally for a spacetime with a 1-parameter group of time-like isometries, where a Hamiltonian exists. There are also other nuclearity conditions allowing to draw similar conclusions, most notably ones involving the modular operator $\Delta$~\cite{buchholz_3,lechner_2} introduced below as a5') in sec.~\ref{sec:upper_bounds}.

\subsection{Examples of algebraic quantum field theories}

Let us now discuss some examples of algebraic quantum field theories which manifestly fit into the operator algebraic setting.

\subsubsection{Free scalar fields}\label{ssec_free}

The free Klein-Gordon (KG) field $\phi$ of mass $m$ on $d+1$-dimensional Minkowski spacetime (or, more generally, on any globally hyperbolic manifold $(\sM,g)$) is the simplest example satisfying all the axioms, including the BW-nuclearity condition, cf.~\cite{buchholz_4,buchholz_5}. Roughly speaking, $\phi(x)$ is an operator for each $x \in \sM$ satisfying the KG-equation $(\Box - m^2) \phi(x) = 0$. If $Q = \phi |_{\C}, P = \partial_n \phi |_\C$ are the restriction of the field and its normal derivative to a Cauchy surface (``time-slice'') $\C$, then the canonical commutation relations $[Q(x), P(y)] = i\delta_\C(x,y)1$ are satisfied, and, informally, the algebra $\A(O_A)$ with $A \subset \C$, should be thought of as being generated by all $P(x),Q(x)$ with $x \in A$. For a  mathematically precise construction one has to pass to bounded operators, which can be done e.g. by treating the field operators $\phi(x)$ as a distribution and by working with suitable exponentials of smeared versions of these fields.

This construction is as follows. We let $K_\RR = C^\infty_0(\C, \RR) \oplus C^\infty_0(\C,\RR)$, and we define a symplectic form, corresponding to the canonical commutation relations, by $\sigma(F,F') = \int_\C [qp'-q'p] \dd V$ for all $F=(q,p), F'=(q',p') \in K_\RR$. For $A \subset \C$, we define $\A(O_A)$ to be the subalgebra of the Weyl algebra $\W(K_\RR,\sigma)$ given by the $C^*$-closure of $\{ W(F) \mid {\rm supp}(F) \subset A\}$.\footnote{Here the braces denote ``generated by, as a $C^*$-algebra'', and ${\rm supp}(F)={\rm supp}(q)\cup{\rm supp}(p)$, where the support ${\rm supp}$ of a function is the closure of the set of all points where it does not vanish.} We think of the Weyl operators $W(F)$ as representing the informal expressions $\exp[i\int_\C (pQ - qP) \dd V]$.

This construction has the disadvantage that we only identify the algebras corresponding to causal diamonds $O_A$ with base $A$ in the fixed Cauchy-surface $\C$, see fig.~\ref{fig:diamond}. However, there is a simple way to get around this by giving an alternative description of $(K_\RR,\sigma)$. For this, we note that $K_{\RR}$ is just the space of initial data of the KG equation, equipped with the natural symplectic form. Since initial data are, on a globally hyperbolic spacetime, in one-to-one correspondence with solutions to the KG equation, we may equally well work with solutions. For this, let $G^A, G^R: C^\infty_0(\sM,\RR) \to C^\infty(\sM,\RR)$ be the unique advanced and retarded propagators to the KG equation uniquely specified by the relations
\ben
(\Box - m^2) G^{A,R} = 1_M \ , \quad
{\rm supp}(G^{A,R}f) \subset J^{-,+}({\rm supp}(f)),
\een
where $J^\pm$ denotes the causal future/past of a set in the spacetime $(\sM,g)$, see e.g. \cite{bar}. Let $G=G^R-G^A$ be the ``causal propagator''. It can be identified with a distributional kernel on $\sM \times \sM$, denoted by $G(x,y)$. This distribution is real-valued, anti-symmetric, a solution to the KG equation in both entries, and vanishes if $x$ is space-like to $y$. Define $\tilde K_{\RR} = C^\infty_0(\sM,\RR)/{\rm Ran}(\Box - m^2)$ (with $(\Box-m^2)$ acting on $C^{\infty}_0(\sM,\RR)$). Elements in this space are equivalence classes $[f]$ of compactly supported, smooth, real-valued functions $f$ on $\sM$ modulo ones in the image of the KG operator. On $\tilde K_\RR$ we define a  symplectic form by $\tilde \sigma([f],[f']) = G(f,f')$. One now shows that $(\tilde K_\RR, \tilde \sigma)$ is a symplectic space isomorphic to $(K_\RR, \sigma)$. The isomorphism is given by assigning to $[f] \in \tilde K_\RR$ the initial data $(q,p) \in K_\RR$ for the solution $Gf$,
\ben\label{iso}
[f] \mapsto
\left(
\begin{matrix}
Gf |_{\C} \\
\partial_n Gf |_{\C}
\end{matrix}
\right)
\equiv
\left(
\begin{matrix}
q \\
p
\end{matrix}
\right) \ .
\een
Consequently, the Weyl algebra $\W(\tilde K_\RR, \tilde \sigma)$ is isomorphic to $\W(K_\RR,\sigma)$. The local algebras for an arbitrary $O \subset \sM$ are now defined as
\ben
\A(O) = \{ \tilde W([f]) \mid {\rm supp}(f) \subset O\} \ .
\een
The obvious analogs of axioms a1), a2) for a Lorentzian spacetime $(\sM,g)$ directly follow from the properties of $G$ and the relation $\tilde W([f]) \tilde W([f]) = e^{-iG(f,f')/2} \tilde W([f+f'])$. Given a representation $\pi$ of the Weyl algebra on a Hilbert space, one may informally think of $\pi(\tilde W([f]))$ as $e^{i\phi(f)}$, where $\phi(f) = \int_M \phi(x) f(x) dV$ is a ``smeared'' quantized Klein-Gordon field. The field $\phi(f)$ actually exists as a self-adjoint, operator valued distribution in general only in certain representations of the Weyl-algebra.

One such representation is the ground state representation of a massive ($m>0$) field on an ultra-static spacetime $\sM=\RR \times \C, g=-dt^2 + h$, where $h$ is a Riemannian metric on $\C$ not depending on $t$, and where $\C$ is assumed to be compact. Let $\nabla^2$ be the Laplacian of the Riemannian metric $h$ on $\C$. $-\nabla^2$ is known to be a positive essentially self-adjoint operator on $L^2(\C)$~\cite{seeley}, so in particular all complex powers of $-\nabla^2+m^2$ are well-defined. Let $C=(-\nabla^2+m^2)^{-1}$. Consider the bilinear form $\mu:  K_\RR \times K_\RR \to \RR$ given by
\ben
\mu(F,F') = \half \Re (C^{1/4} p - iC^{-1/4} q, C^{1/4} p' - iC^{-1/4} q' )
\een
where $F=(q,p)$ and we mean the standard $L^2$-inner product on the right hand side. It follows immediately that \eqref{sigboun} holds, so $\mu$ defines a quasi-free state $\omega_0$ on $\W(K_\RR, \sigma)$, and therefore also one on $\W(\tilde K_\RR, \tilde \sigma)$ via the isomorphism described above in \eqref{iso}. The closure of $K_\RR$ in the topology given by $\mu$ is isomorphic to $W^{+1/2}_\RR(\C) \oplus W^{-1/2}_\RR(\C)$ (here we mean the Sobolev spaces of fractional order $\pm 1/2$) and the corresponding operator $\Sigma$ as defined in eq.(\ref{Sdef}) is given by
\ben\label{Eqn_Sigma}
\Sigma=
i\left(
\begin{matrix}
0 & C^{1/2}\\
-C^{-1/2} & 0
\end{matrix}
\right).
\een
It follows that $\Sigma^2 = 1$, so the state $\omega_0$ is pure. If we compose the map $\sqrt{1+\Sigma}$ with the identification $W^{+1/2}(\C) \oplus W^{-1/2}(\C)\simeq L^2(\C)\oplus L^2(\C)$, determined by $(q,p)\mapsto \frac{1}{\sqrt{2}}(C^{-\quarter}q,C^{\quarter}p)$, we obtain
\ben
K_\RR \owns F = \left(
\begin{matrix}
q\\
p
\end{matrix}
\right)
\mapsto
\frac12
\left(
\begin{matrix}
iC^{1/4} p + C^{-1/4} q\\
C^{1/4} p - iC^{-1/4} q
\end{matrix}
\right)
\equiv
\frac{1}{\sqrt{2}}
\left(
\begin{matrix}
i\kappa(F)\\
\kappa(F)
\end{matrix}
\right)
\in L^2(\C)^{\oplus 2} .
\een
From this it immediately follows that the image of $K_\RR$ under $\sqrt{1+\Sigma}$ is isomorphic to $L^2(\C)$, which may thus be identified with the 1-particle Hilbert space $\H_1$. The map $\kappa:K_\RR \to L^2(\C) = \H_1$ defined in the previous equation is the one-particle structure of Kay \cite{kay_1}. In terms of $\kappa$ one may write the GNS-representation of our state $\omega_0$ as ($F \in K_\RR$)
\ben
\pi_0(W(F)) = \exp i[a^*(\kappa(F)) + a(\kappa(F))] \ ,
\een
where $a, a^*$ are annihilation and creation operators on the bosonic Fock space over $\H_1$. One can show (cf. \cite{kay_1}), that the time translation isometry of $(\sM,g)$ is implemented in the GNS representation of this state by a strongly continuous 1-parameter group with positive generator $H$ (the Hamiltonian). Thus, the state can be considered as a ground state. The same construction works in any ultra-static, globally hyperbolic spacetime even when the slices are not compact (because $\C$ is then complete). In particular, it works in Minkowski space, where the Cauchy surface is given by $\C=\RR^d$, equipped with the flat metric.

In Minkowski space, it is instructive to represent 1-particle wave functions in $\H_1$ in momentum space, using the Fourier
transform\footnote{Our convention for the Fourier transform in one dimension is $\widetilde f(p) = \frac{1}{\sqrt{2\pi}} \int \dd x f(x) e^{-ipx}$.}
to identify $L^2(\RR^d, \dd^d {\bf x})$ with $L^2(\RR^d, \frac{1}{2\omega({\bf k})}\dd^d {\bf k})$, where $\omega({\bf k}):=\sqrt{{\bf k}^2+m^2}$. More precisely, we map $\psi(x)$ to $\sqrt{2\omega({\bf k})}\widetilde{\psi}({\bf k})$, which defines an isometry. For any smooth function $f$ on Minkowski space of compact support, one then writes the smeared quantum KG-field
characterized by $\pi_0(\tilde W([f])) = \exp i\phi(f)$ informally as an integral $\int \phi(x) f(x) \dd^{d+1}x$. Going through the definitions and constructions just given, one can then further write the representer in the familiar form
\ben\label{phidef}
\phi(x) = (2\pi)^{-\frac{d}{2}}\int_{\RR^d} \frac{\dd^d {\bf k}}{2\omega({\bf k})} [a^\dagger ({\bf k}) e^{-ikx} + a({\bf k}) e^{ikx}]
\een
in terms of the creation and annihilation operators\footnote{In our setup, we arrive at the normalization
\ben
[a({\bf k}), a^\dagger({\bf k}')] = 2\omega({\bf k}) \ \delta^d({\bf k}-{\bf k}') \cdot 1 \ , \quad
[a({\bf k}), a({\bf k}')] = 0 .
\een}
on the bosonic Fock space over $\H_1=L^2(\RR^d, \frac{1}{2\omega({\bf k})}\dd^d {\bf k})$, where $xk=x^\mu k_\mu$ and
$(k_\mu)=(-\omega({\bf k}), \bf k)$. After smearing with $f \in C^\infty_0(\RR^{d+1})$, this operator is shown to have a dense set of analytic vectors, so it is essentially self-adjoint by Nelson's analytic vector theorem~\cite{reed}. Thus, we may write the representation of the Weyl-operators by exponentiation in the form $\pi_0(\tilde W([f])) = \exp[i \phi(f)]$, so the relation between Weyl operators and unbounded quantum fields is more than a formal one in this example. The vacuum state satisfies the axioms a3), a4) as well as the BW-nuclearity condition~\cite{buchholz_4}.

\subsubsection{Free fermion fields}\label{sec:freeDirac}
The free spin-$\frac{1}{2}$-field $\psi$ of mass $m$ on $d+1$-dimensional Minkowski spacetime (or, more generally, on any globally hyperbolic spin manifold $(\sM,g)$) is another simple example satisfying all the axioms, including the BW-nuclearity condition in static spacetimes~\cite{dantoni}. Roughly speaking, $\psi(x)$ is an operator for each $x \in \sM$ satisfying the Dirac equation $( \slash \nabla+ m) \psi(x) = 0$.  For a  mathematically precise construction one uses the CAR algebra described in sec.~\ref{ssec_CAR}.

This construction is as follows e.g. for a real (Majorana) field, which exists when $d-1=0,1,2,7 \ {\rm mod} \ 8$, for details see e.g.~\cite{dantoni}. Let $\$$ be the spin-bundle, which is a complex vector bundle over $\sM$ of dimension $2^{\lfloor (d+1)/2 \rfloor}$. At each point, we let $e_\mu$ be a chosen orthonormal $d+1$-bein, so that $\slash \nabla = e_\mu \cdot \nabla_{e_\mu}$ is the Dirac operator, with $e_\mu \cdot$ denoting Clifford multiplication. The action of $e_\mu$ on the conjugate vector bundle\footnote{The complex conjugate, $\bar V$, of a vector space $V$ is identical as a set, but has the scalar multiplication $\lambda \cdot v \equiv \bar \lambda v$.} $\overline{\$}$ is denoted by $\overline{e_\mu} \cdot$. There is then an anti-linear map $C: \$ \to \overline{\$}$ satisfying $\overline{C} C = (-1)^{\frac12(d-1)d}$ which is characterized by $C^{-1} \overline{e}_\mu C = e_\mu$. We also have ``Dirac conjugation'', which is an anti-linear map $B: \$ \to \$^*$ with $B^* B = 1$ (for details on such notions, see e.g.~\cite{fig}). The Hilbert space $K$ needed for the definition of the CAR algebra $\gC(K, \Gamma)$ is the space $K = L^2(\C, \$ |_\C)$ of square integrable spinors on a Cauchy surface $\C$
(unit forward normal: $n$),
with inner product
\ben\label{innerprod}
(k_1, k_2)_K = \int_\C Bk_1(n \cdot k_2) \, \dd V
\een
and the anti-linear involution $\Gamma$ is given by $\Gamma k = \overline{C}\bar k$. If $A \subset \C$ and $O_A$ its causal diamond, we define
\ben
\A(O_A) = \{ \psi(k) \mid k \in K, {\rm supp}(k) \subset A\}^{\rm even} \ ,
\een
where on the right side, curly brackets mean ``generated by, as a $C^*$-algebra'', and the superscript denotes the sub algebra of elements with an even number of generators.

As before, this construction has the disadvantage that only the algebras corresponding to causal diamonds on a fixed Cauchy-surface $\C$ are identified.
However, there is a simple way to get around this by giving an alternative description of $(K,\Gamma)$. For this, we note that $K$ is just the space of initial data of the Dirac equation, equipped with the natural hermitian inner product. Since initial data are, on a globally hyperbolic spacetime, in one-to-one correspondence with solutions to the Dirac equation, we may equally well work with solutions. For this, let $\slash E^A, \slash E^R: C^\infty_0(\sM,\$) \to C^\infty(\sM,\$)$ be the unique advanced and retarded propagators associated with the ``squared''
Dirac operator
\ben
(\slash \nabla + m)(\slash \nabla - m) = \Box - m^2 - \quarter R \ .
\een
The advanced propagator for the Dirac equation is simply $\slash S^A = (\slash \nabla - m) \slash E^A$ (and similarly for the retarded propagator), and we also set
$\slash S = \slash S^A - \slash S^R$. We next equip the complex vector space of all smooth compactly supported spinors with the hermitian sesquilinear form
\ben
(k_1, k_2)_{\tilde K} = \int_\sM Bk_1 (\slash S k_2) \ .
\een
This sesquilinear form is checked to be positive semi-definite, and after factoring with the kernel of $\slash S$, it becomes a pre-Hilbert space $\tilde K= C^\infty_0(\sM, \$)/{\rm Ran} (\slash \nabla + m)$. Now one shows that the closure of this Hilbert space, with the above inner product and conjugation $\tilde \Gamma=\Gamma$ is in fact isometric to $(K, \Gamma)$ as defined before under restriction to $\C$, i.e. under $[k] \in \tilde K \mapsto k |_\C \in K$. Consequently, the CAR algebra $\gC(K, \Gamma)$ is isomorphic to $\gC(\tilde K, \tilde \Gamma)$. The local algebras for an arbitrary region $O$ are then defined as
\ben
\A(O) = \{ \psi([k]) \mid k \in \tilde K, {\rm supp}(k) \subset O\}^{\rm even} \ ,
\een
where ``even'' refers to the subalgebra of even elements w.r.t. the ${\mathbb Z}_2$-grading of the CAR algebra.

We now describe the construction of the ground state representation on an ultra-static spacetime $\sM = \RR \times \C, g=-\dd t^2 + h$ with compact Cauchy surface $\C$, assumed to be spin. Let us assume that the orthonormal frame is chosen in such a manner that $e_0$ is the time-like normal to $\C$. We may write the Dirac equation $(\slash \nabla + m) \psi=0$ as
\ben\label{eq:h1p}
i\partial_t \psi = \sum_{j=1}^d (ie_0 \cdot e_j \cdot \nabla_{e_j} + ie_0 m) \psi \equiv h \psi \ .
\een
The right side defines a ``1-particle Hamiltonian''. By standard theorems, it has a discrete spectrum of eigen-spinors. It is checked that $\Gamma^{-1} h \Gamma = -h$. So, $\Gamma$
exchanges eigen-spinors with positive and negative eigenvalues, and the spectrum is symmetric about $0$.
Around $0$, there is a gap including at least $(-m,m)$.  We let $P$ be the projector in $K$ onto the subspace of eigenspinors with positive eigenvalues, which as a consequence
satisfies $\Gamma P \Gamma = 1-P$. Thus, we can define, as in our general discussion of the CAR algebra in sec.~\ref{ssec_CAR}, a representation of the
CAR algebra $\gC(K, \Gamma)$ associated with $P$ on a fermionic Fock space. The 1-particle Hilbert space is $\H_1 = PK$. On the subspace $\H_n = E_n \H_1^{\otimes n}$ of ``$n$-particle'' states in fermonic Fock space, the Hamiltonian of the QFT is then given by
\ben
H |\Psi_n\rangle = \sum_{i=1}^n (1 \otimes \dots \underbrace{h}_{i{\rm -th \ slot}} \otimes \dots 1) |\Psi_n\rangle \ , \qquad |\Psi_n \rangle \in \H_n \ .
\een
The vacuum state satisfies $H|0\rangle = 0$, so it is a ground state.
The fermionic field operator is represented by \eqref{eq:fieldCAR}.

Rather than unraveling these abstract definitions, we consider as a simple example the Majorana field in $1+1$ dimensional Minkowski spacetime $\RR^{1,1}$.
In this example, $K=L^2(\RR, \dd x; \CC^2)$, eq.~\eqref{innerprod} becomes the standard inner product on this space, and $\Gamma$ becomes component-wise complex conjugation (using a suitable representation of the Clifford algebra). The 1-particle Hamiltonian becomes in this representation
\ben
h= \left(
\begin{matrix}
p & im \\
-im & -p
\end{matrix}
\right)
\een
with $p=i\dd/\dd x$. The projector $P$ onto the positive part of the spectrum can be worked out explicitly by diagonalizing this matrix.
Once this has been done, it is convenient to identify the 1-particle Hilbert space $PK$ with $L^2(\RR, \dd \theta)$ via the isometry
\ben
V: PL^2(\RR, \dd x; \CC^2) \to L^2(\RR, \dd \theta) =: \H_1
\een
defined by
\ben\label{eq:identification}
V
\left(
\begin{matrix}
k_1\\
k_2
\end{matrix}
\right)
=
\frac{1}{\sqrt{2}} [e^{\theta/2 - i\pi/4} \widetilde k_1(m \sinh \theta) + e^{-\theta/2+ i\pi/4} \widetilde k_2(m \sinh \theta)] \ ,
\een
where a tilde means Fourier transform. The Dirac field operator in the representation $\pi_P$ (defined abstractly by \eqref{eq:fieldCAR}) becomes under this identification, and
under the identification of $K$ and $\tilde K$ described in sec.~\ref{ssec_CAR},
the 2-component operator valued  distribution
\ben\label{eq:psiCAR}
\psi(x) = \frac{1}{\sqrt{4\pi}} \int_\RR \dd \theta \left\{
\left(
\begin{matrix}
e^{\theta/2 - i\pi/4}\\
e^{-\theta/2 + i\pi/4}
\end{matrix}
\right) e^{-ip(\theta) x} a^\dagger(\theta) +
\left(
\begin{matrix}
e^{\theta/2 + i\pi/4}\\
e^{-\theta/2 - i\pi/4}
\end{matrix}
\right) e^{ip(\theta) x} a(\theta)
\right\} \ ,
\een
where $x\equiv (x^0,x^1)$, where $(p_\mu(\theta)) = (-m \cosh \theta, m \sinh \theta)$, and where $a(\theta), a^\dagger(\theta)$ satisfy the relations~\eqref{eq:ZF} below with $S_2=-1$.
A wave function $|\Phi_1\rangle \in \H_1$ is created from the vacuum by applying a smeared creation operator, $|\Phi_1\rangle = \int \dd \theta \phi(\theta) a^\dagger(\theta) |0\rangle
\equiv a(\phi)^* |0\rangle$, with $\phi \in L^2(\RR, \dd \theta)$, compare sec.~\ref{ssec_CAR}.

\subsubsection{Integrable models in $1+1$ dimension with factorizing $S$-matrix}\label{Intmodels}

Following an idea of~\cite{schroer_1}, it has been shown in~\cite{lechner_1,lechner_3,buchholz_6}, how to construct a wide class of integrable models in 1+1 dimensional Minkowski space
satisfying the above axioms a1)-a4) and a nuclearity condition [a5) and a5')]. The only input\footnote{Apart from the value $m$ of the mass of the basic particle.} in this construction is a 2-body scattering-matrix. We will not discuss here how in general the concept of a scattering-matrix fits into the algebraic framework, see~\cite{haag_2} for a discussion and references. For the sake of the construction, it is enough to think of the 2-body scattering-matrix as merely a function $S_2$ which is a datum entering the construction of a net $\A(O)$. The properties required of $S_2$ to make this construction work are:

\begin{enumerate}
\item[(s1)] $S_2(\theta)$ is a bounded analytic function on the strip $\{ \theta \in \CC \mid -\epsilon < \Im \theta < \pi + \epsilon\}$ where $0<\epsilon<\pi/2$.
\item[(s2)] $|S_2(\theta)| = 1$ for $\theta \in \RR$, and $S_2(0) = -1$ (it is therefore ``fermionic'' in the terminology of~\cite{lechner_1}).
\item[(s3)] For $\theta$ in $\RR$, $S_2(-\theta) = S_2(\theta)^{-1}$.
\item[(s4)]  For $\theta$ in $\RR$, $S_2(\theta + i\pi) = S_2(\theta)^{-1}$.
\end{enumerate}

s1) corresponds to analyticity, s2) to unitarity, s3) to PCT-invariance, and s4) to crossing symmetry for the full scattering-matrix, if $k = (m \cosh \theta, m \sinh \theta)$ is identified with the incoming momentum of an on-shell particle of mass $m$ in a 2-body collision. For instance, in the Sinh-Gordon model~\cite{abdallah}
\ben
S_2^{\rm ShG}(\theta) = \frac{\sinh \theta - i\sin b}{\sinh \theta + i\sin b} \ , \quad b = \frac{\pi g^2}{1 + g^2}
\een
with $g$ the coupling constant of the Sinh-Gordon potential. $S_2^{\rm ShG}$ satisfies s1)-s4) as long as $0<b<\pi$, as would for instance any product of an odd number of
such factors with different $0<b_i<\pi$.

The construction of the net $O \mapsto \A(O)$ corresponding to a given $S_2$ starts by considering an ``$S_2$-symmetric'' Fock-space over $\H_1=L^2(\RR, \dd\theta)$. This Fock space is a direct sum $\CC \oplus_{n \ge 1} \H_n$ of $n$-particle spaces. By contrast to the case of the bosonic Fock-space, $\H_n$ is not obtained by applying a symmetrization projection to $\H_1^{\otimes n}$. Rather, one applies a projection $E_n$ based on $S_2$. For that, let $\tau_i$ be an elementary transposition of the elements $i$ with $i+1$ in the symmetric group $\mathfrak{S}_n$ on $n$ elements. Define an exchange operator $D_n(\tau_i)$ on $\H_1^{\otimes n}$, identified with unsymmetrized $L^2$-wave functions $\Psi_n$ in $n$-variables, as
\ben
(D_n(\tau_i) \Psi_n)(\theta_1, \dots, \theta_i, \theta_{i+1}, \dots, \theta_n) = S_2(\theta_{i+1} - \theta_{i}) \Psi_n(\theta_1, \dots, \theta_{i+1}, \theta_i, \dots, \theta_n) \ .
\een
It can be shown using s2)-s3) that this exchange operator gives a unitary representation\footnote{I.e., it satisfies the relations of the permutation group.} of $\mathfrak{S}_n$ on $\H_1^{\otimes n}$. Define an $S_2$-symmetric projection $E_n = (1/n!) \sum_{\sigma \in \mathfrak{S}_n} D_n(\sigma)$, define $\H_n = E_n \H_1^{\otimes n}$ ($S_2$-symmetric wave functions), define $\H = \CC \oplus_{n \ge 1} \H_n$ and define creation and annihilation operators $z^\dagger(\Psi), z(\Psi), \Psi \in \H_1$ on $\H$ by analogy with eq.~\eqref{adef}. We write informally\footnote{Informally, $z^\dagger(\theta) = z(\theta)^*$.} $z^\dagger(\Psi) = \int d\theta \Psi(\theta) z^\dagger(\theta)$. These operators satisfy relations called the ``Zamolodchikov-Faddeev (ZF) algebra'', which is
\ben\label{eq:ZF}
z(\theta) z^\dagger(\theta') - S_2(\theta-\theta') z^\dagger(\theta') z(\theta) = \delta(\theta-\theta') \cdot 1 \ , \quad
z(\theta) z(\theta') - S_2(\theta'-\theta) z(\theta') z(\theta) = 0 \ .
\een
Following Schroer and Wiesbrock~\cite{schroer_1}, one then defines a ``field operator'' by putting, with $px = p_\mu x^\mu$ and
$(p_\mu(\theta)) = (-m \cosh \theta, m \sinh \theta)$,
\ben\label{phidef1}
\phi(x) = \frac{1}{\sqrt{4\pi}}\int_\RR \dd\theta[z^\dagger(\theta) e^{-ip(\theta)x} + z(\theta) e^{+ip(\theta)x}] \ .
\een
This field satisfies the KG equation with mass $m$. Its linear structure in creation- and annihilation operators is analogous to that of a KG quantum field in 1+1 dimensions in the vacuum representation (see~\eqref{phidef}), when we identify $z^{\dagger}(\theta)=\frac{1}{\sqrt{2}}a^{\dagger}(k_1(\theta))$. Actually, for $S_2=1$, $\phi(x)$ is exactly equal to the free KG field because the ZF generators then satisfy the standard relations of creation and annihilation operators. In this special case, the field $\phi(x)$ defined by \eqref{phidef1} satisfies space-like commutativity. But in general, it does not have this property. It may, however, be used to define a local quantum field theory in the sense of axioms a1)-a4) by a roundabout route. First, define the following anti-linear operator $J_n$ on $\H_1^{\otimes n}$:
\ben
(J_n\Psi_n)(\theta_1, \dots, \theta_n) = \overline{\Psi_n(\theta_n, \dots, \theta_1)} \ .
\een
The properties of $S_2$ imply that this consistently defines, in fact, an anti-unitary, involutive, operator on the $S_2$-symmetric Fock space $\H$ (it commutes with $E_n$). Call this operator\footnote{$J$ turns out to be equal to the modular conjugation associated with the algebra $\R$ defined below.} $J$, and define $z'(\theta) = J z(\theta) J, z'^\dagger(\theta) = J z^\dagger(\theta) J$. These operators satisfy relations that are identical to the ZF-algebra, except that $S_2(\theta-\theta')$ is replaced by $S_2(\theta'-\theta)$. Define a field $\phi'(x)$ by substituting into \eqref{phidef1} the primed ZF creation and annihilation operators. The key observation~\cite{schroer_1}, which follows from s2)-s4), is:

\begin{lemma}
Assume $f,f'$ are test functions on $\RR^2$ such that the support of $f'$ is space-like and to the left of that of $f$. Then $[\phi(f), \phi'(f')] = 0$ (on a core of vectors in $\H$).
\end{lemma}

One also shows using Nelson's analytic vector theorem~\cite{reed} (just as in the case of the free KG-field) that the operators $\phi(f), \phi'(f')$ are (closable and) essentially self-adjoint. Then their exponentials are well-defined, and the lemma holds also for them, i.e. $[e^{i\phi(f)}, e^{i\phi'(f')}] = 0$. Let $W=\{(t,x) \in \RR^2 \mid x>|t|\}$ be the right wedge in $\RR^2$. Based on the ``half-sided'' locality expressed by the lemma, it is natural to define the wedge algebras
\ben
\R = \{ e^{i\phi(f)} \mid {\rm supp}(f) \subset W\}'', \quad \R' = \{ e^{i\phi'(f')} \mid {\rm supp}(f') \subset W' \}'' \ ,
\een
where $W'$ is the opposite wedge and the double prime is the v. Neumann closure. As the notation suggests, the commutation relations of the lemma not only imply that $\R'$ commutes with $\R$, but even that $\R'$ is in fact the commutant, i.e.~the set of all such operators. To get algebras associated with bounded double cones, one would like to form intersections of appropriately ``translated'' wedge algebras. For this, we first need a representation of translations (and Lorentz-boosts) on $\H$. These are defined by the  unitaries ($a \in \RR^2, \lambda \in \RR$)
\ben\label{eq:Uboost}
(U_n(\lambda,a) \Psi_n)(\theta_1, \dots, \theta_n) = \exp\left(-i\sum_j p(\theta_j)a \right) \Psi_n(\theta_1-\lambda, \dots, \theta_n-\lambda) \ ,
\een
where $\lambda$ is interpreted as the boost-parameter and $a$ as the translation vector of an element $g=(\lambda,a)$ of the 2-dimensional Poincar\'e-group. The properties of $S_2$ once again imply that this consistently defines a strongly-continuous, positive energy representation $U$ of $\rm P$ on $\H$. If $W_a$ denotes the translation of $W$ by the vector $a \in \RR^2$, then any double cone $O$ can be written as an intersection of opposite, translated wedges, $O = W_a^{} \cap W'_b$, for suitable $a,b \in \RR^2$.
Thus, it is natural to define:
\begin{definition}
Let $O = W_a^{} \cap W'_b$, for suitable $a,b \in \RR^2$. Then we define a net by
\ben
O \mapsto \A(O) \equiv U(a) \R U(a)^* \cap U(b) \R' U(b)^* \ .
\een
\end{definition}
It is then straightforward to see that this net $O \mapsto \A(O)$ satisfies axioms a1)-a5) (for v. Neumann algebras), where the vacuum state is just that in the $S_2$-symmetric Fock space. It is not so clear, however, that this net is non-trivial, but this has been established for sufficiently large regions\footnote{Note that~\cite{lechner_1} contained
an error, which has been amended in~\cite{lechner_4}. At present, the arguments only establish a1)-a4),a5') for
regions $O$ of a minimal size, contrary to the claim of~\cite{lechner_1}.} $O$
in~\cite{lechner_1, lechner_4} using a combination of techniques such as modular nuclearity bounds, analyticity methods, and making heavy use of the properties of $S_2$. The methods also show that each $\A(O)$ is of type III$_1$, and that the vacuum vector $|0\rangle$ in our Fock space $\H$ is cyclic and separating.

The surprisingly simple, in principle, construction of a wide class of integrable models just outlined has one serious caveat, though. Unlike for the case of the free KG-field, the operator $\phi(x)$ (and likewise $\phi'(x)$) is {\em not} a local operator. In particular, for instance in the Sinh-Gordon model, $\phi(x)$ must not be interpreted as a quantum version of the classical field appearing in the corresponding Lagrangian~\eqref{shG}, because that would be expected to be local. Local operators at a point $x$ can be characterized in principle as being, in a certain sense~\cite{fredenhagen_1,bostelmann}, elements in  $\cap_{O \owns x} \A(O)$. But in practice, they would in all likelihood have an extremely complicated expression in terms of the non-local operators $\phi$ and $\phi'$, and in this sense one can only say that the model has only been constructed in a very indirect way. Fortunately, for our purposes, it will not be important at all to precisely identify such local operators. Rather, all we need is the information given by the net $\{\A(O)\}$, which has a straightforward and simple definition.

\subsubsection{Chiral CFTs}

Chiral conformal field theories (CFTs) describe ``one chiral half'' of a conformal field theory in $1+1$ dimensions, and are particularly well-investigated.
They are described in the algebraic setting by nets $I \mapsto \A(I)$ parameterized by open intervals of the circle $I \subset S^1$. The axioms are essentially the same as in Minkowski space, with a few fairly evident changes. In the commutativity axiom a2), one replaces the notion of causal complement simply by disjoint intervals, i.e. $[\A(I_1), \A(I_2)]=\{0\}$ if $I_1$ and $I_2$ are disjoint. The covariance axiom now involves the M\" obius group ${\rm G} = {\rm SU}(1,1)/\{1,-1\} = {\rm PSU}(1,1)$, rather than the Poincar\' e group. An element $g \in {\rm PSU}(1,1)$ acts on $z \in S^1$ by
\ben
g\cdot z = \frac{\alpha z+\beta}{\bar \beta z+\bar \alpha} \ , \qquad g=
\left(
\begin{matrix}{}
\alpha & \beta \\
\bar \beta& \bar \alpha
\end{matrix}
\right)  \quad \text{where $|\alpha|^2 - |\beta|^2 = 1$} \ .
\een
In the transcription of the covariance and vacuum requirements a3) and a4), one  requires a  unitary, positive energy representation $U$ of the group ${\rm G}$ (or its
cover $\widetilde{\rm G}$) replacing the group of ``spacetime'' symmetries. A net $\{ \A(I)\}$ over $S^1$ with these properties is called a ``chiral net''.~\footnote{One sometimes requires that the symmetry algebra of the net is the full Virasoro algebra, i.e. that the net contains the algebra of quantized diffeomorphisms as a subnet. Then the split property is automatic~\cite{weiner}.}

Under the Cayley transform $z \mapsto i(z-1)/(z+1)$, the circle gets mapped to the 1-point compactification of $\RR$ and the action of the M\"obius group then corresponds to the action of ${\rm G} \cong {\rm PSL}(2,\RR)$ by the transformations $\RR \owns x \mapsto g \cdot x = \frac{ax + b}{cx +d}$ with real coefficients such that $ad-bc=1$. This action is just the action of the conformal group ${\rm SO}_+(2,2) \cong {\rm G} \times {\rm G}$ on light rays of 2-dimensional Minkowski space. A corresponding 2-dimensional conformal net in 1+1 dimensions can therefore be defined simply as follows. Viewing  2-dimensional Minkowski spacetime as the Cartesian product of two light rays, any double cone $O$ is the Cartesian product of two open intervals $I_L,I_R \subset \RR$. We then simply set $\A_{2d}(O) = \A(I_L) \otimes \A(I_R)$ for $O=I_L\times I_R$. Such a  net on $\RR^2$ is thus the tensor product of chiral theories. It again satisfies analogous versions of the axioms a1)-a4), with the group of spacetime symmetries now replaced by conformal transformations or its cover
$\widetilde{{\rm SO}_+(2,2)}$. (More details can be found e.g. in \cite{kawahigashi_3,rehren_2}.)

\medskip

There are very many examples for chiral nets, for a detailed discussion in the case of a central charge $c<1$ see e.g.~\cite{kawahigashi_1}. Here we only give one such example purely for illustrative purposes. It is an operator algebraic version of the so-called ``minimal models,'' see e.g.~\cite{difrancesco} for the conventional description. A local algebra $\A(I)$ in such a net basically describes ``quantized diffeomorphisms'' on $S^1$ acting non-trivially only in the interval $I \subset S^1$. The construction is more precisely as follows, see e.g.~\cite{fewster_3} for details and references.

Any diffeomorphism $\tilde f$ of $\RR$ satisfying $\tilde f(\theta + 2\pi) = \tilde f(\theta) + 2\pi$ defines an orientation preserving diffeomorphism $f$ of $S^1$ via the formula $f(e^{i\theta}) = e^{i\tilde f(\theta)}$. Call the group of these ${\rm Diff}_+(S^1)$. The ``Bott cocycle'' is the map $B: {\rm Diff}_+(S^1) \times {\rm Diff}_+(S^1) \to \RR$ defined by
\ben
B(f_1,f_2) = -\frac{1}{48\pi} \int_{S^1} \log((f_1 \circ f_2)'(z)) \frac{\dd}{\dd z} \log (f_2'(z)) \, \dd z .
\een
It can be shown that this lifts to a cocycle of the universal covering group $\widetilde{{\rm Diff}_+}(S^1)$, which is shown to be a Fr\'echet Lie-group. A unitary representation $U$ of this group on some Hilbert space $\H$ is called a multiplier representation with central charge $c \in \RR$ if it is strongly continuous with respect to the group topology and
\ben\label{Bott}
U(f_1) U(f_2) = e^{icB(f_1, f_2)} U(f_1 \circ f_2) \ .
\een
The universal covering $\widetilde{\rm G}$ of the M\" obius group is a subgroup of $\widetilde{{\rm Diff}_+}(S^1)$. For $f_1, f_2 \in \widetilde{\rm G}$, the Bott-cocycle vanishes, so $U$ restricts to a bona-fide, ordinary, unitary representation of $\widetilde{\rm G}$. For given $c$, the irreducible multiplier representations can be classified. They correspond to exponentiated versions of the ``highest weight representations'' of the Virasoro-algebra, which is an infinitesimal version of the relation~\eqref{Bott}, the Bott cocycle $B$ corresponding to the central term in the Virasoro-algebra. Such representations only exist for $c>0$, and for $c<1$, the central charge must be quantized according to the rule $c=1-\frac{6(p-p')^2}{pp'}, p,p' \in {\mathbb N}$. The vacuum representation $U_0$ corresponds to the highest weight representation in which the generator $L_0^{\rm vac}$ of rotations of $S^1$ satisfies $L_0^{\rm vac} |0\rangle = 0$, where $|0\rangle \in \H_0$ is the highest weight vector (vacuum). The net $I \subset S^1 \mapsto \A(I)$ is defined by
\ben
\A(I) = \{ U_0(f) \mid \text{$f \in \widetilde{{\rm Diff}_+}(S^1)$ such that $f(z) = z$ for $z \notin I$}\}'' \ .
\een
Using properties of multiplier representations~\cite{toledano}, one shows that this definition satisfies the analogues of axioms a1)-a4) for chiral CFTs.

\section{Entanglement measures in QFT}

In this section we discuss entanglement in a general setting and we review some quantitative measures of entanglement and their properties. (See~\cite{horodecki_2,vidal,vedral, vedral_2}  for more details.)

\subsection{Entanglement}

Let us begin by introducing the basic notion of entanglement. We consider a system, described by a $C^*$ (or v. Neumann)-algebra $\A$, and two subsystems
described by subalgebras $\A_A,\A_B\subset\A$. Furthermore, we let $\omega$ be a state on $\A$ and we wish to characterize and quantify the entanglement in the state $\omega$ between the algebras $\A_A$ and $\A_B$.

Using the GNS-representation $\pi_{\omega}$ on $\H_{\omega}$, we may replace the $C^*$-algebras by the corresponding v. Neumann algebras, which we will again denote by $\A$, $\A_A$ and $\A_B$, respectively. (Note, however, that these algebras may depend in general on the choice of $\omega$.) We will assume that $\A_A$ and $\A_B$ commute, that $\A_A\cap\A_B=\CC 1$, and that $\A_A$ and $\A_B$ are statistically independent, i.e. $\A_A\vee \A_B\simeq \A_A\otimes\A_B$, where the latter algebra acts on $\H_{\omega}\otimes\H_{\omega}$. Since $\omega$ restricts to a normal state on $\A_A\vee \A_B$, we can view this restriction also as a normal state on $\A_A\otimes\A_B$, which is described by a density matrix $\rho$ on $\H_{\omega}\otimes\mathcal{H}_{\omega}$.

We can now introduce the distinction between ``separable states'' and ``entangled states'':
\begin{definition}
A normal state $\omega$ on the tensor product $\A_A \otimes \A_B$ of two v. Neumann algebras is said to be ``separable'' if it can be written as a norm convergent sum $\omega = \sum_j \varphi_j \otimes \psi_j$ for positive normal functionals $\varphi_j, \psi_j$ on $\A_A$ respectively $\A_B$, i.e. $\omega(ab)=\sum_j\varphi_j(a)\psi_j(b)$. A normal state which is not separable is called ``entangled''.
\end{definition}
When $\A_A\otimes\A_B$ is finite dimensional, the set of separable states is norm-closed\footnote{The general case is unclear to us, but one could modify the definition to make the set of separable states norm-closed.}.

The simplest example of a separable state is $\omega=\omega_A\otimes\ \omega_B$ for two vector states
$\omega_A(a)=\langle\Phi|a\Phi\rangle$ and $\omega_A(b)=\langle\Psi|b\Psi\rangle$. By definition, $\omega(ab)=\omega_A(a)\omega_B(b)$, which is a vector state determined by the simple tensor product vector $|\Phi \rangle \otimes |\Psi \rangle$. Alternatively, we can write the states $\omega_A$ and $\omega_B$ in terms of the density matrices $\rho_A$ and $\rho_B$, which are simply orthogonal projections onto $|\Phi \rangle$ and $|\Psi \rangle$, respectively. The state $\omega$ is then determined by the density matrix $\rho=\rho_A\otimes\rho_B$. We remind the reader that in our general setting, vector states need not be pure and density matrices need not be uniquely determined by the state $\omega$.

A general separable state is always a convex combination of such separable vector states. Indeed, for a general separable state $\omega$, decomposed in terms of $\varphi_j$ and $\psi_j$, we have $\varphi_j(a)=\tr(\rho_{A,j}a)$ and $\psi_j(b)=\tr(\rho_{B,j}b)$ for suitable positive trace-class operators $\rho_{A,j},\rho_{B,j}$. We can then write the density matrix as $\rho=\sum_j\rho_{A,j}\otimes\rho_{B,j}$. By diagonalising the operators $\rho_{A,j}$ and $\rho_{B,j}$, decomposing them into one-dimensional projectors, and relabelling indices we can always write $\rho$ as a sum of one-dimensional projectors which project onto vectors of simple tensor product form. Hence, $\omega$ is a convex combination of separable vector states.

Note that a general unit vector $\Theta\in\H_A\otimes\H_B$ can be written as a sum of simple tensor products, $\Theta=\sum_j|\Phi_j \rangle \otimes |\Psi_j \rangle$, but the corresponding positive linear functional is $\langle\Theta,ab\Theta\rangle=\sum_{j,k}\langle \Phi_j|a\Phi_k\rangle\ \langle\Psi_j|b\Psi_k\rangle$, which need not be separable. Analogously, the density matrix $\rho$ of any normal state $\omega$ on $\A_A\otimes\A_B$ can be diagonalised and then decomposed as $\rho=\sum_jx_j\otimes y_j$ with bounded operators $x_j,y_j$, but in general these operators cannot be chosen positive (although they can be chosen self-adjoint). The obstruction is what characterizes entanglement. Another reformulation of this obstruction is that any normal state $\omega$ on $\A_A\otimes\A_B$ can be written as $\omega=\sum_j\varphi_j\otimes\psi_j$ with hermitean normal functionals $\varphi_j,\psi_j$ which may not be positive.

When $\A_A$ and $\A_B$ are in standard form with vectors $|\Omega_A \rangle $ and $|\Omega_B \rangle$, then $\A_A\otimes\A_B$ is also in standard form with vector $|\Omega \rangle :=|\Omega_A \rangle \otimes |\Omega_B \rangle$. Recall that every normal state $\omega$ has a unique vector representative in the natural cone $\mathcal{P}^\sharp \subset \H=\H_A\otimes\H_B$. If $\mathcal{P}^{\sharp}_A,\mathcal{P}^{\sharp}_B$ denotes the natural cone in $\H_A$, $\H_B$ respectively, and if $|\Psi_A \rangle \in\mathcal{P}^{\sharp}_A$ and $|\Psi_B \rangle \in\mathcal{P}^{\sharp}_B$ are unit vectors, then $|\Psi_A \rangle \otimes |\Psi_B \rangle \in\mathcal{P}^{\sharp}$ is a separable state. All pure separable states are necessarily of this form, and using the properties of the natural cones one may show that the norm limit of separable states of the form $\psi\otimes\phi$ is again separable. Note, however, that it is not so easy to recognize when a vector in $\mathcal{P}^{\sharp}$ defines a mixed separable state, because the separable states do not form a cone inside $\mathcal{P}^{\sharp}$.

Depending on the state $\omega$, the outcomes of separate measurements on the two systems $\A_A$ and $\A_B$ can exhibit different kinds of correlations. When $\omega=\omega_A\otimes\omega_B$, there are no correlations at all. For a general separable state, however, there can be correlations, which are of a classical nature. Entangled states exhibit even more general ``quantum'' correlations. For this reason, entanglement has come to be viewed as an experimental resource, which can be enhanced or ``purified'', and subsequently exploited to perform quantum computations, teleportations or other often counter-intuitive experiments.

\subsection{Properties of entanglement measures}\label{sec:properties_measures}

Let us now turn to the question how to quantify the amount of entanglement in a general normal state $\omega$ on $\A_A\otimes\A_B$. We will start in this section by reviewing a number of desirable properties that an entanglement measure $E(\omega)$ could satisfy. In the remainder of this chapter we will then introduce specific examples and discuss the properties that they have.

We start with the following basic properties:
\begin{enumerate}
\item[(e0)] (symmetry) $E(\omega)$ is independent of the order of the systems $A$ and $B$.
\item[(e1)] (non-negative) $E(\omega)\in[0,\infty]$, with $E(\omega)=0$ if and only if $\omega$ is separable, and $E(\omega)=\infty$ when $\omega$ is not a normal state on $\A_A\otimes\A_B$ (e.g. when $\A_A$ and $\A_B$ are not statistically independent).
\item[(e2)] (continuity) Let ${\mathfrak N}_{A 1} \subset {\mathfrak N}_{A 2} \dots \subset {\mathfrak N}_{A i} \dots \subset \A_A$ be an increasing net of type I factors isomorphic to
matrix algebras ${\mathfrak N}_i \cong M_{n_i}(\CC)$, and similarly for $B$. Let $\omega_i, \omega_i'$ be normal states on $\mathfrak{N}_{A i} \otimes \mathfrak{N}_{B i}$  such that
$\lim_{i \to \infty} \| \omega_i' - \omega_i^{} \| = 0$. Then
\ben
\lim_{i \to \infty} \frac{E(\omega_i') - E(\omega_i^{})}{\log n_i} = 0 \ .
\een
\item[(e3)] (convexity) If $\omega = \sum_j \lambda_j \omega_j$ is a convex combination of states $\omega_j$ (with $\lambda_j\ge0, \sum_j\lambda_j =1$), then
\ben
E(\omega) \le \sum_j \lambda_j E(\omega_j),
\een
i.e. $\omega \mapsto E(\omega)$ is convex.
\end{enumerate}
Property (e3) states that entanglement cannot be increased by mixing states. It can be reduced, however: for two independent spin-$\frac12$ systems, one can choose a Bell-basis of four vectors in the tensor product Hilbert space. These vectors define pure, (maximally) entangled states, but an equal mixture of these four states yields the density matrix $\rho
\propto 1=1_A\otimes 1_B$, which is separable.

The next property is based on the idea that certain experimental manipulations cannot increase the amount of entanglement, because they can only introduce classical correlations between measurement results. Before we can formulate this property, we will first review the allowed experimental operations, for which we will use the following terminology:
\begin{definition}
A  linear map $\F: \A_2 \to \A_1$ between two $C^*$-algebras is called {\bf positive} (p) if $\F(a)$ is a positive operator whenever $a$ is. $\F$ is called {\bf completely positive} (cp) if $1_{M_N(\CC)} \otimes \F$ is positive as a map\footnote{Here the tensor product $M_N(\CC) \otimes \A_2$ is algebraic, with no completion required. To obtain the (unique) $C^*$-norm, one may use the fact that there exists a universal representation $\pi_u:\A_2\to\B(\H_u)$, which is faithful and hence isometric. One may then represent $M_N(\CC) \otimes \A_2$ on $\CC^N\otimes\H_u$ and use the operator norm.}
 $M_N(\CC) \otimes \A_2 \to M_N(\CC) \otimes \A_1$ for all $N$. A (completely) positive map is called {\bf normalized} if $\F(1) = 1$. A normalized positive map $\F$ between v. Neumann algebras is called {\bf normal} when $\F^*$ maps normal states to normal states.
\end{definition}

A normalized positive map $\F$ gives rise to a map $\F^*$ from states on $\A_1$ to states on $\A_2$, defined by $(\F^*\omega)(a):=\omega(\F(a))$. (This point of view explains the order of $\A_1$ and $\A_2$ in the definition above.) Conversely, any map $\F^*$ from states on $\A_1$ to states on $\A_2$ arises from a normalized positive linear map in this way. Complete positivity is motivated by the desire to be able to apply the same experimental manipulations independently to $N$ copies of the same system. It is the mathematical characterization of a ``quantum channel'' in the sense of Quantum Information Theory.

In addition to a quantum channel, one could perform measurements and post-select a sub-ensemble according to the results. For a v. Neumann measurement, given by projections $P_k\in\A$ with $\sum_kP_k=1$, we note that the maps $\F_k:\A\to\A$ defined by $a\mapsto P_kaP_k$ are cp, with $0\le\F_k(1)=P_k\le1$. Performing the measurement on a state $\omega$ we obtain the new state
\[
\omega_k:=\frac{\F_k^*\omega}{\omega(P_k)}
\]
with probability $\omega(P_k)$, when $\omega(P_k)>0$. A combination of quantum channels and measurements is called an ``operation''~\cite{Lindblad}. It is described by a family $\F_k:\A_2\to\A_1$ of cp maps with $\sum_k\F_k(1)=1$, which transform a state $\omega$ on $\A_1$ into $\omega_k:=\frac{1}{p_k}\F_k^*\omega$ with probability $p_k:=\omega(\F_k(1))$ when $p_k>0$.

\begin{example}
Let us give some examples of p and cp maps.
\begin{enumerate}[label=(\roman*)]
\item Any (unit preserving) $*$-homomorphism between $C^*$-algebras (and in particular every representation) is a (normalized) cp map. Furthermore, any state of a $C^*$-algebra is a normalized cp map.
\item If $V:\H\to\mathcal{K}$ is a bounded linear map between Hilbert spaces, then $\F:\B(\mathcal{K})\to\B(\H)$ defined by $\F(a):=V^*aV$ is a cp map. It is normalized if and only if $V$ is an isometry.
\item Let $\H$ be a Hilbert space carrying a continuous unitary representation of a finite dimensional compact Lie group $\rm K$. Denote the unitaries representing $g \in {\rm K}$ by $U(g)$, and let $\F: \B(\H) \to \B(\H)$ be the map
\ben\label{Kmean}
\F(a) = \int_{\rm K} {\rm d} g \ U(g) a U(g)^*
\een
where ${\rm d} g$ is the normalized Haar measure. Then $\F$ is normalized and completely positive, because $1 \otimes \F$ can be written in the same form with $U(g)$ replaced by the representation $1 \otimes U(g)$.
\item When $\A_1$ and $\A_2$ are v. Neumann algebras, then $\F:\A_1\to\A_1\otimes\A_2$ defined by $\F(a):=a\otimes 1_2$ defines a normalized cp map. It corresponds to the restriction of states from $\A_1\otimes \A_2$ to $\A_1$.
\item Similarly, given a state $\omega_2$ on $\A_2$ the map $\F:\A_1\otimes\A_2\to\A_1$ defined by $\F(a\otimes b):=a\omega_2(b)$ is a normalized cp map. The corresponding map on states sends $\omega_1$ to $\omega_1\otimes\omega_2$, which corresponds to attaching an ancillary system $\A_2$ in the state $\omega_2$. This map is a right-inverse to the restriction map above.
\item For each $N>1$ the map $\F: \B(\CC^N) \to \B(\CC^N)$ defined by $\F(a) = (\tr\ a) 1_N - a$ is positive. It is normalized only for $N=2$. Interestingly, it is {\em not} completely positive.
\item Let $\psi_j, j=1, \dots, d$ be operators on a Hilbert space $\H$ satisfying the relations of the Cuntz algebra, see sec.~\ref{ssec_Cuntz}. Then $\rho: \B(\H) \to \B(\H)$ given by $\rho(a) = \sum_j \psi_j a \psi_j^*$ is a normalized cp map. It plays a role in the theory of superselection sectors.
\end{enumerate}
\end{example}

A general result due to Stinespring~\cite{stinespring} shows that all completely positive maps $\F:\A\to\B(\H)$ can be written as $\F(a)=V^*\pi(a)V$, where $\pi$ is a representation of $\A$ on some Hilbert space $\mathcal{K}$ and $V:\H\to\mathcal{K}$ is bounded. When $\F$ is normalized, one can choose an isometry $V$. When $\A$ already acts on $\H$ and $\pi(a)=\oplus_ja$ is a (finite or countable) direct sum representation on $\H^{\oplus N}$ one recovers a formulation in terms of Kraus operators:
\ben
\label{kraus}
\F(a)=\sum_j V_j^*aV_j^{},\qquad \sum_jV_j^*V_j^{}=1.
\een
It follows from standard properties of finite type I factors that in this case, all cp maps arise in this way \cite{NielsenChuang}, but this is no longer true for general type, in particular type III.

Returning to properties for entanglement measures, we now consider cp maps $\F: \A_{\hat A}\otimes\A_{\hat B} \to \A_A \otimes \A_B$. We call such a map ``{\bf local}'' if it is of the form
\ben\label{sepop}
\F(a \otimes b) = \F_{A}(a) \otimes \F_{B}(b) \equiv (\F_A \otimes \F_B)(a \otimes b) \ ,
\een
where the $\F_{A}$ and $\F_{B}$ are normal cp maps. More generally, we make the following key definition:

\begin{definition}
A ``{\bf separable operation}'' is by definition a family of normal, local cp maps $\F_j$, which are each
of tensor product form~\eqref{sepop}, satisfying additionally $\sum_j\F_j(1)=1$.
We think of such an operation as mapping a state $\omega$ with probability $p_j:=\omega((\F_{A,j}\otimes\F_{B,j})(1))$
to $\frac{1}{p_{j}}(\F_{A,j}\otimes\F_{B,j})^*\omega$.
\end{definition}

It is clear that separable operations map separable states to separable states. In the literature on Quantum Information Theory, it is argued that an arbitrary combination of local operations and ``classical communication'' (LOCC) between systems $A$ and $B$ is modeled by a separable operation.
However, not all separable operations are actually  LOCC operations~\cite{horodecki_2}. The notion of an LOCC operation is closer to what actually seems experimentally feasible, and thus conceptually superior.  But that notion is also more complicated, and besides, even many LOCC operations may not  be experimentally feasible~\cite{gidke_2}. Moreover, the assumption that $A$ and $B$ can communicate their (classical) measurement results is at any rate inappropriate in a relativistic theory such as QFT, when the regions $A$ and $B$ are spacelike separated, see~\cite{Verch_2} for a discussion. We thus see that it is perhaps overly restrictive to consider all separable operations, especially in QFT. We will nevertheless do
so, since separable operations are rather easy to describe and handle, and we leave a more thorough discussion of this matter to the future.

We can now formulate the idea that on average, no entanglement can be won by performing separable operations:
\begin{enumerate}
\item[(e4)] (monotonicity under separable operations) Consider a separable operation, described by normal cp maps $\F_j=\F_{A,j}\otimes\F_{B,j}$ with $\sum_j\F_j(1)=1$. Then
\ben
\sum_j p_jE\left(\frac{\F_j^*\omega}{p_j}\right) \le E(\omega) \ ,
\een
where we sum over all $j$ with $p_j:=\omega(\F_j(1))>0$.
\end{enumerate}

As examples one can consider some of the cp maps mentioned above. When (e4) holds, $E$ is preserved under the action of local unitaries, i.e. separable operations of the form $\F(a\otimes b):=U_A^*aU_A\otimes U_B^*bU_B$ with unitaries $U_A$ and $U_B$ on $\H_A$ and $\H_B$, respectively, because $\F$ has an inverse which is again a separable operation. Similarly, $E$ is preserved under attaching a local ancillary system, e.g. $\F: (\A_A\otimes\A_{C})\otimes\A_B \to \A_A\otimes\A_B$ with $\F((a\otimes c)\otimes b)=\omega_{C}(c)a\otimes b$, because this separable operation has a left inverse. (Note that we need to choose whether the ancillary system is attached to system A or B in order to view $\A_A\otimes\A_{C}\otimes\A_B$ as a bipartite system and to define separable states and entanglement.) On the other hand, the restriction of states to a subalgebra of the form $\A_{C}\otimes\A_B$ with $\A_{C}\subset\A_A$ may decrease the value of $E$.

Next we consider what happens when the systems $A$ and $B$ themselves are composed of statistically independent subsystems. In that case, one may wish to ask additionally that
(e5) and/or (e6) hold:
\begin{enumerate}
\item[(e5)] (tensor products) Let $\A_A=\A_{A_1}\otimes\A_{A_2}$ and $\A_B=\A_{B_1}\otimes\A_{B_2}$, and let $\omega_{A_jB_j}$, $j=1,2$, be states on $\A_{A_j}\otimes\A_{B_j}$. Then
\ben
E(\omega_{A_1 B_1} \otimes\omega_{A_2B_2})\le E(\omega_{A_1B_1})+E(\omega_{A_2B_2}) \ .
\een
\item[(e6)] (superadditivity) Let $\omega_{AB}$ be a state on $\A_A\otimes\A_B$ with $\A_A=\A_{A_1}\otimes\A_{A_2}$ and
$\A_B=\A_{B_1}\otimes\A_{B_2}$, and let $\omega_{A_i B_i}$ be its restriction to $\A_{A_i}\otimes\A_{B_i}$ (embedded as e.g. $1\otimes\A_{A_2}\otimes 1\otimes\A_{B_2}$). Then
\ben
E(\omega_{A_1 B_1})+E(\omega_{A_2 B_2})\le E(\omega_{AB}).
\een
\end{enumerate}

\subsection{Bell correlations as an entanglement measure}\label{sec:bell}

Historically, the first quantity that was used as a measure of entanglement was the violation of the Bell-inequalities~\cite{bell_1,bell_2}. A convenient formulation is as follows. For commuting subalgebras $\A_A, \A_B$ of some v. Neumann algebra $\A$, and $\omega$ a state on $\A$, we define
\ben\label{Eqn_Bell}
E_B(\omega) := \sup \{ \half \omega(a_1(b_1 + b_2) + a_2(b_1-b_2)) \}
\een
where the supremum is over all self-adjoint elements $a_i, b_i$ such that
\ben\label{opsform}
 a_i \in \A_A, \ \
 -1\le a_i \le 1, \ \
 b_i \in \A_B \ \
 -1 \le b_i \le 1  \ .
\een
The properties of this quantity in the context of algebraic quantum field theory are discussed e.g. in~\cite{summers_2}.
It can be demonstrated that $\sqrt{2} \ge E_B(\omega) \ge 1$~\cite{summers_1,summers_3,tsirelson},
and that the lower bound is achieved for separable states, so no state with $E_B(\omega) > 1$ can be separable. The equality $E_B(\omega)=1$ is equivalent to the Clauser-Horne-Shimony-Holt~\cite{chsh} version of Bell's inequalities.

The measure $E_B(\omega)$ of the Bell correlations has several nice properties, including (e0), (e2), (e3) and (e4), but unfortunately it fails (e1). The normalisation $E_B(\omega)\ge1$ rather than $\ge0$ seems harmless, but the main problem is that there are entangled states $\omega$ with $E_B(\omega)=1$~\cite{gisin}.

There exist many other measures for entanglement and it is impossible to list them all here, but we refer to~\cite{plenio} for an overview (in the type I case). Our main focus will be on the relative entanglement entropy, which will be introduced in the next section. The other measures that we introduce will be useful as convenient tools to derive upper and lower bounds on the relative entanglement entropy.

\subsection{Relative entanglement entropy}\label{sec:ER}

The mother of all notions of entropy is the v. Neumann entropy. It is defined for density matrices $\rho$ on a Hilbert space $\H$ by $H_{\rm vN}(\rho) = -\tr(\rho \log \rho)$. The v. Neumann entropy can be viewed as the lack of information about a system to which one has ascribed the state $\rho$, assuming that the observer has, in principle, access to all operations (observables) in $\B(\H)$. This interpretation is in accord for instance with the facts that $H_{\rm vN}(\rho)\ge0$ and that a pure state $\rho = |\Psi \rangle \langle \Psi|$ has vanishing v. Neumann entropy.

A related notion is that of the relative entropy. It is defined for two density matrices $\rho, \rho'$ by
\ben
\label{drel}
H(\rho,\rho') = \tr(\rho \log \rho - \rho \log \rho') \ .
\een
The relative entropy can be thought of as the expected amount of information we gain when we update our belief about the state of the system from $\rho'$ to $\rho$ \cite{baez}. Like $H_{\rm vN}(\rho)$, $H(\rho,\rho')$ is non-negative, but can be infinite.

It seems hard to generalize the v. Neumann entropy to algebras of arbitrary type, in particular for type III. But a generalization of the relative entropy to v. Neumann algebras of arbitrary type was found by Araki~\cite{araki_3a,araki_3b}. It is formulated using modular theory. One assumes to be given two faithful, normal states $\omega, \omega'$ on a v. Neumann algebra $\A$ in standard form. We choose the vector representatives in the natural cone $\mathcal{P}^\sharp$, called $|\Omega \rangle,|\Omega' \rangle$ (cf. prop.~\ref{prop_cone}). Imitating the construction in sec.~\ref{SSec_algebras}, one defines, following Araki~\cite{araki_1}, $S_{\omega,\omega'} a|\Omega' \rangle = a^* |\Omega \rangle$, and one considers again the polar decomposition $S_{\omega, \omega'} = J \Delta_{\omega,\omega'}^{1/2}$ (the anti-unitary $J$ is seen to coincide with the corresponding $J$ for the state $\omega$). A related object is the Connes-cocycle (Radon-Nikodym-derivative) defined as $[D\omega : D\omega']_t = \Delta_{\omega, \psi}^{it} \Delta_{\psi, \omega'}^{it} \in \A$, where $\psi$ is an arbitrary auxiliary faithful state on $\A'$ (the definition is seen not to depend on it).
\begin{definition}
The relative entropy is
\ben\label{drel1}
H(\omega, \omega') = \langle \Omega | \log \Delta_{\omega, \omega'} \ \Omega\rangle \\
= \lim_{t \to 0} \frac{\omega([D\omega:D\omega']_t - 1)}{it} \ ,
\een
$H$ is extended to positive functionals that are not necessarily normalized by the formula $H(\lambda\omega,\lambda'\omega')=\lambda H(\omega,\omega') + \lambda \log (\lambda/\lambda')$, where $\lambda, \lambda'>0$ and $\omega,\omega'$ are normalized. If $\omega'$ is not normal, then one sets $H(\omega, \omega') = \infty$. When $\omega$ or $\omega'$ are not faithful (such that
$|\Omega\rangle,|\Omega'\rangle$ are not standard), the definition has to be somewhat modified~\cite{ohya_1}.
\end{definition}

\noindent
{\bf Key example} from sec.~\ref{SSec_algebras} continued:
For a type I algebra $\A = \B(\H)$, states $\omega, \omega'$ correspond to density matrices $\rho, \rho'$. The relative modular operator $\Delta_{\omega,\omega'}^{1/2}$ corresponds to $\rho^{1/2} \otimes \rho^{\prime -1/2}$ in the representation of $\A$ on $\H \otimes \bar \H$. In this representation, $\omega$ corresponds to the vector state $|\Omega \rangle =\rho^{1/2} \in \H \otimes \bar{\H}$, and the abstract definition of the relative entropy in~\eqref{drel1} becomes
\ben
\langle \Omega | \log \Delta_{\omega, \omega'} \, \Omega \rangle =  \tr_{\H} \rho^{\frac12}\left(\log\rho \otimes 1 - 1 \otimes \log \rho'\right)\rho^{\frac12} = \tr_\H(\rho \log \rho - \rho \log \rho') \ ,
\een
and therefore reproduces that given in~\eqref{drel} for density matrices.

\medskip
Let us now recall the main properties of $H$ (see~\cite{ohya_1} for a thorough discussion and references).
\begin{enumerate}
\item[(h1)] (positivity) $H(\omega, \omega') \ge 0$, and $H(\omega,\omega') = 0 \Rightarrow \omega=\omega'$ for states $\omega, \omega'$.
If $\omega, \omega'$ are not normal (i.e. their GNS representations are not quasi-equivalent), then $H(\omega, \omega') =\infty$. (The reverse implication is in general false, i.e. the relative entropy can be infinite for normal states).
\item[(h2)] (lower semi-continuity) The map $(\omega,\omega')\mapsto H(\omega,\omega')$ is weakly lower semicontinuous on the space of positive functionals on a $C^*$-algebra.
\item[(h3)] (subadditivity) $H(\sum_j \psi_j, \sum_j \varphi_j) \le \sum_j H(\psi_j, \varphi_j)$ for finite sums of normal positive functionals. (Note that this is equivalent to convexity: $H(\sum_j \lambda_j\psi_j, \sum_j \lambda_j\varphi_j) \le \sum_j \lambda_jH(\psi_j, \varphi_j)$ when $\lambda_j\ge0$ has $\sum_j\lambda_j=1$.)
\item[(h4)] (superadditivity in first argument) $H(\sum_j \omega_j, \omega') \ge \sum_j H(\omega_j, \omega')$ for finite sums of normal positive functionals.
\item[(h5)] (monotonicity) If $\phi\le\omega, \|\phi\|=\|\omega\|$ and $\phi'\le\omega'$ for normal positive functionals $\omega',\phi',\omega,\phi$,
then $H(\omega,\omega')\le H(\phi, \phi')$.
\item[(h6)] (``Uhlmann's monotonicity theorem''~\cite{Lindblad,uhlmann_1}) If $\mathcal{F}: \A_1 \to \A_2$ is a normalized cp map between v. Neumann algebras, then
$H(\F^*\omega,\F^*\omega') \le H(\omega, \omega')$. Equality holds if $\mathcal{F}=\mathcal{E}$ is a faithful, normal, conditional expectation from $\A_1$ to a subalgebra $\A_2$, i.e.
$\mathcal{E}(abc) = a\mathcal{E}(b)c$ for $a,c \in \A_2 \subset \A_1, b \in \A_1$ and there is a faithful normal state $\phi$ on $\A_1$ such that $\phi\circ \cE=\phi$ (such a map is always completely positive).
\item[(h7)] (tensor product) Let $\A = \A_1 \otimes \A_2$ be the (spatial) tensor product of two v. Neumann algebras, let $\omega$ be a normal state on $\A$ with $\omega_i:=\omega|_{\A_i}$ and let $\omega'_i$ be normal states on $\A_i$. Then $H(\omega,\omega_1'\otimes\omega'_2) = H(\omega,\omega_1\otimes\omega_2)+
H(\omega_1^{}, \omega_1') + H(\omega_2^{}, \omega_2')$.
\end{enumerate}

With the help of the relative entropy we can now define two entanglement measures~\cite{vedral}:
\begin{definition}
The ``relative entanglement entropy'' $E_R(\omega)$ of a normal state $\omega$ on the tensor product $\A_A \otimes \A_B$ of two
v. Neumann algebras (in standard form) is given by
\ben
E_R(\omega) := \inf\{ H(\omega,\sigma) \  \mid \ \sigma \ \text{a separable state} \} \ .
\een
The ``mutual information'' $E_I(\omega)$ is given by
\ben
E_I(\omega) = H(\omega, \omega_A \otimes \omega_B) \ ,
\een
where $\omega_A:=\omega|_{\A_A}$ and similarly for $\omega_B$.
\end{definition}

It immediately follows that $E_R(\omega) \le E_I(\omega)$.

As an example, let us consider a bi-partite system with Hilbert space $\H_A \otimes \H_B$ and observable algebra $\A = \B(\H_A) \otimes \B(\H_B)$. A pure state $\omega_{AB}$ on $\A$ corresponds to a density matrix $\rho_{A B} = |\Phi \rangle \langle \Phi|$, where $|\Phi \rangle \in \H_A \otimes \H_B$. One calls $\rho_A = \tr_{\H_B} \rho_{AB}$ the ``reduced density matrix'', which defines a state $\omega_A$ on $\B(\H_A)$ (and similarly for system $B$). The relative entanglement entropy between $A$ and $B$ in the pure state $\omega \equiv \omega_{AB}$ is then~\cite{vedral}
\ben
E_R(\omega) = H_{\rm vN}(\rho_A)  \qquad [=H_{\rm vN}(\rho_B)] \ .
\een
The mutual information, often used in the case when $\omega$ is mixed, i.e. when $\rho_{A B}$ is not a rank 1 projector, is given in our example system by
\ben
E_I(\omega) = H_{\rm vN}(\rho_A) + H_{\rm vN}(\rho_B) - H_{\rm vN}(\rho_{A B}) \ .
\een
When $\omega = \tr(\rho_{AB} \ . \ )$ is pure, then evidently $E_I(\omega) = 2E_R(\omega)$. If $\omega$ is not pure, then $E_I(\omega)$ will be strictly smaller than $2E_R(\omega)$. $E_I$ satisfies (e1) for product states $\omega=\omega_A\otimes\omega_B$, but for general separable states of the
form $\omega = \sum_i p_i \omega_{Ai} \otimes \omega_{Bi}$ (with $\omega_{Ai}, \omega_{Bi}$ states) we can show using (h3), (h5), (h7) only that
$E_I(\omega) \le H_{\rm vN}(\{p_i\})$.  (e3) fails and the status of (e4)-(e6) is the same as for $E_R$, see below.

In the next sections we investigate these quantities in various algebraic quantum field theories, where we will always take
$$
\A_A \cong \pi(\A(O_A))'' , \quad
\A_B \cong \pi(\A(O_B))''
$$
for two space-like separated open sets $A,B$ in some Cauchy surface with finite distance, where $\pi$ is a suitable representation (usually the GNS-representation of
the state $\omega$ considered). We have already noted that these algebras are of type III$_1$, and so {\em never have a normal pure state}. Consequently, in our case, we typically expect a strict inequality $E_R(\omega)  < E_I(\omega)$. It is also essential in this situation that the regions $O_A, O_B$ must have a finite, positive distance. Otherwise standard states, such as the vacuum, will usually not be normal states on $\A(O_A) \otimes \A(O_B)$ (as we have seen in our discussion of the split property, theorem \ref{thm_split}), and hence automatically lead to an infinite relative entanglement entropy by (h1). This phenomenon is indeed encountered in many formal approaches to entanglement entropy in quantum field theory, where one implicitly assumes that the type of the algebra is I.

The properties of the relative entropy directly imply many properties of $E_R(\omega)$, where $\A_A, \A_B$ are two v. Neumann algebras in standard form, as before. In particular we have the properties (e0) (manifest), (e3) (from (h3)), and (e5) (from (h7)). Property (e1) holds due to (h1) with the modification that $E_R(\omega)=0$ implies that $\omega$ is a norm limit of separable states (cf.~\cite{ohya_1} thm.5.5). The continuity (e2) was shown for matrix algebras in~\cite{donald_2}. The key requirement
(e4) does not directly follow from (h4) and (h6), but we can argue as follows adapting somewhat the proof by~\cite{vedral} for type I factors:
First, let $\mathfrak{M}$ be any v. Neumann algebra with $n$ normal cp maps ${\mathcal F}_i$ defined on it such that $\sum_i {\mathcal F}_i(1)=1$. As a technical
simplification, we assume that $\F_i(a) = 0, a \in \A^+$ implies $a=0$ (this assumption can be removed).

Letting $\omega, \omega'$ be two faithful normal states, we first show:
\begin{lemma}
We have $\sum_i H({\mathcal F}_i^* \omega, {\mathcal F}_i^* \omega') \le H(\omega,\omega')$.
\end{lemma}
{\em Proof:}
Define $\hat {\mathfrak M} = {\mathfrak M} \otimes M_n(\CC)$. Denoting by $\{ |i \rangle \}$ an orthonormal basis of $\CC^n$, we define
\ben
\hat {\mathcal F} : \hat {\mathfrak M} \to {\mathfrak M} \ , \qquad a \otimes X \mapsto \sum_i \langle i| X | i \rangle \ {\mathcal F}_i(a) \ ,
\een
which is easily checked to be cp. Using the projections $P_i = 1 \otimes | i \rangle \langle i|$, $i=1, \dots, n$, we also define  the cp
maps ${\mathcal E}_i: \hat {\mathfrak M} \to \hat {\mathfrak M}, {\mathcal E}_i(\hat a) = P_i \hat a P_i$. Using the properties $P_i P_j = \delta_{ij} P_j, \sum_i P_i = 1$,
one can show that for any pair of normal states $\psi, \psi'$ on $\hat {\mathfrak M}$, one has
\ben\label{intermediate}
\sum_i H({\mathcal E}_i^* \psi, {\mathcal E}_i^* \psi') = H\left(\sum_i {\mathcal E}_i^* \psi, \sum_j {\mathcal E}_j^* \psi'\right) \ .
\een
This property is obvious for type I factors, and can be proven in the general case as follows. Let $\alpha_i: {\mathfrak M} \to \hat {\mathfrak M}$ be the $*$-homomorphisms
$\alpha_i(a) = | i \rangle \langle i| \otimes a$. The image is a v. Neumann subalgebra $\hat {\mathfrak M}_i$, and ${\mathcal E}_i$ is obviously a faithful conditional
expectation onto this subalgebra. Now let $\psi_i = \alpha_i^* \psi$ (and similarly $\psi'_i = \alpha^*_i \psi'$), and let $|\Omega_i \rangle, |\Omega_i'\rangle
\in \mathcal{P}^\sharp \subset \H$ be vector representers in a natural cone in a Hilbert space representation of $\mathfrak{M}$ on a Hilbert space $\H$.
Letting $\varphi = \sum_i {\mathcal E}_i^* \psi$, we can describe the associated GNS-representation $\hat \H_\varphi, |\Omega_\varphi \rangle$ as follows.
The Hilbert space is $\hat \H = M_n(\CC) \otimes \H$ with inner product $\langle X' \otimes \Psi' | X \otimes \Psi \rangle = \tr (X^{\prime *} X) \langle \Psi' | \Psi \rangle_{\H}$.
Elements $Y \otimes a \in \hat {\mathfrak M}$ act by $Y \otimes a|X \otimes \Psi \rangle = |YX \otimes a\Psi \rangle$. The GNS vector is
$|\Omega_\varphi \rangle = \sum_i |i\rangle \langle i| \otimes |\Omega_i \rangle$. Analogous statements hold for $\varphi' = \sum_i {\mathcal E}_i^* \psi'$.
If $\psi,\psi'$ are faithful, then both $|\Omega_\varphi \rangle, |\Omega_{\varphi'} \rangle$ are separating, as are $|\Omega_i \rangle, |\Omega_i' \rangle$.
The relative modular operator and modular conjugation are found to be
\ben\label{modularhat}
\Delta(\varphi,\varphi')^{it} (|k \rangle \langle j| \otimes |\Psi \rangle) = |k \rangle \langle j| \otimes \Delta(\psi_k, \psi_j')^{it} |\Psi \rangle \ , \quad
J (|k \rangle \langle j| \otimes |\Psi \rangle) = |j \rangle \langle k| \otimes J |\Psi \rangle
\een
where $J$ on the right side of the last equation is the modular conjugation associated with the natural cone ${\mathcal P}^\sharp$. It immediately follows
from these formulas and the definition of the relative entropy that
\ben
H\left(\sum_i {\mathcal E}_i^* \psi, \sum_j {\mathcal E}_j^* \psi'\right) = \sum_i H(\psi_i, \psi_i') \ .
\een
However, the terms on the right side can also be written as $\sum_i H({\mathcal E}_i^* \psi, {\mathcal E}_i^* \psi')$, since ${\mathcal E}_i$ is a faithful conditional
expectation onto the subalgebra $\hat {\mathfrak M}_i$ and ${\mathcal E}_i^* \psi = \psi |_{\hat {\mathfrak M}_i}$, and similarly for $\psi'$, by (h6). The proof of
\eqref{intermediate} is complete.

If we embed ${\mathfrak M}$ into $\hat {\mathfrak M}$ as $a \mapsto a \otimes 1_n$, then it follows from the definitions
that $\hat {\mathcal F} {\mathcal E}_i |_{\mathfrak M} = {\mathcal F}_i$. It follows from the definitions that $\sum_i \hat {\mathcal F} {\mathcal E}_i = \hat {\mathcal F}$. Using these properties, we have,
for normal states $\omega, \omega'$ on $\mathfrak M$ (noting that $\hat {\mathcal F}^* \omega = \psi, \hat {\mathcal F}^* \omega' = \psi'$ are faithful):
\ben
\begin{split}
\sum_i H({\mathcal F}_i^* \omega, {\mathcal F}_i^* \omega') =&
\sum_i H( (\hat {\mathcal F} {\mathcal E}_i)^* \omega |_{\mathfrak M} , (\hat {\mathcal F} {\mathcal E}_i)^* \omega' |_{\mathfrak M}  ) \\
\le & \sum_i H( (\hat {\mathcal F} {\mathcal E}_i)^* \omega , (\hat {\mathcal F} {\mathcal E}_i)^* \omega'  )\\
= & H( \sum_i (\hat {\mathcal F} {\mathcal E}_i)^* \omega , \sum_j (\hat {\mathcal F} {\mathcal E}_j)^* \omega'  )\\
= & H( \hat {\mathcal F}^* \omega , \hat {\mathcal F}^* \omega'  )
\le  H(\omega, \omega') \ .
\end{split}
\een
To go to the second line, we used (h6) applied to the inclusion of ${\mathfrak M}$ into $\hat {\mathfrak M}$. To go to the third line we used~\eqref{intermediate},
and in the last step we used (h6)
applied to $\hat {\mathcal F}$. \qed

We now have, with $p_i = \omega({\mathcal F}_i(1)), p_i' = \omega'({\mathcal F}_i(1))$:
\ben
\begin{split}
\sum_i p_i H({\mathcal F}_i^* \omega/p_i, {\mathcal F}_i^* \omega'/p_i') =&
\sum_i H({\mathcal F}_i^* \omega, {\mathcal F}_i^* \omega') - \sum_i p_i \log (p_i/p_i') \\
=&  \sum_i H({\mathcal F}_i^* \omega, {\mathcal F}_i^* \omega') - H({\rm diag}\{p_i\}, {\rm diag}\{p_i'\})\\
\le & \sum_i H({\mathcal F}_i^* \omega, {\mathcal F}_i^* \omega') \ ,
\end{split}
\een
using in the first step the scaling properties of the relative entropy, and
using in the last step the property (h1) for the diagonal density matrices ${\rm diag}\{p_i\}, {\rm diag}\{p_i'\}$.
We therefore conclude altogether that
\ben\label{inequ}
\sum_i p_i H({\mathcal F}_i^* \omega/p_i, {\mathcal F}_i^* \omega'/p_i') \le H(\omega, \omega') \ .
\een
To show (e4), one now takes ${\mathfrak M} = \A_A \otimes \A_B$, and for $\omega'$ a separable state $\sigma$
with the property $E_R(\omega) \ge H(\omega, \sigma)- \epsilon$ for an arbitrary but fixed $\epsilon > 0$. The statement then follows immediately
from inequality~\eqref{inequ} since each ${\mathcal F}_i^* \sigma/\sigma({\mathcal F}_i(1))  = \sigma_i$ is again a separable state,
so that $\sum_i p_i E_R({\mathcal F}_i^*\omega/p_i) - \epsilon \le E_R(\omega)$.

\subsection{Logarithmic dominance}

We say that a positive linear functional $\sigma$ on a $C^*$-algebra $\A$ is dominated by a positive linear functional $\sigma'$ when $\sigma'-\sigma$ is positive, i.e. $\sigma'\ge\sigma\ge0$. Using the properties of positive linear functionals we have
\[
\|\sigma'-\sigma\|=\sigma'(1)-\sigma(1)=\|\sigma'\|-\|\sigma\|.
\]
In particular, for two states $\omega$ and $\omega'$ we have $\omega'\ge\omega$ if and only if $\omega'=\omega$.

We will call a positive linear functional $\sigma$ on $\A_A\otimes\A_B$ separable, when $\sigma=r\omega$ with $\omega$ a separable state and $r=\sigma(1)\ge0$.

Using these notions we can now introduce a further entanglement measure, which has been introduced and analyzed in the type I setting in \cite{dutta}, where it was termed ``max entropy''\footnote{We thank Marc M. Wilde for point out this reference to us.}:
\begin{definition}
The ``logarithmic dominance'' $E_N(\omega)$ of a normal state $\omega$ on the tensor product $\A_A \otimes \A_B$ of two v. Neumann algebras (in standard form) is given by
\ben
E_N(\omega) := \inf\{\log(\|\sigma\|) \ \mid \ \sigma\ge\omega, \ \sigma \ \text{separable} \} \ .
\een
If no dominating separable functionals $\sigma$ exist, we set $E_N(\omega):=\infty$.
\end{definition}

$E_N$ satisfies property (e0) in a straightforward way. For a modified version of (e1) we note that $\sigma\ge\omega$ implies that $\|\sigma\|\ge1$ and hence $E_N(\omega)\ge0$. When $\omega$ is separable we have $E_N(\omega)=0$. Conversely, when $E_N(\omega)=0$ there is a sequence $\sigma_n$ of separable positive linear functionals such that $r_n:=\|\sigma_n\|=\|\sigma_n-\omega\|+1$ converges to 1 as $n\to\infty$. Hence, $r_n^{-1}\sigma_n$ converges in norm to $\omega$ and $\omega$ is a norm limit of separable states. Finally, when a separable, and hence normal, functional $\sigma$ dominates $\omega$, then $\omega$ is necessarily normal too. Conversely, if $\omega$ is not normal, then $E_N(\omega)=\infty$.

The validity of property (e2) is unclear, and property (e3) probably fails, because $\log$ is concave rather than convex.

To prove (e4) we consider a separable operation, described by normal cp maps $\F_j=\F_{A,j}\otimes\F_{B,j}$ with $\sum_j\F_j(1)=1$. We let $\omega$ be any state and we set $p_j:=\omega(\F_j(1))$, where we may assume $p_j>0$. We note that each $\F_j$ maps separable positive functionals $\sigma$ to separable positive functionals $\F^*_j\sigma$, and when $\sigma\ge\omega$, then $p_j^{-1}\F^*_j\sigma\ge p_j^{-1}\F^*_j\omega$. Furthermore,
\[
\sum_j\|\F^*_j\sigma\|=\sum_j\sigma(\F^*_j(1))=\sigma(1)=\|\sigma\|.
\]
Using the concavity of $\log$ we therefore find
\begin{eqnarray}
\sum_jp_jE_N(\F^*_j\omega/p_j)&\le&
\sum_jp_j\inf\{\log(\|\F^*_j\sigma/p_j\|) \ \mid \ \sigma\ge\omega, \ \sigma \ \text{separable} \}\nonumber\\
&\le&\inf\left\{\sum_jp_j\log(\|\F^*_j\sigma\|/p_j) \ \mid \ \sigma\ge\omega, \ \sigma \ \text{separable} \right\}\nonumber\\
&\le&\inf\left\{\log\left(\sum_j\|\F^*_j\sigma\|\right) \ \mid \ \sigma\ge\omega, \ \sigma \ \text{separable} \right\}\nonumber\\
&=&\inf\{\log(\|\sigma\|) \ \mid \ \sigma\ge\omega, \ \sigma \ \text{separable} \} = E_N(\omega).\nonumber
\end{eqnarray}

To show (e5) it suffices to note that two separable functionals $\sigma_i$, $i=1,2$, which dominate states $\omega_i$, give rise to a separable functional $\sigma:=\sigma_1\otimes\sigma_2$ with $\sigma\ge \omega_1\otimes\sigma_2\ge\omega_1\otimes\omega_2$ and $\|\sigma\|=\sigma(1)=\|\sigma_1\|\cdot\|\sigma_2\|$. By taking the logarithm and the infimum over the $\sigma_i$ one then finds (e5).

One can also show a weaker version of the superadditivity (e6). Indeed, if $\sigma\ge\omega$, then the restrictions satisfy $\sigma_i\ge\omega_i$ and $\|\sigma_i\|=\|\sigma\|$. Taking the infimum over $\sigma$ then yields the modified estimate
$E_N(\omega_1)+E_N(\omega_2)\le 2E_N(\omega)$.

To conclude this section we show the following useful estimate:
\begin{theorem}\label{Thm_ER<EN}
$E_R(\omega)\le E_N(\omega)$.
\end{theorem}

{\em Proof:} We choose any separable positive linear functional $\sigma\ge\omega$. (If such $\sigma$ does not exist, the inequality is trivially true.) We then find
\begin{eqnarray}
E_R(\omega)&\le&H\left(\omega,\frac{\sigma}{\|\sigma\|}\right)=H(\omega,\sigma)+\log\|\sigma\|\nonumber\\
&\le&H(\omega,\omega)+\log\|\sigma\|=\log\|\sigma\|,\nonumber
\end{eqnarray}
using in the first step the definition of the relative entanglement entropy, using the definition of $H$ in the second step, using
$\sigma\ge\omega$ and the monotonicity (h5) of $H$ in the third step, and using $H(\omega,\omega) = 0$ in the last step. Taking the infimum over separable $\sigma\ge\omega$ then yields the desired estimate.
\qed

\subsection{Modular nuclearity as an entanglement measure}\label{E_Mdef}

Our next entanglement measure makes use of modular operators and is especially convenient in quantum field theories that satisfy a modular nuclearity condition, which is somewhat analogous to the BW-nuclearity condition a5). We will use it below in applications to integrable models in $1+1$ dimensions, as well as for conformal quantum field theories and free
field theories in $d+1$ dimensions.

As usual, we consider a v. Neumann algebra $\A_A\otimes\A_B \cong \A_A \vee \A_B$ represented in standard form on a Hilbert space $\H$ with
fixed natural cone $\cP^\sharp$. This cone is associated with some fixed cyclic and separating reference vector, but the definition of our
entanglement measure will not depend on it. It is natural to define the algebra associated with
the ``environment'' by $\A_E = (\A_A \vee \A_B)'$. The following {\em standing assumption} in this section will be made:

\medskip
\noindent
{\bf Standing assumption:}
All states $\omega$ considered
are such that their vector
representative $|\Omega \rangle \in \cP^\sharp$ in the natural cone is cyclic
for each of $\A_A, \A_B$ and $\A_E$. Due to the Reeh-Schlieder theorem, we are naturally in this situation in the context of quantum field theory on Minkowski space if $A, B$ are open subsets of a Cauchy surface $\cC$ and $E$ is the complement of the closure of $A \cup B$ in $\cC$.
Indeed, if $\pi_0$ is the vacuum representation and
$\A_A = \pi_0(\A(O_A))''$ (and similarly for $B$ and $E$), the standing assumption holds e.g. for normal states $\omega$ with bounded energy if the QFT
satisfies (a1)-(a4). The cyclic property of $|\Omega \rangle$ can in this case be interpreted physically as saying that an observer occupying the environment has
sufficient control over this state.

\medskip
\noindent
Under the standing assumption, $|\Omega \rangle$ is cyclic and separating also for the von Neumann algebras $\A_B'$ and $\A_A'$ and separating for $\A_A$ and $\A_B$\footnote{Actually, this is all that is needed in order to define $E_M$.}. We let $\Delta$ and $J$ be the corresponding modular operator and modular conjugation for $\A_B'$. One can then define the map
\ben\label{Psidefnuclear}
\Psi^A: \A_A \to \H \ , \quad \Psi^A(a) =  \Delta^{\quarter} a | \Omega \rangle \ .
\een
and likewise for $A$ replaced by $B$. Here and in the following, we write $a$ for $a \otimes 1$ and $b$ for $1 \otimes b$
to simplify the notation, and likewise we often make the identification of $\A_A$ with $\A_A \otimes 1$ and of $\A_B$ with $1 \otimes \A_B$.

The {\bf Buchholz partition function} is defined by
\ben
Z(\omega) = {\rm min} (\|\Psi^A\|_1, \|\Psi^B\|_1),
\een
where we use the 1-nuclear norm and the minimum is taken to get a quantity that is manifestly symmetric under an exchange of $A$ with $B$.

\begin{definition}
The modular entanglement measure is defined by
\[
E_M(\omega):=\log(Z(\omega)).
\]
When neither $\Psi^A$ nor $\Psi^B$ is nuclear, we set $Z(\omega)=\infty$ and $E_M(\omega)=\infty$.
\end{definition}

\begin{remark}
The distinguished value of $\quarter$ is due to the formula $\Delta^\alpha a|\Omega \rangle = J \Delta^{\frac12-\alpha} a^*| \Omega \rangle$ for $0 \le \alpha \le \frac12$, implying that the 1-norm of the functional $\Delta^\alpha a |\Omega \rangle$ is symmetric about the value $\alpha=\quarter$. The results given below
also apply to other values of $\alpha \in (0, \frac12)$. We will formulate our proofs in such a way that this should hopefully be evident, but will not make explicit statements.
\end{remark}

\medskip
\noindent
{\bf Key example} from sec.~\ref{SSec_algebras} continued:
Let $\A_A = M_n(\CC) = \A_B$ and $|\Omega\rangle = \sum_i \sqrt{p_i} |i\rangle \langle i| \in \H = \CC^n \otimes \bar \CC^n$, which is a pure state on $\A_A \otimes \A_B$ with corresponding functional $\omega = \langle \Omega | \ . \ | \Omega \rangle$. $|\Omega\rangle$ is therefore not cyclic for $\A_E$, but it is still
cyclic and separating for $\A_B', \A_A'$ if all $p_i > 0$. Thus, $E_M$ can still be defined in this case. Going through the definitions, one finds $\Psi^A(a) = \rho^{1/4} a \rho^{1/4}$ and
$E_M(\omega) = 2 \log \tr \rho^{1/4}$, where $\rho= diag(\{p_i\})$, with the same expressions holding also for $B \leftrightarrow A$. For the maximally entangled
state $\omega^+_n$ defined by $p_i=1/n$, we thereby get $E_M(\omega^+_n) = \tfrac{3}{2} \log n$.
\medskip
\noindent

Our main use of $E_M$ is the following theorem:
\begin{theorem}\label{Thm_EN<Em}
$E_N(\omega) \le E_M(\omega)$.
\end{theorem}

The proof is based on two lemmas\footnote{After this preprint appeared, it was pointed out to the
authors that a similar proof of lemma~\ref{Lem_dominating_sigma} also appears in the unpublished manuscript~\cite{fredenhagen_6}.}:

\begin{lemma}\label{Lem2}
If $\nu:=\| \Psi^A\|_1<\infty$ and $\epsilon>0$, there are sequences of (not necessarily positive) normal linear functionals $\phi_j$ on $\A_A$ and $\psi_j$ on $\A_B$ such that
\ben
\omega(ab)=\sum_j\phi_j(a)\psi_j(b) \ , \qquad a\in\A_A \ , \ b\in \A_B
\een
and $\sum_j\|\phi_j\|\cdot\|\psi_j\|<\nu+\epsilon$.
\end{lemma}

{\em Proof:} Recall that $J|\Omega \rangle=|\Omega\rangle $, and note that $J\Delta^{-\frac12}$ is the Tomita operator for $\A_B$, i.e.
\[
\Delta^{-\frac12}b^*|\Omega \rangle =J(J\Delta^{-\frac12})b^*|\Omega \rangle=Jb|\Omega \rangle=JbJ|\Omega \rangle \ .
\]
Using the commutativity of $\A_A$ and $\A_B$ we then note that
\begin{eqnarray}
\omega(ab)&=&\langle \Omega | ab \Omega \rangle=\langle (\Delta^{\quarter}+\Delta^{-\quarter})^{-1}(1+\Delta^{-\frac12}) b^*\Omega|\Delta^{\quarter}
a \Omega \rangle \nonumber\\
&=&\langle (\Delta^{\quarter}+\Delta^{-\quarter})^{-1}(b^*+JbJ)\Omega| \Psi^A(a) \rangle.\nonumber
\end{eqnarray}

If $\nu<\infty$ and $\epsilon>0$, there are sequences of normal functionals $\phi_j$ on $\A_A$ and vectors $|\chi_j \rangle \in\H$ such that
\[
\Psi^A(a) = \Delta^{\quarter} a| \Omega \rangle = \sum_j| \chi_j \rangle \phi_j(a)
\]
for all $a\in\A_A$, and $\sum_j\|\phi_j\|\cdot\|\chi_j\|<\nu+\epsilon$. Define the normal functionals $\psi_j$ on $\A_B$ by
\[
\psi_j(b) := \langle (\Delta^{\quarter}+\Delta^{-\quarter})^{-1}(b^*+JbJ)\Omega|\chi_j \rangle
\]
and note that $\|\psi_j\|\le \|\chi_j\|$, because $\|(\Delta^{\quarter}+\Delta^{-\quarter})^{-1}\|\le\frac12$ by the spectral calculus.

Putting both paragraphs together we find the conclusion: $\omega(ab)=\sum_j\phi_j(a)\psi_j(b)$ with $\sum_j\|\phi_j\|\cdot\|\psi_j\|<\nu+\epsilon$.
\qed

\begin{lemma}\label{Lem_dominating_sigma}
If there are sequences of (not necessarily positive) normal linear functionals $\phi_j$ on $\A_A$ and $\psi_j$ on $\A_B$ such that
\ben
\omega(ab)=\sum_j\phi_j(a)\psi_j(b) \ , \qquad a\in\A_A \ , \ b\in \A_B
\een
and $\mu:=\sum_j\|\phi_j\|\cdot\|\psi_j\|<\infty$, then there is a separable positive linear functional $\sigma$ such that $\sigma\ge\omega$ and $\|\sigma\|=\mu$.
\end{lemma}

{\em Proof:} By theorem 7.3.2 in~\cite{KadisonRingrose}  there are partial isometries $U_j\in \A_A$ such that $\phi_j(U_j \ . \ )\ge0$ on $\A_A$ and $\phi_j(U_jU_j^* \ . \ )=\phi_j$. It follows in particular that $\phi_j(U_j)=\|\phi_j(U_j \ . \ )\|=\|\phi_j\|$ and
\[
\bar{\phi}_j(a)=\overline{\phi_j(U_jU^*_ja^*)}=\phi_j(U_j(U^*_ja^*)^*)=\phi_j(U_jaU_j)
\]
for all $a\in\A_A$, where we used the fact that $\phi_j(U_j \ . \ )$ is hermitean. (Here $\bar \psi(a) \equiv \overline{\psi(a^*)}$.) Similarly, there are partial isometries $V_j\in \A_B$ such that $\psi_j(V_j \ . \ )\ge0$ and $\psi_j(V_jV_j^* \ . \ )=\psi_j$.

Note that the positive linear functional $\rho_j:=\phi_j(U_j \ . \ )\otimes\psi_j(V_j  \ . \ )$ is separable. Writing $W_j:=U_j\otimes V_j$ we then define
\[
\sigma_j:=\frac12\rho_j+\frac12\rho_j(W^* \ . \ W) \ ,
\]
which is also separable, because $W$ is a simple tensor product. Furthermore,
\[
\|\sigma_j\|=\sigma_j(1)=\rho_j(1)=\|\phi_j\|\cdot\|\psi_j\|
\]
and
\[
0\le \frac12\rho_j((1-W^*)\ .\ (1-W))=\sigma_j-\frac12(\phi_j\otimes\psi_j+\bar{\phi}_j\otimes\bar{\psi}_j) \ .
\]
We conclude that $\sigma:=\sum_j\sigma_j$ is a separable positive linear functional with $\|\sigma\|=\sigma(1)=\sum_j\|\sigma_j\|=\mu$ and
\[
\sigma\ge\frac12\sum_j(\phi_j\otimes\psi_j+\bar{\phi}_j\otimes\bar{\psi}_j)=\frac12(\omega+\overline{\omega})=\omega \ .
\]
\qed

{\em Proof of thm. \ref{Thm_EN<Em}:}
Combining the two lemmas with $\mu=\nu+\epsilon$ we find
$E_N(\omega)\le\log(\|\sigma\|)=\log(\mu)=\log\left(\|\Psi^A\|_1+\epsilon\right)$ for all $\epsilon>0$, and hence $E_N(\omega)\le\log\left(\|\Psi^A\|_1\right)$. Interchanging the roles of $A$ and $B$ we also find
$E_N(\omega)\le\log\left(\|\Psi^B\|_1\right)$ and hence $E_N(\omega)\le\log(Z(\omega))=E_M(\omega)$.
\qed

\medskip

Although we will use $E_M$ only via its relationship to $E_N$  given in thm.~\ref{Thm_EN<Em}, it is perhaps of interest to investigate $E_M$ in its own right.
Here we look at properties (e1)-(e6). (e0) is clearly satisfied by construction. (e1) holds in the restricted sense that $E_M(\omega_p) = 0$ when $\omega$ is a product state  $\omega_A \otimes \omega_B$, which follows immediately from the fact that $\Psi^A(a) = |\Omega\rangle \omega(a)$ in this case. For more general separable states (e1) probably fails. (e2) is unclear to us, but we have the following result, which also implies some sort of continuity different from (e2) for $E_M$:

\begin{proposition}\label{Emdominance}
Let $\omega_i, i=1,2$ be two faithful normal states on $\A_A \vee \A_B \cong \A_A\otimes\A_B$, with GNS representers that are separating for $\A_A'$ and $\A_B'$ and such that, for some $\lambda>0$, $\omega_2 \le \lambda \omega_1$. Then $E_M(\omega_2) - E_M(\omega_1) \le \half \log \lambda$.
\end{proposition}

{\em Proof:}
Let $|\Omega_i\rangle, i=1,2$ be the GNS vector representatives of the states $\omega_i$ in $\cP^\sharp$, see prop.~\ref{prop_cone}, and
let $S_i$ be the Tomita operators for the algebra $\A_B'$ associated with $|\Omega_i\rangle, i=1,2$, with polar decompositions
$S_i = J_i \Delta_i^{1/2}$. Note that $\lambda\ge1$ by the normalisation of states and that $\omega_2\le \mu\omega_1$ for any $\mu>\lambda$. We may therefore write $\omega_1=\frac{1}{\mu}\omega_2+\frac{\mu-1}{\mu}\omega_3$, where $\omega_3$ is a normal state. Because $\omega_3\ge\frac{\mu-\lambda}{\lambda(\mu-1)}\omega_2$ we see that $\omega_3$ is also faithful.

We now need the following important result, which is proved in \cite{araki_3a} (sec. 4). Alternatively, it follows from the ``quadratic interpolations'' of \cite{uhlmann_1} (prop. 8).
\begin{lemma}\label{Lem_WYDL}
Let $\alpha \in (0,\frac12)$, and let $\gM$ be a v. Neumann algebra acting on $\H$ with fixed natural cone $\cP^{\sharp}$.
Then the functional $\omega \mapsto \| \Delta_\omega^\alpha a \Omega \|^2$ on normal faithful states $\omega$ on $\gM$
(with vector representatives $|\Omega\rangle \in \cP^{\sharp}$) is concave.
\end{lemma}

Applying lemma \ref{Lem_WYDL} we find that for every $a\in\A_A$,
\[
\frac{1}{\mu}\|\Delta_2^{\frac14}a\Omega_2\|^2 \le \|\Delta_1^{\frac14}a\Omega_1\|^2,
\]
i.e. $\|\Psi^A_2(a)\| \le \sqrt{\mu}\|\Psi^A_1(a)\|$. This immediately implies that there is an operator $T$ such that $T\Psi^A_1(a)=\Psi^A_2(a)$ and $\|T\|\le\sqrt{\mu}$. We can therefore estimate
\[
\|\Psi^A_2\|_1=\|T\Psi^A_1\|_1\le \|T\| \|\Psi^A_1\|_1\le \sqrt{\mu}\|\Psi^A_1\|.
\]
A similar estimate holds when we swap the roles of $A$ and $B$, so we have
\[
E_M(\omega_2) - E_M(\omega_1)\le \ln(\sqrt{\mu})=\frac12\ln(\mu).
\]
Taking $\mu\to\lambda^+$ yields the result. \qed

\begin{remark}
When $\gM=M_N$ is the algebra of $N$ dimensional matrices, lemma \ref{Lem_WYDL} states that the map $\rho \mapsto \tr (\rho^{1-2\alpha} a^* \rho^{2\alpha} a)$ from density matrices $\rho$ to the reals is concave for any $a \in M_N$. This is a special case of the
Wigner-Yanase-Dyson-Lieb concavity theorem.
\end{remark}

In a similar way we find the converse to (e3), i.e. the concavity of $E_M$.

\begin{proposition}
Let $\omega_i$ be normal states on $\A_A \otimes \A_B$, and let $\omega = \sum_i \lambda_i \omega_i$ with $\lambda_i>0, \sum_i \lambda_i=1$.
Then $\sum_j \lambda_j E_M(\omega_j) \le E_M(\omega)$.
\end{proposition}

{\em Proof:}
Let $|\Omega\rangle$  be the vector representative for the normal state $\omega$ in the chosen natural cone $\cP^{\sharp}$ of $\A_A \otimes \A_B$. Likewise we have vector representatives $|\Omega_i\rangle$ of $\omega_i$ in this cone, and by our standing assumption all these vectors are cyclic for $\A_B'$. Their modular operators are denoted by $\Delta_i$. Applying lemma \ref{Lem_WYDL} to $\gM = \A_B'$ with $\alpha=1/4$ and using the concavity of $x \mapsto \sqrt{x}$ we find
\ben
\sum_{i=1}^n \lambda_i \| \Delta^{\frac{1}{4}}_i a \Omega_i \| \le  \| \Delta^{\frac{1}{4}} a \Omega \|
\een
for all $a \in \A_A$. In terms of the maps $\Psi_i^A(a) = \Delta^{1/4}_i a |\Omega_i \rangle$, and $\Psi^A(a) = \Delta^{1/4} a |\Omega \rangle$, this evidently means $\sum_i \lambda_i^{} \|\Psi_i(a) \| \le \| \Psi(a) \|$. Now let $\Phi^A: \A_A \to \cY = \oplus^n \H$ be defined by $\Phi^A(a) = (\lambda_1 \Psi_1(a), \dots, \lambda_n \Psi_n(a))$, and equip $\cY$ with the Banach space norm $\|Y\|_\cY = \sum_i \| y_i \|$ for $Y=(y_1, \dots, y_n) \in \cY$. Then we obviously have $\|\Phi^A(a) \|_\cY \le \|\Psi^A(a) \|$ for all $a \in \A_A$. It follows that the relation $\{ (\Psi^A(a), \Phi^A(a)) \mid a \in \A_A \} \subset \H \times \cY$ is the graph of a
closed linear operator $T: \H \to \cY$ with the property that $\| T \| \le 1$ and $T \circ \Psi^A = \Phi^A$. Consequently, by the properties of the 1-norm,
$\| \Phi^A \|_1 \le \| \Psi^A \|_1$.

It is elementary to show that $\|\Phi^A\|_1 \ge \sum_i \lambda_i^{} \| \Psi^A_i \|_1$: Suppose $\Phi^A(a) = \sum_\alpha Y_\alpha \varphi_\alpha(a)$ for all $a \in \A_A$, with $Y_\alpha = (y_{1,\alpha}, \dots, y_{n,\alpha}) \in \cY$ and normal functionals $\varphi_\alpha$ chosen such that $\| \Phi^A \|_1 + \epsilon \ge \sum_\alpha \| Y_\alpha \|_\cY \, \|\varphi_\alpha\|$. Obviously, $\lambda_i^{} \Psi_i^\alpha(a)  = \sum_\alpha y_{i,\alpha} \varphi_\alpha(a)$, so $\lambda_i^{} \| \Psi^A_i \|_1 \le \sum_\alpha \| y_{i,\alpha} \| \, \|\varphi_\alpha\|$ and taking the sum over $i$ it follows that $\|\Phi^A\|_1 + \epsilon \ge \sum_i \lambda_i^{} \| \Psi^A_i \|_1$, and from this the claim follows since $\epsilon$ can be made arbitrarily small.

Thus, we know $\|\Psi^A\|_1 \ge \sum_i \lambda_i^{} \| \Psi^A_i \|_1$, and we get the analogous statement for $A$ replaced by $B$.
Taking the $\log$ using its concavity, and taking the minimum over $A,B$ yields the statement.
 \qed

Let us next discuss (e4). Although the next lemma is a special case of the following proposition, we include it here, because the proof
is independent.

\begin{lemma}
\label{lem_embed}
Let $\A_{A_1} \subset \A_{A_2}, \A_{B_1} \subset \A_{B_2}$ let $\omega$ be a normal state on $\A_{A_2} \otimes \A_{B_2}$
satisfying our standing assumption. Then $E_M(\omega \restriction_{\A_{A_1} \otimes \A_{B_1}}) \le E_M(\omega)$.
\end{lemma}

{\em Proof:}
   We let $S_i$ be the Tomita operators for $\A_{B_i}'$ with polar decompositions $S_i = J_i \Delta_i^{1/2}$.  Note that,
since $\A_{B_2}' \subset \A_{B_1}'$, $\dom(S_2) \subset \dom(S_1)$. Let $\lambda>0$.
The set $\dom(S_1)$ is a Hilbert space called $\H_1$ with respect to the inner product
$\langle \Phi | \Psi \rangle_\lambda = \langle \Phi | \Psi \rangle + \lambda^{-1} \langle S_1 \Psi | S_1 \Phi \rangle$, where $\lambda>0$. Letting $I:\H_1 \to \dom(S_1)$
be the identification map, one shows that  $I^{-1} \dom(S_2)$ is a closed subspace $\H_2 \subset \H_1$ with associated orthogonal projection $P_2$.
\cite{fredenhagen_5} shows that $I P_i I^* = (1 + \lambda^{-1}\Delta_i)^{-1}$ (with $P_1=1$) and that $I^* = I^{-1}(1+ \lambda^{-1}\Delta_1)^{-1}$.
It follows for all $b \in \A_{B_2}'$ that
\ben\label{frede}
\langle \Omega | b^* (\lambda + \Delta_1)^{-1} b  \Omega \rangle -  \langle \Omega | b^* (\lambda + \Delta_2)^{-1} b  \Omega \rangle
=  \lambda \, \| (P_1- P_2) I^{-1} (\lambda + \Delta_1)^{-1} b \Omega \|^2 \ .
\een
A standard trick in such a situation is to use the identity ($1 \ge \alpha > 0, t>0$)
\ben
t^\alpha = \frac{\sin \pi\alpha}{\pi} \int_0^\infty \dd \lambda [\lambda^{\alpha-1} - \lambda^{\alpha} (\lambda + t)^{-1} ] \ .
\een
Then, if we multiply~\eqref{frede} by $\lambda^{\alpha}$,  and integrate against $\lambda$, we find via the spectral calculus
\ben
\label{del12}
\begin{split}
&\|\Delta_1^{\frac{\alpha}{2}} b \Omega  \|^2 - \|\Delta_2^{\frac{\alpha}{2}} b \Omega \|^2 \\
=& - \frac{\sin \pi\alpha}{\pi} \int_0^\infty \dd \lambda \, \lambda^{1+\alpha}  \| (1- P_2) I^{-1} (\lambda + \Delta_1)^{-1} b \Omega \|^2 \le 0 \ .
\end{split}
\een
For $\alpha=1/2$, we get\footnote{For an alternative argument, see lemma 2.9 of \cite{lechner_2}, which uses a generalization of the Heinz-L\" owner theorem~\cite{hansen} to unbounded operators.} $\|\Delta_1^{\frac{1}{4}} b \Omega \| \le \|\Delta_2^{\frac{1}{4}} b \Omega \|$ for all $b \in \A_{B_2}'$.
This entails the existence of an operator $T$ with $\|T\| \le 1$ such that $\Delta_1^{\frac{1}{4}} b |\Omega \rangle = T \Delta_2^{\frac{1}{4}} b |\Omega \rangle$
for all $b \in \A_{B_2}'$.

Since $\A_{A_1} \subset \A_{A_2} \subset \A_{B_2}'$ this relation holds for $a \in \A_{A_1}$ and we get from the definition of the map
$\Psi^{A_i}$ given above in eq.~\eqref{Psidefnuclear} that $\|\Psi^{A_1} \|_1 \le \| T \| \| \Psi^{A_2} |_{\A_{A_1}} \|_1 \le \|\Psi^{A_2} \|_1$. Since the same relation also
holds with $A$ replaced by $B$, we get $E_M(\F^* \omega) \le E_M(\omega)$ for the embedding
$\F:\A_{A_1} \otimes \A_{B_1} \to \A_{A_2} \otimes \A_{B_2}$, which is the claimed special case of (e4).
\qed

\medskip
\noindent
A more general, but still special, case of (e4) arises when $\cF$ is a unit-preserving *-homomorphism $\rho$ of $\A_A\otimes\A_B \cong \A_A \vee \A_B$ such that $\rho(\A_A) \subset \A_A$, and likewise for $A$ replaced by $B$. Such ``localized endomorphisms'' arise naturally in the context of the DHR-theory of superselection sectors (charged states) in QFT, see section \ref{sec:charged}. More generally we may consider a homomorphism $\rho: \A_{A_1} \vee \A_{B_1} \to  \A_{A_2} \vee \A_{B_2}$, or finite families thereof.

\begin{proposition}\label{e4em}
$E_M$ satisfies $\sum_i p_i E_M(\rho_i^* \omega) \le E_M(\omega)$ (here $\sum_i p_i = 1, p_i \ge 0$) for localized homomorphisms $\rho_i:
\A_{A_1} \vee \A_{B_1} \to \A_{A_2} \vee \A_{B_2}$ such that each $\omega_i=\rho^*_i \omega$ satisfies our standing assumption for $\A_{A_1} \vee \A_{B_1}$, and $\omega$ that for $\A_{A_2} \vee \A_{B_2}$.
\end{proposition}

{\em Proof:} Consider first a single localized endomorphism $\rho$.  Let $|\Omega_\omega\rangle, |\Omega_{\rho^*\omega}\rangle$ be the vector representatives of $\omega, \rho^*\omega$ in $\cP^\sharp$. It follows from the properties of $\rho$ that the linear operator $V$ defined by $Vx|\Omega_{\rho^*\omega}\rangle=\rho(x) |\Omega_\omega\rangle, x \in \A_{A_1} \vee \A_{B_1}$ is an isometry, $V^* V = 1$.
Next, let $S_\omega, S_{\rho^* \omega}$ be the Tomita operators for $|\Omega_\omega\rangle, |\Omega_{\rho^*\omega}\rangle$ for the v. Neumann algebras $\A_{B_2}, \A_{B_1}$. The trivial calculation
\ben
S_\omega Vb |\Omega_{\rho^*\omega}\rangle = \rho(b)^* |\Omega_\omega\rangle = \rho(b^*) |\Omega_\omega\rangle = Vb^* |\Omega_{\rho^*\omega}\rangle
= VS_{\rho^*\omega} b  |\Omega_{\rho^*\omega}\rangle
\een
for all $b\in\A_{B_1}$ establishes the operator equality $S_\omega V = VS_{\rho^*\omega}$ on the domain of $S_{\rho^*\omega}$. By taking adjoints we find on the form domain of $\Delta_{\rho^*\omega}$
\ben
V^*\Delta_{\omega}V=\Delta_{\rho^*\omega},
\een
where $\Delta_\omega$ is the modular operator for $\A_{B_2}$ and the state $\omega$ and similarly for $\rho^*\omega$. By the
Heinz-L\"owner theorem~\cite{hansen}, the function $\RR_+ \owns x \mapsto x^\alpha$ is operator monotone\footnote{A function $f: \RR \to \RR$ is called operator monotone if $f(A) \le f(B)$ whenever two self-adjoint operators $A,B$ on a Hilbert space $\H$ satisfy $A \le B$ on the form domain of $B$. If $A=B$ on the form domain of $B$ we obtain $f(A)\le f(B)$. Notice especially the asymmetry in the assumption on the form domain.}
% In this case we have $V^*\Delta_{\omega}V=\Delta_{\rho^*\omega}$ on the form domain of $\Delta_{\rho^*\omega}$.}
for $0<\alpha \le 1$, so we get $(V^* \Delta_\omega V)^\alpha \le \Delta_{\rho^*\omega}^\alpha$. Now we need the following result (see e.g. theorem~2.6 and 4.19 of~\cite{Carlen}):

\begin{lemma}
Let $f: \RR \to \RR$ be an operator monotone function, $V$ an operator such that $\| V \| \le 1$, $A$ a positive operator
on $\H$. Then $V^* f(A) V \le f(V^* A V)$ on the form domain of $f(V^* A V)$.
\end{lemma}

If we apply this lemma to $V$ and $A=\Delta_{\omega}$, we get $V^* \Delta_{\omega}^\alpha V \le (V^* \Delta_{\omega} V)^\alpha= \Delta_{\rho^*\omega}^\alpha$. This is the same as saying that
\ben
\Delta^{-\alpha/2}_{\rho^*\omega} V^* \Delta_{\omega}^{\alpha/2} (\Delta^{-\alpha/2}_{\rho^*\omega} V^* \Delta_{\omega}^{\alpha/2})^* \le 1.
\een
We use this with $\alpha=1/2$ and take the norm, which gives
\ben
\|
\Delta_{\rho^*\omega}^{-\frac{1}{4}} V^* \Delta_{\omega}^{\frac{1}{4}} \| \le 1 \ .
\een

The modular operator for $\A_{B_2}'$ required in the definition of $\Psi^A_\omega$ (eq. \eqref{Psidefnuclear}) is
$J_\omega \Delta_\omega J_\omega = \Delta_\omega^{-1}$, and similarly for $\rho^* \omega$. It follows using the definition \eqref{Psidefnuclear} that
\ben
\Psi_{\rho^* \omega}^A(a) = \Delta_{\rho^* \omega}^{-\frac{1}{4}} a |\Omega_{\rho^*\omega}\rangle
=  \Delta_{\rho^* \omega}^{-\frac{1}{4}} V^* V a |\Omega_{\rho^*\omega}\rangle
=  T \Delta_{\omega}^{-\frac{1}{4}} \rho(a) |\Omega_{\omega}\rangle
= T \circ \Psi^A_\omega \circ \rho (a)
\een
for all $a \in \A_{A_1}$, where $T= \Delta_{\rho^* \omega}^{-\frac{1}{4}} V^*  \Delta_{\omega}^{\frac{1}{4}}$. The properties of the 1-norm
then give $\| \Psi_{\rho^* \omega}^A \|_1 \le \|\rho\| \| T \| \| \Psi^A_\omega \|_1 \le \| \Psi^A_\omega \|_1$ and the same for $A$ replaced by
$B$. It follows that $Z(\rho^* \omega) \le Z(\omega)$ and hence that $E_M(\rho^* \omega) \le E_M(\omega)$.

Consider next a finite family of localized endomorphisms $\rho_i$. In this case, the result immediately follows from the previous result and the concavity of $\log$ as (with $\omega_i = \rho_i^* \omega$ and using $Z(\omega_i) \le Z(\omega)$ for the Buchholz partition function):
$
\sum_i p_i E_M(\rho_i^* \omega) = \sum_i p_i \log Z(\omega_i) \le \log \sum_i p_i Z(\omega_i) \le \log Z(\omega) = E_M(\omega).
$
\qed

\medskip
\noindent
This concludes our discussion of (e4). Whether the general case of (e4) holds for families of separable operations is unknown to us. Perhaps one could say that at any rate, the properties expressed by prop. \ref{e4em} and \ref{lem_embed} are the more natural ones in the context of QFTs. Property (e5) is  satisfied since the modular operator behaves functorially under tensor products. Property (e6) is not obvious to us.

\subsection{Distillable entanglement}\label{sec:ED}

The last measure that we will discuss is closely related to ``entanglement distillation''~\cite{rains,Verch_2} and is maybe the most natural of all entanglement measures.
This measure is formulated in terms of {\bf maximally entangled states}, which are defined for finite dimensional type~I factors of the form $\A_A \otimes \A_B$, with
$\A_A = M_n(\CC) = \A_B$. Their density matrix is
\ben
P^+_n = |\Phi^+ \rangle \langle \Phi^+| \ , \quad |\Phi^+ \rangle = \frac{1}{\sqrt{n}} \sum_{i=1}^n |i \rangle \otimes |i \rangle \ ,
\een
where $\{|i \rangle \}$ is a chosen orthonormal basis of $\CC^n$.
The corresponding linear functional is denoted by $\omega^+_n = \tr(P^+_n \ . \ )$ in the following.
In view of (e4) it is justified to think of these states as maximally entangled because one can show~\cite{plenio} that
any other state $\omega$ (pure or mixed) on $\A_A \otimes \A_B$ can be obtained as $\omega = \F^* \omega^+_n$ by a separable operation,
i.e. a normalized cp map $\F$ on $\A_A \otimes \A_B$ that is a convex linear combination of local cp maps of the form~\eqref{sepop}.

The relative entanglement entropy of the maximally entangled state is for instance given by $E_R(\omega^+_n) = \log n$, as one can see using that
$E_R(\omega)=H_{\rm vN}(\omega_A)=H_{\rm vN}(\omega_B)$ for all pure states $\omega$.
Since $E_R(\omega) \le E_N(\omega)$ in general, we also have $E_N(\omega^+_n) \ge \log n$, whereas from lemma~\ref{Lem_dominating_sigma}, we easily get
$E_N(\omega^+_n) \le \log n$, implying equality, $E_N(\omega^+_n) = \log n$. Furthermore, by construction $E_I(\omega^+_n) = 2 \log n$, and $E_B(\omega^+_n) = \sqrt{2}$,
since $\omega_n^+$ can be mapped to a Bell-state by a local operation.

Now let $\A_A, \A_B$ be general v. Neumann algebras and $\omega$
a normal state on $\A = \A_A \otimes \A_B$.  The idea of distillation is to take a large number $N$ of copies of this bipartite system $\omega^{\otimes N}$ and ``distill'' from this ensemble a maximally entangled state $\omega^+_n$ -- for as large an $n=n_N$ depending on $N$ as we can --
by separable operations. More precisely, we consider sequences $\{n_N\}$ of natural numbers and sequences $\{\F_N\}$ of separable operations, i.e. normalized cp maps $\F_N: M_{n_N}(\CC) \otimes M_{n_N}(\CC) \to \A_A^{\otimes N} \otimes \A_B^{\otimes N} = \A^{\otimes N}$  that are
each convex linear combinations of cp maps of the form~\eqref{sepop} and have $\F_N(1)=1$, such that
\ben
\| \F_N^*\omega^{\otimes N} - \omega^+_{n_N} \| \to 0 \quad \text{as $N \to \infty$.}
\een
If such sequences exist, then we call $\omega$ ``{\bf distillable}'' and the sequence $\{ \F_N \}$ a ``distillation protocol''. (This and the following notions clearly do not depend on the choice of basis made above since basis rotations can be implemented by separable operations).

The notion of distillable entropy captures the efficiency of this process. Since $\log n_N$ is the relative entanglement entropy of the reference state $\omega_{n_N}^+$, the distillation process would  be considered as rather inefficient if $\log n_N \ll N$ asymptotically, while we would consider the distillation process to possess a finite rate if $\log n_N \propto N$ asymptotically, and the rate itself would be the proportionality constant. The entanglement achieved by the distillation processes $\{ \F_N \}$ is defined to be this  rate, i.e. we set\footnote{We may always pass to a new protocol whose rate is arbitrarily close, so that the $\limsup$ is actually a $\lim$.}
\ben
E_{\{ \F_N \}}(\omega) = \limsup_{N \to \infty} \frac{\log n_N}{N} \ .
\een
The distillable entropy is the optimum rate achievable by any such process, i.e. one defines:
\begin{definition}
The distillable entropy is defined by $E_D(\omega) = \sup_{\{\F_N\}} E_{\{ \F_N \}}(\omega)$.
\end{definition}

Some general properties of $E_D(\omega)$ are immediately clear from the definition. For instance, we clearly have (e0) and also $E_D(\omega)\ge0$. (e2) is fairly obvious from the definition, too. We prove (e4) as a lemma:

\begin{lemma}
$E_D(\omega)$ satisfies (e4), i.e. for any separable operation $\{\cE_i\}$ with $\sum_i \cE_i(1)=1$, $\omega(\cE_i(1))=p_i>0$, we have
$\sum_ip_iE_D(p_i^{-1}\cE_i^*\omega)\le E_D(\omega)$.
\end{lemma}

{\em Proof:}
For local operations, i.e. normalized cp maps $\cE$ of the form~\eqref{sepop}, this is immediate. For general separable operations, let
$\{ \F_{i,N} \}$ be near optimal distillation protocols for $\omega_i = p_i^{-1} \cE^*_i \omega$ with rates $r_i$ within $\epsilon$ of $E_D(\omega_i)$. For any $N$ let $N_i=\lfloor p_i N \rfloor$ and $M:=N-\sum_iN_i\ge0$. Define the following protocol for $\omega$:
\[
\hat \F_N = 1^{\otimes M}  \otimes (p_1^{-1} \cE_1)^{\otimes N_1} \F_{1,N_1} \otimes (p_2^{-1} \cE_2)^{\otimes N_2} \F_{2,N_2} \otimes \cdots \quad ,
\]
where $1^{\otimes M}$ is the map $M_1(\CC)\ni 1\mapsto 1\in \A_A^{\otimes M}$. Because $\omega_N^+=\omega_M^+\otimes\omega_{N_1}^+\otimes\omega_{N_2}^+\otimes\ldots$ it follows straightforwardly that $\{ \hat \F_N \}$ is a distillation protocol for $\omega$ with rate $\sum_i r_i$. (e4) then follows. \qed

Instead of (e5) we have
\ben
\frac{1}{N} E_D(\omega^{\otimes N}) = E_D(\omega) \ ,
\een
as is obvious from the definition. $E_D(\omega)$ is not in general convex in the sense of property (e3), although it is convex on pure states (this property is not so obvious). Since there are no pure states in the type III case relevant for quantum field theory, this is at any rate not helpful for us, and we have no analogue of (e3).

% New version:
For us, it is most important that the distillable entropy has the superadditivity property (e6), as remarked without proof e.g. in~\cite{wolf_1}.

\begin{lemma}
$E_D(\omega)$ satisfies (e6), i.e. for any state $\omega$ on $\A = (\A_{A_1} \otimes \A_{A_2})\otimes(\A_{B_1}\otimes\A_{B_2})$
the restrictions $\omega_i$ to $\A_{A_i}\otimes\A_{B_i}$ satisfy $E_D(\omega_1)+E_D(\omega_2)\le E_D(\omega)$.
\end{lemma}

{\em Proof:}
Let $\{ \F_{N,1} \}$ be a distillation protocol for $\omega_1$, i.e. a sequence of cp maps such that $\| \F_{N,1}^* \omega_1^{\otimes N} - \omega_{n_{N,1}}^+ \| \to 0$. The rate of this protocol is $\lim (\log n_{N,1})/N = r_1$, and similarly for $\omega_2$. We claim that $\{ \F_{N,1} \otimes \F_{N,2} \}$ is a distillation protocol for $\omega$ with rate $r=r_1+r_2$. Letting $\sigma_N = (\F_{N,1} \otimes \F_{N,2})^* \omega^{\otimes N}$, and $X \in M_{n_{N,1}} (\CC) \otimes M_{n_{N,2}}(\CC) $ and $1_i \equiv 1_{n_{N,i}}, P^+_i \equiv P^+_{n_{N,i}}$, we have
\ben\label{project}
\begin{split}
|\sigma_N ( (1_1 - P^+_{1}) \otimes 1_2 \cdot X  ) |  & \le  \sigma_N \Big( (1_1 - P^+_{1}) \otimes 1_2 \cdot X X^* \cdot (1_1 - P^+_{1}) \otimes 1_2 \Big)^{\frac12} \\
& \le \| XX^* \|^{\frac12} \sigma_N \Big( (1_1 - P^+_{1}) \otimes 1_2 \Big)^{\frac12} \\
&= \|X\| \ \Big| \F_{N,1}^* \omega_1^{\otimes N}(1_1 - P^+_{1}) - \omega_{n_{N,1}}^+(1_1 - P^+_{1}) \Big|^{\frac12}   \\
&\to 0 \quad \text{as $N \to \infty$}
\end{split}
\een
uniformly for $\|X\| \le 1$. The same conclusion can be drawn, with similar proof, for $\sigma_N ( X \cdot (1_1 - P^+_{1}) \otimes 1_2)$ and for $(1 \leftrightarrow 2)$. We have
\ben
\omega^+_{n_{N,1} n_{N,2}}(X)  = \omega_{n_{N,1}}^+ \otimes \omega_{n_{N,2}}^+(X)
=  \frac{\sigma_N(P_1^+ \otimes P_2^+ \cdot X \cdot P_1^+ \otimes P_2^+)}{\sigma_N(P_1^+ \otimes P_2^+)} \ ,
\een
and therefore in view of~\eqref{project} (and the analogous relations for $(1 \leftrightarrow 2)$),
\ben
\begin{split}
&|(\F_{N,1} \otimes \F_{N,2})^* \omega^{\otimes N}(X) - \omega^+_{n_{N,1} n_{N,2}}(X) | \\
&=
|\sigma_N(X) - \omega_{n_{N,1}}^+ \otimes \omega_{n_{N,2}}^+(X) | \\
& \le |\sigma_N(X) - \sigma_N(P_1^+ \otimes P_2^+ \cdot X \cdot P_1^+ \otimes P_2^+)| + \left|1-\sigma_N(P_1^+ \otimes P_2^+)^{-1}\right| \|X\| \to 0
\end{split}
\een
uniformly for $\| X\| \le 1$, which immediately gives the claim. Thus, we see that $\{ \F_{N,1} \otimes \F_{N,2} \}$ is a distillation protocol, whose rate is evidently $r_1+r_2$.
Choosing $r_i$ arbitrarily close to $E_D(\omega_i)$, we see that there is a protocol for $\omega$ whose rate is at least $E_D(\omega_1) + E_D(\omega_2) - \epsilon$ for
any $\epsilon>0$, which implies superadditivity (e6). \qed

It might be guessed from the involved variational characterization of $E_D(\omega)$ that this quantity is difficult to calculate in practice even for the simplest examples, and this expectation turns out to be correct. For us, the usefulness of this quantity lies in the fact it has the very convenient property (e6), and that it is a lower bound for a large class
of entanglement measures, in particular the relative entropy of entanglement\footnote{We even get the statement for the ``asymptotic'' relative entropy of entanglement defined by
$E_R^\infty(\omega) =
\lim_{n \to \infty} \frac{1}{n} E_R(\omega^{\otimes n})$. Note that the limit exists: Use (e5) and
lemma~12 of~\cite{donald_2}.}:

\begin{theorem}\label{Thm:dominance}
For any normal state on $\A = \A_A \otimes \A_B$, and any entanglement measure satisfying (e2), (e4), (e5) and normalization
$E(\omega_n^+) = \log n$, we have $E(\omega) \ge E_D(\omega |_{\mathfrak N})$ and in particular
$E_R(\omega) \ge E_D(\omega |_\mathfrak{N})$, where $\mathfrak{N}$ is any type I subfactor of $\A$ of the form $\mathfrak{N} = \mathfrak{N}_A \otimes \mathfrak{N}_B$,
with $\mathfrak{N}_A \subset \A_A,   \mathfrak{N}_B \subset \A_B$ intermediate type I subfactors.
\end{theorem}

\begin{remark}
The existence of many such intermediate type I subfactors exhausting $\A$ is guaranteed by the split property.
\end{remark}

The proof of this theorem is given in~\cite{donald_2} for the case of finite-dimensional type~I algebras, where
it is shown more precisely that $\frac{1}{N} E(\omega^{\otimes N}) \ge E_D(\omega) - \epsilon$ for sufficiently large $N$ depending on $\epsilon>0$. This immediately implies
the theorem in view of (e5) for finite-dimensional type~I algebras.  Inspection of the proof~\cite{donald_2} shows that it can be generalized fairly easily to the
case of infinite-dimensional type~I algebras by a straightforward approximation argument.
The general case then follows trivially in view of (e4), $E_R(\omega) \ge E_R(\omega |_\mathfrak{N})$. \\

\subsection{Summary of entanglement measures}\label{SSec_summarymeasures}

We summarize the various entanglement measures and some of their properties and relationships in the following table. Unlisted properties may either be false or unknown to the authors.\footnote{Property (e4) for $E_M$ has been proven only in a restricted sense, and instead of (e3) we have the opposite: concavity
$\overline{{\rm (e3)}}$.}

\begin{center}
    \begin{tabular}{| l | l  | l | l | l |}
    \hline
    Measure & Properties & Relationships & $E(\omega_n^+)$ \\ \hline
     $E_B$ & (e0), (e2), (e3), (e4) & &  $\sqrt{2}$ \\ \hline
     $E_D$ & (e0), (e1), (e2), (e4), (e6) & $E_D \le E_R, E_N, E_M, E_I$ &  $\log n$ \\ \hline
     $E_R$ & (e0), (e1), (e2), (e3), (e4), (e5) & $E_D \le E_R \le E_N, E_M, E_I$ &  $\log n$ \\ \hline
     $E_N$ & (e0), (e1), (e4), (e5) & $E_D, E_R \le E_N \le E_M$ &  $\log n$ \\ \hline
     $E_M$ & (e0), $\overline{{\rm (e3)}}$, (e4), (e5) & $E_D, E_R, E_N \le E_M$ &  $\tfrac{3}{2} \log n$ \\ \hline
     $E_I$   & (e0), (e2), (e4), (e5) &  $E_D, E_R \le E_I$ &  $2 \log n$ \\ \hline
    \end{tabular}
\end{center}

\section{Upper bounds for $E_R$ in QFT}\label{sec:upper_bounds}

In this section, we derive some upper bounds on the relative entanglement entropy in quantum field theories. To illustrate the idea, let us first consider the spatial slice $\C=\{t=0\}$ in $d+1$-dimensional Minkowski spacetime, and let $A$ and $B$ be two disjoint open regions in $\C$. Let $O_A$ and $O_B$ be the domains of dependence of these regions and $\A(O_A)$ and $\A(O_B)$ the associated algebras, see fig.~\ref{fig:diamond}. The ground state (vacuum) $\omega_0$ of the QFT gives rise to a GNS-triple $(\pi_0,\H_0,|0\rangle)$, which yields the von Neumann algebras $\A_A:=\pi_0(\A(O_A))''$ and $\A_B:=\pi_0(\A(O_B))''$. Note that $\A_A$ and $\A_B$ commute, due to causality. We want to investigate when
$\omega_0$ defines a normal state on $\A_A\otimes\A_B$, i.e. when $\A_A$ and $\A_B$ are statistically independent, and what its relative entanglement entropy is.

When the theory satisfies the BW-nuclearity condition a5), the statistical independence of $\A_A$ and $\A_B$ follows from the assumption that the distance ${\rm dist}(A, B)>0$ is positive. Moreover, we will show in section \ref{SSec_Hamiltonian} below that $E_R(\omega_0|_{\A_A\otimes\A_B})$ can be estimated in terms of the 1-nuclear norm of the operator $\Theta$ which appears in the definition of the BW-nuclearity condition. Using similar methods one can also estimate the relative entanglement entropy of thermal (KMS) states.

The illustrative example above can easily be generalised as follows. We can choose two disjoint regions $A$ and $B$ in a Cauchy surface $\C$ of a globally hyperbolic spacetime $(\sM,g)$, and we let $O_A$ and $O_B$ denote their domains of dependence, see fig.~\ref{fig:diamond}. We let $\omega$ be a state on the QFT on the entire spacetime with GNS-triple $(\pi,\H,|\Omega \rangle)$, and we introduce the von Neumann algebras $\A_A:=\pi(\A(O_A))''$ and $\A_B:=\pi(\A(O_B))''$. Once again we can ask whether $\omega$ defines a normal state on $\A_A\otimes\A_B$ and what its relative entanglement entropy is. In general, however, the BW-nuclearity condition is no longer available, due to the lack of a Hamiltonian operator. For this reason we will consider a modular nuclearity condition instead.

Let us assume for simplicity\footnote{A modular operator can still be defined when these assumptions are not met~\cite{lechner_2}, and our estimates, e.g. in the proof of lemma \ref{Lem2}, still hold.}
that $|\Omega \rangle$ is cyclic and separating for $\A_A$ and for $\A_B$. (This will be the case e.g. if $\omega = \omega_0$ on Minkowski spacetime, or if $\omega$ is any other state with finite energy, by the same argument as in the Reeh-Schlieder theorem. For results in curved spacetimes, see e.g.~\cite{sanders_1}.) In particular, $|\Omega \rangle$ is then cyclic and separating for $\A_{B'}=\pi(\A(O_{B'}))''$, where $B':=\C\setminus \overline{B}$. Let $\Delta_{B'}$ be the corresponding modular operator, and note that $\A_{B'} \subset \A_B'$. Instead of the BW-nuclearity condition a5),  we are going to impose/use the following \textbf{modular nuclearity condition}:

\begin{enumerate}
\item[a5')]
The operator
\[
\Phi^A: \A_A \to \H \ , \quad \Phi^A(a) =  \Delta^{\quarter}_{B'} a| \Omega \rangle \
\]
has $\|\Phi^A\|_1<\infty$ when ${\rm dist}(A, B)>0$ is positive.
\end{enumerate}

As shown in~\cite{buchholz_3}, a5') implies a5) in Minkowski spacetime without the bounds on the nuclear norms.
The nuclearity condition a5') again suffices to prove the statistical independence of $\A_A$ and $\A_B$, and we can estimate the relative entanglement entropy of $\omega$ between $\A_A$ and $\A_B$ using $\|\Phi^A\|_1$, making use of theorems \ref{Thm_ER<EN} and \ref{Thm_EN<Em}. Indeed, noting that $\A_{B'} \subset \A_B'$, it follows arguing as in the proof of
prop. \ref{Emdominance} that $\| \Psi^A \|_1 \le \| \Phi^A \|_1$, where $\Psi^A$ is the map~\eqref{Psidefnuclear} appearing
in the definition of $E_M$.

In the first section below, we point out some general upper bounds that follow from the BW-nuclearity condition. (The corresponding results for modular operators already follow from the theorems \ref{Thm_ER<EN} and \ref{Thm_EN<Em}.) In the sections after that we apply the various nuclearity conditions to obtain concrete upper bounds for free fields, 2-dimensional integrable models with factorizing scattering-matrices and CFTs.

\subsection{General upper bounds from BW-nuclearity}\label{SSec_Hamiltonian}

The first type of general bound is for the ground (vacuum) state, $\omega_0$, and holds for a theory on Minkowski spacetime
with mass gap satisfying the BW-nuclearity condition a5). Our result is:
\begin{theorem}\label{Thm_m>0}
1) Assume that the Hamiltonian $H=P^0$ in the vacuum representation has a mass gap, ${\rm spec}(H) \subset \{0\} \cup [m, \infty)$, with $m>0$. Let $A$ and $B$ be contained in balls of radius $r$ in a $t=0$ time slice, separated by the distance $R$. As usual we set $\A_A = \pi_0(\A(O_A))''$, where $O_A$ is the causal diamond with base $A$, and similarly for $B$.
Assume that the BW-nuclearity condition~\eqref{BW} holds with constants $c,n$.
Then the relative entanglement entropy in the vacuum state between $A$ and $B$ satisfies, for any $k<1$, an upper bound of the form
\ben
E_R(\omega_0) \lesssim C \exp[-(mR)^k] \ ,
\een
for sufficiently large $R/c \gg 1$, where $C$ is a constant depending on $c,k,n$.

2) Under the same assumptions as in 1) but not necessarily $m>0$, we have for
$R/c \ll 1$ an upper bound of the form
\ben
E_R(\omega_0) \lesssim  \sqrt{2} \left( \frac{c \ctg\frac{\pi}{4n}}{R  } \right)^n .
\een
\end{theorem}

Thus, we see from part 1) that the relative entanglement entropy decays almost exponentially as the separation $R$ between the regions tends to infinity in the presence of a mass gap. Since free bosons~\cite{buchholz_4} and fermions~\cite{dantoni} in $d+1$ dimensions are known to satisfy the BW-nuclearity condition, one immediately gets almost exponential decay in those models. The upper bound expressed by part 2) for short distances shows that the growth of $E_R$ is not faster than an inverse power of $R$, which depends on the constant $n$ in the nuclearity condition. For free fields, $n=d, c \propto r$, giving thus for $R \ll r$
\ben
E_R(\omega_0) \lesssim c_d \left( \frac{r}{R } \right)^d \ ,
\een
so this upper bound falls short of the expected ``area law''. For a better bound qualitatively consistent with the area law in the case of Dirac fields, see sec.~\ref{sec:areaDirac}.

The second theorem is concerned specifically with thermal states (KMS states) $\omega_\beta$. Again, we take $A$ and $B$ as balls of radius $r$ separated by the distance $R$ in some slice $\RR^d$  of Minkowski spacetime for simplicity. We go to the GNS-representation $(\H_\beta, \pi_\beta, |\Omega_\beta \rangle)$ of $\omega_\beta$, in which this state is represented by the vector $|\Omega_{\beta}\rangle$. The state $\omega_\beta$ is not pure and the representation is highly reducible. The time-translation subgroup is implemented by the unitary $U_\beta(t) = e^{itH_\beta}$. Unlike in the vacuum representation, the generator $H_\beta$ always has as its spectrum the entire real line ${\rm spec}(H_\beta) = \RR$, even if the theory has a mass gap in the vacuum sector. As a replacement of the BW-nuclearity condition in this case, we assume that
1) the point $\{0\}$ is a non-degenerate eigenvalue of $H_\beta$ associated with the vector $|\Omega_\beta\rangle $, 2) Letting $P_+$ be the spectral projection of $H_\beta$ corresponding to the set $(0, \infty)$ and $P_-$ that corresponding to the set $(-\infty, 0)$, the maps $\Theta_{z}^\pm: \A_B \to \H_\beta$
\ben\label{Eqn_Theta_rz}
\Theta^\pm_{r,z}(b) := P_\pm e^{\pm iz H_\beta} b |\Omega_\beta \rangle \ , \qquad \Im(z) > 0,
\een
are assumed to be nuclear. By invariance under spatial translations, the same then obviously holds for the region $A$.
\begin{theorem}\label{Thm_KMS}
Assume that $A$ and $B$ are as in theorem \ref{Thm_m>0} and assume
\ben\label{thbound}
\| \Theta^\pm_{r,z}\|_1 \le
  |\Im(z)|^{-\alpha}
\exp (c/|\Im(z)|)^n \een
 for a constant $c$ depending on $r,\beta$ and $n>0,\alpha>1$. Then
 \ben
 E_R(\omega_\beta)  \le C R^{-\alpha+1}
 \een
 for sufficiently large $R=dist(A,B)$ and a constant $C$ depending on $r,\beta$.
\end{theorem}

The theorem leaves the possibility
that the relative entanglement entropy of two regions in a thermal state may decay more slowly than that in a vacuum state.
The precise rate of the upper bound
is related to spectral information, which in our case is encoded in the assumption about the nuclear norms. The last theorem concerns massless theories.

\medskip

\begin{theorem}\label{Thm_m=0}
In a massless theory on Minkowski spacetime with vacuum $\omega_0$, assume that the 1-norm of the map $\Theta_{r,z}: a \mapsto e^{izH} a |0\rangle$ fulfills
\eqref{thbound} for $\Im z > 0$. Then
\ben
E_R(\omega_0) \le C R^{-\alpha}
\een
for sufficiently large $R=dist(A,B)$ and a constant $C$ depending on $r$.
\end{theorem}

The strategy to prove these theorems is as follows: we show that we can write the relevant state $\omega$ on $\A_A\otimes\A_B$ in the form $\omega=\sum_j\phi_j\otimes\psi_j$, with normal linear functionals $\phi_j$ and $\psi_j$ satisfying a bound on $\nu:=\sum_j\|\phi_j\|\cdot\|\psi_j\|$, which is controlled by the respective nuclearity condition assumed in each theorem. The desired bound on  $E_R(\omega)$ is then obtained in conjunction with lemma \ref{Lem_dominating_sigma} and theorem \ref{Thm_ER<EN}, which imply that
\ben\label{masterbound}
E_R(\omega) \le \log \nu.
\een
We now turn to the detailed estimation in each case.

\subsubsection{Proof of thm. 6}

The basic idea of all three proofs is to consider suitable correlation functions
$f_{a,b}(z)$ whose analyticity properties in $z$ encode the commuting nature of $\A_A$ and $\A_B$. Part 1) of the proof of the present theorem relies on an argument due to~\cite{fredenhagen_2,fredenhagen_3} (up to a slightly better control on the bounds), which we repeat here merely for the convenience of the reader and to set the stage for part 2).

\medskip
\noindent
{\em Proof of thm. 6, Part 1):}  Following~\cite{fredenhagen_2,fredenhagen_3}, we consider the function
\ben\label{f1def}
f_{a,b}(z) =
\begin{cases}
\langle 0| a P e^{+izH} b |0\rangle & \text{for $0< \Im (z)$,}\\
\langle 0| b P e^{-izH} a |0\rangle & \text{for $0> \Im (z)$,}
\end{cases}
\een
where $a \in \A_A, b \in \A_B$, and where $P=1-|0\rangle \langle 0|$. $z \mapsto f_{a,b}(z)$ is by construction analytic for $\Im(z) \neq 0$, i.e. away from the real axis. For the jump across the real axis, one finds
\ben
f_{a,b}(t+i0) - f_{a,b}(t-i0) = \langle 0 | [a, e^{itH} b e^{-itH}] |0\rangle = 0
\een
as long as $|t|<R$, because then the time-translated region $O_B$ (by $t$) remains space-like to $O_A$. By the edge-of-the-wedge theorem
(see appendix~\ref{eow}),
the function $f_{a,b}(z)$ may thus be extended to an analytic function in the doubly cut plane $\CC \setminus \{z \in \CC \mid \Im(z) = 0, |\Re(z)| \ge R\}$.
We consider next for $|t|<R$ the mapping $w \mapsto z=2t/(w+w^{-1})$ which maps the open disk $\{ |w|<1 \}$
into the doubly cut plane. The image of the contour $C_\rho: \varphi \mapsto \rho e^{i\varphi}, 0< \rho<1$ in the doubly cut $z$-plane under this mapping is illustrated in the following fig.~\ref{fig:doubly_cut}. Applying Cauchy's formula to this contour gives
\ben
\begin{split}
\langle 0 | ab | 0 \rangle - \langle 0 | a | 0 \rangle \langle 0 |b | 0 \rangle &= \int_{C_\rho} \frac{\dd w}{2\pi i \ w} f_{a,b}\left( \frac{2t}{w+w^{-1}} \right) \\
&\to
\frac{1}{2\pi} \int_0^{2\pi} \dd \varphi \ f_{a,b}\left(\frac{t}{\cos \varphi} \right)  \ ,
\end{split}
\een
sending $\rho \to 1$ in the second line.

\begin{figure}[h!]
\begin{center}

\begin{tikzpicture}
	%\tkzInit[xmin= -3.5, ymin= -3.5, xmax=3.5, ymax=3.5]
	%\tkzDrawXY[noticks]
	\tkzDefPoint(0,0){O}
	\tkzDefPoint(0.5, 0.2){B}
	\tkzDefPoint(-0.5, 0.2){BB}
	\tkzDefPoint(0.5, 0-.2){C}
	\tkzDefPoint(-0.5, 0-.2){CC}
	\tkzDefPoint(2.64, 1.45){D}
	\tkzDefPoint(-2.64, 1.45){DD}
	\tkzDefPoint(2.64, -1.45){E}
        \tkzDefPoint(-2.64, -1.45){EE}
        %\tkzDrawArc[color=red,line width=1pt](O,B)(C)
        %\draw[color = green, line width = 1 pt] (B) -- (C);
        %\draw[color = green, line width = 1 pt] (BB) -- (CC);
        %\draw[color = blue, line width = 1 pt] (DD) -- (BB);
        %\draw[color = blue, line width = 1 pt] (CC) -- (EE);
        %\begin{scope}[decoration={markings,
 	   % mark=at position .5 with {\arrow[scale=2]{>}};}]
	    %\tkzDrawSegments[postaction={decorate},color=red,line width=1pt](B,D E,C)
        %\end{scope}
        \begin{scope}[decoration={markings,
	    mark=at position .20 with {\arrow[scale=1.5]{latex}},
	    mark=at position .70 with {\arrow[scale=1.5]{latex}};}]
	    \tkzDrawArc[postaction={decorate}, color=black, line width=1pt](O,D)(DD)
        \end{scope}
        \begin{scope}[decoration={markings,
	    mark=at position .20 with {\arrow[scale=1.5]{latex}},
	    mark=at position .70 with {\arrow[scale=1.5]{latex}};}]
	    \tkzDrawArc[postaction={decorate}, color=black, line width=1pt](O,EE)(E)
        \end{scope}
        % deformation of contour1
        \draw [black, line width=1pt] plot [smooth] coordinates {(2, 0.72) (0.5, 0) (2, -0.72)};
        \draw [black, line width=1pt] plot [smooth] coordinates {(-2, 0.72) (-0.5, 0) (-2, -0.72)};
        \draw [black, line width=1pt] plot [smooth] coordinates {(2, 0.72) (2.7, 1)  (2.64, 1.45)};
        \draw [black, line width=1pt] plot [smooth] coordinates {(-2, 0.72) (-2.7, 1)  (-2.64, 1.45)};
        \draw [black, line width=1pt] plot [smooth] coordinates {(-2, -0.72) (-2.7, -1)  (-2.64, -1.45)};
        \draw [black, line width=1pt] plot [smooth] coordinates {(2, -0.72) (2.7, -1)  (2.64, -1.45)};
        % contour2
         \draw [black, line width=1pt] plot [smooth] coordinates {(3.0, 0.1) (1.3, 0) (3.0, -0.1)};
         \draw [black, line width=1pt] plot [smooth] coordinates {(-3, 0.1) (-1.3, 0) (-3, -0.1)};

        \filldraw (1.5, 0) circle (1pt);
        \filldraw (-1.5, 0) circle (1pt);
        % axis
        \draw[dashed] (- 1.5, 0) -- (1.5, 0);
        \draw (-3.5,0) -- (-1.5,0);
        \draw[%width=0.2mm,
        >={Triangle[length=2mm, width=1mm]}, ->] (1.5, 0) -- (3.5, 0);
        \draw[%width=0.2mm,
        >={Triangle[length=2mm, width=1mm]}, ->] (0, -3.5) -- (0, 3.5);
        % labelling
        \node[right] at (0, 3.5) {$\Im(z) $};
        \node[below] at (3.5, 0) {$\Re(z) $};
        \node[below] at (1.5, 0) {$ R $};
        \node[below] at (-1.5, 0) {$ - R $};
    \end{tikzpicture}

\end{center}
\caption{The image of the contour $C_\rho$ for two different values of $\rho$.}
\label{fig:doubly_cut}
\end{figure}
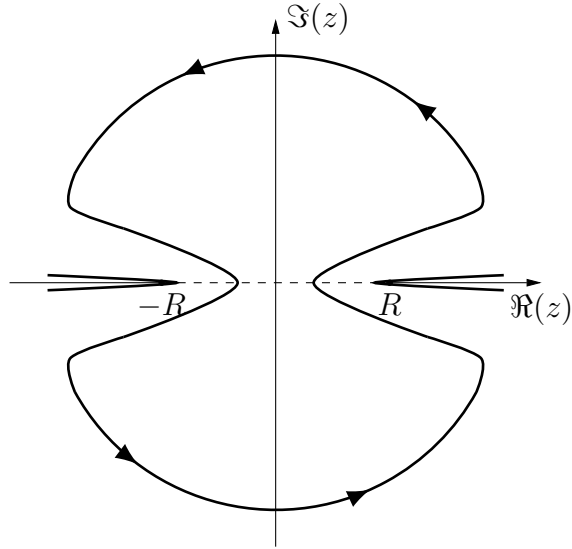

For the part above resp. below the real axis we can use the respective representations of $f_{a,b}$,
and this gives
\ben
\begin{split}
\langle 0 | ab | 0 \rangle - \langle 0 | a | 0 \rangle \langle 0|b | 0 \rangle
&=
\frac{1}{2\pi} \int_0^\pi \dd \varphi \, \langle 0| a P \exp(+itH/\cos \varphi) b | 0 \rangle \\
&+ \frac{1}{2\pi} \int_\pi^{2\pi} \dd \varphi \, \langle 0| b P \exp(-itH/\cos \varphi) a | 0 \rangle \
\end{split}
\een
for all $|t|<R$. Following~\cite{fredenhagen_3}, we next proceed as follows.
We multiply this identity with a smooth test function of the form $g_k(t/R)/R$ where the support of $g_k(t)$ is contained in $(-1,1)$.
For convenience, we normalize $g_k(t)$ so that $\int_\RR g_k(t) \dd t = 1$ and furthermore make a choice such that $|\tilde g_k(E)| \le C_k \exp(-|E|^k)$, for $k<1$
for $|E| \to \infty$, and for some $C_k>0$.
It is well-known that such a choice is possible, see e.g.~\cite{jaffe,ingham}.\footnote{A permissible non-normalized choice of $g_k(t)$ is for instance
$g_k(t) = h_k(1+t)h_k(1-t)$, where $h_k(t)$ has the Fourier Laplace transform
$\int_0^\infty h_k(t) e^{-Et} \, dt = e^{-E^k}$.
For further discussion and explicit formulas, see e.g.~\cite{fewster_4}, where one finds
\ben
h_k(t) = -\frac{1}{\pi} \sum_{j\ge 1} \frac{(-1)^j \Gamma(jk+1) \sin(\pi jk)}{j!t^{jk+1}} \ .
\een
}
This gives a representation of the form
\ben
\begin{split}
\langle 0 | ab | 0 \rangle - \langle 0 | a | 0 \rangle \langle 0|b | 0 \rangle
&= \langle 0 | a G(-H)P b | 0 \rangle + \langle 0 |  b G(H)P a | 0 \rangle \\
&=
\langle 0 | b \Xi^+(a) | 0 \rangle + \langle 0 | a\Xi^-(b) | 0 \rangle \ ,
\end{split}
\een
where $\sqrt{2\pi} G(E) = \int_0^\pi \tilde g(RE/\cos \varphi) \ d\varphi$. In particular, it follows that $|G(E)|$ is of order $\exp(-|RE|^k)$ for sufficiently large $|RE|$.
The terms on the right side define maps $\Xi^+: \A_A \to \H$ resp. $\Xi^-: \A_B \to \H$ by
 $\Xi^+(a) = P G(H) a |0 \rangle$ respectively
$\Xi^-(b) =  P G(-H) b |0 \rangle$. Let $E_j$ be the spectral projector for $H$ onto the interval $[mj, m(j+1))$, where $j=0,1,2, \dots$. We may
then write
\ben
\Xi^+(a) =
G(H)P a | 0 \rangle = \sum_{j=1}^\infty e^{\beta_j H} G(H) E_j \Theta_{\beta_j,r}(a) \ ,
\een
where the $\beta_j>0$ are to be chosen. (There is no $j=0$ term in the sum  because $P$ projects out the vacuum state and $H$ has a mass gap.)
Using this formula, we estimate:
\ben
\begin{split}
\| \Xi^+ \|_1 &\le \sum_{j=1}^\infty \| G(H) E_j \| e^{\beta_j(j+1) m} \| \Theta_{\beta_j, r} \|_1 \\
&\le C_k \sum_{j=1}^\infty e^{-(Rmj)^k} e^{\beta_j m(j+1)}   e^{(c/\beta_j)^n} \\
&\le C_k \sum_{j=1}^\infty e^{-(Rmj)^k} e^{3(cmj)^{n/(n+1)}}
\end{split}
\een
using the properties of $G$ and the nuclearity assumption to go to line two, and making the choice
$\beta_j = c(cmj)^{-1/(n+1)}$ to go to line three. Choosing any $k>n/(n+1)$, there follows the bound
$\|\Xi^+ \|_1 \lesssim C e^{-(mR)^k}$ (for a new
constant $C=C(k,n,c)$),
and we can get the same type of estimate for $\Xi^-$. Combining these, we find that there exist functionals $\psi_j, \varphi_j$
on $\A(O_B), \A(O_A)$ respectively such that
\ben\label{part1}
\omega_0(ab) = \omega_0(a) \omega_0(b) + \sum_{j=1}^\infty \varphi_j(a) \psi_j(b)
\een
and such that  $\sum_j \| \varphi_j \| \ \| \psi_j \| \le C e^{-(mR)^k}$ for large $R$. Thus, we are in the situation of lemma \ref{Lem_dominating_sigma} with $\nu = 1 + C e^{-(mR)^k}$.
Lemma \ref{Lem_dominating_sigma} and theorem \ref{Thm_ER<EN} together with the elementary bound $\log (1+x) \le x$ now implies the statement.

\medskip
\noindent
{\em Proof of thm. 6, Part 2):} We view $f_{a,b}(z)$ as
defining a function $\Xi_z$ from the doubly cut plane $\CC \setminus \{z \in \CC \mid \Im(z) = 0, |\Re(z)| \ge R\}$ into the linear maps $\B = {\mathfrak B}(\A_B, (\A_A)_*)$
by the formula $[\Xi_z(b)](a) = f_{a,b}(z)$. The BW-nuclearity assumption implies the nuclearity
of this map for $|\Im z|>0$, with 1-norm bounded from above by $\|\Xi_z\|_1 \le e^{(c/|\Im z|)^n}$. Without loss of generality, we may assume
that $n>1/2$, since increasing $n$ makes the bound less tight (in practicle, one expects $n \sim d$ anyhow).

\begin{figure}[h!]
\begin{center}

\begin{tikzpicture}
%axis
        \draw[dashed] (- 1.5, 0) -- (1.5, 0);
        \draw[line width=0.2mm, >={Triangle[length=2mm, width=1mm]}, ->] (0, -2.5) -- (0, 2.5);
        	\draw[line width=0.2mm, >={Triangle[length=2mm, width=1mm]}, ->] (1.5, 0) -- (4, 0);
	\draw[line width=0.2mm] (-1.5, 0) -- (-4, 0);
        % contour
        \draw[line width=0.5mm, >={Triangle[length=1.5mm, width=2mm]}, ->] (-1.5, 0) -- (0, .5);
        \draw[line width=0.5mm, >={Triangle[length=1.5mm, width=2mm]}, ->] (0, .5) -- (1.5, 0);
        \draw[line width=0.5mm, >={Triangle[length=1.5mm, width=2mm]}, ->] (1.5, 0) -- (0, -.5);
        \draw[line width=0.5mm, >={Triangle[length=1.5mm, width=2mm]}, ->] (0, -0.5) -- (-1.5, 0);
        	\draw (-1.5, 0) circle (2pt);
	\draw (1.5, 0) circle (2pt);
        % others	
	\node[right] at (0, 2.5) {$\Im(z)$};
	\node[below] at (4, 0) {$\Re(z)$};
	\node[right] at (1.5, -0.3) {$R$};
	\node[below] at (-2, 0) {$-R$};
	\node[below] at (0.9, -.5) {$-i R \tan \frac{\pi}{4n}$};
	\node[above] at (0.9, .5) {$i R \tan \frac{\pi}{4n}$};
	\node[above] at (-1.0, .2) {$C_R$};
    \end{tikzpicture}

   \end{center}
    \caption{The contour $C_R$.}
\label{fig:betacontour}
    \end{figure}
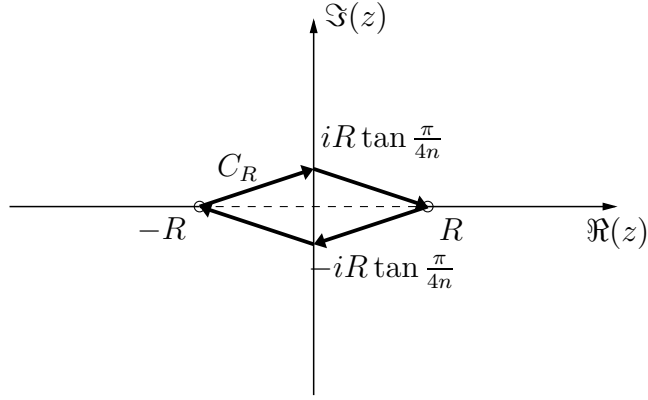

Let $C_R$ be the contour shown in fig. \ref{fig:betacontour}. Consider the map $z \mapsto \Upsilon_z$ from the interior of  $C_R$
to ${\mathfrak B}$ defined by
\ben
\Upsilon_z = \varphi(z) \cdot \Xi_z
\een
where
\ben\label{varphidef}
\varphi(z) = C(n,R,c)
\exp \left[ -\sqrt{2} \left( \frac{c}{(R+z) \sin\frac{\pi}{4n} } \right)^n \right]
\exp \left[ -\sqrt{2} \left( \frac{c}{(R-z) \sin\frac{\pi}{4n} } \right)^n \right] ,
\een
and where
$$
C(R,n,c) =
\exp \left[ 2\sqrt{2} \left( \frac{c}{R \sin\frac{\pi}{4n} } \right)^n \right] \ .
$$
Since $C_R$ is contained in the doubly cut plane $\CC \setminus \{\Im(z) = 0, |\Re(z)| \ge R \}$
where $\Xi_z$ is analytic, and since $\varphi$ is analytic in the interior of $C_R$
it follows that $z \mapsto \Upsilon_z$ is holomorphic in the interior of $C_R$.
Using the explicit form of the function
$\varphi(z)$ and the fact that $\|\Xi_z\|_1 \le e^{(c/|\Im z|)^n}$ in particular along the contour $C_R$,
we see that
\ben
\|\Upsilon_z\|_1 \le |\varphi(z)| e^{(c/|\Im z|)^n} \le \exp \left[ \sqrt{2} \left( \frac{c\ctg\frac{\pi}{4n}}{R } \right)^n \right] \quad \text{for all $z \in C_R$ .}
\een
Also, by definition, $\Xi_0=\Upsilon_0$.

Inside the contour, $\Upsilon_z$ has, for fixed $a \in \A_A, b \in \A_B$, the contour integral representation
\ben\label{contoured}
[\Upsilon_z(b)](a) = \int_{C_R} \frac{\dd w}{2\pi i \ (w-z)} [\Upsilon_w(b)](a) \ .
\een
Making an argument of the same kind as in~\cite{borchers_1} based on this contour integral formula,
we can see that $z \mapsto \Upsilon_z$ is in fact a strongly holomorphic map from the interior of the contour $C_R$ to the nuclear (not just bounded) operators from
$\A_B \to (\A_A)_*$ with uniformly bounded 1-norm inside the contour $C_R$. By lemma~\ref{maxlemma} in appendix~\ref{eow}, the maximum of $\|\Upsilon_z\|_1$ is achieved on the boundary $C_R$, and thus
$
\|\Upsilon_z\|_1 \le \exp [ \sqrt{2} ( \tfrac{c\ctg\frac{\pi}{4n}}{R } )^n ]
$
also for $z$ inside -- and not just on -- the contour $C_R$. Thus, from $\Xi_0=\Upsilon_0$,
we conclude in particular that $\|\Xi_0\|_1 \le \exp [ \sqrt{2} ( \frac{c\ctg\frac{\pi}{4n}}{R } )^n ]$. Arguing now as in Part 1),
we get \eqref{part1} with $\sum_j \| \varphi_j \| \ \| \psi_j \| \le \exp [ \sqrt{2} ( \frac{c\ctg\frac{\pi}{4n}}{R } )^n ]$ which holds for all $R$, in particular for $R/c \ll 1$. Thus, we are in the situation of lemma \ref{Lem_dominating_sigma} with $\nu = 1 +
\exp [ \sqrt{2} ( \frac{c\ctg\frac{\pi}{4n}}{R } )^n ]$.

Lemma \ref{Lem_dominating_sigma} and theorem \ref{Thm_ER<EN} imply that $E_R(\omega) \le \log \nu$, which gives the statement.
\qed

\subsubsection{Proof of thm. 7 and thm. 8}

{\em Proof of thm. 7:}
As in the previous proof, we omit the reference to the representation $\pi_\beta$ and simply write $a$ for $\pi_\beta(a)$. We also write
$J$ for the modular conjugation associated with\footnote{Note that $\pi_\beta(\A)$ has a non-trivial commutant since the
the representation $\pi_\beta$ is not irreducible.} $\A$ (the v. Neumann closure of $\cup_O \A(O)$) -- {\em not} with $\A_A$ --  and the state $\omega_\beta$, acting on the GNS-Hilbert space with implementing vector $|\Omega_\beta \rangle$. Fix $a \in \A_A \vee J\A_AJ,
b \in \A_B$, and define the function $F_{a,b}(z)$ as
\ben
F_{a,b}(z) =
\begin{cases}
\langle \Omega_\beta| a P_+ e^{izH_\beta} b \Omega_\beta \rangle -
\langle \Omega_\beta| b P_- e^{-izH_\beta} a \Omega_\beta \rangle
& \text{for $0< \Im (z)$}\\
- \langle \Omega_\beta| a P_- e^{izH_\beta} b \Omega_\beta \rangle
+ \langle \Omega_\beta| b P_+ e^{-izH_\beta} a \Omega_\beta \rangle
& \text{for $0> \Im (z)$} ,
\end{cases}
\een
which coincides with a function considered also by~\cite{jaekel_1}.
Here $H_\beta$ is the generator of time translations in the representation $\pi_\beta$ (``Liouvillean'').
For $-R < t < R$ one finds for the jump across the real axis:
\ben
F_{a,b}(t+i0) - F_{a,b}(t-i0) = \langle \Omega_\beta | [a, e^{itH_\beta} b e^{-itH_\beta}] \Omega_\beta \rangle = 0
\een
using space-like commutativity and the fact that the time-translated region $O_B$ remains space-like to $O_A$ as long as the time translation parameter
$t$ stays in the range $t \in (-R, R)$.
By the edge-of-the-wedge theorem (see appendix~\ref{eow}), $F_{a,b}(z)$ therefore defines an analytic function in
the doubly cut plane $\CC \setminus \{\Im(z) = 0, |\Re(z)| \ge R \}$. Using $N$ times the
KMS-condition (applying the last item in prop.~\ref{Prop1} to the pair $\A, \omega_\beta$) for $\omega_\beta$, one can derive the ``image sum'' formula
\ben\label{rep1}
\begin{split}
\langle \Omega_\beta | ab \Omega_\beta \rangle - \langle \Omega_\beta | a \Omega_\beta \rangle \langle \Omega_\beta| b \Omega_\beta\rangle
&= \sum_{k=-N}^N F_{a,b}(i\beta k) + \sum_{k=-(N-\half)}^{N-\half} F_{Ja^*J, b}(i\beta k) \\
&+ \langle \Omega_\beta | b P_- e^{\beta N H_\beta} a \Omega_\beta \rangle + \langle \Omega_\beta | a P_- e^{\beta NH_\beta} b \Omega_\beta \rangle .
% \\ &\to \sum_{k=-\infty}^\infty f_{a,b}(i\beta k) + \sum_{k=-\infty}^{\infty} f_{Ja^*J, b}(i\beta(k+\half))
\end{split}
\een
Next, we view $F_{a,b}(z)$ as
defining a function $\Xi_z$ from the doubly cut plane into the linear maps $\B = {\mathfrak B}(\A_B, (\A_A \vee J\A_A J)_*)$
by the formula $[\Xi_z(b)](a) = F_{a,b}(z)$. The definition of $F_{a,b}(z)$, implies that for $|\Im(z)| >0$, the nuclear 1-norm
of $\Xi_z$ satisfies the upper bound
\ben
\| \Xi_z\|_1 \le \| \Theta^+_{r,z} \|_1 +  \| \Theta^-_{r,z}\|_1 \le 2|\Im(z)|^{-\alpha}
\exp (c/|\Im(z)|)^n
\een
using in the second step the assumption of the theorem.

We also  consider the map $z \mapsto \Upsilon_z$ from the doubly cut plane $\CC \setminus \{\Im(z) = 0, |\Re(z)| \ge R \}$
to ${\mathfrak B}$ defined by
\ben
\Upsilon_z = R^{-\alpha} (R-z)^{\alpha} (R+z)^{\alpha}  \cdot \varphi(z) \cdot \Xi_z
\een
where $\alpha>0$ is as in the assumptions of the theorem, and where
$\varphi(z)$ is the function defined previously in eq.\eqref{varphidef}.
It follows that also $z \mapsto \Upsilon_z$ is strongly
analytic as a map from the doubly cut plane to $\B$.

Now let $C_R$ be the contour of figure \ref{fig:betacontour}, which is entirely within the doubly cut plane.
Inside this contour, $\Upsilon_z$ has, for fixed $a \in \A_A \vee J \A_A J, b \in \A_B$, a contour integral representation
as in \eqref{contoured}. Using the definition of $\varphi(z)$ and our previous bound on $\| \Xi_z\|_1$, we find that the nuclear 1-norm of $\Upsilon_z$ is bounded by
\ben\label{Ubound}
\begin{split}
\| \Upsilon_z \|_1 &\le  R^{-\alpha} \sup_{w \in C_R} \left\{ |w-R|^{\alpha} |w+R|^{\alpha}
 |\varphi(w)| \cdot \| \Xi_w \|_1 \right\}\\
& \le 2 \exp [ \sqrt{2} ( \tfrac{c\ctg\frac{\pi}{4n}}{R } )^n ]\ (\tfrac{1}{2} \sin\tfrac{\pi}{4n})^{-\alpha}
\end{split}
\een
for all $z$ {\em on} the contour $C_R$. Making again an argument of the same kind as in~\cite{borchers_1} based on the contour integral formula,
we can see that $\|\Upsilon_z\|_1$ must remain bounded also {\em inside} the contour $C_R$,
and that $z \mapsto \Upsilon_z$ is in fact a holomorphic map from the interior of the contour $C_R$ to the nuclear (not just bounded) operators from
$\A_B \to (\A_A \vee J\A_A J)_*$ with uniformly bounded 1-norm inside the contour $C_R$.

By lemma~\ref{maxlemma} in appendix~\ref{eow} the map $z \mapsto \| \Upsilon_z \|_1$ from the interior of $C_R$ in fact assumes its maximum on the boundary, $C_R$, so
\eqref{Ubound} also holds for $z$ inside $C_R$. For $k \in \half \ZZ$ and
$|k| \le N := \lfloor \beta^{-1}R\sin\frac{\pi}{4n} \rfloor$,
the points $i\beta k$ are inside the contour $C_R$.
Going back from $\Upsilon_z$ to $\Xi_z$, it follows that, for such $k$, and for $R/c \gg 1$,
\ben
\| \Xi_{i\beta k} \|_1 \le C_1  R^{\alpha} (R^2 + (\beta k)^2)^{-\alpha}
\een
for some constant $C_1$. Therefore
\ben\label{sumkbound}
\sum_{k \in \half \ZZ, |k| \le N } \| \Xi_{i\beta k} \|_1 \le C_2 R^{-\alpha+1}
\een
using our assumption that $\alpha>1$. On the other hand, using \eqref{rep1}
and our definitions of $\Xi_z, \Theta^\pm_z$, we may write
\ben
\begin{split}
&\langle \Omega_\beta | ab \Omega_\beta \rangle
- \langle \Omega_\beta | a \Omega_\beta \rangle \langle \Omega_\beta| b \Omega_\beta\rangle\\
=
&\sum_{k\in \ZZ, |k| \le N} [\Xi_{i\beta k}(b)](a) + \sum_{k \in \ZZ + \half, |k| \le N}
[\Xi_{i\beta k}(b)](Ja^*J) \\
&
+ \langle \Omega_\beta | b \Theta^-_{\beta N, r}(a) \Omega_\beta \rangle + \langle \Omega_\beta | a \Theta^-_{\beta N, r}(b) \Omega_\beta \rangle
\ .
\end{split}
\een
As a consequence of the nuclearity bound \eqref{sumkbound}, the bound $\| \Theta_{\beta N,r}^\pm\|_1 \le
(N\beta)^{-\alpha} \exp[(c/N\beta)^n]$ assumed in the theorem, and
since $a \mapsto Ja^*J$ is bounded, we can write
\ben
\langle \Omega_\beta | ab \Omega_\beta \rangle
- \langle \Omega_\beta | a \Omega_\beta \rangle \langle \Omega_\beta| b \Omega_\beta\rangle
= \sum_j \varphi_j(a) \psi_j(b) \ ,
\een
where the linear functionals $\varphi_j$ on $\A_A$ and $\psi_j$ on $\A_B$
satisfy $\sum_j \| \varphi_j \| \ \| \psi_j \|  \le C_3 R^{-\alpha+1}$, for a new constant $C_3$.
Thus, we can apply the lemma \ref{Lem_dominating_sigma} and theorem \ref{Thm_ER<EN}
with $\nu := 1 + C_3R^{-\alpha+1}$.
The bound~\eqref{masterbound} together with $\log(1+x) \le x$ then gives the statement. \qed

\medskip

{\em Proof of thm. 8:} The proof is very similar to that of the previous theorem
and is now based on the correlation function \eqref{f1def}.  We omit the details. \qed

\subsection{Upper bounds for free  quantum field theories in $d+1$ dimensions}

\subsubsection{Free scalar fields}

The BW-nuclearity condition is well-established for free scalar fields in $d+1$ dimensional Minkowski spacetime both in the massive ($m>0$) and the massless ($m=0$) case. Thus, thm.~\ref{Thm_m>0} respectively thm.~\ref{Thm_m=0} apply and provide upper bounds on the entanglement entropy $E_R(\omega_0)$ of two diamonds of size $r$ separated by a distance $R$.

In fact, in the massive case, an explicit nuclearity bound is given in~\cite{buchholz_4}.\footnote{To be specific, \cite{buchholz_4} show that in the massive case, $\| \Theta_{\beta, r} \|_1 \le \exp[(c/\beta)^d |\log(1-e^{-m\beta/2})|]$
for $r>m^{-1}$ and $0<\beta\le r$, where $c=c_0 r$.} Therefore thm.~\ref{Thm_m>0} can be applied to infer sub-exponential decay of the entanglement entropy for large separation $R$.
In the massless case, it can be extracted from the work of \cite{buchholz_5}
that $\|\Theta_{\beta, r}\|_1 \le \beta^{-(d-2)} \exp(c/\beta)^d$ for
$\beta>0, r>0$ and $d> 2$. Applying thm.~\ref{Thm_m=0}
then immediately gives
\begin{proposition}
Let $\omega_0$ be the vacuum state of a free $m=0$ Klein-Gordon field in Minkowski spacetime $\RR^{d,1}$. Let $A,B \subset \RR^d$ be two balls of
radius $r$ separated by a distance $R$. Then
$$
E_R(\omega_0) \lesssim CR^{-(d-2)}
$$
when $R/r \gg 1$, where $C$ is a constant depending on $r$ and $d>2$.
\end{proposition}

This upper bound is consistent with a general bound for conformal quantum field theories given below in the remark after thm.~\ref{confddim}.
More general, and somewhat better bounds, can be obtained from the modular nuclearity bound $E_R(\omega) \le E_M(\omega)$ in thms.~\ref{Thm_EN<Em} and \ref{Thm_ER<EN}, using results of \cite{lechner_2} in the case $m>0$. We formulate them first in the general setting of open regions $A$ and $B$ in a Cauchy surface $\C$ of a globally
hyperbolic spacetime $(\sM,g)$. Furthermore, we let $\omega$ be any quasi-free state on the Weyl algebra of a free scalar field on $(\sM,g)$.

Recall from section \ref{ssec_free} that the space of real initial data $K_{\RR}:=C_0^{\infty}(\C,\RR)\oplus C_0^{\infty}(\C,\RR)$ carries a symplectic form
\[
\sigma((f_0,f_1),(h_0,h_1))=\int_{\C}(f_0h_1-f_1h_0) \dd V.
\]
Following the notations of section \ref{ssec_Weyl}, the quasi-free state $\omega$ corresponds to an inner product $\mu$ on $K_{\RR}$, and we let $\text{clo}_\mu K$ denote the complex Hilbert space completion of the complexification $K$ of $K_{\RR}$. We let $\Gamma$ denote the complex conjugation on $\text{clo}_\mu K$ and $\Sigma$ the operator which implements the symplectic form. The one-particle Hilbert space can then be identified with $\H_1:=\overline{\text{ran}(1+\Sigma)}$, whereas the full GNS-representation space is the bosonic Fock space $\F_+(\H_1)$ with GNS vector $|\Omega \rangle$.

If $V\subset\C$ is any open region, the real initial data in $V$ generate a symplectic subspace $K_{\RR}(V)$ of $K_{\RR}$, and the complex initial data in $V$ generate a closed subspace $\K(V)\subset\text{clo}_{\mu}K$. Let $P_V$ be the orthogonal projection onto $\K(V)$ and note that $\Sigma_V:=P_V\Sigma P_V$ implements the symplectic form on $\K(V)$. The local Weyl algebra $\W(K_{\RR}(V),\sigma)$ gives rise to a von Neumann algebra $\A_V:=\pi_{\mu}(\W(K_{\RR},\sigma))''$, and we may associate to $(\A_V,\Omega)$ a modular operator $\Delta_V$ and a modular conjugation $J_V$. These operators are the second quantizations of operators $\delta_V$ and $j_V$ on $\H_1$, which we now describe.

First we let $\H_1(V):=\overline{\text{ran}(P_V+\Sigma_V)}$ denote the one-particle space corresponding to the restriction of $\omega$ to $\A_V$, and we note that $\H_1(V)\subset\K(V)$. Then we let $\tilde{R}_V$ be the projection onto $\overline{\text{ran}(P_V-|\Sigma_V|)}$, which is a subspace of $\H_1(V)$. Recall from section \ref{ssec_Weyl} that $\Omega$ is separating for $\A_V$ if and only if $\tilde{R}_V=P_V$, which also entails that $\H_1(V)=\K(V)$. Using these projections we may write the one-particle modular operator $\tilde{\delta}_V$ and modular conjugation $\tilde{\jmath}_V$ on $\H_1(V)$ as
\begin{eqnarray}
\tilde{\delta}_V&=&\frac{P_V-\Sigma_V}{P_V+\Sigma_V}\ \tilde{R}_V\nonumber\\
\tilde{\jmath}_V&=& \Gamma \tilde{R}_V.\nonumber
\end{eqnarray}
One easily verifies that $\tilde{\jmath}_V\tilde{\delta}_V^{\frac12}=s_V$, where $s_V \tilde{R}_V\sqrt{P_V+\Sigma_V}F=\tilde{R}_V\sqrt{P_V+\Sigma_V}\ \Gamma F$ for complex initial data $F$ in $V$.

Note that $\H_1(V)$ is isometric to a subspace of $\H_1$. Restricting the domain of this isometry to $\tilde{R}_V\H_1(V)$, it takes the form
\begin{equation}\label{Eqn_Uv}
U_V=\sqrt{1+\Sigma}\ (P_V+\Sigma_V)^{-\frac12}\tilde{R}_V.
\end{equation}
This satisfies $U_V^*U_V=\tilde{R}_V$ and we denote the range projection of $U_V$ by $R_V:=U_VU_V^*=U_V\tilde{R}_VU_V^*$. We may then introduce the modular operator $\delta_V$ and the modular conjugation $j_V$ on $\H_1(V)$, defined by
\begin{eqnarray}\label{Eqn_delta}
\delta_V&:=&U_V\tilde{\delta}_VU_V^*=\sqrt{1+\Sigma}\ \frac{P_V-\Sigma_V}{(P_V+\Sigma_V)^2}\ \tilde{R}_V\sqrt{1+\Sigma}\\
j_V&:=&U_V\tilde{\jmath}_VU_V^*=\Gamma \sqrt{1-\Sigma}\ (P_V-\Sigma_V^2)^{-\frac12}\tilde{R}_V\sqrt{1+\Sigma}.\nonumber
\end{eqnarray}
Note that $\delta_VR_V=\delta_V$ and that
\[
\delta_V^{\frac12}:=U\tilde{\delta}_V^{\frac12}U^*=\sqrt{1+\Sigma}\ \frac{(P_V-\Sigma_V)^{\frac12}}{(P_V+\Sigma_V)^{\frac32}}\ \tilde{R}_V\sqrt{1+\Sigma}.
\]

We now apply these notations to the case $V=B'=\C\setminus\overline{B}$, i.e. we consider the modular operator $\Delta_{B'}$ associated to the algebra $\A_{B'}$. The 1-nuclear norm of the operator
$
\Phi^A: \A_A \to \H \ , \quad \Phi^A(a) =  \Delta_{B'}^{\quarter} a| \Omega \rangle \
$
as in a5') can then be estimated in terms of one-particle operators as follows:
\begin{theorem}\label{thm_qf}
For open regions $A\subset B'$ in $\C$ we define
\begin{eqnarray}
\alpha&:=&(1+\delta_A)^{-\frac12}R_A\delta_{B'}^{\frac12}R_A(1+\delta_A)^{-\frac12}\nonumber\\
&=&\frac12U_A(1+\Sigma)\frac{(P_{B'}-\Sigma_{B'})^{\frac12}}{(P_{B'}+\Sigma_{B'})^{\frac32}}\tilde{R}_{B'}(1+\Sigma)U_A^*.\nonumber
\end{eqnarray}
Then $\|\Phi^A\|_1\le \det(1-\sqrt{\alpha+j_A\alpha j_A})^{-4}$ and
\[
E_R(\omega)\le\log(\|\Phi^A\|_1)\le-4\tr\log(1-\sqrt{\alpha+j_A\alpha j_A})\, .
\]
\end{theorem}

{\em Proof:}
The equality of both formulae for $\alpha$ can be verified using equations (\ref{Eqn_delta},\ref{Eqn_Uv}) and the fact that $U_A$ intertwines the functional calculus of $\delta_A$ and $\tilde{\delta}_A$, which yields in particular that
\[
(1+\delta_A)^{-\frac12}R_A=\frac{1}{\sqrt{2}}U_A\sqrt{1+\Sigma}.
\]
The main part of the proof is then a particular case of theorem 3.3 and of eq.(3.15) of~\cite{lechner_2}  (the latter is originally due to~\cite{buchholz_5}). Instead of reproducing the entire argument, we only give the following remarks. If $\omega$ is not cyclic and separating for $\A_{B'}$, the modular operator $\delta_{B'}$ and conjugation $j_{B'}$ vanish on a certain subspace. However, we may extend $j_{B'}$ to a complex conjugation $\Gamma_{B'}$ on the entire Hilbert space $\H_1$, and we use this conjugation in the doubling procedure in the proof of theorem 3.3 in~\cite{lechner_2}. On $\H_1\oplus\H_1$ we then use the one-particle operator $X_1:=\delta_{B'}^{\quarter}\oplus\delta_{B'}^{-\quarter}$.

Let $H_A$ be the closed real-linear subspace in $\H_1$ generated by $\sqrt{1+\Sigma}F$ with real initial data $F$ supported in $A$, and let $H_A^{\circ}$ be its symplectic complement. The complex subspaces $\underline{\K}_{\pm}$ constructed in~\cite{lechner_2} are then the sums $\underline{\K}_{\pm}=\underline{K}_0\oplus\underline{\K}'_{\pm}$ of the spaces
\begin{eqnarray}
\underline{K}_0&=&(H_A^{\circ})^{\perp}\oplus \Gamma_{B'}(H_A^{\circ})^{\perp}\nonumber\\
\underline{K}'_{\pm}&=&\left\{
\left(\begin{array}{c}
h_1+ih_2\\
\pm\Gamma_{B'}(h_1-ih_2)
\end{array}\right)
\mid h_1,h_2\in R_AH_A\right\}.\nonumber
\end{eqnarray}
Because these summands are orthogonal, the orthogonal projections $E_{\pm}$ onto $\underline{\K}_{\pm}$ can be decomposed into sums of orthogonal projections, $E_0+E'_{\pm}$, where $E_0$ projects onto $\underline{K}_0$ and $E'_{\pm}$ onto $\underline{K}'_{\pm}$. We then have $X_1E_{\pm}=X_1E'_{\pm}$, and one may compute that
\[
E'_{\pm}=\left(\begin{array}{cc}
(1+\delta_A)^{-1}R_A&\pm(1+\delta_A)^{-1}\delta_A^{\frac12}j_A\Gamma_{B'}\\
\pm\Gamma_{B'}j_A(1+\delta_A)^{-1}\delta_A^{\frac12}&\Gamma_{B'}(1+\delta_A)^{-1}\Gamma_{B'}
\end{array}\right).
\]

It turns out that $|X_1E_{\pm}|=|X_1E'_{\pm}|$ is unitarily equivalent to the matrix
\[
\left(\begin{array}{cc}
\alpha+j_A\alpha j_A&0\\
0&0
\end{array}
\right)
\]
for both signs. The estimate~(2.24) of~\cite{buchholz_5} then yields $\|\Phi^A\|_1\le \det(1-\sqrt{\alpha+j_A\alpha j_A})^{-4}$ (cf. (3.15) in~\cite{lechner_2}), and the estimate on $E_R(\omega)$ follows from our theorems \ref{Thm_ER<EN} and \ref{Thm_EN<Em}, together with the well-known formula $\log \det(X) = \tr \log(X)$ for positive operators $X$.
\qed

\medskip
The results of~\cite{lechner_2} show that for quasi-free Hadamard states with relatively compact $A$ and ${\rm dist}(A,B) > 0$, the relative entanglement entropy $E_R(\omega)$ between $A$ and $B$ is finite.

The expression for the operator $\alpha$ simplifies when the state is separating for $\A_{B'}$ (e.g. when it is cyclic for $\A_B$). In this case $\tilde{R}_A=P_A\le P_{B'}=\tilde{R}_{B'}$ and
\[
\alpha=\frac12U_A\sqrt{P_{B'}-\Sigma_{B'}^2}U_A^*=j_A\alpha j_A \ .
\]
Using the estimate $(P_A(P_{B'}-\Sigma_{B'}^2)^{\frac12}P_A)^2\le P_A(P_{B'}-\Sigma_{B'}^2)P_A$ and taking
quartic roots we find that
$|(P_{B'}-\Sigma_{B'}^2)^{\quarter}P_A|\le|(P_{B'}-\Sigma_{B'}^2)^{\frac12}P_A|^{\frac12}$ and hence that
\[
\sqrt{\alpha+j_A\alpha j_A}=U_A|(P_{B'}-\Sigma_{B'}^2)^{\quarter}P_A|U_A^*\le U_A|(P_{B'}-\Sigma_{B'}^2)^{\frac12}P_A|^{\frac12}U_A^*
\]
and
\ben\label{Eqn_EstforRS}
E_R(\omega)\le-4\tr_{\K(A)} \log(1-|(P_{B'}-\Sigma_{B'}^2)^{\frac12}P_A|^{\frac12})  \ .
\een

Let us now specialise to an ultra-static spacetime $\sM=\RR\times\C$ with a Cauchy surface $\C$ and metric $g=-\dd t^2+h$ and to a massive $m>0$ minimally coupled scalar field. Recall from section \ref{ssec_free} that in the ground state $\K$ can be identified with $W^{(\frac12)}(\C)\oplus W^{(-\frac12)}(\C)$, where the Sobolev spaces are defined using the operator $C^{-1}:=-\nabla^2+m^2$, and that $\Sigma$ takes the form given in equation (\ref{Eqn_Sigma}). Analogously, we may identify $\K(A)=W^{(\frac12)}(A)\oplus W^{(-\frac12)}(A)$ and similarly for $B$ and $B'$, where $W^{\pm\frac12}(A)$ is the closed subspace of $W^{\pm\frac12}(\C)$ generated by test-functions supported in $A$. We denote by $P_{A\pm}$ the orthogonal projections in $W^{\pm\frac12}(\C)$ onto $W^{\pm\frac12}(A)$, so that $P_A=P_{A+}\oplus P_{A-}$, and similarly for $B'$.

Note that the ground state has the Reeh-Schlieder property, so we may use the estimate (\ref{Eqn_EstforRS}). Furthermore, using the facts that $\Sigma^2=1$ (which is true for any pure state) and $P_A\le P_{B'}$ we find from the definition of $\Sigma_{B'}$ that
\begin{eqnarray}\label{prematrix}
|(P_{B'}-\Sigma_{B'}^2)^{\frac12}P_A|^2&=&P_A\Sigma(1-P_{B'})\Sigma P_A=|(1-P_{B'})\Sigma P_A|^2\nonumber\\
&=&\left|\left(\begin{array}{cc}
0&(1-P_{B'+})C^{\frac12}P_{A-}\\
-(1-P_{B'-})C^{-\frac12}P_{A+}&0
\end{array}\right)\right|\nonumber\\
&=&\left(\begin{array}{cc}
|(1-P_{B'-})C^{-\frac12}P_{A+}|&0\\
0&|(1-P_{B'+})C^{\frac12}P_{A-}|
\end{array}\right) \ .
\end{eqnarray}
We may use the unitary operators $U_{\pm}:W^{\pm\frac12}(\C)\to L^2(\C)$ with $U_{\pm}f:=C^{\mp\quarter}f$ and
$U^*_{\pm}f:=C^{\pm\quarter}f$ to rewrite this operator in terms of the $L^2$-inner product and the operator $C$. For this purpose we introduce the projections $Q_{A\pm}=U^*_{\pm}P_{A\pm}U_{\pm}$ in $L^2(\C)$ and similarly for $Q_{B'\pm}$. Note that $Q_{A\pm}$ projects onto the subspace
\ben\label{subspace}
\overline{\{C^{\mp\quarter}f \mid f\in C_0^{\infty}(A) \}} \subset L^2(\C) \ .
\een
We then find from (\ref{Eqn_EstforRS})\footnote{For ground states one may in fact omit the doubling procedure of \cite{lechner_2} and directly apply the results of \cite{buchholz_5}, which improves this estimate by a factor $\half$.}:

\begin{proposition}
On an ultra-static spacetime $\sM=\RR\times\C$ with metric $g=-\dd t^2+h$, the ground state $\omega_0$ of a massive KG field fulfills
\ben\label{Eqn_Estground}
E_R(\omega_0)\le-4\sum_{\pm} \tr\log(1-|(1-Q_{B'\mp})Q_{A\pm}|^{\frac12})\ .
\een
Here the trace is in $L^2(\C)$ and $Q_{A\pm}$ are the orthogonal projectors onto the closed subspaces~\eqref{subspace}, and similarly for
$Q_{B'\pm}$.
\end{proposition}

{\em Proof:}
We use $Q_{A\pm}=U^*_{\pm}P_{A\pm}U_{\pm}$ and $U_{\pm}f:=C^{\mp\quarter}f$ in eqs.~\eqref{prematrix} and~\eqref{Eqn_EstforRS}, and
the facts that $|UX|=|X|$ and $|XU|=U^*|X|U$ for all closed operators $X$ and unitary operators $U$. Using the fact that $C$ is bounded by $m^{-2}$ one may show that $|(1-Q_{B'\mp})Q_{A\pm}|<1$ for both signs, so the right-hand side of (\ref{Eqn_Estground}) is well-defined. Note that the estimate clearly only depends on the geometry of $(\C,h)$ and the regions $A$ and $B$. \qed

\medskip
We close this section proving our main estimate, which is an application of the previous proposition and some other techniques.
This estimate shows that the entanglement entropy falls off exponentially with the distance between the regions $A$ and $B$, in the following precise sense:
\begin{theorem}
Consider an ultra-static spacetime $\sM=\RR\times\C$ with metric $g=-\dd t^2+h$ and the ground state $\omega_0$ of a minimally coupled free scalar field with mass $m>0$. Let $A\subset\C$ be a bounded open region and for any $r>0$ let $B_r:=\{x\in\C \mid \dist(x,A)>r\}$.
For any $R>0$ and all $r\ge R$ the ground state $\omega_0$ has an entanglement entropy between the regions $A$ and $B_r$ which satisfies
\[
E_R(\omega_0)\le c e^{-\frac12mr}
\]
where $c>0$ is independent of $r\ge R$.
\end{theorem}

{\em Proof:}
By the results of~\cite{lechner_2}, the operators $|(1-Q_{B'_r\mp})Q_{A\pm}|$ are compact, and their eigenvalues are in $[0,1)$. It follows that the largest eigenvalue is less than 1, i.e. $\|(1-Q_{B'_r\mp})Q_{A\pm}\|<1$. For $t\in[0,a]$ with $a<1$ we have the elementary estimate
$-\log(1-t)<\frac{t}{1-a}$, which we may apply to $t=|(1-Q_{B'_r\mp})Q_{A\pm}|^{\frac12}$. It then follows from (\ref{Eqn_Estground}) that
\[
E_R(\omega_0)\le\sum_{\pm} \frac{4}{1-\|(1-Q_{B'_r\mp})Q_{A\pm}\|^{\frac12}}
\tr \ |(1-Q_{B'_r\mp})Q_{A\pm}|^{\frac12}\ ,
\]
where the trace is over $L^2(\C)$. As $r$ increases, the projections $1-Q_{B'_r\mp}$ decrease, so the prefactors $4(1-\|(1-Q_{B'_r\mp})Q_{A\pm}\|^{\frac12})^{-1}$ attain their maximum at $r=R$. This value is still finite, because $R>0$. We let $c_0$ denote the maximum of this value over the choices $+$ and $-$.

To estimate the traces we will use the fact that for all bounded linear operators $X,Y$  we have $|XY|\le \|X\|\cdot|Y|$ and $|XY|=V^*|Y^*X^*|V\le \|Y^*\|\cdot V^*|X^*|V=\|Y\|\cdot V^*W|X|W^*V$, where $V$ and $W$ are the partial isometries appearing in the polar decompositions $XY=V|XY|=|(XY)^*|V$ and $X=W|X|=|X^*|W$.

Following~\cite{lechner_2} we choose a test-function $\chi_A$ with support in $B'_{\frac12R}$ such that $\chi_A\equiv 1$ on $A$. We then choose test-functions $\chi_i$, $i=1,\ldots,4$, such that $\chi_1\equiv 1$ on $\mathrm{supp}(\chi_A)$ and $\chi_{i+1}\equiv 1$ on $\mathrm{supp}(\chi_i)$. We fix an $l\in\mathbb{N}$ such that $l>\frac12+\frac34\mathrm{dim}(\C)$ and we introduce
\[
X_i:=\chi_iC^{(5-i)l}\chi_iC^{-(4-i)l}
\]
for $i=1,\ldots,4$. The operators $X_i$ are Hilbert-Schmidt (see~\cite{lechner_2} theorem 4.2) and hence $X:=X_1X_2X_3X_4$ has $\tr|X|^{\frac12}<\infty$. Note that $C^{-1}$ is a partial differential operator, so considering the supports of the $\chi_i$ we have
\[
\chi_AC^{-4l-b}X=\chi_AC^{-b}
\]
for any $b\in\mathbb{N}$. We fix $b$ such that $b\pm\quarter\ge0$ and we use the fact that $Q_{A\pm}=C^{\mp\quarter}\chi_AC^{\pm\quarter}Q_{A\pm}$ in order to find
\begin{eqnarray}
|(1-Q_{B'_r\mp})Q_{A\pm}|&=&|(1-Q_{B'_r\mp})C^{\mp\quarter}\chi_AC^{-b}C^{\pm\quarter+b}Q_{A\pm}|\nonumber\\
&=&|(1-Q_{B'_r\mp})C^{\mp\quarter}\chi_AC^{-4l-b}XC^{\pm\quarter+b}Q_{A\pm}|\nonumber\\
&\le&\|(1-Q_{B'_r\mp})C^{\mp\quarter}\chi_AC^{-4l-b}\|\cdot\|C^{\pm\quarter+b}Q_{A\pm}\|\cdot |X|\nonumber
\end{eqnarray}
and because $C^{\pm\quarter+b}$ and $Q_{A\pm}$ are both bounded we have
\[
\tr |(1-Q_{B'_r\mp})Q_{A\pm}|^{\frac12}\le c_1\|(1-Q_{B'_r\mp})C^{\mp\quarter}\chi_AC^{-4l-b}\|^{\frac12}
\]
for some $c_1>0$ which is independent of $r$ and of the sign $\pm$. Combining this estimate with the first paragraph of this proof we have
\ben\label{Eqn_est01}
E_R(\omega_0)\le c_0c_1\sum_{\pm}\|(1-Q_{B'_r\mp})C^{\mp\quarter}\chi_AC^{-4l-b}\|^{\frac12}.
\een

For any test-function $\chi_B$ supported in $B'_r$ and identically 1 on $B'_{\frac12 R}$ we have $(1-Q_{B'_r\mp})=(1-Q_{B'_r\mp})C^{\pm\quarter}(1-\chi_B)C^{\mp\quarter}$ and hence
\begin{eqnarray}
\|(1-Q_{B'_r\mp})C^{\mp\quarter}\chi_AC^{-4l-b}\|&=&\|(1-Q_{B'_r\mp})C^{\pm\quarter}(1-\chi_B)C^{\mp\frac12}\chi_AC^{-4l-b}\|\nonumber\\
&\le&\|C^{\pm\quarter}(1-\chi_B)C^{\mp\frac12}\chi_AC^{-4l-b}\|\nonumber\\
&=&\|C^{-4l-b}\bar{\chi}_AC^{\mp\frac12}(1-\bar{\chi}_B)C^{\pm\quarter}\| \ .\nonumber
\end{eqnarray}
Using lemma 4.4 in~\cite{lechner_2} we may find $\eta_0,\ldots,\eta_{4l+b}\in C_0^{\infty}(B'_{\frac12R})$ such that
\[
\|C^{-4l-b}\bar{\chi}_A\psi\|\le\left(\sum_{k=0}^{4l+b}\|\bar{\eta}_kC^{-k}\psi\|^2\right)^{\frac12}
\le\sum_{k=0}^{4l+b}\|\bar{\eta}_kC^{-k}\psi\|
\]
for all $\psi$ in the domain of $C^{-4l-b}$. Applying this result to $C^{\mp\frac12}(1-\bar{\chi}_B)C^{\pm\quarter}f$ with test-functions $f$ and using the fact that $\bar{\eta}_kC^{-k\mp\frac12}(1-\bar{\chi}_B)C^{\pm\quarter}$ is bounded (\cite{lechner_2} theorem 4.5) we find
\ben\label{Eqn_est2}
\|(1-Q_{B'_r\mp})C^{\mp\quarter}\chi_AC^{-4l-b}\|\le\sum_{k=0}^{4l+b}\|C^{\pm\quarter}(1-\chi_B)C^{-k\mp\frac12}\eta_k\|.
\een
Note that the $\eta_k$ are independent of $r$, as is the number of functions $4l+b+1$.

For given $r\ge R$ we now choose a real-valued $\chi_B$ with support in $B'_r$ such that $0\le\chi_B\le 1$, $\chi_B\equiv 1$ on $B'_{r-\frac12R}$ and such that $|\nabla\chi_B|\le R$. For the $+$ sign it then follows immediately from $\|C\|\le m^{-2}$ and proposition 4.3 of~\cite{lechner_2} that
\ben\label{Eqn_est+}
\sum_{k=0}^{4l+b}\|C^{\quarter}(1-\chi_B)C^{-k-\frac12}\eta_k\|\le c_2^2e^{-mr}\ ,
\een
because the supports of the $\eta_k$ and $1-\chi_B$ are separated by a distance $\ge r-R$. For the $-$ sign we note that
\[
(1-\chi_B)C^{-1}(1-\chi_B)\le C^{-1}\frac{(1-\chi_B)^2}{2m^2}C^{-1}+\frac{(1-\chi_B)^2m^2}{2}+|\nabla\chi_B|^2 \ ,
\]
where we estimated the term $-\nabla^2\frac{(1-\chi_B)^2}{2m^2}\nabla^2\le 0$. Combining this estimate with the fact that
$\|C^{-\quarter}\psi\|\le m^{-\frac12}\|C^{-\frac12}\psi\|$ for all $\psi$ in the domain of $C^{-\frac12}$, it then follows that
\[
\|C^{-\quarter}(1-\chi_B)\psi\|^2\le \frac{1}{2m^2}\|(1-\chi_B)C^{-1}\psi\|^2+\frac{m^2}{2}\|(1-\chi_B)\psi\|^2+\|\ |\nabla\chi|\ \psi\|^2
\]
for all $\psi$ in the domain of $C^{-1}$, where $|\nabla\chi|=\sqrt{\nabla^{i}\chi\cdot\nabla_{i}\chi}$ is a positive continuous function. We choose $\psi=C^{-k+\frac12}\eta_k$ and applying proposition 4.3 of~\cite{lechner_2} again, noting that its result also holds when
the left function in the product is only continuous. We then find that
\ben\label{Eqn_est-}
\sum_{k=0}^{4l+b}\|C^{-\quarter}(1-\chi_B)C^{-k+\frac12}\eta_k\|\le c_3^2e^{-mr}\ ,
\een
for some $c_3\ge0$, which is independent of $r$.

Putting together the estimates (\ref{Eqn_est2},\ref{Eqn_est+},\ref{Eqn_est-}) (and taking square roots) yields
\[
\|(1-Q_{B'_r\mp})C^{\mp\quarter}\chi_AC^{-4l-b}\|^{\frac12}\le\max\{c_2,c_3\}e^{-\frac12mr}.
\]
Inserting this into (\ref{Eqn_est01}) and setting the constant in the theorem $c:=2c_0c_1\max\{c_2,c_3\}$ completes the proof.
\qed

\subsubsection{Free Dirac fields}\label{sec:areaDirac}

The BW-nuclearity condition is well-established for free massive Dirac fields in $d+1$ dimensional static spacetimes~\cite{dantoni}. Thus, thm.~\ref{Thm_m>0}  applies and provides upper bounds on the entanglement entropy $E_R(\omega_0)$ of two diamonds of size $r$ in Minkowski space separated by a distance $R$ (or more generally, two bounded regions $A$ and $B$ in a static time-slice $\C$ separated by a distance $R$). One can again get better bounds for large $R$ using techniques from modular theory in a similar way as for
scalar fields~\cite{onirban}.

Here we give an upper bound in the opposite regime when the distance, $\epsilon$, between $A$ and $B$ goes to zero. Our reason for doing so is that this bound is qualitatively better than the general bound presented in thm.~\ref{Thm_m>0} -- in fact it is of ``area law'' type. For simplicity, we focus on a Majorana field [cf. sec.~\ref{sec:freeDirac}] on a static spacetime of the form $\sM = \RR^{1,1} \times \Sigma$, where $(\Sigma, \gamma)$ is a compact $d-1$-dimensional spin manifold, and where the metric on $\sM$ is $g=-\dd t^2 + \dd x^2 + \gamma_{AB}(y) \dd y^A \dd y^B$. But our results presumably hold more generally when $O_A$ is the interior of a black hole region in a spacetime with bifurcate Killing horizon\footnote{See e.g.~\cite{wald_2} for an explanation of this concept.} and $O_B$ is a subset of the exterior, separated from the horizon by a corridor of diameter $\epsilon$.

Our theorem is the following.

\begin{theorem}
Let $B=\{ x<0, t= 0\}$ and $A =\{ x> \epsilon, t=0\}$, and let $\omega_0$ be the
ground state of a free Majorana field on $\sM=\RR^{1,1} \times \Sigma$ of mass $m>0$. Then for sufficiently small $\epsilon > 0$, we have for every fixed
$N \in \mathbb{N}, \delta>0$ the upper bound
\ben
E_R(\omega_0) \le C |\log(m'\epsilon)| \sum_{j=d-1}^{-N} \epsilon^{-j} \int_{\partial A} a_j(y)  \ ,
\een
where $C_0>0$ is a constant, $m'=2m/[(1+\delta)(1+\delta^{-1})]^{\frac12}$ and where $a_j$ are the heat kernel expansion coefficients of the operator
$(1+\delta)^{-\frac12}(-\nabla_\Sigma^2+\quarter R_\Sigma^{})^{\frac12}$ (see proof).
\end{theorem}

\begin{remark}
The first heat kernel coefficient $a_{d-1}$ is constant (see e.g.~\cite{gilkey}), so we get, to leading order for $\epsilon \to 0$, the ``area law'',
\ben
E_R(\omega_0) \lesssim c_0 |\log(m\epsilon)| \frac{|\partial A|}{\epsilon^{d-1}} \ .
\een
\end{remark}

{\em Proof:}  Spacetime dimension $d+1=2$:  The essence of the proof is already seen in the case $d=1$. The regions $A$ and $B$ and their
causal completion $O_A$ and $O_B$ are in this case shown in fig.~\ref{fig:wedges} below (where $R$ is replaced by $\epsilon$ here); they are left and right wedges. The construction of the algebra for the free Majorana field in $1+1$ dimensions was given above in sec.~\ref{sec:freeDirac}, which we use here.

Let $|0\rangle$ be the vector representative of the vacuum state $\omega_0$ in the GNS-Hilbert space $\H$.
As usual, we consider $\Phi^A: \A_A \to \H, a \mapsto \Delta^{1/4}_{B'} a |0\rangle$.
We wish to apply theorems~\ref{Thm_ER<EN} and \ref{Thm_EN<Em}, which together with $\| \Psi^A\|_1 \le \| \Phi^A \|_1$ ($\Psi^A$ given by
\eqref{Psidefnuclear}) give $E_R(\omega_0) \le \log \| \Phi^A\|_1$.

So we need to estimate the nuclear 1-norm $\| \Phi^A \|_1$. The key point is that $\Delta^{it}_{B'}$ implements boosts on
the Hilbert space $\H$, by the Bisognano-Wichmann theorem~\cite{bisognano}. More precisely, let $U(\lambda)$ be the unitary implementer of the boost
$
\left( \begin{matrix}
\cosh \lambda & \sinh \lambda\\
\sinh \lambda & \cosh \lambda
\end{matrix}
\right)
$
given by eq.~\eqref{eq:Uboost} on fermonic Fock space. Then $\Delta_{B'}^{it} = U(-2\pi t)$.
Now if $b' \in \A_{B'}$, then since translations act geometrically on the algebras,
$e^{-i\epsilon P^1} b' e^{i\epsilon P^1} = a \in \A_A$ [here $e^{-i\epsilon P^1}$ is the implementer~\eqref{eq:Uboost} for a
translation by $(0,\epsilon)$ which maps the wedge $O_B'$ to $O_A$ by construction].

Therefore, the 1-norm of the map $\Phi^A$ is the same as the 1-norm of the map
\ben
\A_{B'} \owns b' \mapsto \Xi^{B'}(b') = \Delta_{B'}^{\quarter} e^{-i\epsilon P^1} b'|0\rangle \in \H \ .
\een
Due to the geometrical action of boosts, we have the identity
\ben
\Delta_{B'}^{it} e^{-i\epsilon P^1} |\Psi \rangle =  e^{-i\epsilon \sinh (2\pi t) P^0 - i\epsilon \cosh(2\pi t) P^1} \Delta_{B'}^{it} |\Psi \rangle \ .
\een
for $|\Psi \rangle \in {\rm dom} \Delta_B'$. If we formally set
$t=-i/4$, and put $|\Psi\rangle = b' |0\rangle$, we obtain
\ben\label{Xidefmod}
\Xi^{B'}(b')  = e^{-\epsilon P^0} \Delta^{\quarter}_{B'} b' |0\rangle = e^{-\epsilon P^0} U(i\pi/2) b' |0\rangle \ ,
\een
where $U(i\pi/2)$ is the representer of a boost with parameter $\lambda$ analytically continued to $\lambda = i\pi/2$.
A rigorous proof may be given using a similar argument as in sec.~\ref{sec:CFT1}; alternatively see~\cite{buchholz_6}.

Thus, at this stage, we have managed to show that $E_R(\omega_0) \le \log \| \Xi^{B'}\|_1$, where $\Xi^{B'}$ is
given by \eqref{Xidefmod}. We must thus understand this map. Since the operator
$e^{-\epsilon P^0} U(i\pi/2)$ is of ``second quantized form'' in a free field theory such as ours-- it is given by eq.~\eqref{eq:Uboost} in general --
it is plausible that that nuclear norm can be estimated
using operators acting only on the 1-particle Hilbert space
$\H_1 = L^2(\RR, \dd \theta)$. That this is indeed the case can be be shown using results of~\cite{lechner_2}, which are in turn
based on results of~\cite{buchholz_6,buchholz_5}.

One first defines the {\em real} Hilbert space [recall $B'=(0, \infty)$]
\ben
H_{B'} = \{ e^{\theta - i\pi/4} \widetilde k_1(m \sinh \theta) + e^{-\theta + i\pi/4} \widetilde k_2(m \sinh \theta)  \mid k_i \in C^\infty_0(B', \RR) \} \subset L^2(\RR, \dd \theta)
\een
which is obtained by acting with $\psi(k)$ on $|0\rangle$ for {\em real}-valued test-spinors $k=(k_1,k_2)$ supported in $O_{B'}$, compare eq.~\eqref{eq:psiCAR}. On $\H_1$, we next introduce an anti-unitary ``time reversal'' operator $\tilde T$. To this end, we first define, on $K=L^2(\RR, \dd x; \CC^2)$ the anti-linear operator $T: (k_1, k_2) \mapsto (\bar k_2, \bar k_1)$, which clearly satisfies $T^2=1$ (involution property).
It follows that $T$ commutes with the 1-particle hamiltonian $h$ [see eq.~\eqref{eq:h1p}], and therefore also with the projector $P$ onto the positive part of the spectrum of $h$.
Therefore, $T$ restricts to an operator on $PK$. Since, according to our discussion in sec.~\ref{sec:freeDirac}, the 1-particle Hilbert space $\H_1=L^2(\RR, \dd \theta)$ is
identified as $VPK$ under the isometry~\eqref{eq:identification}, it follows that $VTV^*=\tilde T$ is an anti-unitary involution on $\H_1$. Next, we define the
{\em complex} subspaces
\ben
H^\pm_{B'} = \CC \cdot (1 \pm \tilde T) H_{B'} \subset \H_1 \ .
\een
Concretely, using the definition of $V$ in eq.~\eqref{eq:identification} and of $T$, we have
\ben
\begin{split}
H^+_{B'} &= \{ \Psi_k(\theta) = \cosh(\theta/2 - i\pi/4) \widetilde{k}(m \sinh \theta) \mid \supp(k) \subset B', k \in L^2(\RR) \} \ ,\\
H^-_{B'} &= \{ \Psi_k(\theta) = \sinh(\theta/2 - i\pi/4) \widetilde{k}(m \sinh \theta) \mid \supp(k) \subset B', k \in L^2(\RR) \} \ .
\end{split}
\een
The wave functions in $H^\pm_A$ are in the domain of the operator $e^{-\epsilon P^0} U(i\pi/2)$, and the action is in fact given by
\ben
[e^{-\epsilon P^0} U(i\pi/2) \Psi_k](\theta) = e^{-m\epsilon \cosh \theta} \Psi_k(\theta -i\tfrac{\pi}{2}) \equiv (X_1 \Psi_k)(\theta) \ ,
\een
where we note that the analytic continuation on the right side is indeed possible due to the support of $k$,
and where we used eq.~\eqref{eq:Uboost} on fermonic Fock space with $n=1$, $\lambda = i\pi/2$, and $a = (\epsilon, 0)$. We omit the straightforward calculation.
Theorem~3.11 of~\cite{lechner_2} now tells us that, if $E^\pm_{B'}$ are the projection operators onto the subspaces $H^\pm_{B'} \subset \H_1$, then
we have the upper bound
\ben\label{112}
\| \Xi^{B'} \|_1 \le \exp \{ 2\| X_1 E^+_{B'} \|_1 + 2\| X_1 E^-_{B'} \|_1 \} \ ,
\een
where the 1-norm on the right side is in $\H_1$. Similarly as in~\cite{buchholz_6}, one may use contour integration to rewrite the operator $X_1$ in the form
\ben
(X_1 \Psi_k)(\theta) = \frac{-1}{2\pi i} e^{-m\epsilon \cosh \theta} \int \dd \theta' \left\{
\frac{1}{\theta'-\theta+i\pi/2} + \frac{1}{\theta'+\theta-i\pi/2}
\right\} \Psi_k(\theta') \
\een
when $\Psi_k \in H^\pm_{B'}$. For $\kappa,s \in \RR, \kappa \neq 0, s > 0$, let us define an operator $T_{\kappa, s}$
on $L^2(\RR, \dd \theta)$ given by the following kernel:
\ben\label{Tkadef}
T_{\kappa,s}(\theta, \theta') = -{\rm sign}(\kappa) \frac{e^{-\frac12 s \cosh \theta}}{2\pi i(\theta'-\theta+i\kappa/2)} \ .
\een
In terms of this operator, we immediately get
\ben\label{113}
\|X_1 E^\pm_{B'} \|_1\le  \| T_{ \pm \pi, 2m\epsilon} \|_1 \ .
\een
The following lemma describes properties of the operator $T_{\kappa,s}$ needed to
estimate the right side and also for later purposes:
\begin{lemma}\label{Tlemma}
\begin{enumerate}
\item
In terms of the momentum operator $p=i\dd/\dd \theta$ on $L^2(\RR, \dd \theta)$, we can write
\ben
T_{\kappa,s} = e^{-\frac12 s \cosh \theta} \Theta(\kappa p) e^{-|\kappa p|/2}
\een
where $\Theta$ is the Heaviside step function (characteristic function of the set $\RR_+$).
\item
The operator $A_{\kappa,s}=T_{+\kappa,s}^{} T_{+\kappa,s}^* + T_{-\kappa,s}^{} T_{-\kappa,s}^*$ has the integral kernel
\ben
A_{\kappa,s}(\theta,\theta') = \frac{|\kappa|}{\pi} \frac{e^{-\frac12 s \cosh \theta} e^{-\frac12 s \cosh \theta'}}{(\theta-\theta')^2 + \kappa^2} \ .
\een
\item
\ben
\| T_{\kappa,s} \|_1 \lesssim
\begin{cases}
c_1 e^{-s/2} & \text{for $s \gg |\kappa|$}\\
c_2 \log s & \text{for $s \ll |\kappa|$}
\end{cases}
\een
where $c_i$ are numerical constants diverging no worse than $c_1 \lesssim |\log \kappa|/|\kappa|^{2}$ respectively
$c_2 \lesssim 1/|\kappa|^{2}$ for $|\kappa| \to 0$.
\end{enumerate}
\end{lemma}

Combining now item 3) of lemma~\ref{Tlemma} with~\eqref{112},~\eqref{113} gives, for $\epsilon \ll m^{-1}$,
\ben
\log \|\Phi^A\|_1 = \log \| \Xi^{B'} \|_1  \le C |\log(2m\epsilon)| \ ,
\een
and since we have already noted that $E_R(\omega_0) \le \log \|\Phi^A\|_1$, this proves the theorem when $d=1$. \\

\medskip
\noindent
Spacetime dimension $d+1>2$:  Let $\lambda_j$ be the (real) eigenvalues of the elliptic operator $e_0 \cdot \sum_{A=2}^d e_A \cdot \nabla_{e_A}$ on the compact manifold
$\Sigma$ (enumerated without multiplicities), where $e_A$ is a frame field for $\Sigma$.
By decomposing a general spinor on $\C=\RR \times \Sigma$ into the corresponding eigenmodes, one can
easily show using the product structure $\sM = \RR^{1,1} \times \Sigma$ and eq.~\eqref{eq:functor}
that the algebra $\A_A$ is isomorphic to the tensor product $\otimes_j \A_{A,j}$, where $\A_{A,j}$ is isomorphic to the algebra
for a Dirac field on $1+1$-dimensional Minkowski spacetime with mass $m_j = \sqrt{m^2 + \lambda_j^2}$. The analogous statement holds for $B$.
If $\omega_{0,j}$ is the vacuum state for such a Dirac field on $\RR^{1,1}$, property (e5) tells us that
\ben
E_R(\omega_0) \le \sum_j E_R(\omega_{0,j}) \ .
\een
Now let $\delta>0$.
Using our previous results in $d=1$, and the trivial relation
$m_j \sqrt{1+\delta}  \ge \lambda_j \sqrt{1+\delta}^{-1} + m \sqrt{1+\delta^{-1}}^{-1}$
we also have
\ben
\begin{split}
E_R(\omega_{0,j}) &\le 4\| T_{ \pi, 2m_j\epsilon} \|_1 \\
&\le 4e^{-\epsilon |\lambda_j|/(1+\delta)} \| T_{ \pi, m'\epsilon} \|_1 \le
Ce^{-\epsilon |\lambda_j|/(1+\delta)} \log (\epsilon m') \ ,
\end{split}
\een
for $\epsilon \ll m'$.
Now we take the sum over $j$ and use the relation
\ben
\sum_j e^{-t |\lambda_j|} = \tr_{L^2(\Sigma, \$|_\Sigma)} e^{-t \sqrt{-\nabla_\Sigma^2+\quarter R_\Sigma^{}}} \ .
\een
The result then follows using well-known results on the heat kernel of elliptic pseudo-differential operators, see e.g.~\cite{gilkey}.
\qed

\medskip
\noindent
{\em Proof of lemma}~\ref{Tlemma}:
1) and 2) follow by taking Fourier transforms.
3) Consider first $s \gg \kappa$, and take $\kappa>0$ for definiteness (the other case is similar).
We have using 1)
\ben
\begin{split}
T_{\kappa,s}
&= e^{-\frac12 s \cosh \theta} e^{-\kappa p/2} \Theta(p) \\
&= e^{-\frac12 (s-\kappa) \cosh \theta} \cdot e^{-\quarter \kappa \cosh \theta} (p^2+1)^{-\frac12} \cdot (p^2+1)^{\frac12} e^{-\quarter \kappa \cosh \theta} (p^2+1)^{-1} \cdot
(p^2+1) e^{-\kappa p/2} \Theta(p) \ . \nonumber
\end{split}
\een
Now we apply the standard inequalities $\| XY \|_1 \le \|X\|_2 \|Y\|_2$ and $\|XY\|_1 \le \|X\| \|Y\|_1$ and use that $\| e^{-\frac12 (s-\kappa) \cosh \theta} \|
\le e^{-\frac12(s-\kappa)}$, $\| (p^2+1) e^{-\kappa p/2} \Theta(p) \| \le c_4 \kappa^{-2}$ (for $|\kappa|<2$). We get
\ben
\| T_{\kappa,s} \|_1 \le c_4 \kappa^{-2} e^{-\frac12(s-\kappa)} \, \| e^{-\quarter \kappa \cosh \theta} (p^2+1)^{-\frac12} \|_2 \| (p^2+1)^{\frac12} e^{-\quarter \kappa \cosh \theta} (p^2+1)^{-1} \|_2 \ .
\een
Using $\frac{1}{4\pi} e^{-|\theta|}= \frac{1}{2\pi} \int_\RR \dd p (1+p^2)^{-1} e^{ip\theta}$ for the integral kernel of $(1+p^2)^{-1}$, we have
\ben
\| e^{-\quarter \kappa \cosh \theta} (p^2+1)^{-\frac12} \|_2^2 = \frac{1}{4\pi} \int \dd \theta \dd \theta' e^{-\quarter \kappa \cosh \theta -\quarter \kappa \cosh \theta'-|\theta-\theta'|}
 \le c_5(1+|\log \kappa|)
\een
for some $c_5$ independent of $\kappa$.
The norm $\| (p^2+1)^{\frac12}e^{-\quarter \kappa \cosh \theta} (p^2+1)^{-1} \|_2$ can be estimated in a similar way noting that $p^2+1$ is just a differential operator. This proves the lemma in
the case $s \gg \kappa$. The other case is treated similarly. \qed

\subsection{Upper bounds for integrable models}

Here we apply the general upper bounds presented in theorems~\ref{Thm_ER<EN} and \ref{Thm_EN<Em} to the case of certain
integrable models in 1 + 1 dimensions with factorizing $S$-matrix described in sec.~\ref{Intmodels}.

\begin{figure}[h!]
\begin{center}
\begin{tikzpicture}
        \coordinate (a) at (1, 0);
        \coordinate (b) at (4, 3);
        \coordinate (c) at (4, -3);
        \coordinate (d) at (-1, 0);
        \coordinate (e) at (-4, 3);
        \coordinate (f) at (-4, -3);

        \filldraw[color=red!60, fill=red!15, very thick] (b) -- (a) -- (c);
        \filldraw[color=green!60, fill=green!15, very thick] (f) -- (d) -- (e);

        \draw[line width=1pt, >={Triangle[length=1mm,width=2mm]}, ->] (-4, 0) -- (4, 0);
        \draw[line width=1pt, >={Triangle[length=1mm,width=2mm]}, ->] (0, -3) -- (0, 3);

        \node[right] at (0, 2.7) {$t$};
        \node[below] at (4, 0) {$x$};
        \node[below] at (a) {$ R $};
        \node[below] at (d) {$ 0 $};
        \node[below] at (2.75, 0) {$A$};
        \node[below] at (-2.75, 0) {$B$};
        \node[left] at (3, 2.5) {$ O_{A} $};
        \node[right] at (- 3, 2.5) {$ O_{B} $};
    \end{tikzpicture}
\end{center}
\caption{The wedge regions $O_A,O_B$.}
\label{fig:wedges}
\end{figure}
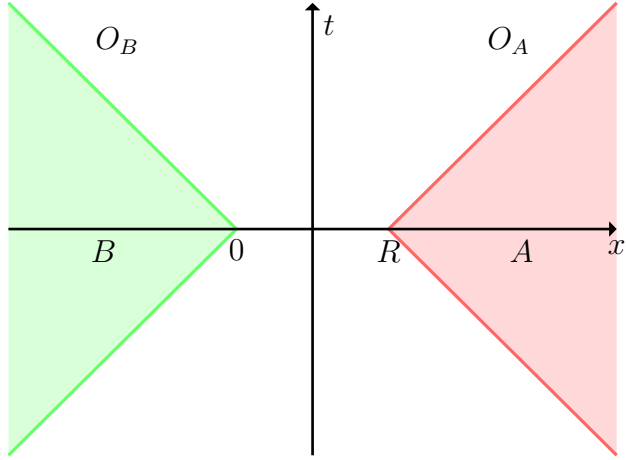

We denote points in $1 + 1$ dimensional Minkowski space by $(t, x)$. For $A$ we choose $A = \{t=0, x>R\}$ and for $B$ be choose
$B=\{t=0, x<0\}$. The regions $O_A$ and $O_B$ are obviously left and right wedges translated by $R>0$, i.e.
$O_A = \{x-R > |t|\}$ and $O_B = \{-x > |t|\}$. The distance between $A$ and $B$ is $R$, see fig.~\ref{fig:wedges}.

As usual, we set
$\A_A = \pi_0(\A(O_A))''$, and similarly for $\A_B$,
where $\pi_0$ is the GNS-representation of the vacuum state, $\omega_0$, of the model.
For definiteness, we consider integrable models defined via a factorizing 2-body $S$-matrix of the form
\ben\label{S2def}
S_2(\theta) = \prod_{k=1}^{2N+1} \frac{\sinh \theta - i\sin b_k}{\sinh \theta + i\sin b_k}  \ ,
\een
where each $0< b_i < \pi/2$. This $S_2$-matrix has the required properties s1)-s4) and therefore defines a net of v. Neumann algebras as described
in sec.~\ref{Intmodels}.
Our theorem is:

\begin{theorem}
Let $\omega_0$ be the vacuum state of the integrable quantum field theory of mass $m$ defined by $S_2$, and let $A=\{t=0, x>R\}$ and $B = \{t=0, x<0\}$.
For $mR \gg 1/(\kappa \delta)$,  we have for any $\kappa$ with $\min\{b_i\} > \kappa > 0$ and any $\delta>0$
\ben
E_R(\omega_0)  \lesssim \frac{4ec}{\kappa \sqrt{\pi mR}} \, e^{-mR(1-\delta)}
\een
where $c=\sup\{ |S_2(\zeta)|  : -\kappa < \Im \zeta < \pi+\kappa\}^{1/2}$ is a constant depending on $S_2,\kappa$.
\end{theorem}

\begin{remark}
1) For $mR\ll 1$, our estimations do not produce a bound so far. Looking at the proof, it is clear that an improved estimation of the quantity $\tr \wedge^n A_{\kappa,s}$
appearing below in eq.~\eqref{Aint} for $\kappa \ll 1$ is necessary. We conjecture that this would lead for $mR \ll 1$
to a bound of the type
\ben
E_R(\omega_0)  \lesssim C |\log mR|^\alpha
\een
for some constants $\alpha,C$ depending on $S_2$. One may also guess that $E_R$ is perhaps asymptotic to the v. Neumann entropy for $A$ with a UV-cutoff $\sim R$. This v. Neumann entropy has been computed using the replica trick (and also approximations) in \cite{Cardy}. From that work we
thus perhaps expect the sharp upper bound to be of the form $\sim (c_{\rm eff}/6) \log (mR)$, where $c_{\rm eff}$ is an effective UV central charge defined in that work. \\
2)  Our upper bounds
also hold for any pair of diamond regions $O_{A/B}$
that are space-like separated by a Lorentz invariant distance $R = {\rm dist}(O_A,O_B)$, due to the monotonicity property of the relative
entanglement entropy, (e4), and the
Lorentz-invariance of $\omega_0$.
\end{remark}

\medskip
\noindent
{\em Proof:}
Let $|0\rangle$ be the vector representative of the vacuum state $\omega_0$ in the GNS-Hilbert space $\H$.
As usual, we consider $\Phi^A: \A_A \to \H, a \mapsto \Delta^{1/4}_{B'} a |0\rangle$.
We wish to apply theorems~\ref{Thm_ER<EN} and \ref{Thm_EN<Em}, which together with $\| \Psi^A\|_1 \le \| \Phi^A \|_1$ ($\Psi^A$ given by
\eqref{Psidefnuclear}) give $E_R(\omega_0) \le \log \| \Phi^A\|_1$.
So we need to estimate the nuclear 1-norm
$\| \Phi^A \|_1$, which is also equal to $\| \Phi^B \|_1$, since the setup is clearly symmetric in $A,B$. These estimates can
to a large extent be extracted from~\cite{lechner_4}, which in turn improve and correct estimations in~\cite{lechner_1}.
Some of our arguments differ somewhat and may lead to important improvements of~\cite{lechner_4}, so we give
some details.

First, we observe that $\| \Phi^A\|_1 = \| \Xi^{B'}\|_1$, where $\Xi^{B'}: \A_{B'} \to \H$ is the map defined already above in eq.~\eqref{Xidefmod}.
The argument is precisely the same as there and relies again on the fact that $\Delta_{B'}^{it}$ are the generators of boosts of
the wedge $O_{B'}$.

Let $\Xi_n = P_n \Xi^{B'}$, where $P_n: \H \to \H_n$ is the orthogonal
projector onto the $n$-particle subspace of the $S_2$-symmetric Fock space, see sec.~\ref{Intmodels}. Using eq.~\eqref{Xidefmod}
and using the action of boosts and translations on wave functions in $\H_n$ described in sec.~\ref{Intmodels}, eq.~\eqref{eq:Uboost},
we immediately get
\ben
\Xi_n(b') = \prod_{j=1}^n e^{- mR \cosh \theta_j} \cdot (P_n b' |0 \rangle)(\theta_1 - i\tfrac{\pi}{2}, \dots, \theta_n -i\tfrac{\pi}{2}) \ .
\een
Following~\cite{lechner_1,lechner_4},
we now consider the decomposition
\ben
\Xi_n = E_n \circ X_n \circ \Upsilon_n \ .
\een

a) $\Upsilon_n: \A_{B'} \to H^2(\RR^n + iC_n)$ is the map into the Hardy space defined by ($\delta>0$ small)
\ben
\Upsilon_n: b' \mapsto \prod_{j=1}^n e^{- i\delta mR \sinh \zeta_j} \cdot (P_n b' |0\rangle)(\zeta_1, \dots, \zeta_n) \ .
\een
Here, $(\zeta_1, \dots, \zeta_n) \in \RR^n + iC_n \subset \CC^n$ is an $n$-tuple of complex numbers, $C_n$ is a suitable
open polyhedron in $\RR^n$, and the Hardy space is defined to be the Banach space of those holomorphic functions $h(\zeta_1, \dots, \zeta_n)$
having finite Hardy norm
\ben
\| h \|^2_{H^2} = \sup_{(\lambda_1, \dots, \lambda_n) \in C_n} \int_{\RR^n} \dd^n \theta \, |h(\theta_1 + i\lambda_1, \dots, \theta_n+i\lambda_n)|^2 \ .
\een
It can be shown using $b' \in \A_{B'}$ that the analytic continuation of the $n$-particle wave function
$(P_n b' |0\rangle)(\theta_1, \dots, \theta_n)$ to $\RR^n + iC_n$ is
possible e.g. for the choice $C_n=(-\frac{\pi}{2}, \dots, -\frac{\pi}{2}) + (-\frac{\kappa}{2n},\frac{\kappa}{2n})^{\times n}$, with $\kappa$ as in the
hypothesis of the theorem.

b) $X_n: H^2(\RR^n + iC_n) \to L^2(\RR^n)$ is the map defined by
\ben
X_n: h(\zeta_1, \dots, \zeta_n) \mapsto \prod_{j=1}^n e^{-(1-\delta) mR \cosh \theta_j} \cdot h(\theta_1 - i\tfrac{\pi}{2}, \dots, \theta_n -i\tfrac{\pi}{2}) \ ,
\een
and $E_n$ is the projector onto the $S_2$-symmetric wave functions described in sec.~\ref{Intmodels}. Next, we need bounds on the
norms of $E_n \circ X_n$ respectively $\Upsilon_n$:

a) Combining prop. 3.8 and lemma 5.1 of~\cite{lechner_4}, and 1), 2) of lemma~\ref{Tlemma}, one gets the following upper bound. There is
a constant $c$ depending only on $\kappa$ and $S_2$ such that
\ben\label{Xnbound}
\| E_n \circ X_n \|_1 \le
c^n \sum_{\sigma_1, \dots, \sigma_n = \pm 1} \tr \left( \bigwedge_{j=1}^n R_{\sigma_j, \kappa/2n, (1-\delta) mR} \right)
= c^n \tr (\wedge^n A_{\kappa/2n, (1-\delta)mR}) \ ,
\een
where $R_{\pm , \kappa, s} = T_{\pm \kappa, s}^{} T_{\pm \kappa,s}^{*}$ with $T_{\kappa,s}$ the operator in lemma~\ref{Tlemma},
and where $A_{\kappa, s}=R_{+,\kappa,s}+R_{-,\kappa,s}$ is equal to
the operator defined in item 2) of lemma~\ref{Tlemma}. The $n$-th exterior power
$\wedge^n A$ of an operator $A$ on $L^2(\RR, \dd \theta)$ by definition
means the restriction of $\otimes^n A$ to the subspace of totally {\em anti-symmetric} ({\em not} $S_2$-{\em symmetric!}) wave functions in $L^2(\RR^n, \dd^n \theta)$.
It is also shown in lemma 5.7 of~\cite{lechner_1} that the constant $c$ may be chosen as
$c=\|S_2\|^{1/2}_\kappa$, using the shorthand $\| S_2 \|_\kappa = \sup \{|S_2(\zeta)|\}$, with the supremum taken in the strip $-\kappa < \Im \zeta < \pi + \kappa$.

b) Furthermore, it is shown in prop. 4.5 of~\cite{lechner_4} that
\ben\label{115}
\| \Upsilon_n \| \le {\rm max} \left\{
1, \| S_2 \|_\kappa^n \left( \frac{2}{\pi \kappa}  \int_{0}^\infty \dd \theta \, e^{-mR\delta \sin \kappa \cosh \theta} \right)^{\half}
\right\} \ ,
\een
These upper bounds will now be applied to the right side of
\ben\label{114}
\|\Phi^A\|_1 = \| \Xi^{B'}\|_1 \le \sum_{n=0}^\infty \|\Xi_n\|_1 \le \sum_{n=0}^\infty \| E_n \circ X_n\|_1 \| \Upsilon_n \| \ .
\een
To get a better handle on the right side of the bound~\eqref{Xnbound}, we next use the following two lemmas:

\begin{lemma}\label{Alemma}
Let $A$ be a positive trace-class operator on $L^2(\RR, \dd \theta)$ with smooth integral kernel $A(\theta, \theta')$. Then there holds
\ben
\tr (\wedge^n A) = \frac{1}{n!} \int_{\RR^n} \dd^n \theta \, \det[A(\theta_i, \theta_j)]_{1 \le i, j \le n} \ .
\een
\end{lemma}
\noindent
{\em Proof:} Obvious generalization of well-known formula in statistical mechanics, see e.g. sec.~7.2 in \cite{kardar}. \qed

\begin{lemma}\label{Gramlemma}
Let $T$ be a complex, positive definite $n \times n$ matrix. Then
\ben
\det T \le \prod_{i=1}^n T_{ii}\ .
\een
\end{lemma}
\noindent
{\em Proof:} Well-known. Follows e.g. from ``Gram's'' or ``Hadamard's inequality''. \qed

We apply these two lemmas to the operator $A = A_{\kappa, s}$ defined in item 2) of lemma~\ref{Tlemma}.
Applying first lemma~\ref{Alemma} gives
\ben\label{Aint}
\tr \wedge^n A_{\kappa,s} = \frac{\kappa^n}{\pi^n n!}
\int_{\RR^n} \dd^n \theta \,
\det [(\theta_i-\theta_j)^2+\kappa^2]^{-1}
\prod_{j=1}^n e^{-s\cosh \theta_j} \ .
\een
The integral kernel $[(\theta-\theta')^2+\kappa^2]^{-1}$ is positive definite (i.e. gives a positive operator)
because its Fourier transform is a positive constant times the function $e^{-\kappa|p|}>0$.
It follows from standard characterizations of positive definite kernels that $T_{ij} = [(\theta_i-\theta_j)^2+\kappa^2]^{-1}$ is a positive $n\times n$
matrix for any choice of $\{\theta_j\}$. Therefore lemma~\ref{Gramlemma} gives $| \det T| \le \kappa^{-2n}$. This immediately results in
\ben
\tr \wedge^n A_{\kappa,s} \le  \frac{1}{\pi^n \kappa^n n!} \int_{\RR^n} \dd^n \theta \, \prod_{j=1}^n e^{-s\cosh \theta_j}
= \frac{1}{n!} \left( \frac{2}{\kappa \pi} K_0(s) \right)^n
\ ,
\een
using the representation $\int_{0}^\infty \dd\theta \, e^{-x\cosh \theta} = K_0(x), x>0$ of the Bessel function $K_0$.
Hence, \eqref{Xnbound} gives us:
\ben
\| E_n \circ X_n \|_1 \le \frac{1}{n!} \left( \frac{4nc}{\kappa \pi} K_0[(1-\delta)mR] \right)^n \le  \left( \frac{4ec}{\kappa \pi} K_0[(1-\delta)mR] \right)^n
\een
for $n \ge 1$ (using Stirling's approximation $n^n \le e^n n!$). From \eqref{114}, \eqref{115}, we then get the bound
\ben\label{199}
\| \Xi^{B'}\|_1 \le \sum_{n=0}^\infty \{c_1 K_0[(1-\delta)mR]\}^n  {\rm max} \left\{
1, c_2^n [ K_0 (mR\delta \sin \kappa)]^{\half}
\right\}
\een
where $c_1=\frac{4ec}{\kappa \pi}, c= \|S_2\|_{\kappa}^{1/2}, c_2$ are  constants depending on $S_2$ and $\kappa$ which will diverge when $\kappa \to 0$.
For the Bessel function $K_0$ it is well-known that
\ben
 K_0(x) \sim
\begin{cases}
-\log x& \text{for $x \to 0^+$}\\
\sqrt{\pi/x} e^{-x} & \text{for $x \to \infty$.}
\end{cases}
\een
We can now discuss the asymptotic behavior of $\| \Xi^{B'}\|_1$.
For $0< mR \ll 1$, the right side of \eqref{199} does not converge, so we are unable to obtain a bound on $\| \Xi^{B'}\|_1$ in that case.
For $mR \gg 1/\kappa\delta$, instead we get convergence, and in fact,
\ben
(E_R(\omega_0) \le ) \log \| \Xi^{B'} \|_1 \lesssim \frac{4e}{\kappa} \sqrt{\frac{\|S_2\|_{\kappa}}{\pi mR}} \, e^{-mR(1-\delta)}  \ ,
\een
as claimed. \qed

\subsection{Upper bounds for conformal QFTs in $d+1$ dimensions}\label{sec:CFT1}

Here we apply our methods to derive a general bound on $E_R(\omega_0)$ for the vacuum state of a conformal field theory (CFT)
in $d+1$ dimensions, where $d > 1$ (the case when $d=1$ is somewhat special and is treated in subsection \ref{ssec_CFTchiral}).
In the case of conformal quantum field theories, the axioms a3),a4) are suitably extended to the conformal group, ${\rm G} = {\rm SO}_+(d+1,2)/{\mathbb Z}_2$ of $d+1$-dimensional Minkowski spacetime. The action of ${\rm G}$ on points $x \in \RR^{d,1}$ can be efficiently described in the well-known ``embedding formalism'',
wherein one considers first the action of ${\rm G}$ on the cone $C=\{ \xi \in \RR^{d+3}  \mid \tilde \eta(\xi, \xi) = 0 \}$, where $\tilde \eta=diag(-1,1,\dots, 1,-1)$ is the metric of signature $(-2,d+1)$ on $\RR^{d+1,2}$. The projective cone $C/\RR_\times = \overline{\RR^{d,1}}$ is the Dirac-Weyl compactification (see e.g.~\cite{guido})
of Minkowski space, and on this compactification,
the action of ${\rm G}$ is defined {\em globally}. Uncompactified Minkowski space can be identified with the subset of points for which
$x^\mu = \xi^\mu/(\xi^{d+1} + \xi^{d+2}), \mu=0,1, \dots, d$ is finite. The transformation $\xi \mapsto g\xi$ induces a {\em local} action of ${\rm G}$ on $\RR^{d,1}$ via this identification, which we write $x \mapsto g \cdot x$, where local means that it is not defined for all pairs $g,x$.
The geometric significance of the various subgroups of ${\rm G}$ can be described as follows:
\begin{enumerate}
\item[(i)] The subgroup leaving $\xi^{d+1}, \xi^{d+2}$ fixed given by $\xi^{\prime \mu} = \Lambda^\mu{}_\nu \xi^\nu$, where $\Lambda$ is
a proper orthochronous Lorentz transformation.

\item[(ii)] The $d+1$-parametric subgroup of transformations $\xi^{\prime \mu} = \xi^\mu + \xi^+  a^\mu, \xi^{\prime -} = \xi^-, \xi^{\prime +} = \xi^+ + 2 \xi_\mu a^\mu + \xi^- a_\mu a^\mu$
corresponding to translations by $a^\mu$. Here $\xi^\pm = \xi^{d+1} \pm \xi^{d+2}$.

\item[(iii)]
The $d+1$-parametric subgroup of transformations $\xi^{\prime \mu} = \xi^\mu - \xi^- c^\mu, \xi^{\prime +} = \xi^+, \xi^{\prime -} = \xi^- - 2 \xi_\mu c^\mu + \xi^+ a_\mu c^\mu$
corresponding to special conformal transformation with parameters $c^\mu$.

\item[(iv)] The dilations by $\lambda >0$ correspond to $\xi^{\prime \mu} = \xi^\mu, \xi^\pm = \lambda^{\mp 1} \xi^\pm$.
\end{enumerate}
Since, by contrast with the action of the Poincar\'e group, conformal transformations cannot be globally defined for all pairs $g,x$, when stating
the axioms of covariance in conformal field theory, one can at best require covariance only for orbits of points which do not pass trough
infinity\footnote{Alternatively, one may pass to a net on the Dirac-Weyl compactification, as described in~\cite{guido}.}:

\begin{enumerate}
\item[a3')] (Conformal invariance) For any finitely extended region $U \subset \RR^{d,1}$ there exists a neighborhood ${\rm N} \subset {\rm G}$ of the identity such that for any
$g \in {\rm N}$, one has an algebraic isomorphism $\alpha_g$ respecting the net structure in the sense that $\alpha_g \A(O) = \A(g \cdot O)$ for all causal diamonds $O \subset U$.
For $g,g' \in {\rm N}$ such that also $gg' \in {\rm N}$, there holds  $\alpha_g \alpha_{g'} = \alpha_{gg'}$.
\end{enumerate}

The vacuum axiom becomes:
\begin{enumerate}
\item[a4')] (Vacuum) There is a unique state $\omega_0$ on $\A$ such that $\omega_0(\alpha_g(a))=\omega_0(a)$ whenever $\alpha_g(a)$ is defined.
On its GNS-representation $(\pi_0, \H_0, |0 \rangle)$, $\alpha_g$ is implemented by a strongly continuous projective positive energy\footnote{This means as usual that
$P^0$ has non-negative spectrum. It can be shown that this implies that also the ``conformal Hamiltonian'' $\half(P^0+K^0)$ appearing below has non-negative spectrum~\cite{mack}.
The relation $e^{isD} (P^0+K^0) e^{-isD} = e^{-s} P^0 + e^{+s}K^0$ then implies the same for $K^0$ (letting $s \to \infty$).}
representation $U$ of the
covering group $\widetilde{{\rm G}}$
%$= \widetilde{{\rm SO}_+(d+1,2)}$
in the sense that
\ben
U(g) \pi_0(a)|0 \rangle = \pi_0(\alpha_g (a))|0 \rangle \ ,
\een
whenever $\alpha_g(a)$ is defined.
\end{enumerate}

Denoting the generators of the Lie algebra $\mathfrak{so}(d+1,2)$ by $M_{AB}, A,B = 0, \dots, d+2$ with relations
\ben
[M_{AB}, M_{CD}] = 2(\tilde \eta_{A[C} M_{D]B} - \tilde \eta_{B[C} M_{D]C}),
\een
we define the following self-adjoint
generators on $\H_0$:
\ben\label{gendef}
\begin{split}
P^0 &= \frac{1}{i} \frac{\dd}{\dd t} U(\exp t(M_{0(d+2)} + M_{0(d+1)})) \Big|_{t=0} \ , \\
K^0 &= \frac{1}{i} \frac{\dd}{\dd t} U(\exp t(M_{0(d+2)} - M_{0(d+1)})) \Big|_{t=0} \ , \\
D &= \frac{1}{i} \frac{\dd}{\dd t} U(\exp tM_{(d+1)(d+2)} ) \Big|_{t=0} \ ,
\end{split}
\een
which are the generators of time-translations (ii), special conformal transformations (iii) in the time-direction, and dilations (iv), respectively.

We can now state
the first result of this subsection. Let $R>r>0$ and let $A$ be a ball of radius $r$, $B$ be the complement
a ball of radius $R$ centered at the origin in a time slice $\RR^d$ of $d+1$-dimensional Minkowski spacetime, see fig.~\ref{fig:concentric}.

As usual, let $O_A, O_B$ be the corresponding domains of dependence, and $\A_A = \pi_0(\A(O_A))''$ and
$\A_B = \pi_0(\A(O_B))''$ the corresponding v. Neumann algebras of observables acting on the vacuum Hilbert space $\H_0$.

\begin{theorem}\label{confddim1}
Let $\omega_0$ be the vacuum state. We have
\ben\label{nubound}
E_R(\omega_0) \le \log \ \tr \left( \frac{r}{R} \right)^{\frac{1}{2}(P^0 + K^0)} \ .
\een
\end{theorem}

{\em Proof:}
Since the theory is invariant under dilations, it is clearly sufficient to prove the theorem in the special case $R=1>r>0$.
We would like to apply theorems~\ref{Thm_ER<EN} and \ref{Thm_EN<Em}. We note again that $\pi(\A(O_{B'}))''\subset \A_B'$, it follows
$\Delta\le\Delta_{B'}$, where $\Delta$ is the modular operator for $\A_B'$ considered before in condition~\eqref{Psidefnuclear}, so that $\| \Psi^A \|_1 \le \| \Phi^A \|_1$.
This shows that
$E_R(\omega_0) \le \log \|\Phi^A\|_1$, where $\Phi^A: \A_A \to \H_0$ is
defined as usual by $a \mapsto \Delta_{B'}^{\quarter} a |0 \rangle$. Here $B'= \RR^d \setminus \overline{B}$, as usual. So
we need to estimate $\|\Phi^A\|_1$. Inspired by~\cite{buchholz_1}, we define
an operator $T$ by
\ben\label{Tdef}
T = \Delta_{B'}^{\quarter} \Delta_A^{-\quarter} \ .
\een
We would like to show that
\ben\label{Tid}
T = r^{\frac{1}{2}(P^0 + K^0)} r^{iD} \
\een
on a dense core of vectors in $\H_0$.
Let us assume this has been done. Then, since $D$ is self-adjoint, we clearly have
\ben
\sqrt{ TT^* } = r^{\frac{1}{2}(P^0 + K^0)} \ .
\een
It follows that
\ben
\| T \|_1 = \| \sqrt{ TT^* } \|_1 = \tr \ r^{\frac{1}{2}(P^0 + K^0)} \ .
\een
Now,
as we will argue momentarily, the map $\Xi^A: \A_A \to \H_0$ given by $\Xi^A(a) = \Delta^{\quarter}_A a |0 \rangle$ has norm
$\|\Xi^A\| \le 1$, so the above operator identity for $T$ gives us,
\ben\label{bound1}
\| \Phi^A\|_1 = \| T   \Xi^A \|_1 \le \|T  \|_1  \cdot \| \Xi^A \| \le \tr \ r^{\frac{1}{2}(P^0 + K^0)}  \ ,
\een
and therefore $E_R(\omega_0) \le \log \tr \ r^{\frac{1}{2}(P^0 + K^0)}$, thereby showing the theorem if we can demonstrate $\| \Xi^A \| \le 1$. This
can be demonstrated by the following standard argument. Consider the function appearing in the KMS-condition,
$f_a(z) = \langle 0 | a^* \Delta_A^{-iz} a | 0 \rangle$. By proposition~\ref{Prop1}, it is analytic on the strip
$0< \Im(z) < 1$ and continuous at the boundary of the strip. For the boundary value at $\Im(z) = 0$ one finds $|f_a(t)| \le \| a |0\rangle \|^2 \le \|a\|^2$.
For the boundary value at $\Im(z) = 1$, one can use the KMS-condition to find $|f_a(t+i)| = |f_{a^*}(-t)| \le \|a\|^2$. We can now apply the three line-theorem
and conclude that $|f_a(z)| \le \|a\|^2$ also in the interior of the strip. The value $z = i/2$ gives the result because $f_a(i/2) = \| \Xi^A(a) \|^2$, so indeed
$\|\Xi^A(a)\| \le \|a\|$.

We still need to demonstrate the operator identity~\eqref{Tid}. For this, we first formulate a lemma:

\begin{lemma}\label{bdformula}
For $s,t \in \RR$, we have the operator identity
\ben
\Delta_{B'}^{it} e^{isD} \Delta_{B'}^{-it} = \exp \bigg[ -\frac{is}{2} \sinh(2\pi t) (P^0+K^0) + is \cosh(2\pi t) D \bigg]
\een
on $\H_0$.
\end{lemma}

\noindent
{\em Proof:} To prove this formula, we use that the modular operators of double cones act in a geometrical way
in conformal quantum field theories according to the Hislop-Longo theorem~\cite{hislop, guido}\footnote{The proof was given in~\cite{hislop}
for a free massless scalar field. It was subsequently shown by Brunetti, Guido and Longo~\cite{guido} that the theorem
generalizes to theories that fit into our axiomatic setting for CFTs.}. The precise result is as follows. Let $L(t)$ be the 1-parameter family of conformal transformations
defined by
\ben\label{Ldef}
L(t) = \exp(-2\pi t M_{0(d+1)}) =
\left(
\begin{matrix}
\cosh (2\pi t) & 0 & \dots & -\sinh(2\pi t) & 0 \\
0                   & 1 & \dots & 0                  & 0\\
\vdots            &  \vdots  &          &   \vdots        & \vdots \\
 -\sinh(2\pi t)  & 0          & \dots &    \cosh(2\pi t) & 0 \\
 0 & 0 & \dots & 0 & 1
\end{matrix}
\right) \ .
\een
Then the theorem~\cite{hislop,guido} is that $\Delta_{B'}^{it} = U(L(t))$. Recalling that the generator of dilations in the Lie algebra
$\mathfrak{so}(d+1,2)$ is given by $M_{(d+1)(d+2)}$, we compute in ${\rm G}$
\ben
{\rm Ad}(L(t)) M_{(d+1)(d+2)} = - \sinh(2\pi t) M_{0(d+2)}+ \cosh(2\pi t) M_{(d+1)(d+2)} \ .
\een
Applying the unitary representation $U$, and recalling the defining relations \eqref{gendef} for $D,P^0,K^0$ this immediately gives
the statement of the lemma. \qed

\medskip

Next, since $D$ generates dilations, and since region $A$ is obtained from region $B'$ by
shrinking $B'$ by $r$, it is geometrically clear (and can easily be proven) that
\ben\label{ABid}
\Delta_A = e^{-i(\log r) D} \Delta_{B'} e^{i(\log r) D} \ ,
\een
so taking $s=-\log r$ in the lemma and multiplying the formula in the lemma by $e^{i(\log r) D}$ from the right, we get
\ben
\Delta_{B'}^{it} \Delta_{A}^{-it} = \exp \bigg[ +\frac{i}{2} \log (r) \sinh(2\pi t) (P^0 + K^0) - i \log (r) \cosh(2\pi t) D \bigg] \ e^{ i(\log r) D }
\een
in $\H_0$. The desired formula~\eqref{Tid} now formally follows by setting $t=-i/4$.

Justifying this last step occupies the
remainder of this proof. Generalizing~\eqref{Tdef}, let us set $T_s(z) = \Delta_{B'}^{z} \Delta_A^{-z}$, where $z$ is in the
strip $0 \le \Re(z) \le \frac12$. We include a subscript ``$s$'' here to emphasize the dependence on the regions $A,B$, since their relative positions
are fixed by $r=e^{-s}<1$. Since $\A_{B'} \supset \A_A$, it follows that (lemma 2.9 of \cite{lechner_2}, which uses a generalization of the Heinz-L\" owner theorem~\cite{hansen} to unbounded operators) $\Delta_{B'}^{\alpha} \le \Delta_A^{\alpha}$ for $0\le \alpha \le 1$.
Therefore, $T(\alpha/2)^* T(\alpha/2) \le 1$ for this range of $\alpha$.
It then also follows that $T_s(z)$ is holomorphic in the strip $0< \Re(z)<\frac{1}{2}$ and continuous on its closure, and that
$\| T_s(z) \| = \| \Delta_{B'}^{i\Im z} T_s(\Re z) \Delta_A^{-i\Im z}\|\le 1$ for $0\le \Re(z)\le \frac{1}{2}$.

Now let
\ben
\begin{split}
f_1(z) &=
\Delta_{B'}^{z} e^{isD} \Delta_{B'}^{-z} |\chi \rangle = T_s(z) e^{isD} |\chi \rangle \ , \\
f_2(z) &=
\exp \bigg[ \frac{is}{2} \sinh(2\pi iz) (P^0+K^0) + is \cosh(2\pi i z) D \bigg] |\chi \rangle \ ,
\end{split}
\een
defined first for $z$ such that $\Re(z) = 0$. By what we have just said about $T_s(z)$, the function $f_1(z)$ has
an analytic continuation to the strip $0< \Re(z)<\frac{1}{4}$ that is continuous on its closure. $f_2$ has an analytic continuation
to an open complex neighborhood of the form $\{z \mid |z|<\frac{1}{2}\}$ in the complex plane for a dense set of (``smooth'')
vectors $|\chi \rangle \in \H_0$~\cite{nelson}, provided $|s|$ is sufficiently small. By lemma~\ref{bdformula},
$f_1$ and $f_2$ agree on the imaginary axis. It follows by the edge of the wedge theorem (see Appendix A) that $f_1$ and $f_2$ coincide in the
open neighborhood of $\{ z \in \CC \mid 0< \Re(z)< \frac{1}{2}, |z| < \frac{1}{2} \}$ and by continuity on its closure. Thus we may take $z=\quarter$, and it follows
\ben\label{bprimedbprime}
\Delta_{B'}^{\quarter} e^{isD} \Delta_{B'}^{-\quarter} |\chi \rangle = \exp \bigg[ -\frac{s}{2} (P^0+K^0)  \bigg] |\chi \rangle
\een
for sufficiently small $|s|$ and all smooth vectors $|\chi \rangle$. Since the operators are bounded for $s>0$, the formula in fact holds for
all $|\chi \rangle \in \H_0$ when $s>0$ is sufficiently small.
Furthermore, both sides define 1-parameter semi-groups in $s$, so the identity holds also for all $s >0$.
The desired operator identity~\eqref{Tid} now follows setting $s=-\log(r)>0$ in view of~\eqref{ABid}. \qed

\medskip

Our next aim is to relate the ``partition function'' on the right side of our bound~\eqref{nubound} to the ``spectrum of operator dimensions'' in the given CFT.
To state our result, we need to describe our conformal field theory in terms of quantum fields $\O(x)$, which are unbounded operator-valued distributions.
Given an algebraic quantum field theory described by a net of observables algebras satisfying a1)-a5),~\cite{fredenhagen_1} have shown
how to define a set of linearly independent operator valued distributions
\ben
\Phi = \Bigg\{ \O: f \mapsto \O(f) = \int \O(x) f(x) \dd^{d+1} x \in {\mathfrak L}({\mathcal D}, \H_0) \bigg\}
\een
defined on a common, dense, invariant, domain $\mathcal D$ given by the subspace of vectors $|\chi\rangle \in \H_0$ such that $\|(1+ P^0)^\ell \chi  \| < \infty$ for all $\ell>0$.
These operator valued distributions are unbounded but  each field in this collection satisfies a bound of the form
$\| (1+P^0)^{-\ell} \O(x) (1+P^0)^{-\ell} \| < \infty$ for some sufficiently large number $\ell$, i.e. field operators exist point-wise if we
damp them appropriately.
The smeared fields $\O(f)$ associated with test-functions $f \in C^\infty_0(O)$ localized in a causal diamond $O$ are ``affiliated'' with the
local v.Neumann algebra $\pi_0(\A(O))''$ in the sense that their spectral projections are elements of this algebra. (They cannot of course themselves be in
the local v.Neumann algebra because they are unbounded.) The fields $\O$ can furthermore be arranged into multiplets transforming naturally
under Poincar\'e transformations in the sense that $U(\Lambda, a) \O(x) U(\Lambda,a)^* = D(\Lambda^{-1}) \O(\Lambda x + a)$, where $D$ is
some  irreducible representation of the covering of the Lorentz group $\widetilde{{\rm SO}_+(d,1)}$. We will assume for simplicity that all of these
representations $D$ are finite dimensional, i.e. that the multiplets have a finite number of components.

If the underlying net is even conformally invariant in the sense of a3'), a4'), then it is natural
to assume that the fields can be suitably re-organized into (larger) multiplets of the conformal group $\widetilde{\rm G}$. By this we shall mean that among all fields $\O \in \Phi$
there is a countable subset of linearly independent ``primary fields''. These by definition should transform as (in the sense of operator-valued distributions)~\cite{mack_1}
\ben\label{transf}
U(g) \O(x) U(g)^* = N(g,x)^{d_\O} D[\Lambda(g,x)^{-1}] \O(g \cdot x)
\een
for all pairs $(x,g)$ of points $x$ and conformal transformations $g \in \widetilde{\rm G}$ such that $g \cdot x$ can be deformed to $x$ for a path of
conformal transformations $t \mapsto g(t)$ without passing through the point at infinity. Here, $N(g,x)$ is the conformal
factor of the transformation, i.e. $N(g,x)^2 \eta_{\mu\nu} = g^* \eta_{\mu\nu}$. $d_\O \ge 0$ is called the ``dimension'' of the
primary field. $D(\Lambda(g,x))$ implements the tensorial transformation behavior of the field,
where $\Lambda(g,x) = N(g,x)^{-1} \partial(g \cdot x)/\partial x \in \widetilde{{\rm SO}_+(d,1)}$ is a Lorentz transformation associated with $g,x$, and $D$ is an irreducible, finite dimensional,
representation of $\widetilde{{\rm SO}_+(d,1)}$, see~\cite{mack,mack_1} for more explicit expressions. Besides primary fields,
there are ``descendants'', which are by definition fields of the form $\O_{\mu_1 \dots \mu_k} = [P_{\mu_1}, [ \dots, [P_{\mu_k}, \O]]]$,
where $\O$ is a primary field. The dimension of such a descendant is then defined to be
$d_\O + k$. We assume that the set of all fields $\Phi$ is spanned by the countably many primary fields and their (countably many) descendant
fields.\footnote{It would be interesting to see whether such an assumption can be derived from the basic axioms a1),a2),a3'),a4'),a5'). Partial progress in this
direction has been made by~\cite{bostelmann}.} We also assume that
\ben\label{HF}
{\rm span} \bigg\{ \O(f) | 0 \rangle \mid f \in C^\infty_0, \O \in \Phi \bigg\} \ \ \ \text{is dense in $\H_0$,}
\een
i.e. we may approximate in norm, with arbitrary precision, any vector in $\H_0$ by applying a suitable combination of smeared field $\O(f)$ to the vacuum.
Under these assumptions we now show:

\begin{theorem}\label{confddim3}
Let $A$ be a double cone whose base is a ball of radius $r$, and let
$B$ be the causal complement of a double cone whose base is a concentric ball of
radius $R>r$. Let
$\omega_0$ be the vacuum state. Under the assumptions on our conformal field theory just described, we have
\ben\label{nubound3}
E_R(\omega_0) \le \log \ \sum_{\O \in \Phi} \left( \frac{r}{R} \right)^{d_\O} \ .
\een
\end{theorem}

\begin{remark}
A corollary of the theorem is that if $\O$ is the primary field with the smallest non-zero dimension $d_\O$, then for large $R \gg r$, we have
\ben
E_R(\omega_0) \lesssim N_\O \left( \frac{r}{R} \right)^{d_\O}
\een
where $N_\O$ is the number of independent components of $\O$. \\
\end{remark}

\medskip
\noindent
{\bf Example:} For a free hermitian massless scalar field $\phi$ in $3+1$ dimensions, a basis of fields $\O$ for the set $\Phi$ can be chosen to be the
Wick monomials $\O(x) = :D^{(n_1)} \phi(x) \cdots D^{(n_k)} \phi(x):$, where the double dots denote
normal ordering, i.e. all creation operators are put to the left of all annihilation operators upon inserting relation~\eqref{phidef}.
The derivative operators $D^{(n)}$ are defined as
\ben
D^{(n)}_{\mu_1\dots\mu_n} = P^{\nu_1 \dots \nu_n}_{\mu_1\dots\mu_n} \partial_{\nu_1} \cdots \partial_{\nu_n} \ ,
\een
with $P^{\nu_1 \dots \nu_n}_{\mu_1\dots\mu_n}$ denoting the projection onto tensors which are trace free with respect
to any pair of indices (upon contraction with $\eta^{\mu_i\mu_j}$). The trace free condition arises from
the fact that $\partial^\mu \partial_\mu \phi=0$. The dimension is given by $d_{\O} = k + n_1 + \dots + n_k$.
The dimension of the space of trace free tensors of rank $n$
is given by $(n+1)^2$ in $3+1$ spacetime dimensions. From this, the conformal partition function is found to be
\ben
\log \sum_{\O \in \Phi} \left( \frac{r}{R} \right)^{d_\O}  = \log \prod_{n=1}^\infty \left( \frac{1}{1-(r/R)^n} \right)^{n^2} \lesssim  \frac{\pi^4}{45} \tau^{-3} \ ,
\een
as $\tau \to 0^+$, where $r/R = e^{-\tau}$, so according to our theorem
$E_R(\omega_0) \lesssim \frac{\pi^4}{45} \tau^{-3}$ as $r \to R$. On the other hand,
the field with the smallest dimension which is not the identity is $\phi$ itself, and $d_\phi =1$.
From this one finds $E_R(\omega_0) \lesssim r/R$ for $R \gg r$. \\
\medskip
\noindent

{\em Proof:} By conformal invariance, we may again assume without loss of generality that
$R=1>r>0$. The idea of the proof is to define the vectors
\ben\label{vectors}
|\O\rangle = \Delta^{\quarter} \O(0) | 0 \rangle \ ,
\een
where $\O$ runs through some basis of $\Phi$, and where here and in the rest of the proof, $\Delta=\Delta_{B'}$ is the modular
operator for the region $O_{B'} = O_B'$. Evaluating the operator identity of lemma~\ref{bdformula} for $t=-i/4$ formally gives for $s>0$
\ben
\exp \bigg[ -\frac{s}{2}  (P^0+K^0) \bigg] |\O \rangle = \exp \bigg[ -\frac{s}{2}  (P^0+K^0) \bigg] \Delta^{\quarter} \O(0) |0 \rangle = \Delta^{\quarter} e^{isD} \O(0) |0 \rangle \ .
\een
On the other hand, from the relation $e^{isD} P_\mu e^{-isD} = e^{-s} P_\mu$
of the conformal algebra, the fact that the conformal factor for a dilation by $\lambda$ is $\lambda$, and the invariance
of the vacuum,  we have
$e^{isD} \O(0) |0 \rangle = e^{isD} \O(0) e^{-isD} |0 \rangle = e^{-sd_\phi} \O(0) |0 \rangle$ for any primary or descendant field $\O$.
It follows that
\ben\label{eigenvalue}
\exp \bigg[ -\frac{s}{2}  (P^0+K^0) \bigg] |\O \rangle = e^{-sd_\O} |\O \rangle \ .
\een
If we can show that $\{ |\O \rangle \ \mid \ \O \in \Phi\}$ forms a basis of $\H_0$, then the vectors with fixed $d_\O$ span an eigenspace of $\exp\left[ -\frac{s}{2}  (P^0+K^0) \right]$. Putting $s=-\log(r)$ we therefore find
\ben\label{maineq}
\tr \, r^{\frac{1}{2}(P^0 + K^0)} = \sum_{\O \in \Phi} r^{d_\O} \ .
\een
In view of thm.~\ref{confddim1}, this would complete the proof. We now make the above somewhat formal arguments
rigorous. In order to do this, we first note that $\O(x)|0 \rangle$ is a $\H_0$-valued distribution that is the boundary value of a strongly holomorphic
$\H_0$-valued (see appendix~\ref{eow}) function in
the domain $\RR^{d,1} + iV^+$, where $V^+$ is the interior of the future lightcone. Indeed, this holomorphic extension may be defined as
(here $Pz=P_\mu z^\mu$)
\ben
\O(z) |0 \rangle = e^{-iPz} \O(0) | 0 \rangle := e^{-iPz} (1+P^0)^{\ell} \cdot [(1+P^0)^{-\ell} \O(x) (1+P^0)^{-\ell}] | 0 \rangle \ ,
\een
noting that
\ben
\| \O(z) | 0 \rangle \| \le
\| e^{-iPz} (1+P^0)^{\ell} \| \ \| (1+P^0)^{-\ell} \O(x) (1+P^0)^{-\ell} \| \le
C_\ell (\Im(z^0)-|\Im({\bf z})|)^{-\ell}.
\een
Hence, by a simple generalization of thm.~3.1.15 \cite{hormander} to Hilbert-space valued distributions, $\O(x)|0 \rangle$ is indeed the
distributional boundary value in the strong sense\footnote{This means that
\ben
\O(f) |0\rangle = \lim_{y \to 0, y \in V^+} \int d^{d+1} x \, \O(x+iy) |0\rangle f(x) \ .
\een
where the limit is understood in the norm topology on $\H_0$.}
of the holomorphic function $\RR^{d,1} + iV^+ \owns z \mapsto \O(z) |0\rangle \in \H_0$.
Next we use again the Hislop-Longo theorem~\cite{hislop} stating that the modular group $\Delta^{it}_{B'}$
is equal to $U(L(t))$, where $L(t)$ was given above by~\eqref{Ldef}. Let $x(t) = L(t) \cdot x$ (for $x \in O_{B'}$), and let $x_\pm = x^0 \pm |{\bf x}|$. Then it follows that
\ben
x_{\pm}(t) = \frac{
(1+x_\pm) - e^{2\pi t}(1-x_\pm)
}{
(1+x_\pm) + e^{2\pi t}(1-x_\pm)
}
\een
while ${\bf x}(t)/|{\bf x}(t)| = {\bf x}/|{\bf x}|$ for all $t \in \RR$. It is easy to check from this expression that, for fixed $t_0>0$ and $x$ in a sufficiently small neighborhood $O$ of the origin, the
complex points $x(t-is), 0< s < \quarter, |t| < t_0$ remain within $\RR^{d,1} + iV_+$. Thus, $\O(x(t-is)) |0 \rangle$ is a well-defined vector
in $\H_0$ for all $x \in O$. The conformal factor $N(x,t) \equiv N(L(t),x)$, and the associated Lorentz transformation,
$\Lambda(x,t) \equiv \Lambda(L(t),x)$ of the conformal transformations $L(t)$~\eqref{Ldef} appearing in the transformation law~\eqref{transf} are found to be\footnote{The second relation can be found by integrating the explicit infinitesimal versions of the transformation law given e.g. in~\cite{mack,mack_1}. }
\ben
\begin{split}
N(x,t) &=  \bigg( \cosh(\pi t) - x_+ \sinh(\pi t) \bigg)^{-1} \bigg( \cosh(\pi t) - x_- \sinh(\pi t) \bigg)^{-1} \ , \\
\Lambda(x,t) &= \exp \bigg(
\frac{2{\bf x} \cdot {\bf C}}{x_+-x_-} \, \log
\frac{\cosh(\pi t) - x_+ \sinh(\pi t)}{\cosh(\pi t) - x_- \sinh(\pi t)}
\bigg) \ ,
\end{split}
\een
where ${\bf C}=(M_{01},\dots,M_{0d})$ are the generators of boosts in $\mathfrak{so}(d+1,2)$.
It can be seen from these expressions that the analytic continuation $N(x,t-is)$ avoids the negative real axis for $|t|<t_0, 0 \le s < \quarter$ as long as $x \in O$
and as long as $O$ is a sufficiently small neighborhood of the origin (depending on $t_0$). Similarly, $\Lambda(x,t-is)$ remains single-valued in this range.
These facts imply that, for a primary field $\O$,
\ben\label{rigorous}
\Delta^{s + it} \O(x) |0 \rangle= N(x,t-is)^{d_\O} D[\Lambda(x,t-is)^{-1}] \O(x(t-is)) |0 \rangle
\een
pointwise for all $x \in O$, $|t|<t_0, 0<s<\quarter$, as both sides have the same distributional boundary value (in $x$) when $s \to 0^+$ by
the transformation law~\eqref{transf} for primary fields, and hence must coincide by the edge-of-the-wedge theorem, see appendix~\ref{eow}.
We may now set in this equation $x=0, t=0$
and let $s \to \tfrac{1}{4}^-$. Using $x_\pm(-\tfrac{1}{4}i) = i$, we find $x(-\tfrac{1}{4} i) = ie_0=(i,0,0\dots,0)$, $\Lambda(0,-\tfrac{1}{4} i)=1$
and $N(0,-\tfrac{1}{4} i)=1$, and we arrive at the formula
\ben
|\O \rangle = \Delta^{\quarter} \O(0) |0 \rangle = %2^{-d_\O} \,
\O(ie_0)|0 \rangle \ .
\een
Since $e_0$ is clearly inside the forward lightcone, the right side is a well-defined, non-zero vector in $\H_0$ (finite norm). Thus, we have
shown that $|\O \rangle$ is a well-defined vector when $\O$ is a primary field. By applying a suitable number of commutators with $[P_\mu, \ . \ ]$ to
\eqref{rigorous}, we can easily reach a similar conclusion for descendant fields $\O_{\mu_1 \dots \mu_k} = [P_{\mu_1}, [ \dots, [P_{\mu_k}, \O]]]$, namely
\ben\label{dectransf}
|\O_{\mu_1 \dots \mu_k} \rangle = i^k
\ \partial_{\mu_1} \dots \partial_{\mu_k} [N(x;-\tfrac{1}{4}i)^{d_\O} D[\Lambda(x,-\tfrac{1}{4}i)^{-1}]\,
 \O(x(-\tfrac{1}{4}i ))]| 0 \rangle \ \bigg|_{x=0} \ .
\een
That the set $\{|\O \rangle \mid \O \in \Phi\}$ forms a basis of $\H_0$ can now be seen as follows. Assume that $|\chi \rangle$ is orthogonal to
all $|\O \rangle, \phi \in \Phi$. It follows from~\eqref{dectransf} that for any $k$, we  have $\langle \chi | (\partial_{\mu_1}  \dots \partial_{\mu_k}
 \O)(ie_0) | 0 \rangle = 0$. Since $\O(z) |0 \rangle$ is holomorphic in an open neighborhood in $\CC^{d+1}$ of $z=ie_0$,
it follows that $ \langle \chi| \O(z) | 0 \rangle = 0$ for all $z$ in such a neighborhood. By the edge-of-the-wedge theorem, it follows
that $\langle \chi| \O(x) | 0 \rangle = 0$ in the distributional sense (i.e. after smearing with $f(x)$). Since this holds for all fields $\O \in \Phi$,
we conclude that $|\chi \rangle$ is in the orthogonal complement of the set~\eqref{HF}. Since that set is by assumption dense in $\H_0$, we conclude
that $|\chi \rangle=0$, i.e. we learn that $\{|\O \rangle \mid\O \in \Phi \}$ spans a dense subset of $\H_0$.

We next show that the elements in the set  $\{|\O \rangle \mid \O \in \Phi\}$ are linearly independent. Suppose that there exists a vanishing finite linear combination
$\sum_i c_i |\O_i \rangle = 0$ for a set of linearly independent fields $\O_i \in \Phi$.
Using~\eqref{dectransf}, we can also write this as $\sum_i c_i' \O_i'(ie_0) |0 \rangle = 0$, where $c_i'$ is a new set of
complex numbers and $\O_i'$ a new set of linearly independent fields in $\Phi$.  Let $\psi = \sum_i c_i' \O_i'$.
We conclude that $\psi(x+i\epsilon e_0)|0 \rangle=e^{-iP \cdot x} e^{(1-\epsilon) P^0} \psi(ie_0) |0 \rangle=0$ for any sufficiently small $\epsilon>0$ and all $x \in \RR^{d,1}$.
Thus, by the edge-of-the wedge theorem (see appendix~\ref{eow}), $f \mapsto \psi(f)|0\rangle=0$ in the distributional sense. Since $\psi(f)$ is affiliated with $\A(O)$,
we have $[\psi(f), a]=0$ in the strong sense for $a \in \A(O')$, so $\psi(f) a |0\rangle = 0$.
By the Reeh-Schlieder theorem, the set $a| 0 \rangle, a \in \A(O')$ is dense in $\H_0$, therefore we see that $\psi(f)=0$ for all test functions $f$, or in other
words, $\sum_i c_i' \O_i' = 0$ as an identity between quantum fields (i.e. when smeared with any test function $f$). Thus, we see that $c_i'=0$, and this is also easily seen
to imply that all $c_i=0$. This completes the proof that $\{ |\O\rangle \mid \O \in \Phi \}$ forms a basis of $\H_0$.

The rest of the argument leading to eq.~\eqref{maineq} can now also be made rigorous using eq.~\eqref{rigorous}
repeating the above formal steps with this equation for $0<s<\quarter$, and taking $s \to \quarter^-$ in the end.
\qed

\subsection{Upper bounds for CFTs in $3+1$ dimensions}

In the previous section, we have treated the case when $O_A$ is a diamond whose base is a ball, $A$, and where $O_B$ is the complement of a concentric diamond. It is of interest to obtain also a bound when the diamonds are in arbitrary position (i.e. not concentric), but still of course
$O_A \subset O_B'$ is in the causal complement, see fig.~\ref{fig:diamond1}. Upper bounds can be obtained in this case by essentially the same method as in the previous subsection,
but the formula for the upper bound becomes somewhat more complicated. To keep the complications at a minimum, we will only consider the case when $d=3$, i.e. $3+1$ dimensional CFTs.

The key point is basically to understand the finite dimensional irreducible representations of $\widetilde{{\rm SO}_+(3,1)} \cong {\rm SL}_2(\CC)$, $D$, in the transformation
formula~\eqref{transf} for the quantum fields. These are best described in spinorial form. The inequivalent $D$'s are labelled by
two natural numbers $s,s'$ and act on the vector space
\ben
V_{s,s'} = E_s(\CC^{2 \ \otimes s}) \otimes E_{s'}(\bar \CC^{2 \ \otimes s'})
\een
where $E_s$ projects onto the subspace of symmetric rank $s$ tensors. The action of $D_{s,s'}(g), g \in {\rm SL}_2(\CC)$ on a tensor $T \in V_{s,s'}$ is given by
\ben
(D_{s,s'}(g)T)_{A_1 \dots A_sB_1' \dots B_{s'}'} =
g_{A_1}{}^{C_1} \dots g_{A_s}{}^{C_s} \bar g_{B_1'}{}^{D_1'} \dots \bar g_{B_{s'}'}{}^{D_{s'}'} T_{C_1 \dots C_sD_1' \dots D_{s'}'} \ .
\een
Tensors over $\RR^{3,1}$ correspond to elements of $V_{s,s'}$ by the rules explained in detail e.g. in~\cite{wald_3}.
For instance, an anti-symmetric tensor $T_{\mu\nu}=-T_{\nu\mu}$ decomposes into one complex
component $T_{AB}$ in $V_{2,0}$ and another one $\bar T_{A'B'}$ in $V_{0,2}$.
The ``spin'' of the finite dimensional representation is $S=\frac12 s + \frac12 s'$, and we can also define
the left and right chiral spins by $S^L=\frac12 s, S^R= \frac12 s'$. The transformation behavior~\eqref{transf} of
a quantum field $\O$ under Lorentz transformations is described by $S^L_\O,S^R_\O$, and the transformation behavior under dilations by its
dimension, $d_\O$. We can now state our result.

\begin{theorem}\label{confddim}
Let $O_A$ be a double cone which is the intersection of the past of a point $x_{A+}$ and the future of
a point $x_{A-}$. Similarly, let $O_B$ be the complement of a double cone which is the intersection of the past of a point $x_{B+}$ and the future of
a point $x_{B-}$. It is required that $O_A$ is properly contained in the other double cone, i.e. the causal complement of $O_B$.
Define the conformally invariant cross-ratios by
\ben\label{xr}
\begin{split}
u &=\frac{
(x_{B+}-x_{B-})^2(x_{A+} - x_{A-})^2
}{
(x_{A-}-x_{B-})^2(x_{A+} - x_{B+})^2
} > 0\\
v &= \frac{
(x_{B+}-x_{B-})^2(x_{A+} - x_{A-})^2
}{
(x_{A-}-x_{B+})^2(x_{A+} - x_{B-})^2
} > 0,
\end{split}
\een
and let $\tau, \theta$ be defined by
\ben\label{thtau}
\theta = \cosh^{-1} \left( \frac{1}{\sqrt{v}} - \frac{1}{\sqrt{u}} \right) \ , \quad
\tau = \cosh^{-1} \left( \frac{1}{\sqrt{v}} + \frac{1}{\sqrt{u}} \right) \ .
\een
 Let
$\omega_0$ be the vacuum state. Under the assumptions on our conformal field theory described in the previous subsection, we have
in $3+1$ dimensions:
\ben\label{nubound1}
E_R(\omega_0) \le \log \ \sum'_{\O \in \Phi} e^{-\tau d_\O}
[2S_\O^R+1]_\theta [2S_\O^L+1]_\theta \ ,
\een
with $[n]_\theta = (e^{n\theta/2} - e^{-n\theta/2})/(e^{\theta/2} - e^{-\theta/2})$.
For $\tau \sim |\theta| \gg 1$ this gives
\ben
E_R(\omega_0) \lesssim N_{\O} \cdot e^{-\tau (d_\O-S_\O)}
 \ ,
\een
where $\O$ is the operator with the smallest ``twist'' $d_\O-S_\O$ and $N_{\O}$ its multiplicity.
\end{theorem}

\begin{remark}
Note that, unlike in thm.~\ref{confddim3}, the sum $\sum'$ over $\O$ is over all different independent field {\em multiplets} under ${\rm SL}_2(\CC)$, not
their individual operator {\em components}. Thus, for instance, a hermitian tensor field operator $\O_{\mu\nu}$ satisfying
$\O_{\mu\nu}=-\O_{\nu\mu}$ would
correspond under the identification $\epsilon_{AB} \O_{AB} + \bar \epsilon_{A'B'} \O_{A'B'}^*$ to $2$ multiplets,
namely $\O_{AB}$ and $\O_{A'B'}^*$, one having $S_\O^L=1, s_\O^R=0$, and the other $S_\O^L=0, s_\O^R=1$, and not $6$ real component fields.
\end{remark}

{\em Proof:} By applying a conformal transformation to the double cones, one can achieve that
\ben
x_{B\pm} = \pm (1, 0 , 0 , 0 ) \ , \quad x_{A\pm}= (\pm e^{-\tau} \cosh \theta, e^{-\tau} \sinh \theta, 0, 0) \
\een
for some $\tau, \theta$ satisfying $\tau>|\theta|$ (the last statement uses the assumptions on the relative position of the diamonds).
Computing the cross ratios $u,v$ for these points, one finds precisely the
relations~\eqref{thtau}. Since the CFT is conformally invariant in the sense of $a3'), a4')$ it suffices to prove the
theorem for this special configuration. As in the previous subsection, let $\Delta_A, \Delta_{B'}$ be the modular operators
for the diamonds $O_A, O_B'$. Define the operator $T$ as before in~\eqref{Tdef}. As before, it follows that $E_R(\omega_0) \le \log \tr |T|$.
Thus, we need to compute this trace.

Define
\ben\label{gendef1}
\begin{split}
P^1 &= -\frac{1}{i} \frac{\dd}{\dd t} U(\exp t(M_{15} + M_{14})) \Big|_{t=0} \ , \\
K^1 &= -\frac{1}{i} \frac{\dd}{\dd t} U(\exp t(M_{15} - M_{14})) \Big|_{t=0} \ , \\
L_{01} &= \frac{1}{i} \frac{\dd}{\dd t} U(\exp tM_{01} ) \Big|_{t=0} \ ,
\end{split}
\een
which are the generators of translations/special conformal transformations in the $1$-direction, and boosts in the $01$-plane.
By conformal invariance, we can write (compare~\eqref{ABid})
\ben
T= \Delta_{B'}^{\quarter} e^{-i\tau D + i\theta L_{01}} \Delta_{B'}^{-\quarter} e^{i\tau D - i\theta L_{01}}  \equiv
X e^{i\tau D - i\theta L_{01}} \ .
\een
Thus, $\tr |T| = \tr |X|$. In basically the same way as in the previous subsection, one derives the operator identity (compare~\eqref{bprimedbprime})
\ben
X = \exp \left[
-\frac12 \tau(P^0 + K^0) + \frac12 \theta(P^1-K^1)
\right] \ ,
\een
so $\tr |X| = \tr \exp [
-\frac12 \tau(P^0 + K^0) + \frac12 \theta(P^1-K^1) ]$.  Define next the vectors $| \O \rangle$ as in~\eqref{vectors}. By the same arguments as there
(compare \eqref{eigenvalue}), we find
\ben
X|\O \rangle = e^{-\tau d_\O} D_{s_\O,s_\O'}[\exp (-\theta M_{01})] | \O \rangle \ .
\een
We must now determine how the finite dimensional matrix $D_{s_\O,s_\O'}(\exp \theta M_{01})$ acts on an operator $\O$ transforming in the
representation $V_{s_\O, s_\O'}$ of ${\rm SL}_2(\CC)$. This is conveniently done by choosing a basis $e_0, e_1$ in $\CC^2$ such that, relative to this basis
\ben
(D_{s_\O,s_\O'}(\exp \theta M_{01})T)_{A_1 \dots A_sB_1' \dots B_{s'}'} = e^{\frac12\theta(n_0-n_1)+\frac12 \theta(n_0'-n_1')} T_{A_1 \dots A_sB_1' \dots B_{s'}'}
\een
where $n_0$ is the number of times an $A_i$ assumes the value $0$, $n_1$ the number of times an $A_i$ assumes the value $1$,
where $n_0'$ is the number of times a $B_i'$ assumes the value $0$, and $n_1'$ the number of times a $B_i'$ assumes the value $1$.
The above expression follows from the way in which the isomorphism $\widetilde{{\rm SO}_+(3,1)} \cong {\rm SL}_2(\CC)$ is set up. We may now
compute the trace $\tr |X|$ in the basis $\{|\O \rangle \}$. A straightforward computation using $n_0 + n_1 = s_\O, n_0'+n_1'=s_\O'$ gives the right side of the
formula~\eqref{nubound1}, and the proof is complete. \qed

\subsection{Upper bounds for chiral CFTs}\label{ssec_CFTchiral}

Here we apply the general upper bounds provided by
theorems~\ref{Thm_ER<EN} and \ref{Thm_EN<Em} to the case of a chiral CFT described by a net of v. Neumann algebras
$\{\A(I)\}$ over the circle $S^1$. We  consider the entanglement entropy $E_R(\omega_0)$ of two disjoint open intervals $A,B \subset S^1$
in the vacuum state $\omega_0$. By abuse of notation, we denote these intervals by $A=(a_1 ,a_2), B = (b_2, b_1)$, where\footnote{We adopt the convention that points on $S^1$ are labelled clockwise.} $a_1, a_2, b_1, b_2 \in S^1$.
We denote the GNS-representation of $\omega_0$ on $\H$ by $\pi_0$ and the vacuum vector by $|0\rangle$.
The unitary projective
positive energy representation of the covering $\widetilde{\rm G}$ of the
conformal group ${\rm G}={\rm SU}(1,1)$ on $\H$ is denoted by $U_0(g), g \in \widetilde{\rm G}$; invariance of the vacuum
means that $U_0(g) |0\rangle = |0 \rangle$ for all $g \in \widetilde{\rm G}$. The infinitesimal generator of
rotations of $S^1$ in the representation $\pi_0$ is denoted by $L_0=\dd/\dd t \, U_0({\rm diag}(e^{it/2}, e^{-it/2}))|_{t=0}$. The following result can be obtained in exactly the same way as those in the previous two subsections.

\begin{theorem}
Let $\omega_0$ be the vacuum state. We have
\ben\label{nubound4}
E_R(\omega_0) \le \log \left( \tr \exp \left\{ - 2\sinh^{-1} \sqrt{\xi} \cdot L_0 \right\} \right)
\een
where $\xi$ is the conformally invariant cross ratio associated with the pair of intervals $A=(a_1,a_2), B=(b_2,b_1) \subset S^1$ given by
\ben
\xi = \frac{(a_2-b_2)(b_1-a_1)}{(a_2-a_1)(b_2-b_1)}  \ .
\een
\end{theorem}

\begin{remark}
1) The upper bound is obviously only non-trivial if $e^{-\tau L_0}$ has finite trace for $\tau>0$, which is the case e.g. in all rational conformal field theories.
In many such theories, there are concrete formulas for the character $\tr \ e^{-\tau L_0}$ leading to an explicit bound under the substitution
$\tau = 2\sinh^{-1} \sqrt{\xi}$, see e.g.~\cite{difrancesco}.

2) As before in thms.~\ref{confddim1},~\ref{confddim}, our bound is  an upper bound on $E_M(\omega_0) \ge E_R(\omega_0)$ as well, where $E_M$ is the modular entanglement measure.

\end{remark}

{\em Proof}:
Define again an operator $T$ as above in~\eqref{Tdef}, so that, as argued in the proof of
thm.~\ref{confddim1}, $E_R(\omega_0) \le \log \tr |T^*|$, where $|T^*|=\sqrt{TT^*}$. \cite{buchholz_1}
show
\ben
\sqrt{ T T^* } = e^{- \ell(A,B) L_0} \ .
\een
Here $\ell(A,B)$ is an ``inner distance'' associated with the inclusion $(a_1, a_2) \subset (b_1, b_2)$. Going through the implicit definitions
given by these authors, one finds the explicit formula
$
\ell(A,B) = 2 \sinh^{-1} \sqrt{\xi}
$.
 \qed

\medskip

Via the Cayley transform
$z \in S^1 \mapsto i(z-1)/(z+1) \in \RR$ we get a corresponding bound for the theory on the light ray $\RR$, which, in fact, has the same form
(with $a_i, b_i$ now in $\RR$ rather than $S^1$), since the cross ratio $\xi$ retains its form under the Cayley transform.

We can also get asymptotic formulas for the entanglement
entropy (e.g. in the lightray picture) using the known behavior of the character $\tr e^{-\tau L_0}$ for small respectively large $\tau>0$. These bounds are conveniently expressed in terms of the dimensionless ratio
$
r = \frac{{\rm dist}(A,B)}{\sqrt{|A|\cdot|B|}},
$
where $A,B \subset \RR$ have lengths $|A|$, $|B|$. Thus, widely separated intervals have $r\gg 1$ and $r\simeq\sqrt{\xi}$, while intervals separated by a short distance compared to their length have $r \ll 1$. Using that the smallest non-zero eigenvalue of $L_0$ is $1$ with some multiplicity $n_1$ in the vacuum sector, the theorem immediately implies
\ben
E_R(\omega_0) \lesssim \frac{n_1}{4r^2}  \quad \text{for $r \gg 1$.}
\een
We can also get a bound in the opposite regime $r \ll 1$ if we have an asymptotic bound on
$\tr e^{-\tau L_0}$ for $\tau \to 0^+$. For a rational conformal chiral net, \cite{kawahigashi_2} have shown for instance that
\ben
\log \tr \ e^{-\tau L_0} = \frac{c\pi^2}{6 \tau} - \frac{1}{2} \log \mu_{\A} + O(\tau) \ ,
\een
where $\mu_\A$ is the so-called $\mu$-index of the net $\{\A(I)\}$, given by the sum of the square of the statistical dimensions
$\mu_\A = \sum \dim (\rho)^2$ over all irreducible sectors (see the next subsection for further explanations concerning this notion).
This implies
$
E_R(\omega_0) \lesssim \frac{c\pi^2}{6 r},
$
for $r \ll 1$,
which falls short of the expected~\cite{calabrese} logarithmic behavior of $E_R(\omega_0)$ for $0<r \ll 1$.
We should remember, however, that our bound is also an upper bound on $E_M(\omega_0)$, since the proof really estimates this quantity and
since $E_M \ge E_R$ in general. If we use instead the tighter bound $E_I \ge E_R$ in terms of the mutual information, then using the
exact result $E_I(\omega_0) = -\frac{c}{3} \log \xi$ recently obtained by \cite{longo_6} (building on earlier work by \cite{casini_3}) for free fermions where
$c=1/2$, one gets this logarithmic behavior at least in that case.

\subsection{Charged states}\label{sec:charged}

According to the general philosophy, the algebras of observables $\A(O)$ contain the observables of the theory accessible to an observer in $O$.
They are not, however, supposed to contain non-observable fields, such as e.g. charge carrying fields, where ``charge'' is understood here in a rather broad sense.
For instance, in a theory containing fermionic fields, only bosonic combinations (such as bi-linears in such fields) would be in $\A(O)$.
Similarly, fields that are charged under some group $G$ of internal symmetries, would not be in $\A(O)$, only combinations which are invariant under $G$,
i.e. `singlets'. In the algebraic approach, these objects arise ``through the back door''
when considering the GNS-representation of ``charged states'', or simply ``charged representations''.
It is outside the scope of this article to review this beautiful theory, initiated by~\cite{doplicher_1,doplicher_2,doplicher_4}, see~\cite{fredenhagen_2,longo_1,longo_2}
for the case of quantum field theories in 1+1 dimensions, where qualitatively new phenomena are possible.

Here we only give the basic framework and make some comments. For more comprehensive expositions besides the original papers see~\cite{haag_2,fredenhagen_4,araki_6}.
For simplicity, we will work in Minkowski spacetime, and we let $\pi_0$ be the vacuum
representation. Using $\pi_0$, we pass to the corresponding net $\pi_0(\A(O))''$ of v. Neumann algebras on $\H_0$, which by abuse of notation is
denoted again by $\A(O)$.  A localized, charged representation $\pi_\rho$ is one such that there exists a double cone $O$ such that $\pi_0$ is unitarily equivalent to $\pi_\rho$ when restricted to $\A(O')$, where $O'$ is the causal complement, i.e. $\pi_0(a) = V \pi_\rho(a) V^*$ for all $a \in \A(O')$. In order to make possible a general analysis, one assumes for technical reasons the so-called ``Haag duality'', i.e. $\A(O)' = \A(O')$ [a strengthened version of a2)] for all causal diamonds, which is satisfied in many models.
By identifying the Hilbert spaces $\H_0$ and $\H_\rho$ with the isometry $V$,
one can then easily show that there exists an algebra homomorphism $\rho$ of $\A=\cup_O \A(O)$ such that
\ben
\pi_\rho \circ \rho = \pi_0 \ , \quad \rho(a) = a \quad \text{for all $a \in \A(O')$} .
\een
Since we identify the net with its representation under $\pi_0$, the vacuum representation is effectively the identity, and we may thus drop the symbol $\pi_0$.
Because of the last property, $\rho$ is a ``localized endomorphism'', i.e. it acts non-trivially only on observables localized within $O$,
from which it follows that $\rho(\A(O)) \subset \A(O)$. The study
of charged representations is thereby reduced to the study of such localized endomorphisms and the associated inclusions of v. Neumann algebras.

One may ask what this notion of charge has to do with the notion of charge carrying field alluded to above.
This is clarified by the famous Doplicher-Roberts (DR)-reconstruction theorem~\cite{doplicher_3}.
Its basic content is the following. Assume the number of spatial dimensions $d$ is greater than one.
Then there exists a ``field net'' $\{ \mathfrak{F}(O) \}$ represented on a larger Hilbert space $\H$ which decomposes as\footnote{Here $[\rho]$ is the equivalence class of all $\rho$ under the natural notion of unitary equivalence. It is meant that there is one summand for each class of irreducible $\rho$'s.}
\ben\label{decomp}
\H = \bigoplus_{[\rho]} \H'_\rho \otimes \H_\rho^{} \ ,
\een
and a compact group $\rm G$
acting by automorphisms $\alpha_g, g \in {\rm G}$ on the field net, such that $\A(O)$ consists precisely of those elements $F \in \F(O)$ that are invariant
under all $\alpha_g$, i.e. $\alpha_g(F) = F$ for all $g \in {\rm G}$. The operators $F \in \mathfrak{F}(O)$ which do not have this property are the ``charge carrying fields''
localized in $O$. Each $\alpha_g$ is implemented by a unitary representation of the form $U(g) = \oplus_{[\rho]} U_\rho(g) \otimes 1_{\H_\rho}$,
where each $U_\rho$ is an irreducible, unitary representation of $G$ on $\H'_\rho$. The charged vectors for irreducible $\rho$ correspond to the vectors in
the subspace $\H_\rho' \otimes \H_\rho^{}$, which is often called the ``superselection sector'' (of $\rho$).

DR~\cite{doplicher_3} have shown that for each localized endomorphism $\rho$ in $O$ the field algebra $\F(O)$
contains a copy of the Cuntz algebra $\mathcal{O}_{\dim(\rho)}$ (see sec.~\ref{ssec_Cuntz}),
where $\dim (\rho) = {\rm dim}(\H'_\rho)$ is called the ``statistical dimension'' of $\rho$. If $F_i, i=1, \dots, \dim(\rho)$ is a collection of operators
(`multiplet') in $\F(O)$ transforming under
$U_\rho$, i.e. $\alpha_g(F_i) = \sum_j U_\rho(g)_{ij} F_j$, then it can be shown that $F_i$ can be written as $F_i = a \psi_i$, where $a \in \A(O)$ and $\psi_i$
is the generator of a Cuntz algebra sitting inside $\F(O)$. For $a \in \A$, the action of the endomorphism $\rho$ is
\ben\label{rhodef}
\rho(a) = \sum_{i=1}^{\dim(\rho)} \psi_i^{} a \psi_i^* \ .
\een
Thus, if $\omega$ is a state represented by a vector $|\Psi \rangle$ in $\H_0$, it has no net charge. The state $\omega \circ \rho \equiv \rho^* \omega$ corresponds to adding one unit of
charge and can be represented by the vector $\sqrt{\dim(\rho)} \psi_i^* |\Psi \rangle$. In $d=1$ spatial dimensions, the DR reconstruction theorem does not necessarily hold.
However, the notion of localized endomorphism still makes sense. Indeed, one of the most attractive
features of the DHR theory of superselection sectors is that the theory can be formulated intrinsically in terms of these. In particular, even in $d=1$, one can still
give an intrinsic definition of the statistical dimension $\dim (\rho)>0$, which still has many properties of a ``dimension'', even though it no longer needs to be an integer.

\medskip

After this brief review, we now make a connection between statistical dimensions and entanglement entropies.
As before, we set $\A_A = \pi_0(\A_A(O_A))''$ and similarly for $B$. By abuse of language we say that $\rho$ is localized in $A$ or $B$
if $\rho$ is an endomorphism localized in $O_A$ or $O_B$. We then have $\rho(\A_A) \subset \A_A$, and similarly for $B$.

\begin{proposition}\label{propdim}
Let $\omega$ be any faithful normal state in the vacuum representation,
and $\rho = \prod_i \rho_i^{n_i}$ a product of
finitely many irreducible sectors $\rho_i$ with statistical dimensions $\dim (\rho_i)$ localized in $A$ or in $B$, so that $\rho^* \omega = \omega\circ \rho$ can be thought of as
containing for each $i$ precisely $n_i$ additional units of charge of type $[\rho_i]$ relative to $\omega$. Then
\ben\label{rhodef1}
0\le  E_R(\omega) - E_R(\rho^*\omega)  \le \log \prod_i \dim (\rho_i)^{2n_i} \ .
\een
\end{proposition}
\begin{remark}
1) A state is faithful if $\omega(a) = 0$ for $a \in \A^+$ implies $a=0$. By the same argument as in the Reeh-Schlieder theorem, this will hold for instance if
$\omega$ is implemented by a vector in $\H_0$ with finite energy. \\
2) Our formula reminds one of results by \cite{Caputa} and \cite{Nozaki}, where the difference between the v. Neumann entropy for $A$ of the vacuum state
and a state obtained by applying a charge-carrying field to the vacuum is computed.
\end{remark}

{\em Proof:}
To save writing, we put $\mathfrak{M} = \A_A \vee \A_B \cong \A_A \otimes \A_B$, and $\mathfrak{N} = \rho(\mathfrak{M})$.
We denote by $\sigma'$ a separable state on $\mathfrak{N}$, and by $\omega'$ the restriction of $\omega$ to $\mathfrak{N}$.
 By (e4) it immediately follows that
$
E_R(\omega') \le E_R(\omega) \ .
$
In order to get a lower bound on $E_R(\omega')$
we recall the so-called ``left-inverse'' of $\rho$, which is a standard ingredient in DHR-theory~\cite{doplicher_1,doplicher_2}. The left inverse is a linear map $\Psi_\rho: \mathfrak{M} \to \mathfrak{M}$
such that $\Psi_\rho \rho = id$ and such that $\Psi_\rho(\rho(a)b\rho(c)) = a\Psi_\rho(b)c$ (it is not an endomorphism in general).
A canonical (called ``standard'')
left inverse always exists if $\rho$ is irreducible. [In case we are in $>1$ spatial dimensions the DHR reconstruction theorem applies, as described above. In terms of the Cuntz-generators $\psi_i$, $i=1, \dots, \dim (\rho)$,
the standard left inverse is then given by the explicit formula $\Psi_\rho(a) = \dim (\rho)^{-1}\sum_i \psi_i^* a \psi_i$, where $\dim(\rho) \in {\mathbb N}$ is the statistical dimension. The left inverse property then follows manifestly from~\eqref{rhodef} and the relations of the Cuntz algebra \eqref{cntz}.]
In that case $\mathcal{E} = \rho \Psi_\rho$ is shown to be a faithful conditional expectation from
$\mathfrak{M} \to \mathfrak{N}$. The smallest constant $c>0$ such that $\mathcal{E}(a^*a) \ge c^{-1} a^*a$ is the Jones-index $[\mathfrak{M}:\mathfrak{N}]$ associated with the inclusion $\mathfrak{N} \subset \mathfrak{M}$ (Pimsner-Popa inequality~\cite{pim}).
By the index-statistics theorem~\cite{longo_1,longo_2},
\ben
[\mathfrak{M} : \mathfrak{N}]^{\frac12} = \dim(\rho) \ ,
\een
so ${\mathcal E}^* \omega \ge \dim(\rho)^{-2} \omega$, implying in particular that ${\mathcal E}^* \omega$ is faithful.

For an arbitrary $\epsilon>0$ let $\sigma'$ be a separable state on $\mathfrak{N}$ such that $E_R(\omega') \ge H(\omega', \sigma') - \epsilon$. Then $\sigma := \mathcal{E}^* \sigma'$ is a separable state on $\mathfrak{M}$ because, due to the localization properties of
$\Psi_\rho$, $\mathcal{E}$ preserves tensor products in the sense that
$\mathcal{E}(a \otimes b) = \mathcal{E}(a) \otimes \mathcal{E}(b)$ for $a \in \A_A, b \in \A_B$. Thus, due to the infimum in the definition of $E_R(\omega)$, we get
$E_R(\omega) \le H(\omega, \sigma)$. On the other hand, using the chain rule for the Connes-cocycle (and the definitions of $\sigma'$ and $\omega'$)
\ben\label{chain}
\begin{split}
[D\omega:D\sigma]_t
&= [D\omega:D(\mathcal{E}^*\omega)]_t [D(\mathcal{E}^*\omega):D\sigma]_t \\
&= [D\omega:D(\mathcal{E}^*\omega)]_t [D(\mathcal{E}^* \omega'):D(\mathcal{E}^*\sigma')]_t \\
&= [D\omega:D(\mathcal{E}^*\omega)]_t [D\omega':D\sigma' ]_t ,
\end{split}
\een
using in the last line the fact that
$[D(\mathcal{E}^*\omega' ):D(\mathcal{E}^*\sigma')]_t = [D\omega':D\sigma' ]_t$, which follows since there exists  a
faithful, $\mathcal{E}$ invariant state (namely ${\mathcal E}^* \omega$) on $\mathfrak{M}$, see e.g. sec.~4 of~\cite{ohya_1} for a discussion. We get
\ben
\begin{split}
H(\omega, \sigma) &= \lim_{t \to 0} \frac{ \omega([D\omega:D\sigma]_t-1) }{it} \\
&= \lim_{t \to 0} \frac{\omega([D\omega':D\sigma']_t -1)}{it} +
\lim_{t \to 0} \frac{\omega([D\omega:D({\mathcal E}^*\omega)]_t -1)}{it}\\
&+ \lim_{t \to 0} \frac{\langle ([D\omega:D(\mathcal{E}^*\omega)]_t -1)^* \Omega| ([D\omega':D\sigma']_t-1) \Omega \rangle}{it} \\
&= \lim_{t \to 0} \frac{\omega'([D\omega':D\sigma']_t -1)}{it} +
\lim_{t \to 0} \frac{\omega([D\omega:D({\mathcal E}^*\omega)]_t -1)}{it}\\
&= H(\omega' , \sigma' ) + H(\omega, \mathcal{E}^* \omega)  \le H(\omega', \sigma') + \log \dim (\rho)^2 \ .
\end{split}
\een
In the first step we used the alternative definition of $H$ in terms of the Connes-cocycle~\eqref{drel1}. In the second step we used the chain rule
for the Connes-cocycle and~\eqref{chain}. In the third step we used that $[D\omega':D\sigma']_t \in \mathfrak{N}$ so that
$\omega([D\omega':D\sigma']_t)= \omega'([D\omega':D\sigma']_t)$,
as well as
\ben
\begin{split}
& \left|
\frac{1}{it}
\langle ([D\omega:D(\mathcal{E}^*\omega)]_t -1)^* \Omega| ([D\omega':D\sigma']_t-1) \Omega \rangle
\right| \\
\le &
\frac{1}{|t|}
\| ([D\omega:D(\mathcal{E}^*\omega)]_t -1)^* \Omega \|
\
\| ([D\omega':D\sigma']_t-1) \Omega \| \\
= &2 \bigg\{
\frac{\Re \omega([D\omega:D(\mathcal{E}^*\omega)]_t -1)}{t} \
\frac{\Re \omega'([D\omega':D\sigma']_t-1)}{t}
\bigg\}^{\frac12} \\
\to & 0 \quad \text{as $t\to 0$,}
\end{split}
\een
since $\omega([D\omega:D(\mathcal{E}^*\omega)]_t -1)/t \to iH(\omega,\mathcal{E}^* \omega)$ and
$\omega'([D\omega':D\sigma']_t-1)/t \to iH(\omega',\sigma')$ and since $\infty >  H(\omega',\sigma')$.
 In the fourth
step, we used again the definition of $H$ in terms of the Connes-cocycle.
In the last step we  used  $ \mathcal{E}^*\omega \ge \dim (\rho)^{-2} \ \omega$,
the monotonicity of the relative entropy in the second entry, (h5), and $H(\omega,\omega)=0$.
We therefore get
\ben
E_R(\omega) \le E_R(\omega') + \epsilon + \log \dim(\rho)^2 \ .
\een
From (h2), one also knows that $E_R(\omega') = E_R(\rho^* \omega)$ (since $\rho: \mathfrak{M} \to \mathfrak{N}$ is faithful).
Since $\epsilon$ was arbitrary, the proof is complete for irreducible $\rho$. In case $\rho = \prod \rho^{n_i}_i$, we proceed by
iterating the above argument treating the irreducible endomorphisms $\rho_i$ in the product one by one from the right. \qed

\medskip
\noindent
{\bf Example:} (Real $N$-component free KG-field in 3+1 dimensions). The quantum field theory is a simple variant of a 1-component KG theory, the algebraic formulation of
which has been described in sec.~\ref{ssec_free}. The
symplectic space $K_\RR$ for the theory with one component is replaced now by $N$ copies $K_\RR^N=K_\RR \oplus \dots \oplus K_\RR$ corresponding to the $N$
components of the field, i.e. the smearing functions $f$ now have $N$ components $f=(f_I)_{I=1, \dots, N}$. The vacuum state $\omega_0$
and its GNS-triple $(\H, \pi, |0\rangle)$ are only modified in a trivial way.
The field algebra (in the sense described above) is $\mathfrak{F}(O) = \pi(\{ W(f) \mid \supp(f) \subset O \})''$. An element $g \in {\rm O}(N)$ acts
on a test function by $(g.f)_I = \sum_J g_{IJ} f_J$, and this gives a symplectic map on $K_\RR^N$. By the general theory of the Weyl algebra,
it corresponds to an automorphism on the field net
characterized by $\alpha_g (W(f)) = W(g.f)$. The Hilbert space $\H$ on which the field net acts is the Fock-space
of the standard vacuum and it carries a unitary representation $g \mapsto U(g)$ of ${\rm O}(N)$ implementing $\alpha_g$
in the sense that $U(g) \pi(W(f)) U(g)^* = \pi(\alpha_g(W(f)))$. The defining representation $\pi$ of the field net
$\{ \F(O)\}$ decomposes as in~\eqref{decomp}, where the labels $[\rho]$ correspond to the irreducible representations of ${\rm O}(N)$, which in turn are
well-known to be characterized by Young tableaux. $\H_0$ is the subspace of ${\rm O}(N)$ invariant vectors and corresponds to the trivial representation
of the net $\{ \A(O) \}$.
It is precisely the closure of $\{a |0 \rangle \mid a \in \A(O) \}$ (for any causal diamond $O$).

Consider now a
tensor $T^{I_1 \dots I_k}$ whose symmetry properties under index permutations are
characterized by a Young-tableau  $\mathbf{\lambda}=(\lambda_1,...,\lambda_s)$ with $k$ boxes.
Next, take functions $f_I \in C^\infty(\RR^{4})$ with support in a causal diamond $O_A$ with base $A \subset \RR^3$ in a time-slice. Define
\ben
F(T) = \sum_{I_1, \dots, I_k=1}^N T^{I_1 \dots I_k} \phi_{I_1}(f_1) \dots \phi_{I_k}(f_k)  \ ,
\een
where $\phi_I(f)=\int \phi_I(x) f(x) d^4 x$ are the smeared KG quantum fields (so that, $\pi(W(f)) = \exp i\sum_I \phi_I(f_I)$). We assume that our
test functions have been chosen so that $F(T) \neq 0$. The transformation law gives
$U(g) F(T) U(g)^* = F(g.T)$, where $g.T$ is the action of $g$ on the tensor $T$.  Let $\dim(\mathbf{\lambda})$ be the dimension
of this representation and let $\{T_i\}_{i = 1, \dots, \dim(\lambda)}$ be an orthonormal basis of  tensors with Young-tableau symmetry $\mathbf{\lambda}$.

By DR theory, there exist corresponding elements $\psi_i \in \mathfrak{F}(O_A), i = 1, \dots, \dim (\mathbf{\lambda})$
satisfying the relations of a Cuntz-algebra and $a$ affiliated with $\A(O_A)$ such that $F(T_i) = a \psi_i$, and this $a$ can be chosen
to satisfy $a^*=a$. $\rho$ defined by
\eqref{rhodef} with $\dim (\rho)=\dim (\mathbf{\lambda})$  is an endomorphism localized in $O_A$.
As one may verify, the corresponding charged state for the net $\{\A(O)\}$ can be written
\ben
\rho^*\omega_0(b) = \omega_0(\rho(b)) = \langle \Phi| b |\Phi \rangle \ ,
\een
where the vector representer of the charged state, $|\Phi \rangle$, is
\ben
|\Phi \rangle = F(T)^*[{\mathcal F}(F(T) F(T)^*)]^{-\frac12} |0\rangle \ ,
\een
where $\mathcal F$ is the mean over the group ${\rm K} = {\rm O}(N)$ already defined above in~\eqref{Kmean}.
$\dim (\mathbf{\lambda})$ is by the general theory equal to
the statistical dimension of the charged state $\rho^* \omega_0$.  It is given
by a standard formula in terms of the shape of the Young tableau, so we obtain
in this example,
\ben
0 \le E_R(\omega_0) - E_R(\rho^* \omega_0) \le \log \dim (\mathbf{\lambda})^2 = 2 \log \prod_{i,j \in \lambda} \frac{(N+j-i)}{h(i,j)} \ ,
\een
where the ``hook length'' parameter $h(i,j)$ of a box with coordinates $(i,j)$  ($i$-th row and $j$-th column)
of the Young tableau is the number of the boxes to the right plus
the number of boxes below, plus one, equal to the numbers written in the following example diagram $\mathbf{\lambda}: \young(865421,5321,1)$.
For this diagram and $N=10$ the right side is $2 \log(5,945,940)$.

\medskip
\noindent
{\bf Example:} (minimal model of type $(p+1,p)$ in 1+1 dimensions) The irreducible inequivalent representations are labeled by a pair $(m,n)$ of natural numbers.
It is discussed in~\cite{kawahigashi_1} how these representations can be implemented by localized endomorphisms. The statistical
dimensions of the corresponding endomorphisms are
\ben
\dim(\rho_{(m,n)}) = (-1)^{n+m}
\frac{
\sin \left( \frac{\pi(p+1)m}{p} \right)
\sin \left( \frac{\pi pn}{p+1} \right)
}{
\sin \left( \frac{\pi(p+1)}{p} \right)
\sin \left( \frac{\pi p}{p+1} \right)
} \ .
\een

It is interesting that a similar bound as in prop.~\ref{propdim} can be obtained for the entanglement measure $E_M$
defined in sec.~\ref{E_Mdef}. To set things up, we consider the vacuum representation of the
quantum field theory. The vacuum vector $|0\rangle$ is cyclic and separating for $\A_A \vee \A_B$ by the Reeh-Schlieder theorem and
therefore defines a natural cone $\cP^\sharp$. Any state of the QFT
$\omega$ with finite energy has a vector representative in $\cP^\sharp$ that is cyclic for $\A_A, \A_B$
and for $\A_E = (\A_A \vee \A_B)'$, again by the Reeh-Schlieder theorem.
It follows that for such states, the standing assumption made in sec.~\ref{E_Mdef} holds. We now consider
a state $\omega$ with finite energy and a localized endomorphism $\rho$ such that $\rho^*\omega$ has finite energy.

\begin{proposition}
Under the same hypothesis as in prop.~\ref{propdim}, if $d+1>2$ we have
\ben
0\le E_M(\omega)-E_M(\rho^*\omega) \le  \log \prod_i \dim(\rho_i)^{5n_i/2} \ .
\een
\end{proposition}

{\em Proof:} Consider first an irreducible $\rho$.
In $d+1>2$ dimensions the DR reconstruction theorem applies and the left-inverse of $\rho$ [see eq.~\eqref{rhodef}] is given by $\Psi_\rho(x) = N^{-1}\sum_i \psi_i^* x \psi_i$ where $N=\dim (\rho) \in \mathbb{N}$ and where $\psi_i$ are the generators of the Cuntz algebra \eqref{cntz}. By prop.~\ref{e4em}, we have $E_M(\rho^* \omega) \le E_M(\omega)$, which is the first inequality. As in the proof of prop.~\ref{propdim}, we also
have $N^2 \Psi_\rho^* \rho^* \omega \ge \omega$ from the Pimsner-Popa inequality. By prop.~\ref{Emdominance},
\ben\label{emmm}
E_M(\omega) \le E_M(\Psi_\rho^* \rho^* \omega) + \log N .
\een
Now consider the linear map $\phi_N: \A_A \vee \A_B \to M_N(\CC)^{\otimes 2} \otimes (\A_A \vee \A_B)$ defined by
\ben
\phi_N(x) = \sum_{i,j=1}^N | i \rangle \langle j| \otimes 1_N \otimes \psi_i^* x \psi_j.
\een
We get $\phi_N(x)^* = \phi_N(x^*)$, and the relations of the Cuntz algebra \eqref{cntz} furthermore give
$\phi_N(1) = 1_N \otimes 1_N \otimes 1$ and
\ben
\phi_N(x) \phi_N(y) = \sum_{i,j=1}^N | i \rangle \langle j| \otimes 1_N \otimes \psi_i^* x \left( \sum_{k=1}^N \psi_k^{} \psi_k^* \right) y \psi_j = \phi_N(xy),
\een
so $\phi_N$ is a unital *-homomorphism. Next, let $\varphi = \rho^* \omega$ and let $\omega_N^+$ be the maximally entangled state on
$M_N(\CC)^{\otimes 2}$. The definitions imply $(\omega^+_N \otimes \varphi)(\phi_N(x)) = \varphi(\Psi_\rho(x))$. Prop.~\ref{e4em}
now gives $E_M(\phi^*_N(\omega^+_N \otimes \varphi)) \le E_M(\omega^+_N \otimes \varphi)$. On the other hand, the tensor product property (e5) of
$E_M$ together with $E_M(\omega_N^+) = \frac{3}{2} \log N$ gives $E_M(\omega^+_N \otimes \varphi) \le E_M(\varphi) + \frac{3}{2} \log N$. Putting this together gives
$E_M(\Psi_\rho^* \rho^* \omega) \le E_M(\rho^* \omega) + \frac{3}{2} \log N$. Combining with \eqref{emmm}, we thus get $E_M(\omega) \le E_M(\rho^* \omega) + \frac{5}{2} \log N$, which is the claim of the proposition for irreducible $\rho$. The general case follows by iterating the argument. \qed

\section{Lower bounds}

\subsection{Lower bounds of area law type}\label{ssec:arealaw}

As one may guess from the definition of $E_R$ (infimum over separable comparison states), it is not evident how to obtain lower bounds.
In fact, it is not even entirely obvious that $E_R(\omega)>0$, say, in the vacuum. We first settle this question.

\begin{corollary}\label{cor1}
Let $\omega$ be any state such that the conclusions of the Reeh-Schlieder theorem hold, such as the vacuum $\omega_0$, a KMS-state $\omega_\beta$,
or any state with bounded energy in a Minkowski quantum field theory. Let $A,B$ be open non-empty regions with ${\rm dist}(A,B)>0$.
Then $E_R(\omega)>0$.
\end{corollary}

{\em Proof:} Our proof is rather similar to that of ~\cite{narnhofer_1}, which in turn is based on the works of
\cite{summers_1,summers_2}. As usual, we represent our net $\{\A(O)\}$ on a Hilbert space
via the GNS-construction, which gives a representation $\pi$ on $\H$ such that $\omega$ is represented by a vector $|\Omega \rangle$.
We write $\A_A = \pi(\A(O_A))''$, where $O_A$ is the causal diamond with base $A$, and similarly for $B$.

Assume that $E_R(\omega) = 0$. By corollary~\ref{corI} below there exists, for each $\delta>0$, a separable state
$\omega' = \sum_j \varphi_j \otimes \psi_j$ with positive normal functionals $\varphi_j, \psi_j$ such that $\| \omega - \omega'\| < \delta$.
Using the split property, one can choose a type I subalgebra $\W_A$ of the type III$_1$-algebra $\A_A$. This subalgebra may be chosen to be
a factor on some Hilbert space $\H_A$ and may be realized as the v. Neumann closure of a Weyl-algebra for one degree of freedom (``Cbit''),
i.e. we may think of $\W_A$ as being isomorphic to $\W_A \cong \W(\RR^2, \sigma_2)''$, where $\sigma_2 = \left(
\begin{matrix}
0 & 1 \\
-1 & 0
\end{matrix}
\right)$ is the standard symplectic form on $\RR^2$. The same construction can of course be made for $B$.
We now choose a state $\eta$ on $\W_A \otimes \W_B \cong \W(\RR^2 \oplus \RR^2, \sigma_2 \oplus \sigma_2)''$ such that $E_B(\eta)$
(the entanglement measure defined in sec.~\ref{sec:bell}) satisfies $E_B(\eta) \ge \sqrt{2}- \epsilon$ for some small $\epsilon > 0$, i.e.
 a state in which the Bell-inequality is nearly maximally violated. Extend $\eta$ to a state on $\A_A \vee \A_B$ via the Hahn-Banach theorem\footnote{This theorem gives a bounded extension
$\hat \eta$ with norm $\| \hat \eta\| \le \|\eta\| = \eta(1)$. Since $1 \in \W_A \otimes \W_B$, we have $\hat \eta(1) = 1$, and we may also
take $\hat \eta$ to be hermitian. If not, we take instead $\Re \hat \eta$. It follows that $\Re \hat \eta$ is also positive, i.e. a state.}. The extended state, called again $\eta$,
need not be normal to $\omega$. But by Fell's theorem (see e.g. \cite{haag_2}), we can choose a normal state $\psi$ which approximates $\eta$
arbitrarily well in the weak topology on the subalgebra $\W_A \otimes \W_B$. In particular, by choosing suitable operators in \eqref{Eqn_Bell} we can achieve that $E_B(\psi)\ge E_B(\eta)-\epsilon\ge\sqrt{2}-2\epsilon$ for arbitrarily small $\epsilon$.
Let $|\Psi \rangle \in \H$
be the unique representer of this $\psi$ in the natural cone of $|\Omega \rangle \in \H$ (the
GNS-representative of $\omega$). Since we are assuming the Reeh-Schlieder property, $|\Omega \rangle$ is both cyclic and separating for
$\A_A$, say, and so we can find an $a$ from this algebra such that $a|\Omega \rangle$ approximates $|\Psi \rangle$
arbitrarily well and such that $\| a  |\Omega \rangle \| = 1$. Let $\varphi = \omega(a^* \ . \ a)$ be the
corresponding positive functional on $\A_A \vee \A_B$. In particular, we may choose $a$ such that $E_B(\varphi) \ge \sqrt{2} - 3\epsilon$. Next,
consider $\varphi' = \omega'(a^* \ . \ a) = \sum_j \varphi_j(a^* \ . \ a) \otimes \psi_j$. Clearly, $\varphi'$ is separable (so $E_B(\varphi') = 1$), and
\ben
\| \varphi - \varphi' \| \le \|a\|^2 \ \| \omega - \omega'\| < \delta \| a \|^2   \ .
\een
Therefore, by choosing $\delta$ sufficiently small, we can achieve that $E_B(\varphi) \le 1 + \epsilon$ (the invariant $E_B$ is norm continuous).
This is in contradiction with $E_B(\varphi) \ge \sqrt{2} - 3\epsilon$ for sufficiently small $\epsilon$. \qed \\

\medskip

The lower bound we have just derived is of course not satisfactory and only serves to confirm our expectation that the invariant $E_R$ is non-trivial in
the context of quantum field theory.  To get $E_R(\omega)>0$ in the previous proof, we
employed a pair of type I subalgebras $\W_A \subset \A_A, \W_B \subset \A_B$, each isomorphic to the algebra of one continuous quantum mechanical
degree of freedom (``Cbit''). We showed that for a large class of states such as the vacuum $\omega_0$, the restriction to $\W_A \otimes \W_B$, i.e. our Cbit pair, is entangled.

To obtain a better lower bound, we now pass to a large number $N$ of Cbits embedded into disjoint subregions $A_i \subset A$ and $B_i \subset B$, where $i=1, \dots, N$.
The idea is that each of these $N$ copies will contribute at least one Cbits' worth of entanglement, and thus give us a much better lower bound.
This will work as stated if the entanglement measure $E$ satisfies the strong superadditivity property (e6). In this situation, we thus expect an
entanglement at least proportional to $N$ (because each of the $N$ Cbit pairs is expected to contribute one unit), while $N$ itself is restricted only
by the requirement that the regions $A_i, B_i$ should be non-intersecting, i.e. a geometrical property. For an entanglement measure $E$ that does not
fulfill (e6) -- like for instance $E_R$ -- we will argue via an auxiliary measure -- like $E_D$ -- which does.

Our reasoning will work
most straightforwardly for a conformal field theory in $d+1$ dimensions, and for simplicity we will stick to these theories here.
We are interested primarily in the case when there is only a ``thin corridor'' of size $\epsilon$ between $A$ and $B$. To formalize this,
we take $A$ to have a smooth boundary $\partial A$ and outward unit normal $n$.
We can ``flow'' the boundary outwards along the geodesics tangent to $n$ by a small proper distance $\epsilon>0$. In this way, we obtain a slightly larger region $A_\epsilon\supset A$, and we let $B \subset \RR^{d} \setminus \overline A_\epsilon$. The proof of the following simple theorem was inspired by conversations with J. Eisert~\cite{eisert_2}.

\begin{theorem}
Given a  net in spatial dimensions $d \ge 1$ with invariance group consisting of Poincar\'e-transformations and dilations, and
with invariant vacuum state $\omega_0$,  and given regions $A,B$ separated by a thin corridor of size $\epsilon$ we have for $\epsilon \to 0$
\ben\label{arealaw}
E_R({\omega_0}) \gtrsim
\begin{cases}
D_2 \cdot \frac{|\partial A|}{\epsilon^{d-1}} & \text{when $d\ge 2$,} \\
D_2' \cdot \log \frac{{\rm min}(|A|,|B|)}{\epsilon} & \text{when $d=1$,}
\end{cases}
\een
where $|\partial A|$ is the surface area of the boundary when $d \ge 2$, where
$|A|, |B|$ denote the lengths of the intervals when $d=1$, where $D_2$ is the distillable entropy of one Cbit pair (defined more precisely in the proof) and $D_2' = D_2 {\rm log}_3 e$.
\end{theorem}

\begin{remark}
Instead of $E_R$, one can obtain the same result obviously for any other entanglement measure dominating $E_D$ which
satisfies (e4), or with any entanglement measure obeying (e4) for automorphisms and (e6), such as an appropriate generalization of the
``squashed entanglement'' $E_S$~\cite{christandl} for type III factors.
\end{remark}

{\em Proof:} $(d\ge 2)$: We consider a pair consisting of a unit cube $A_0=c = (0,1)^{d}$ at the origin in a spatial slice $\cong \RR^d$, and a unit cube $B_0$ obtained from $A_0$ by
a translation in some arbitrarily chosen coordinate direction. We fix the distance between $A_0$ and $B_0$ to be, say, one. As in the previous proof, we embed
a Cbit pair (i.e. pair of type I algebras $\W_{A_0}, \W_{B_0}$ each isomorphic to the v. Neumann closure of the Weyl algebra of one continuous quantum mechanical degree of freedom)
into $\A_{A_0}, \A_{B_0}$, respectively, and call $D_2$ the
distillable entropy of this pair in the restriction of the state $\omega_0$.

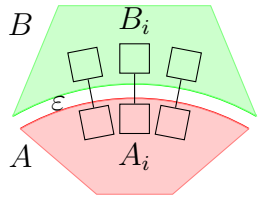
\begin{figure}[h!]
\begin{center}
\begin{tikzpicture}[square/.style={regular polygon,regular polygon sides=4}]
            \draw[thick,red] ([shift=(60:1cm)] 0, 0) arc (60:120:3cm);
            \draw[thick,green] ([shift=(60:1.2cm)] 0, 0) arc (60:120:3.2cm);

        \filldraw[fill=red!20!white, draw=red!50]
        ([shift=(60:1cm)] 0, 0) arc (60:120:3cm) -- (-1.5,0) -- (-0.5, 0) -- cycle;
        \filldraw[fill=green!20!white, draw=green!50]
        ([shift=(60:1.2cm)] 0, 0) arc (60:120:3.2cm) -- (-2.0,2.5) -- (0.0, 2.5) -- cycle;

        \node at (-1.0, 1.0) [square,draw, rotate=0] (a0) {};
        \node at (-1.5, 0.95) [square,draw, rotate=11] (a1) {};
        \node at (-0.5, 0.95) [square,draw, rotate=169] (a2) {};

        \node at (-1.0, 1.8) [square,draw, rotate=0] (b0) {};
        \node at (-1.65, 1.72) [square,draw, rotate=11] (b1) {};
        \node at (-0.35, 1.72) [square,draw, rotate=169] (b2) {};

        \draw (a0) -- (b0);
        \draw (a1) -- (b1);
        \draw (a2) -- (b2);

        \node at (-2, 1.21) {$\varepsilon$};
        \node at (-1.0, 2.3) {$B_{i}$};
        \node at (-2.5, 2.25) {$B$};
        \node at (-1.0, 0.5) {$A_{i}$};
        \node at (-2.5, 0.5) {$A$};
\end{tikzpicture}

\end{center}
\caption{The the sets $A_i,B_i$ in $d+1 > 2$ spacetime dimensions.}
\label{fig:aibi1}
\end{figure}

We can apply to this pair $(A_0, B_0)$ group elements $\{g_i\}$ of the invariance group generated by dilations, rotations, and spatial translations
so that each $(A_i, B_i) = g_i \cdot (A_0, B_0)$ is a pair of cubes of size $2\epsilon$ lying on opposite sides of the corridor separating $A$ from $B$, see fig.~\ref{fig:aibi1}.
We assume that $1/\epsilon$ is much larger than the maximum of the extrinsic curvature $(K_{ij} K^{ij})^{1/2}$ along $\partial A$, so that the boundary is essentially flat on the scale $\epsilon$.
If we demand that the cube pairs do not intersect with each other, then it is clear that
we can fit in $N \gtrsim |\partial A|/\epsilon^{d-1}$ cube pairs (asymptotically for $\epsilon \to 0$).
Defining $\W_{A_i} = \alpha_{g_i} \W_{A_0} $ (and similarly for $B_i$),
we then have an inclusion $\iota_N: \otimes_i \W_{A_i} \to \A_A$ (and similarly for $B$). Let $\omega_i$ be the restriction of $\omega$ to $\W_{A_i} \otimes \W_{B_i}$ under this inclusion.
The properties of $E_R, E_D$ imply:
\ben\begin{split}
E_R(\omega_0)  \ge &E_R\left(\omega_0 \restriction \bigvee_{i=1}^N  \A_{A_i} \otimes  \A_{B_i}\right) \\
\ge & E_R\left(\omega_0 \restriction \bigvee_{i=1}^N \W_{A_i} \otimes  \W_{B_i}\right) \\
\ge & E_D\left(\omega_0 \restriction \bigvee_{i=1}^N \W_{A_i} \otimes  \W_{B_i}\right) \\
\ge & \sum_{i=1}^N E_D(\omega_0 \restriction \W_{A_i} \otimes \W_{B_i}) \\
= & \sum_{i=1}^N E_D(\alpha_{g_i}^* \omega_0 \restriction \W_{A_0} \otimes \W_{B_0}) =  N \cdot D_2 \gtrsim D_2  \cdot \frac{|\partial A|}{\epsilon^{d-1}} \ .
\end{split}
\een
In the first and second step, we used (e4) (letting ${\mathcal F}$ be the inclusion map $\vee  \A_{A_i} \otimes \vee \A_{B_i} \to \A_A \otimes \A_B$ in the first,
and  $\vee  \W_{A_i} \otimes \vee \W_{B_i} \to \vee  \A_{A_i} \otimes \vee \A_{B_i}$ the second step). In the
third step we used that $E_R$ dominates $E_D$ for type I algebras, by thm.~\ref{Thm:dominance}. In the fourth step we used (e6) for $E_D$.
In the fifth step we used that $\omega_0$ is invariant under $\alpha_{g_i}$ (conformal invariance of the vacuum), and that $E_D(\omega_0 \restriction \W_{A_0} \otimes \W_{B_0}) = D_2$ by definition.

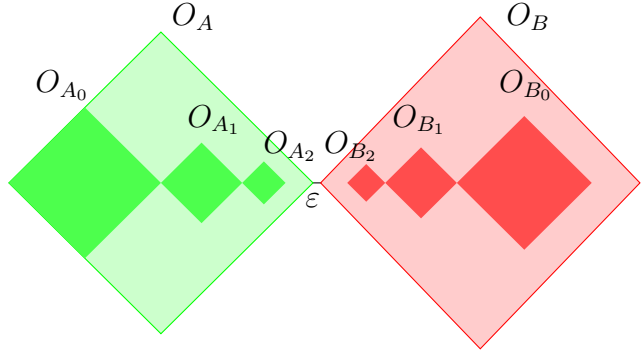
\begin{figure}[h!]
\begin{center}
\tikzset{
	bigger/.style={decoration={shape start size=5mm, shape end size=24mm}},
	smaller/.style={decoration={shape start size=20mm, shape end size=3mm}},
	decoration={shape backgrounds, shape=diamond,
	    shape sep={0.0cm, between borders},shape scaled}
    }
    \begin{tikzpicture}
	\draw (-4, 0) -- (4.3,0);
	\filldraw[fill=green!20, draw=green] (- 4, 0) -- (-2, 2) -- (0, 0) -- (-2, -2) --cycle;
	\filldraw[fill=red!20, draw=red] (4.3, 0) -- (2.2, 2.2) -- (0.1, 0) -- (2.2, -2.2) --cycle;
	\fill [decorate, smaller, green!70] (-3,0) -- (-0.2,0);
	\fill [decorate, bigger, red!70] (0.7,0) -- (3.8,0);
	% A
	\node[right] at (-2, 2.2) {$O_A$};
	\node[left] at (-2.8, 1.3) {$O_{A_{0}}$};
	\node[above] at (-1.3, 0.5) {$O_{A_{1}}$};
	\node[above] at (-0.3, 0.15) {$O_{A_{2}}$};
	% \epsilon
	\node[below] at (0, 0) {$\varepsilon$};
	% B
	\node[right] at (2.4, 2.2) {$O_B$};
	\node[left] at (3.3, 1.3) {$O_{B_{0}}$};
	\node[above] at (1.4, 0.5) {$O_{B_{1}}$};
	\node[above] at (0.5, 0.15) {$O_{B_{2}}$};
    \end{tikzpicture}
\end{center}
\caption{The sets $A_i,B_i$ in $d+1=2$ spacetime dimensions.}
\label{fig:aibi}
\end{figure}

$(d=1)$: By dilation invariance, we may assume without loss of generality that ${\rm min}(|A|,|B|)=1$. A cube is now an interval, and we consider the interval pairs
\ben
A_i = (-(\tfrac{1}{3})^i, -(\tfrac{1}{3})^{i+1}), \quad  B_i =  ((\tfrac{1}{3})^{i+1}, (\tfrac{1}{3})^{i}) \ ,
\een
see fig.~\ref{fig:aibi}. These intervals are obviously disjoint and they
satisfy $A_i \subset A$ respectively $B_i \subset B$ as long as $i +1 \le \lfloor {\rm log}_3 \epsilon^{-1} \rfloor$. The number $N$ of $(A_i, B_i)$-pairs is thus
$\sim {\rm log}_3 \epsilon^{-1} $ when $\epsilon \to 0$. The rest of the proof then follows the same argument as in the case $d \ge 2$.
\qed

\medskip
\noindent
That generic states satisfying the Reeh-Schlieder property are distillable across a pair of spacelike regions has been shown in a rather general setting by~\cite{Verch_2}.
Here, we would like to ensure that the distillation {\em rate} for the vacuum state is in fact non-zero -- or more precisely that $D_2>0$ -- which is a stronger statement.
We now present an argument that this must be the case at least for the
massless free KG field which defines a conformal net: Since this theory
satisfies the Reeh-Schlieder property, we can argue just as in the proof of corollary~\ref{cor1} that $\omega$ restricted to  $\W_{A_0} \vee \W_{B_0}$ cannot be separable. Since this algebra is
isomorphic to the weak closure of $\W(\RR^2, \sigma_2) \otimes \W(\RR^2, \sigma_2) \cong \W(\RR^4, \sigma_4)$ in the restriction of the vacuum representation to this subalgebra, and since the vacuum state is quasi-free, its restriction to $\W(\RR^4, \sigma_4)$ must also be a quasi-free state. For such states it is known~\cite{werner_1,duan,simon_1} that they cannot have a property called ``positive partial transpose''. Using this it is shown in~\cite{gidke_1} that such states satisfy a ``reduction criterion''
which in~\cite{hor_1} was shown to imply a finite distillable entropy. Hence $D_2>0$ for the free massless KG field.\footnote{See also~\cite{eisert_1} for further discussion
on the distillation of quasi-free states.}

\medskip
\noindent

\begin{remark}
Looking at the proof, one sees that one could replace the elementary Cbit pair by $N$ continuous quantum mechanical degrees of freedom, i.e.
by replacing $\W_{A_0}$ resp. $\W_{B_0}$ with a v. Neumann algebra isomorphic to the weak closure of $\W(\RR^{2N}, \sigma_{2N})$ sitting inside
$\A_{A_0}$ resp. $\A_{B_0}$, and then defining $E_D(\omega_0 \restriction \W_{A_0} \otimes \W_{B_0}) = D_N$. We can then maximize over the parameter $N$, and in the lower bound~\eqref{arealaw}. $D_2$  is then replaced by the maximum possible $D_N$.
For a free scalar field with $N$ components, this yields an improvement of the lower bound by the factor $N$ since $D_N = N D_2$. More generally, the conclusions of the theorem are likely to be true for any state $\omega$ that is asymptotically dilation invariant on small scales (e.g. states with finite energy)
and for any theory which approaches a free field theory on small scales, i.e. any asymptotically free theory. Thus, it is highly plausible that the following bound holds for
an asymptotically free theory and any state with finite energy:
\ben
E_R({\omega}) \gtrsim N \cdot D_2 \cdot \frac{|\partial A|}{\epsilon^{d-1}} \ ,
\een
where $N$ is the number of independent free fields in the scaling limit.
\end{remark}

\subsection{General lower bounds}

One can use the variational definition of $E_R$ to obtain some (rather indirect) lower bounds in terms of the norm distance of
$\omega$ to the subspace of separable states. We now
explain these--essentially well-known--bounds.  Returning to the general situation, let $\A$ be a v. Neumann algebra, and $\omega, \omega'$ two
faithful normal states. Then $\varphi(a) = \omega(a) - \omega'(a)$ is a linear, hermitian, continuous, non-positive functional on $\A$. For any such functional, one can define its ``range projection'', $e=e(\varphi) \in \A$.\footnote{This follows by applying the arguments in the proofs of thms.~7.3.1 and 7.3.2 in \cite{
KadisonRingrose} to the self-adjoint part of the unit ball.}
 For instance, if $\varphi$ is the functional defined by $\varphi(a) = \tr(Fa)$ on a matrix v. Neumann algebra $\A = M_n(\CC)$ in terms of some self-adjoint matrix $F=F^*$, the range projection $e$ would be given by the projection onto the non-negative eigenvalues of $F$.
It follows from this definition that the norm of any hermitian linear functional on $\A$
is given in this case by
\ben
\| \varphi \| = \sup_{a \in \A, \| a \| \le 1} | \varphi(a) | = \varphi(e) - \varphi(1-e) \ ,
\een
as one can easily prove by showing that the left side is neither bigger nor smaller than the right side. These formulas generalize to general continuous, hermitian, linear functionals on arbitrary v. Neumann algebras $\A$, and one can show that $e=e^*=e^2$ is always
an element of $\A$. For the norm of $\omega-\omega'$ we therefore get
\ben\label{difference}
\| \omega - \omega' \| = (\omega - \omega')(e) - (\omega - \omega')(1-e) = 2(p-q) \ ,
\een
where $0 < q \le  p < 1$ have been defined by $p = \omega(e), q = \omega'(e)$. Consider now
the subalgebra $\D$ of $\A$ generated by $\{e, 1\}$. It is obvious that this subalgebra
is abelian and isomorphic to the trivial v. Neumann algebra of diagonal complex 2 by 2 matrices, and
under this isomorphism, the restrictions $\omega|_{\D}, \omega'|_{\D}$ correspond to the diagonal density matrices
\ben
\rho_\D = \left(
\begin{matrix}
p & 0 \\
0 & 1-p
\end{matrix}
\right)
\ ,
\quad
\rho'_\D = \left(
\begin{matrix}
q & 0 \\
0 & 1-q
\end{matrix}
\right).
\een
Combining this with (h2) (for the inclusion map ${\mathcal F}: \D \to \A$), we get
\ben\label{Hred}
H(\omega, \omega') \ge H(\omega |_{\D}, \omega' |_{\D}) = p \log \frac{p}{q} + (1-p) \log \frac{1-p}{1-q} \ .
\een
\cite{hiai} estimate the right side as $\ge 2(p-q)^2$, which in view of~\eqref{difference}
immediately gives the well-known result, also stated by these authors,
\ben
H(\omega, \omega') \ge \half \|\omega - \omega'\|^2 .
\een
By a trivial modification of the argument, one can obtain a  tighter lower bound. Define $s(x)$ to be the infimum of
the right hand side of \eqref{Hred} under the constraint $p-q=x\ge0$, i.e.
\ben\label{sdef}
s(x) \equiv \inf_{p,q: p-q=x, 0 < q \le p <1}  \left[ p \log \frac{p}{q} + (1-p) \log \frac{1-p}{1-q} \right] \ .
\een
In view of $x = \half \| \omega - \omega'\|$, \eqref{Hred} actually gives the improved lower bound
\ben
\label{ineq1}
H(\omega, \omega') \ge s(\half \| \omega - \omega' \|) \ .
\een
The function $s:(0,1) \to \RR$ is monotonically increasing, strictly convex, positive, and has the asymptotic behavior~\cite{audenaert}
\ben
s(x) \sim
\begin{cases}
2x^2  + \tfrac{4}{9} x^4 + \tfrac{32}{135} x^6 + \dots  & \text{for $x \to 0$,}\\
-\log(1-x) & \text{for $x \to 1$.}
\end{cases}
\een
From the second line it is seen that the improvement of the lower bound is most drastic when $x \to 1$, i.e. when $\|\omega - \omega'\| \to 2$ (note that $2$ is
the maximum value since $\omega, \omega'$ are functionals of norm one).
For matrix algebras $\A = M_N(\CC)$, where the states $\omega, \omega'$ can be identified with density matrices $\rho_\omega, \rho_{\omega'}$, the norm distance is $\|\omega - \omega'\|
= \| \rho_\omega- \rho_{\omega'} \|_1$, the $1$-norm of an operator being defined by definition~\ref{def_1norm}. Our inequality \eqref{ineq1} thereby reduces to an inequality found
by~\cite{audenaert} using a more involved method.

As an aside we note that instead of using the norm $\|\omega - \omega'\|$, one can also obtain a lower bound directly in terms of suitable vector representatives $|\Omega \rangle, |\Omega' \rangle$
in the GNS representation of, say, $\omega$, using prop.~\ref{prop_cone}. Using also the monotonicity of $s$, we immediately arrive at:

\begin{theorem}\label{thm1}
Let $\omega, \omega'$ be faithful normal states on a v. Neumann algebra $\A$, with vector
representatives $|\Omega \rangle, |\Omega' \rangle \in {\mathcal P}^\sharp$ in the natural cone, so that $1 \ge \langle \Omega'| \Omega \rangle>0$. Then we have
\ben
H(\omega, \omega') \ge s\bigg(1 - \langle \Omega'| \Omega \rangle \bigg) \ ,
\een
where $s: (0,1) \to \RR$ is the universal positive monotonic function  defined by \eqref{sdef}.
\end{theorem}

This lower bound is useful in the context of Gaussian states for free fields, as $\langle \Omega | \Omega' \rangle$ can be
expressed in terms of the operators $\Sigma, \Sigma'$ defining these states. \\

Returning from these general considerations to quantum field theory, consider a local net $O \mapsto \A(O)$, and let $O_A$ and $O_B$ be two causal diamonds with disjoint bases $A$ and $B$ on some Cauchy surface $\C$. As in the
description of the split construction above, we assume that there is a safety distance
${\rm dist}(A,B)  >0$ between the two bases. Let $\omega$ be a faithful
normal state on the algebra $\A_A \vee \A_B$ (e.g. the vacuum state for the entire net).
Then the decoupled state $\omega'(ab) = (\omega \otimes \omega)(ab) := \omega(a) \omega(b)$ is well-defined
by the split property. Obviously
\ben
\| \omega - \omega' \| \ge \frac{(\omega - \omega')(ab)}{\|ab\|}
\ge \frac{\omega(ab) - \omega(a)\omega(b)}{\| a\| \cdot  \|b\|}
\een
From the definitions of the mutual information and entanglement entropy of the pair $A,B$ and
the monotonicity of $s$, we immediately get
\begin{corollary}\label{corI}
Let $O_A$ and $O_B$ be causal diamonds with bases $A$ and $B$ on some Cauchy surface
such that ${\rm dist}(A,B) > 0$. Then
\ben
E_I(\omega)  \ge \sup
s \left(
 \frac{\omega(ab) - \omega(a)\omega(b)}{2\| a\| \cdot \|b\|}
\right),
\een
the supremum being over all nonzero $a \in \A_A, b \in \A_B$. Similarly
\ben\label{secondineq}
E_R(\omega) \ge \inf_\sigma s\left( \half
\| \omega - \sigma \|
\right)
\een
where the infimum is over all separable states on $\A_A \vee \A_B$.
\end{corollary}

It is possible to see form the second inequality~\eqref{secondineq} and the asymptotic behavior of $s$
that, if $B = \RR^d \setminus A_\epsilon$ as above, then $E_R(\omega)$ must diverge
as $\epsilon \to 0$, for any normal state $\omega$.

\medskip
\noindent
{\bf Acknowledgements:} S.H. likes to thank J. Eisert, P. Grangier and R. Longo for discussions, and O. Islam for help with figures. K. S. would like to thank Leipzig University,
where this project was carried out, and U. Rome II (``Tor Vergara'') for funding a visit in November 2016 and for the opportunity to present
some of the results in this volume in a seminar. He also gratefully acknowledges an invitation to MFO where some of the results relevant to this volume were
presented during the workshop ``Recent mathematical developments in quantum field theory'' (ID1630). We thank Y. Tanimoto for pointing out an error
concerning the ordering of entanglement measures in an earlier version of this volume and K.-H. Rehren for comments on our manuscript.

\appendix

\section{The edge of the wedge theorem}\label{eow}

In the body of the volume, we used several times the edge-of-the wedge theorem. For the convenience of the reader, we give a statement of this theorem and make some remarks.
In its most basic form, the
theorem deals with the following situation. $U=(x_1,x_2)$ is an open interval in $\RR$, $F_1$ a function that is holomorphic on the upper half plane of $\CC$, $F_2$ a function holomorphic on the lower half plane, both $F_1$ and $F_2$ have the same bounded, continuous limit on $U$. Then there exists a function $F$, holomorphic in the cut plane
$\CC \setminus [(-\infty, x_1] \cup [x_2, \infty)]$, which is a joint extension of $F_1, F_2$.

A more general version of the theorem applies to analytic functions $F_1, F_2$ holomorphic on a domain of $\CC^n$ of the form $U + iC$ resp. $U-iC$, where $U \subset \RR^n$ is an open domain, and where $C \subset \RR^n$ is the intersection of some open, convex cone with an open ball.  It is assumed that
\ben\label{boundaryvalue}
T_1(f) = \lim_{y \in C, y \to 0} \int \dd^n x F_1(x + iy) f(x) , \quad
T_2(f) = \lim_{y \in C, y \to 0} \int \dd^n x F_2(x - iy) f(x)
\een
define distributions on $U$ such that, actually, $T_1=T_2$. The edge of the wedge theorem is (see e.g.~\cite{wightman_1}):

\begin{theorem}
There exists a function $F$ which is holomorphic on an open complex neighborhood $N \subset \CC^n$ containing $U$ such that
$F$ extends both $F_1, F_2$ where defined.
\end{theorem}

One often applies the theorem to the case when a holomorphic function $F_1$ on $U +iC$ is given with distributional boundary value $T_1=0$.
Then choosing $F_2\equiv 0$, one learns that also $F_1=0$ where defined.

The edge of the wedge theorem has a straightforward generalization to the case when $F_1, F_2$ take values in a Banach space, $\cX$, which we also use
in this volume.
A function $F$ valued in $\cX$ is called (weakly) holomorphic near $z_0$ if $\psi(F(z))$ is holomorphic near $z_0$ for any linear functional
$\psi$ in the topological dual $\cX^*$.
It is easy to see (see e.g.~\cite{feldman}) that a weakly holomorphic function is in fact even strongly holomorphic in the sense that it
has a norm-convergent expansion $F(z) = \sum_{n \ge 0}  x_n (z-z_0)^n, x_n \in \cX$ near $z_0$ (and of course vice versa).
By going through the proof of the edge of the wedge-theorem in the
$\CC$-valued case, one can see as a consequence that an $\cX$-valued version holds true, too: if $F_i$ are  holomorphic $\cX$-valued
functions in $U \pm iC$ such that their distributional boundary values~\eqref{boundaryvalue} (limit in the norm topology on $\cX$)
on $U$ coincide as distributions valued in $\cX$, then there is a holomorphic extension
$F$ on $N$. For a related discussion, see also~\cite{strohmaier}.

We also use in this volume the following (related) lemma about $\cX$-valued holomorphic functions.

\begin{lemma}\label{maxlemma}
Let $U$ be an open domain in $\CC$ and let $F: U \to \cX$ be a holomorphic function with continuous limit on $\partial U$.
Then $\overline U \owns z \mapsto \| F(z) \|_{\cX}$ assumes its maximum on the boundary $\partial U$.
\end{lemma}

{\em Proof:} The norm $u(z) \equiv \| F(z) \|_{\cX}$ is continuous on $\overline U$ and for each continuous linear map $l:\cX\to\CC$ the scalar function
$l\circ F$ is continuous on $\overline U$ and holomorphic on the interior. If $\cX^*_1$ denotes the unit ball of the dual space $\cX^*$, then
\ben\begin{split}
\max_{z\in\overline U} u(z) =& \max_{z\in\overline U} \max_{l\in\cX^*_1} |l(F(z))|
= \max_{l\in\cX^*_1} \max_{z\in\overline U} |l(F(z))|\\
=& \max_{l\in\cX^*_1} \max_{z\in\partial U} |l(F(z))|
=\max_{z\in\partial U} \max_{l\in\cX^*_1} |l(F(z))|
= \max_{z\in\partial U} u(z),\\
\end{split}
\een
where we applied the maximum principle to $l\circ F$ to get to the second line. \qed

\end{document}